\documentclass[12pt,a4paper]{article}

\usepackage{comment}

\usepackage[font=small]{caption}
\usepackage[text={450pt,650pt},headheight={14.5pt},centering]{geometry}

\usepackage{hyperref}
\usepackage{graphicx}
\usepackage{booktabs}
\usepackage{subfig}
\usepackage{tikz} 
\usetikzlibrary{calc}
\usetikzlibrary{arrows}

\usepackage[compress]{cite}

\usepackage{amsmath,amssymb}
\usepackage{mathrsfs}
\usepackage{mathtools}
\usepackage{bbm}
\usepackage{slashed}
\usepackage{braket}

\usepackage{dsfont}

\usepackage{accents}

\DeclareMathAlphabet{\mathsfit}{\encodingdefault}{\sfdefault}{m}{sl}

\usepackage{xspace}

\numberwithin{equation}{section}

\makeatletter
 \let\old@startsection=\@startsection
 \let\oldl@section=\l@section
 \renewcommand{\@startsection}[6]{\old@startsection{#1}{#2}{#3}{#4}{#5}{#6\mathversion{bold}}}
 \renewcommand{\l@section}[2]{\oldl@section{\mathversion{bold}#1}{#2}}
\makeatother

\renewcommand{\leq}{\leqslant}

\def\XXint#1#2#3{{\setbox0=\hbox{$#1{#2#3}{\int}$}
    \vcenter{\hbox{$#2#3$}}\kern-.5\wd0}}

\newcommand{\ce}{\text{c.e.}}

\newcommand{\stL}{\tilde{\text{\tiny L}}}

\newcommand{\AdS}{\textup{AdS}}
\newcommand{\CFT}{\textup{CFT}}
\newcommand{\Sphere}{\textup{S}}
\newcommand{\Torus}{\textup{T}}
\newcommand{\CP}{\textup{CP}}

\newcommand{\Smat}{\mathcal{S}}

\newcommand{\action}{S}
\newcommand{\lagr}{L}
\newcommand{\ham}{H}

\newcommand{\de}{\textup{d}}

\newcommand{\alg}[1]{\mathfrak{#1}}
\newcommand{\grp}[1]{\mathrm{#1}}

\newcommand{\suA}{\bullet}
\newcommand{\suB}{\circ}

\newcommand{\algSL}{\alg{sl}}

\newcommand{\algSU}{\alg{su}}

\newcommand{\algU}{\alg{u}}

\newcommand{\algPSU}{\alg{psu}}

\newcommand{\algSO}{\alg{so}}
\newcommand{\grpSO}{\grp{SO}}

\newcommand{\gen}[1]{\mathbf{#1}}

\newcommand{\su}{\alg{su}}
\newcommand{\psu}{\alg{psu}}

\newcommand{\1}{\mathbf{1}}

\newcommand{\ie}{\textit{i.e.}\xspace}

\newcommand{\cf}{\textit{cf.}\xspace}

\newcommand{\comm}[2]{[#1,#2]}
\newcommand{\acomm}[2]{\{#1,#2\}}

\newcommand{\commPB}[2]{[#1,#2]_{\scriptscriptstyle \text{PB\strut}}}
\newcommand{\acommPB}[2]{\{#1,#2\}_{\scriptscriptstyle \text{PB\strut}}}

\newcommand{\cchi}{\underline{\chi}{}}
\newcommand{\cbchi}{\underline{\widetilde{\chi}}{}}

\newcommand{\sL}{\mbox{\tiny L}}
\newcommand{\sR}{\mbox{\tiny R}}

\newcommand{\smallL}{\sL}
\newcommand{\smallR}{\sR}

\newcommand{\st}{\text{st}}

\newcommand{\wh}{\hat{\omega}}
\newcommand{\wc}{\check{\omega}}

\newcommand{\pri}[1]{\accentset{\prime}{#1}}
\newcommand{\ppri}[1]{\accentset{\prime\prime}{#1}}

\begin{document}

\thispagestyle{empty}

\begin{flushright}\footnotesize\ttfamily
Imperial-TP-OOS-2014-03\\
HU-Mathematik-2014-11\\
HU-EP-14/19\\
ITP-UU-14/17\\
SPIN-14/15
\end{flushright}
\vspace{1em}

\begin{center}
\textbf{\Large\mathversion{bold} The complete $\AdS_3\times \Sphere^3\times \Torus^4 $ worldsheet S-matrix}

\vspace{2em}

\textrm{\large Riccardo Borsato${}^1$, Olof Ohlsson Sax${}^2$, Alessandro Sfondrini${}^{3}$\\
and Bogdan Stefa\'nski jr.${}^4$ } 

\vspace{2em}

\begingroup\itshape
1. Institute for Theoretical Physics and Spinoza Institute, Utrecht University,\\ Leuvenlaan 4, 3584 CE Utrecht, The Netherlands\\[0.2cm]

2. The Blackett Laboratory, Imperial College,\\  SW7 2AZ, London,
U.K.\\[0.2cm]

3. Institut f\"ur Mathematik und Institut f\"ur Physik, Humboldt-Universit\"at zu Berlin\\
IRIS Geb\"aude, Zum Grossen Windkanal 6, 12489 Berlin, Germany\\[0.2cm]

4. Centre for Mathematical Science, City University London,\\ Northampton Square, EC1V 0HB, London, U.K.\par\endgroup

\vspace{1em}

\texttt{R.Borsato@uu.nl, o.olsson-sax@imperial.ac.uk, Alessandro.Sfondrini@physik.hu-berlin.de, Bogdan.Stefanski.1@city.ac.uk}


\end{center}

\vspace{3em}

\begin{abstract}\noindent
We derive the non-perturbative worldsheet S~matrix for fundamental excitations of Type IIB superstring theory on $\AdS_3\times\Sphere^3\times\Torus^4 $ with Ramond-Ramond flux. 
To this end, we study the off-shell symmetry algebra of the theory and its representations.
We use these to determine the S~matrix up to scalar factors and we derive the crossing equations that these scalar factors satisfy. Our treatment automatically includes fundamental massless excitations, removing a long-standing obstacle in using integrability to study the $\AdS_3/\CFT_2 $ correspondence. The present paper contains a detailed derivation of results first announced in~arXiv:1403.4543.
\end{abstract}

\newpage

\tableofcontents

\section{Introduction}
\label{sec:introduction}
The $\AdS/\CFT$ correspondence is a remarkable equivalence between quantum gauge and gravity theories. In its simplest form it posits a strong/weak duality between superstring theories on $\AdS_{d+1}\times \mathcal{M}_{9-d}$, where~$\mathcal{M}_{9-d}$ is a $(9\!-\!d)$-dimensional compact space, and $d$-dimensional Conformal Field Theories (CFTs) on the boundary of $\AdS_{d+1}$~\cite{Maldacena:1997re,Witten:1998qj,Gubser:1998bc}. 
 This conjecture has inspired important advances in our understanding of quantum gravity and Quantum Field Theory (QFT).
An intriguing feature of the $\AdS/\CFT$ duality is the emergence of integrable structures in the 't Hooft, or planar, limit~\cite{'tHooft:1973jz} of certain classes of dual theories. The prototypical example is the case of type IIB strings on~$\AdS_5\times \Sphere^5$  and the dual $\mathcal{N}=4$ Supersymmetric Yang-Mills (SYM) theory, see~\cite{Arutyunov:2009ga,Beisert:2010jr} for a review. Following the discovery of the ABJM Chern-Simons theory~\cite{Aharony:2008ug}, integrability was found also to underpin the duality between this CFT and Type IIA string theory on $\AdS_4\times \CP^3$ in the planar limit.\footnote{See~\cite{Klose:2010ki} for a review and a more complete list of references.} The key role of integrability in providing a quantitative handle on both the $\AdS_5/\CFT_4$ and  $\AdS_4/\CFT_3$ dualities is rather striking. It hints very strongly that, for certain classes of dual pairs, integrability provides the right set of tools with which to investigate the $\AdS/\CFT$ correspondence. As a result, identifying other dual pairs where integrable methods may be applicable is an important challenge in developing a detailed understanding of the $\AdS/\CFT$ correspondence.

Another set of classes where integrability emerges are strings on~$\AdS_3\times \mathcal{M}_7$ backgrounds with 16 real supersymmetries. The $\AdS_3/\CFT_2$ correspondence is a particularly important example of gauge/string duality. Historically, gravity on~$\AdS_3$ gave rise to an early example of holography~\cite{Brown:1986nw}. The gravity theory was found to have an (infinite-dimensional) conformal symmetry on the boundary whose central charge could be calculated. Further, black hole solutions could be constructed in the gravitational theory~\cite{Banados:1992wn,Banados:1992gq} and their entropy was understood using holography~\cite{Strominger:1997eq}. Moreover, the D1-D5 brane system, whose near-horizon limit gives rise to the $\AdS_3/\CFT_2$ correspondence, has played a central role in the string theory derivation of the black-hole entropy formula~\cite{Strominger:1996sh}. At low energy, such a brane construction gives rise to a $1+1$ dimensional supersymmetric Yang-Mills theory with matter multiples in the fundamental and adjoint representations, adding new features with respect to $\mathcal{N}=4 $ SYM and ABJM theories.

In the context of string theory, it is natural to first consider $\AdS_3$ backgrounds with
maximal supersymmetry. Such backgrounds have 16 real supersymmetries and come in two distinct types. String theory on~$\AdS_3\times \Sphere^3\times \Torus^4 $ gives rise to the small $\mathcal{N}=(4,4)$ superconformal algebra~\cite{Maldacena:1997re,Seiberg:1999xz},\footnote{String theory on $\AdS_3\times \Sphere^3\times \textup{K3} $ also leads to a small $\mathcal{N}=(4,4)$ superconformal algebra. From the point of view adopted in this paper this background can be viewed  as a blow-up of an orbifold of $\AdS_3\times \Sphere^3\times \Torus^4 $.}
while string theory on~$\AdS_3\times \Sphere^3\times \Sphere^3\times \Sphere^1 $ 
leads to the large $\mathcal{N}=(4,4)$ superconformal algebra~\cite{Boonstra:1998yu}. Both types of backgrounds can be supported by a mixture of Ramond-Ramond (R-R) and Neveu-Schwarz-Neveu-Schwarz (NS-NS) fluxes.
In the case of pure NS-NS flux, much progress was made by studying the worldsheet theory with two-dimensional CFT techniques~\cite{Giveon:1998ns, deBoer:1998pp,Elitzur:1998mm,Aharony:1999ti,Maldacena:2000hw, Maldacena:2000kv, Maldacena:2001km}. 
These results can be mapped onto the D1-D5 system via S~duality \footnote{In fact, S~duality acts on mixed-flux background by swapping R-R with NS-NS fluxes.}, which however acts in a non-perturbative and non-planar way. It is then natural to ask if backgrounds involving R-R fluxes can be studied more directly~\cite{Berkovits:1999im}. In particular, developing a quantitative understanding of the pure R-R  string theory is essential in understanding generic unprotected properties of the D1-D5 system, and a starting point to tackling more general $\AdS_3/\CFT_2$ dualities.

With this motivation in mind, it was realised  that the equations of motion of type II string theory on the pure R-R background are integrable~\cite{Babichenko:2009dk}\footnote{Integrable structures were also recently found from studying the  Gubser-Klebanov-Polyakov ``spinning string''~\cite{Sundin:2013uca}.
}
 and that this extends to mixed fluxes as well~\cite{Cagnazzo:2012se}. This prompted an extensive investigation of the quantum integrability properties of these backgrounds~\cite{Babichenko:2009dk,Zarembo:2010sg, Sundin:2012gc, Pakman:2009mi, OhlssonSax:2011ms,Sax:2012jv, Borsato:2012ud,Borsato:2012ss,  Borsato:2013qpa,Borsato:2013hoa,Rughoonauth:2012qd, Beccaria:2012kb,Beccaria:2012pm,Abbott:2012dd, Sundin:2013ypa,Sundin:2013uca,Abbott:2013ixa, Hoare:2013pma,Hoare:2013ida,Hoare:2013lja, Engelund:2013fja,Sundin:2014sfa,Babichenko:2014yaa,Bianchi:2014rfa}, mainly by means of the S-matrix approach that proved successful in the case of~$\AdS_5/\CFT_4$, see also~\cite{Sfondrini:2014via} for a review.

A new feature of the~$\AdS_3$ backgrounds is the presence of massless fundamental excitations on the worldsheet. Because massless modes are notoriously difficult to incorporate into integrability constructions~\cite{Zamolodchikov:1992zr,Fendley:1993wq,Fendley:1993xa}, this presented an early challenge to fully understanding the $\AdS_3/\CFT_2$ correspondence using integrable methods.
On the other hand, massive S-matrices and Bethe ansatz equations of~$\AdS_3\times \Sphere^3\times \Torus^4 $~\cite{Borsato:2013qpa,Borsato:2013hoa} and~$\AdS_3\times \Sphere^3\times \Sphere^3\times \Sphere^1 $~\cite{Borsato:2012ud, Borsato:2012ss} are relatively well-understood in the pure R-R case;%
\footnote{%
The S-matrix for mixed R-R and NS-NS fluxes has also been studied~\cite{Hoare:2013pma,Hoare:2013ida}, but remains somewhat more puzzling, see~\cite{Hoare:2013lja}. Other integrable aspects of the mixed flux backgrounds have been investigated in~\cite{David:2014qta,Ahn:2014tua,Banerjee:2014gga}.} the giant magnon associated to the massive modes was also understood some time ago~\cite{David:2010yg,David:2008yk}. In~\cite{Sax:2012jv} massless modes were incorporated in the weakly-coupled spin-chain picture. On the string side, only very recently it has been shown how massless modes can be included  in the classical integrability machinery~\cite{Lloyd:2013wza}. Both of these results demonstrate that the real intricacies involved in understanding massless modes occur away from the weakly-coupled string and spin-chain regimes.
 
The aim of this paper is to present in detail how massless excitations can be included in the non-perturbative integrability picture, and how the non-perturbative asymptotic worldsheet S~matrix for all fundamental particles can be found in the case of pure R-R $\AdS_3\times \Sphere^3\times \Torus^4 $ background. These results were first presented in~\cite{Borsato:2014exa}.

Our analysis starts from the determination of the off-shell symmetry algebra~$\mathcal{A} $ of the theory. Before light-cone gauge fixing, the symmetries of $\AdS_3\times \Sphere^3\times \Torus^4 $ are given by the~$\AdS_3\times \Sphere^3 $ superisometries\footnote{%
The two copies of $\alg{psu}(1,1|2)$ carry labels ``L'' (left) and ``R'' (right) corresponding to chiralities in the dual $\CFT_2 $.}
$\alg{psu}(1,1|2)_{\sL}\oplus\alg{psu}(1,1|2)_{\sR} $ together with the $\Torus^4 $ isometries. Fixing light-cone gauge breaks some of these symmetries, and in particular halves the supersymmetries. We are interested in the symmetry generators that are linearly realised after gauge fixing, as the S~matrix will have to commute with them. Such generators will sit in~$\mathcal{A} $, together with some additional central charges  which are expected from the case of~$\AdS_5\times \Sphere^5 $ \cite{Frolov:2006cc, Arutyunov:2006ak,Arutyunov:2006yd}. We will determine the form of these and find, as it should be,  that they have a non-trivial action only on states that do not satisfy the level-matching constraint (\textit{i.e.} off shell). Once the off-shell algebra of the theory is determined we will use it to constrain the non-perturbative $2\to2$ S~matrix, which will then satisfy the Yang-Baxter equation. 

Unlike what happened in~$\AdS_5/\CFT_4$~\cite{Frolov:2006cc, Arutyunov:2006ak,Arutyunov:2006yd}, we cannot use the coset action~\cite{Rahmfeld:1998zn,Park:1998un, Metsaev:2000mv,Babichenko:2009dk} for our calculations. The coset action requires the use of a particular kappa gauge~\cite{Babichenko:2009dk}, which does not allow for a straightforward quantization of the massless modes; see~\cite{Sundin:2012gc} for a discussion of the coset kappa gauge. We will therefore work with the Green-Schwarz action~\cite{Grisaru:1985fv}, in light-cone gauge. Furthermore, we take the decompactification limit, whereby the world-sheet cylinder becomes a plane and the asymptotic states can be defined.
It is interesting to note that our results give an example of integrability where the fermionic degrees of freedom do not enter the dynamics through a coset action; similar observations have recently been made in integrable $\AdS$ backgrounds which preserve even less supersymmetry~\cite{Wulff:2014kja}.

In this way, we are able to establish the off-shell symmetry algebra, including the non-linear momentum-dependent central extension reminiscent of~\cite{Arutyunov:2006ak}. As expected, the light-cone-gauge worldsheet theory is non-relativistic. Massive and massless excitations will then have periodic dispersion relations, with the energy of the latter being linear in the momentum for small values of it.
Using these results, the two-body S~matrix will follow immediately by symmetry arguments, and is fixed up to some dressing factors, for which crossing equations can be written down. As expected, the massive-sector S~matrix of~\cite{Borsato:2013qpa}, including the crossing-invariant dressing factors of~\cite{Borsato:2013hoa} can be consistently embedded in the full S~matrix of the present paper.

This paper is structured as follows. In section~\ref{sec:off-shell-algebra} we consider the type IIB superstring action for $\AdS_3\times\Sphere^3\times\Torus^4$ in light-cone gauge, and derive its conserved supercurrents. This is done at leading order in the fermions and at subleading order in the bosons.
In section~\ref{sec:algebra} we study the symmetry algebra~$\mathcal{A}$ and the representations that emerge from the supercurrent analysis. We find three short irreducible representations of  the centrally extended $\alg{psu}(1|1)^4\oplus\alg{so}(4)$ algebra\footnote{%
The symbol~$\oplus$ here and later indicates the direct sum of vector spaces, and not necessarily of (super)algebras. To avoid  introducing non-standard notation we will always explicitly detail the non-vanishing (anti-)commutation relations of the (super)algebras considered.%
}: two massive representations of dimension four, and one massless one of dimension eight. In section~\ref{sec:representations-allloop} we deform the representations found perturbatively in order to reproduce the correct non-linear central extension and shortening condition. We also comment on the possibility of quantum corrections to the massless dispersion relation, arguing that they would break part of~$\mathcal{A} $.
Using those exact representations, in section~\ref{sec:smat} we construct an invariant S~matrix for all of the superstring's excitations, including the massless ones, up to some dressing factors which we constrain by crossing symmetry. We conclude in section~\ref{sec:conclusion}. We relegate the more technical aspects of our results to the appendices.

\section{The off-shell symmetry algebra of superstrings on \texorpdfstring{$\AdS_3\times \Sphere^3\times \Torus^4$}{AdS3 x S3 x T4}}
\label{sec:off-shell-algebra}

In this section we compute the algebra $\mathcal{A}$ of off-shell symmetries for classical Type IIB superstring theory on $\AdS_3\times \Sphere^3\times \Torus^4$. At first sight it may appear that the natural setting for this would be the coset action~\cite{Rahmfeld:1998zn,Park:1998un, Metsaev:2000mv,Babichenko:2009dk}, since one can use the algebraic structure of the coset to facilitate the computations. The coset action is obtained from a Green-Schwarz action~\cite{Grisaru:1985fv} by fully fixing the kappa symmetry to the so-called coset gauge. While it is useful in the study of the classical integrability of this theory, the coset gauge leads to a kinetic term for the massless fermions which contains no quadratic piece. As a result, computing $\mathcal{A}$ using Poisson brackets is not straightforward in the coset gauge. Instead, we will perform the calculations using the Green-Schwarz action in the BMN light-cone kappa gauge. Explicit expressions up to quartic order in fermions have been recently found~\cite{Wulff:2013kga}, but we will only work up to quadratic order in fermions and so will use the component action~\cite{Cvetic:1999zs}.

This section is divided into four parts. In section~\ref{sec:Killing-spinors} we find the Killing spinors of the background in the metric~\eqref{eq:metric}. In section~\ref{sec:action} we write down explicitly the action for Type IIB superstrings on $\AdS_3\times \Sphere^3\times \Torus^4$, both before and after imposing the kappa gauge along the  BMN light-cone coordinates. In section~\ref{sec:charges} we write down the super-currents for the $\mathcal{A}$ charges and in section~\ref{sec:A-computation} we compute the off-shell algebra $\mathcal{A}$ of the classical theory. In appendix~\ref{app:index-conventions} we establish our conventions.

\subsection{Killing spinors for type IIB supergravity on  \texorpdfstring{$\AdS_3\times \Sphere^3\times \Torus^4$}{AdS3 x S3 x T4}}
\label{sec:Killing-spinors}

In this sub-section we construct the Killing spinors for type IIB supergravity on  $\AdS_3\times \Sphere^3\times \Torus^4$. Expressions for these are well-known in the literature~\cite{Lu:1996rhb,Lu:1998nu}. We adapt these well-known calculations to the metric
\begin{equation}
\label{eq:metric}
  ds^2 = ds_{\AdS_3}^2 + ds_{\Sphere^3}^2 + dX_i dX_i \,,
\end{equation}
where
\begin{equation}
\label{eq:s3metric}
  ds^2_{\Sphere^3} = +\Bigl(\frac{1 - \frac{y_3^2 + y_4^2}{4}}{1 + \frac{y_3^2 + y_4^2}{4}}\Bigr)^2 d\phi^2 + \Bigl(\frac{1}{1 + \frac{y_3^2 + y_4^2}{4}}\Bigr)^2 ( dy_3^2 + dy_4^2 )
\end{equation}
and
\begin{equation}
\label{eq:ads3metric}
  ds^2_{\AdS^3} = -\Bigl(\frac{1 + \frac{z_1^2 + z_2^2}{4}}{1 - \frac{z_1^2 + z_2^2}{4}}\Bigr)^2 dt^2 + \Bigl(\frac{1}{1 - \frac{z_1^2 + z_2^2}{4}}\Bigr)^2 ( dz_1^2 + dz_2^2 ) \,,
\end{equation}
since this metric is well suited for expansion around the BMN ground state. 

The ten-dimensional Killing spinor equations of Type IIB supergravity on $\AdS_3 \times \Sphere^3 \times \Torus^4$ with R-R flux are
\begin{equation}\label{eq:AdS3-S3-T4-Killing-spinor-eq}
  D_m \varepsilon^1 + \frac{1}{24} \slashed{F} \slashed{E}_m \varepsilon^1 = 0 ,
  \qquad
  D_m \varepsilon^2 - \frac{1}{24} \slashed{F} \slashed{E}_m \varepsilon^2 = 0 ,
\end{equation}
where the covariant derivative is given by
\begin{equation}
  D_m\varepsilon^I = (\partial_m + \frac{1}{4} \slashed{\omega}_m)\varepsilon^I ,
\end{equation}
and the R-R field strength by
\begin{equation}
  \slashed{F} = \Gamma^{ABC} F_{ABC} = 6(\Gamma^{012} + \Gamma^{345}) .
\end{equation}
As is shown in more detail in appendix~\ref{app:Killing-spinors}, these equations are solved by
\begin{equation}
\label{eq:full-ks-soln}
    \varepsilon^1
    =
    \hat{M} \varepsilon_0^1,
    \,,\qquad\qquad
    \varepsilon^2
    =
    \check{M} \varepsilon_0^2,
\end{equation}
where $\varepsilon_0^I$ are constant 9+1 dimensional Majorana-Weyl spinors\footnote{Our spinor and gamma matrix conventions are given in appendix~\ref{sec:gamma-matrix-conventions}.}, which further satisfy
\begin{equation}
  \frac{1}{2} ( 1 + \Gamma^{012345} ) \varepsilon^I = 
    \frac{1}{2} ( 1 + \Gamma^{012345} ) \varepsilon_0^I  = 0\,.
\end{equation}
The matrices  $\hat{M}$ and $\check{M}$ depend on the  $\AdS_3\times \Sphere^3$ coordinates and for later convenience we seperate out the dependence on  $t$ and $\phi$ from the other coordinates by writing
\begin{equation}\label{eq:MN-10d-expressions}
  \hat{M} = M_0 M_t , \qquad
  \check{M} = M_0^{-1} M_t^{-1} ,
\end{equation}
where
\begin{equation}
  \begin{aligned}
    M_0 &= \frac{1}{\sqrt{\bigl( 1 - \frac{z_1^2 + z_2^2}{4} \bigr) \bigl( 1 + \frac{y_3^2 + y_4^2}{4} \bigr)}}
    \Bigl( {1} - \frac{1}{2} z_{\underline{i}} \Gamma^{\underline{i}} \Gamma^{012} \Bigr)
    \Bigl( {1} - \frac{1}{2} y_{\underline{i}} \Gamma^{\underline{i}} \Gamma^{345} \Bigr) ,
    \\
    M_0^{-1} &= \frac{1}{\sqrt{\bigl( 1 - \frac{z_1^2 + z_2^2}{4} \bigr) \bigl( 1 + \frac{y_3^2 + y_4^2}{4} \bigr)}}
    \Bigl( {1} + \frac{1}{2} z_{\underline{i}} \Gamma^{\underline{i}} \Gamma^{012} \Bigr)
    \Bigl( {1} + \frac{1}{2} y_{\underline{i}} \Gamma^{\underline{i}} \Gamma^{345} \Bigr) ,
  \end{aligned}
\end{equation}
and
\begin{equation}
  M_t = e^{-\frac{1}{2} ( t \, \Gamma^{12} + \phi \, \Gamma^{34} )} , \qquad
  M_t^{-1} = e^{+\frac{1}{2} ( t \, \Gamma^{12} + \phi \, \Gamma^{34} )} .
\end{equation}

\subsection{Type IIB superstring action on  \texorpdfstring{$\AdS_3\times \Sphere^3\times \Torus^4$}{AdS3 x S3 x T4}}
\label{sec:action}

In this sub-section we write down the action for Type IIB superstring action on  $\AdS_3\times \Sphere^3\times \Torus^4$. In section~\ref{sec:vielbeins-bos-eoms} we begin by introducing a set of bosonic vielbeins, particularly adapted to the analysis in the remainder of this section, and expressing the bosonic equations of motion in terms of these. In section~\ref{sec:action-before-lc} we write down the action to quadratic order in fermions. By picking suitably defined fermionic fields, our action realises the 16 unbroken supersymmetries of the background via linear shifts of the massive fermionic fields. In section~\ref{sec:action-after-lc} we write down the BMN light-cone kappa gauge-fixed action to quadratic order in fermions. Just as was done in~\cite{Alday:2005ww}, we find it useful to redefine the fermions further so that they are neutral under the $\algU(1)$ charges associated with $t$ and $\phi$ translations. The action is then re-expressed in first-order formalism and fully gauge-fixed in the uniform light-cone gauge in section~\ref{sec:first-order-uniform-lc}.

\subsubsection{A suitable vielbein and bosonic equations of motion}
\label{sec:vielbeins-bos-eoms}

The Lagrangian for the bosonic sigma model is given by
\begin{equation}
  \lagr_B = - \frac{1}{2} \gamma^{\alpha\beta} E_\alpha{}^A E_\beta{}^B \eta_{AB} ,
\end{equation}
where $E_\alpha{}^A = E_m{}^A \partial_\alpha X^m$ denotes the pullback of the vielbein. $\lagr_B$ is invariant under $\grpSO(1,9)$ rotations in tangent space. As a result, all vielbeins that describe the same metric will lead to the same bosonic equations of motion, up to field redefinitions. Nevertheless, picking a suitable vielbein may reduce substantially the computational complexity of the analysis. Since we will be working with the metric~\eqref{eq:metric}, one seemingly natural choice is to pick diagonal vielbeins $E_m{}^A$  given in equations~\eqref{eq:diag-s3-vielbein} and~\eqref{eq:diag-ads3-vielbein}. It turns out that, for the purpose of understanding the realisation of supersymmetry in the Green-Schwarz action, it is instead more conventient to use vielbeins $\hat{K}_m{}^A$  and $\check{K}_m{}^A$, which are related to the $E_m{}^A$ by orthogonal transformations 
\begin{equation}
\label{eq:def-of-hat-check-vielbeins}
  \hat{K}_m{}^A = \hat{\mathcal{M}}^A{}_B E_m{}^B ,
  \qquad
  \check{K}_m{}^A = \hat{\mathcal{M}}^A{}_B E_m{}^B \,.
\end{equation}
The matrices $\hat{\mathcal{M}}$ and $\check{\mathcal{M}}$ are defined in equations~\eqref{eq:orth-rot-ads3-s3},~\eqref{eq:orth-rot-ads3} and~\eqref{eq:orth-rot-s3}. They follow from considering bilinears formed out of the Killing spinors $\varepsilon^I$, (\cf equation~\eqref{eq:MN-identities}). As a result, as shown in equations~\eqref{eq:s3-isometries} and~\eqref{eq:ads3-isometries}, $\hat{K}_m{}^A$  and $\check{K}_m{}^A$ satisfy the Killing vector equation~\eqref{eq:killing-vector-s3} and generate the $\algSO(2,2) \oplus \algSO(4) = \algSL(2) \oplus \algSL(2) \oplus \algSU(2) \oplus \algSU(2)$ isometry algebra of $\AdS_3\times \Sphere^3$.

The bosonic equations of motion that follow from $\lagr_B$ are
\begin{equation}\label{eq:bosonic-eom-I}
    \begin{aligned}
    0 &= \eta_{AB} \Bigl[
    \partial_\alpha ( \gamma^{\alpha\beta}  E_m{}^A E_n{}^B \partial_\beta X^n )
    - \frac{1}{2} \gamma^{\alpha\beta} \partial_m ( E_n{}^A E_k{}^B ) \partial_\alpha X^n \partial_\beta X^k
    \Bigr]
    \\
    &=
    \gamma^{\alpha\beta} \Bigl[
    - \frac{1}{2} \bigl ( \omega_{kAB} E_n{}^A + \omega_{nAB} E_k{}^A \bigr) E_m{}^B
    + \eta_{AB} E_m{}^A \partial_n E_k{}^B
    \Bigr ] \partial_\alpha X^n \partial_\beta X^k
    \\ &\qquad
    + \eta_{AB}   E_m{}^A E_n{}^B \partial_\alpha ( \gamma^{\alpha\beta} \partial_\beta X^n ) \,,
 \end{aligned}
\end{equation}
where in the second line we have used the fact that $E_m{}^A$ is covariantly constant.
For a generic vielbein the first term above is nonvanishing. However, the vielbeins $\hat{K}_m{}^A$ and $\check{K}_m{}^A$ satisfy the Killing vector equation which makes it vanish, see~\eqref{eq:symmetric-relation-E-omega}. Hence the equations of motion written in terms of the worldsheet pullbacks $\hat{K}_\alpha{}^A$ and $\check{K}_\alpha{}^A$ are simply
 \begin{equation}
  \partial_\alpha ( \gamma^{\alpha\beta} \hat{K}_\beta{}^A ) = 0 , \qquad
  \partial_\alpha ( \gamma^{\alpha\beta} \check{K}_\beta{}^A ) = 0 \,.
\end{equation}
This form of the equations of motion is not only particularly simple, but will prove to be very useful in analysing the supersymmetries of string theory on this background.

\subsubsection{Green-Schwarz action before kappa gauge fixing}
\label{sec:action-before-lc}

In this sub-section we write down the Green-Schwarz action for a superstring propagating in $\AdS_3 \times \Sphere^3 \times \Torus^4$ up to quadratic order in fermions and construct supercharges preserving the non-gauge-fixed action. The Green-Schwarz action for Type IIB superstrings in a generic supergravity background was constructed in terms of superfields in~\cite{Grisaru:1985fv}, and explicit expressions in terms of fields are known to quadratic~\cite{Cvetic:1999zs} and quartic order~\cite{Wulff:2013kga} in the fermions. We will perform a field redefinition of the conventional fermions~\cite{Cvetic:1999zs} so that the 16 real supersymmetries of this background are realised as linear shifts of the massive fermions.\footnote{In a background described by a super-coset conventional fermions~\cite{Cvetic:1999zs} correspond to picking the super-coset element $g=g_{\text{\scriptsize bos}}g_{\text{\scriptsize ferm}}$---see for example appendix B in~\cite{Babichenko:2009dk}. The field redefinition we perform would amount to picking a super-coset element of the form $g=g_{\text{\scriptsize ferm}}g_{\text{\scriptsize bos}}$. Such changes of variable were discussed in the context of the maximally supersymmetric type IIB plane-wave background in~\cite{Metsaev:2001bj}.}

The Green-Schwarz Lagrangian can be written as
\begin{equation}
  \lagr = \lagr_B + \lagr_{\text{\scriptsize kin}} + \lagr_{\text{WZ}} .
\end{equation}
The bosonic Lagrangian $\lagr_B$ was discussed in the previous sub-section. We have split the fermionic Lagrangian into two terms: a term dependent on the worldsheet metric, $\lagr_{\text{\scriptsize kin}}$, and the Wess-Zumino term $\lagr_{\text{WZ}}$. In the background we are considering, the former term is~\cite{Cvetic:1999zs}
\begin{equation}
  \lagr_{\text{\scriptsize kin}} = 
  -i\gamma^{\alpha\beta} \bar{\theta}_I \slashed{E}_\alpha \bigl( \delta^{IJ} D_\beta +\frac{1}{24} \sigma_3^{IJ} \slashed{F} \slashed{E}_\beta \bigr) \theta_J \,,
\end{equation}
where $\bar{\theta}_I = \theta_I^\dag \Gamma^0$ and we have redefined the fermions compared to Cveti\v{c}, L\"u, Pope and Stelle~\cite{Cvetic:1999zs}
\begin{equation}
  \theta_{1\text{\scriptsize \,CLPS}} = \frac{\theta_1 + \theta_2}{\sqrt{2}} ,
  \qquad
  \theta_{2\text{\scriptsize \,CLPS}} = \frac{\theta_1 - \theta_2}{\sqrt{2}} ,
\end{equation}
so that they enter diagonally in $\lagr_{\text{\scriptsize kin}}$.

Next we define new fermions $\vartheta^\pm_I$ which are related to $\theta_I$ by
\begin{equation}\label{eq:theta-chi-eta-def}
  \begin{aligned}
    \theta_1 &= \frac{1}{2} ( 1 + \Gamma^{012345} ) \hat{M} \vartheta^+_1 + \frac{1}{2} ( 1 - \Gamma^{012345} ) \hat{M} \vartheta^-_1 , \\
    \theta_2 &= \frac{1}{2} ( 1 + \Gamma^{012345} ) \check{M} \vartheta^+_2 + \frac{1}{2} ( 1 - \Gamma^{012345} ) \check{M} \vartheta^-_2 ,
  \end{aligned}
\end{equation}
where the matrices $\hat{M}$ and $\check{M}$ were given in~\eqref{eq:MN-10d-expressions}.
Inserting this into the Lagrangian and using the relations in section~\ref{app:Killing-spinors} we find
\begin{equation}
  \begin{aligned}
    \lagr_{\text{\scriptsize kin}}
    = -i \gamma^{\alpha\beta} \Bigl[
    &\bar{\vartheta}^-_I \slashed{\hat{K}}_\alpha \partial_\beta \vartheta^-_I
    + 2\bar{\vartheta}^+_I \slashed{\dot{E}}_\alpha \partial_\beta \vartheta^-_I
    + \bar{\vartheta}^+_I \slashed{\hat{K}}_\alpha \partial_\beta \vartheta^+_I
    \\
    &- \frac{1}{2}\sigma^3_{IJ} \bar{\vartheta}^+_I \Gamma^{012} \vartheta^+_J 
    ( \hat{K}_\alpha^a \hat{K}_\beta^b \eta_{ab} + \dot{E}_\alpha^{\dot{a}} \dot{E}_\beta^{\dot{b}} \eta_{\dot{a}\dot{b}} )
    \Bigr]\,.
  \end{aligned}
\end{equation}
The definitions of the vielbeins appearing above are given in equations~\eqref{eq:def-of-hat-check-vielbeins} and~\eqref{eq:various-vielbeins}.
The Lagrangian $\lagr_B + \lagr_{\text{\scriptsize kin}}$ is invariant under the supersymmetry transformations
\begin{equation}\label{eq:susy-transformations}
  \delta\vartheta^-_I = \epsilon_I \,, 
  \qquad 
  \delta\vartheta^+_I = 0 \,,
  \qquad
  \delta\hat{K}_\alpha{}^A = -i \bar{\epsilon}_I \Gamma^A \partial_\alpha \vartheta^-_I \,, 
  \qquad
  \delta \dot{E}_\alpha =0\,,
\end{equation}
where in the above equation the index $A=0,\dots,5$. 
By imposing the Majorana condition on the fermions this gives us the expected 16 real supersymmetries of the background.

We now consider the Wess-Zumino term\footnote{%
We set~$\epsilon^{\tau\sigma}=+1$.
}
\begin{equation}
  \lagr_{\text{WZ}} = +i\epsilon^{\alpha\beta} \Bigl(
  \bar{\theta}_2 \slashed{E}_\alpha \bigl( D_\beta + \frac{1}{24} \slashed{F} \slashed{E}_\beta \bigr) \theta_1
  +
  \bar{\theta}_1 \slashed{E}_\alpha \bigl( D_\beta - \frac{1}{24} \slashed{F} \slashed{E}_\beta \bigr) \theta_2
  \Bigr)
\end{equation}
After introducing the rotated fermions we find
\begin{equation}
  \begin{aligned}
    \lagr_{\text{WZ}} = +i \epsilon^{\alpha\beta} \Bigl(
    &\bar{\vartheta}^-_2 \check{M}^{-1} \hat{M} \slashed{\hat{K}}_\alpha \partial_\beta \vartheta^-_1
    + 2\bar{\vartheta}^+_2 \check{M}^{-1} \hat{M} \slashed{\dot{E}}_\alpha \partial_\beta \vartheta^-_1
    + \bar{\vartheta}^+_2 \check{M}^{-1} \hat{M} \slashed{\hat{K}}_\alpha \partial_\beta \vartheta^+_1
    \\
    &+ \bar{\vartheta}^-_1 \hat{M}^{-1} \check{M} \slashed{\check{K}}_\alpha \partial_\beta \vartheta^-_2
    + 2\bar{\vartheta}^+_1 \hat{M}^{-1} \check{M} \slashed{\dot{E}}_\alpha \partial_\beta \vartheta^-_2
    + \bar{\vartheta}^+_1 \hat{M}^{-1} \check{M} \slashed{\check{K}}_\alpha \partial_\beta \vartheta^+_2
    \\ 
    &- \frac{1}{2} \bar{\vartheta}^+_2 \check{M}^{-1} \hat{M} ( \slashed{\hat{K}}_\alpha \slashed{\hat{K}}_\beta + \slashed{\dot{E}}_\alpha \slashed{\dot{E}}_\beta ) \Gamma^{012} \vartheta^+_1
    \\
    &+ \frac{1}{2} \bar{\vartheta}^+_1 \hat{M}^{-1} \check{M} ( \slashed{\check{K}}_\alpha \slashed{\check{K}}_\beta + \slashed{\dot{E}}_\alpha \slashed{\dot{E}}_\beta ) \Gamma^{012} \vartheta^+_2
    \Bigr) .
  \end{aligned}
\end{equation}
This term is also invariant to quadratic order in the fermions under the supersymmetry transformations~\eqref{eq:susy-transformations}. To see this we can use the identity
\begin{equation}\label{eq:epsilon-NME}
  \epsilon^{\alpha\beta} \partial_\alpha \bigl(
  \check{M}^{-1} \hat{M} \slashed{\hat{K}}_\beta
  \bigr)
  (1 - \Gamma^{012345}) = 0
\end{equation}
to show that
\begin{equation}
  \epsilon^{\alpha\beta} \bar{\vartheta}^-_2\check{M}^{-1} \hat{M} \slashed{\hat{K}}_\alpha \partial_\beta \vartheta^-_1
  =
  \epsilon^{\alpha\beta} \bar{\vartheta}^-_1 \hat{M}^{-1} \check{M} \slashed{\check{K}}_\alpha \partial_\beta \vartheta^-_2\,,
\end{equation}
up to a total derivative. In appendix~\ref{app:proof-wz-identity} we prove~\eqref{eq:epsilon-NME}. Together with an obvious extension of the above argument to exressions involving $\check{K}$ instead of $\hat{K}$, the Lagrangian $\lagr_{\text{WZ}}$ can therefore be written in a form where $\vartheta^-_I$ only appears with a partial derivative acting on it, making the symmetry under shifts of that fermion manifest.

\subsubsection{Neutral fermions and the kappa gauge-fixed action}
\label{sec:action-after-lc}

In this sub-section we impose the BMN light-cone kappa gauge on the Lagrangian obtained in the previous sub-section. In addition, we will further redefine the fermions. Recall that the tangent space rotations~\eqref{eq:theta-chi-eta-def} introduced in the previous section were useful for obtaining the supersymmetry transformations before fixing kappa gauge. However, $\hat{K}_m{}^A$ and $\check{K}_m{}^A$, and therefore also the fermions $\eta_{\!K}$ and $\chi_{\!K}$, transform nontrivially under shifts of the coordinates $t$ and $\phi$. When imposing uniform light-cone gauge it is useful to work with fermions that are uncharged under these shifts~\cite{Alday:2005ww}, which motivates the further re-definition of the fermions.\footnote{In a super-coset background this redefinition amounts to picking a coset representative of the form $g=g_{t,\phi}g_{\text{\scriptsize ferm}}g_{\text{\scriptsize bos}'}$, where $g_{\text{\scriptsize bos}'}$ involves the (eight) bosonic coordinates transverse to $t$ and $\phi$.}

To perform this field redefinition, recall that the rotation matrices $\hat{M}$ and $\check{M}$ can be written in terms of the matrices $M_0$ and $M_t$ (see equation~\eqref{eq:MN-10d-expressions}), where $M_0$ is independent of $t$ and $\phi$ while $M_t$ only depends on those two coordinates. In order to have fermions that are uncharged under shifts of $t$ and $\phi$ one needs to multiply the fermions $\vartheta^\pm_1$ by $M_t^{-1}$ and the fermions $\vartheta^\pm_2$  with $M_t$. In other words, we define
\begin{equation}\label{eq:theta-chi-eta-def-t}
  \begin{aligned}
    \theta_1 &= 
    \frac{1}{2} ( 1 + \Gamma^{012345} ) \mathrlap{M_0 \chi_1}\hphantom{M_0^{-1} \chi_2} 
    + \frac{1}{2} ( 1 - \Gamma^{012345} ) M_0 \eta_1
    \\
    \theta_2 &= \frac{1}{2} ( 1 + \Gamma^{012345} ) M_0^{-1} \chi_2 + \frac{1}{2} ( 1 - \Gamma^{012345} ) M_0^{-1} \eta_2 .
  \end{aligned}
\end{equation}
We also need to perform the corresponding rotation on the vielbeins defining new vielbeins ${\hat{E}}$ and ${\check{E}}$,
\begin{equation}
  \slashed{\hat{K}} = M_t^{-1} \slashed{\hat{E}} M_t , \qquad
  \slashed{\check{K}} = M_t \slashed{\check{E}} M_t^{-1} .
\end{equation}
The components of the inverse vielbeins can easily be read off from equations~\eqref{eq:E-components-S3} and~\eqref{eq:E-components-AdS3} by dropping the first $t$- and $\phi$-dependent factor. 

It is useful to introduce light-cone coordinates
\begin{equation}
  E^{\pm} = \frac{1}{2} ( E^5 \pm E^0 ) , \qquad
  x^{\pm} = \frac{1}{2} ( \phi \pm t ).
\end{equation}
This leads to
\begin{equation}
  E_{x^+}^{+} = E_{x^-}^{-} = \frac{1}{2} \bigl( E_{\phi}^5 + E_t^0 \bigr) , \qquad
  E_{x^+}^{-} = E_{x^-}^{+} = \frac{1}{2} \bigl( E_{\phi}^5 - E_t^0 \bigr) .
\end{equation}
The light-cone components of the tangent space metric are given by
\begin{equation}
  \eta^{+-} = \eta^{-+} = + \frac{1}{2} , \qquad
  \eta_{+-} = \eta_{-+} = + 2 .
\end{equation}
The bosonic Lagrangian then takes the form
\begin{equation}
  \lagr_B = - \frac{1}{2} \gamma^{\alpha\beta} \bigl(
  4 E^+_\alpha E^-_\beta + E^{\underline{i}}_\alpha E^{\underline{i}}_\beta + 
  E^i_\alpha E^i_\beta 
  \bigr) .
\end{equation}

We will work in the BMN light-cone kappa gauge
\begin{equation}\label{eq:bmn-lc-kappa-gauge}
  \Gamma^+ \eta_I = 0 , \qquad
  \Gamma^+ \chi_I = 0 , \qquad
  \Gamma^{\pm} = \frac{1}{2} \bigl( \Gamma^5 \pm \Gamma^0 \bigr) .
\end{equation}
The kappa gauge-fixed Lagrangian then takes the form
\begin{align}\label{eq:lkingfix}
  \!\!\!\!\!\!\!\!\!
  \lagr_{\text{\scriptsize kin}}
  = -2i \gamma^{\alpha\beta} \Bigl(
  &\bar{\eta}_1 \hat{E}_\alpha^+ \Gamma^- \partial_\beta \eta_1 - \bar{\eta}_1 \Gamma^{012} \eta_1 \hat{E}_\alpha^+ \partial_\beta x^+
  + \bar{\eta}_2 \check{E}_\alpha^+ \Gamma^- \partial_\beta \eta_2 + \bar{\eta}_2 \Gamma^{012} \eta_2 \check{E}_\alpha^+ \partial_\beta x^+
  \nonumber \\
  + &\bar{\chi}_1 \hat{E}_\alpha^+ \Gamma^- \partial_\beta \chi_1 
  - \frac{1}{4} \bar{\chi}_1 \Gamma^{012} \chi_1 
  \bigl( \sum_{A,B=0}^5\hat{E}_\alpha^A \hat{E}_\beta^B \eta_{AB} + \dot{E}_\alpha^i \dot{E}_\beta^i - 4 \hat{E}_\alpha^+ \partial_\beta x^-\bigr)
  \nonumber \\
  + &\bar{\chi}_2 \check{E}_\alpha^+ \Gamma^- \partial_\beta \chi_2 
  + \frac{1}{4} \bar{\chi}_2 \Gamma^{012} \chi_2
  \bigl( \sum_{A,B=0}^5\check{E}_\alpha^A \check{E}_\beta^B \eta_{AB} + \dot{E}_\alpha^i \dot{E}_\beta^i  - 4 \check{E}_\alpha^+ \partial_\beta x^-\bigr)
  \Bigr).
\end{align}
\begin{align}\label{eq:lwzgfix}
  \lagr_{\text{WZ}} = +i \epsilon^{\alpha\beta} \Bigl(
  &\bar{\eta}_2 \slashed{\check{E}}_\alpha M_0^2 \partial_\beta \eta_1
  + \bar{\eta}_2 \slashed{\check{E}}_\alpha M_0^2 \Gamma^{12} \eta_1 \partial_\beta x^+
  \nonumber \\
  + &\bar{\eta}_1 \slashed{\hat{E}}_\alpha M_0^{-2} \partial_\beta \eta_2
  - \bar{\eta}_1 \slashed{\hat{E}}_\alpha M_0^{-2} \Gamma^{12} \eta_2 \partial_\beta x^+
  \nonumber \\
  + &\bar{\chi}_2 \slashed{\check{E}}_\alpha M_0^2 \partial_\beta \chi_1
  - \bar{\chi}_2 \slashed{\check{E}}_\alpha M_0^2 \Gamma^{12} \chi_1 \partial_\beta x^-
  - \frac{1}{2} \bar{\chi}_2 ( \slashed{\check{E}}_\alpha \slashed{\check{E}}_\beta + \slashed{\dot{E}}_\alpha \slashed{\dot{E}}_\beta ) M_0^2 \Gamma^{012} \chi_1
  \nonumber \\
  + &\bar{\chi}_1 \slashed{\hat{E}}_\alpha M_0^{-2} \partial_\beta \chi_2
  + \bar{\chi}_1 \slashed{\hat{E}}_\alpha M_0^{-2} \Gamma^{12} \chi_2 \partial_\beta x^-
  + \frac{1}{2} \bar{\chi}_1 ( \slashed{\hat{E}}_\alpha \slashed{\hat{E}}_\beta + \slashed{\dot{E}}_\alpha \slashed{\dot{E}}_\beta ) M_0^{-2} \Gamma^{012} \chi_2
  \nonumber \\
  + &2 \bar{\chi}_2 \slashed{\dot{E}}_\alpha M_0^2 \partial_\beta \eta_1
  + 2 \bar{\chi}_2 \slashed{\dot{E}}_\alpha M_0^2 \Gamma^{12} \eta_1 \partial_\beta x^+
  \nonumber \\
  + &2 \bar{\chi}_1 \slashed{\dot{E}}_\alpha M_0^{-2} \partial_\beta \eta_2
  - 2 \bar{\chi}_1 \slashed{\dot{E}}_\alpha M_0^{-2} \Gamma^{12} \eta_2 \partial_\beta x^+
  \Bigr) .
\end{align}

\subsubsection{First-order action and uniform light-cone gauge}
\label{sec:first-order-uniform-lc}

To fix the bosonic gauge we will impose uniform light-cone gauge~\cite{Arutyunov:2005hd}.
The simplest way to introduce this gauge is to rewrite the action in a first-order formalism by introducing coordinates $x^M$
\begin{equation}
  p_M = \frac{\delta \action}{\delta \dot{x}^M} .
\end{equation}
From the definition of the light-cone coordinates $x^\pm$ we then have
\begin{equation}
  p_+ = p_\phi + p_t , \qquad
  p_- = p_\phi - p_t .
\end{equation}
The isometries generated by shifts in $t$ and $\phi$ lead to the conservation of the energy $E$ and angular momentum $J$
\begin{equation}
  E = - \int_{-r}^{+r} d\sigma \, p_t , \qquad
  J = + \int_{-r}^{+r} d\sigma \, p_\phi .
\end{equation}
For the light-cone momenta we then find
\begin{equation}
  P_+ = \int_{-r}^{+r} d\sigma \, p_+ = J - E , \qquad
  P_- = \int_{-r}^{+r} d\sigma \, p_- = J + E .
\end{equation}
The uniform light-cone gauge fixing is now obtained by setting\footnote{%
  Here we only consider string states with zero winding number. For more general states the gauge fixing condition becomes $x^+ = \tau + \frac{1}{2} m \sigma$, where $m$ is the integer winding number along the angle~$\phi $.%
}%
\begin{equation}
  x^+ = \tau , \qquad
  p_- = 2 .
\end{equation}
The above gauge condition sets $p_-$ to $2$. To make the origin of various expressions more clear we generally still write out factors of $p_-$, unless this clutters our formulae excessively. In any case, the correct factors of $p_-$ can be restored from dimensional considerations.

To see how this gauge works let us consider the bosonic first-order action, which takes the form
\begin{equation}
  \action_B = \int_{-r}^{+r} d\sigma \, d\tau \Bigl(
  p_+ \dot{x}^+ + p_- \dot{x}^- + p_{\underline{i}} \dot{x}^{\underline{i}} + p_i \dot{x}^i + \frac{\gamma^{01}}{\gamma^{00}} C_1 + \frac{1}{2\gamma^{00}} C_2
  \Bigr) ,
\end{equation}
where
\begin{equation}
  C_1 = p_+ \pri{x}^+ + p_- \pri{x}^- + p_{\underline{i}} \pri{x}^{\underline{i}} + p_i \pri{x}^i
\end{equation}
and
\begin{equation}
  \begin{split}
    C_2 &=
    G^{++} p_+ p_+ + 2 G^{+-} p_+ p_- + G^{--} p_- p_- + G^{\underline{ij}} p_{\underline{i}} p_{\underline{j}} + G^{ij} p_i p_j
    \\ &\qquad
    + G_{++} \pri{x}^+ \pri{x}^+ + 2 G_{+-} \pri{x}^+ \pri{x}^- + G_{--} \pri{x}^- \pri{x}^- + G_{\underline{ij}} \pri{x}^{\underline{i}} \pri{x}^{\underline{j}} + G_{ij} \pri{x}^i \pri{x}^j \,.
  \end{split}
\end{equation}
The equations of motion for the worldsheet metric leads to the Virasoro constraints $C_1 = 0$ and $C_2 = 0$. Since $\pri{x}^+ = 0$ we can solve the first constraint by
\begin{equation}
  \pri{x}^- = -\frac{1}{p_-} \bigl( p_{\underline{i}} \pri{x}^{\underline{i}} + p_i \pri{x}^i \bigr) \,.
\end{equation}
Inserting this into the expression for $C_2$ we can solve the second constraint for $p_+$. The gauge-fixed action can then be written as\footnote{%
  We have omitted the total derivative term $p_- \dot{x}^-$.%
}%
\begin{equation}
  \action_B = \int_{-r}^{+r} d\sigma \bigl( p_{\underline{i}} \dot{x}^{\underline{i}} + p_i \dot{x}^i - \ham_B \bigr) ,
\end{equation}
with
\begin{equation}
  \ham_B = - p_+ .
\end{equation}
For the transverse fields we impose periodic boundary conditions\footnote{%
  Here we also ignore possible winding modes along the $\Torus^4$ directions. See below for a further discussion of these modes.%
} %
$x^{\underline{i}}(+r) = x^{\underline{i}}(-r)$ and $x^i(+r) = x^i(-r)$. Since we further assume there is no winding along the angle $\phi$, we find that a physical state should satisfy the level matching condition
\begin{equation}
  \Delta x^- = x^-(+r) - x^-(-r) = \int_{-r}^{+r} d\sigma \, \pri{x}^- = 0 .
\end{equation}
The gauge-fixed action is invariant under worldsheet translations, which leads to the conservation of the worldsheet momentum
\begin{equation}
  p_{\text{ws}} = -\int_{-r}^{+r} d\sigma \, \bigl( p_{\underline{i}} \pri{x}^{\underline{i}} + p_i \pri{x}^i \bigr) = p_- \Delta x^-.
\end{equation}
From the level matching constraint we then find that a physical string in the zero winding sector has to have vanishing total worldsheet momentum
\begin{equation}
  p_{\text{ws}} = 0 .  
\end{equation}

In order to study the worldsheet S~matrix we need to be able to create well-defined asymptotic states to scatter. To do this we will from now on work in the decompactification limit by sending the parameter $r$, which gives the circumference of the worldsheet cylinder, to infinity. Note that after gauge fixing, the light-cone momentum $P_-$ is given by
\begin{equation}
  P_- = \int_{-r}^{+r} d\sigma \, p_- = 4 r.
\end{equation}
Hence, in the large-$r$ limit the light-cone momentum becomes infinite.

By imposing periodic boundary conditions on the $\Torus^4$ coordinates $x^i$ we are ignoring winding modes on the torus. This is justified since we study local properties of the field theory on the worldsheet and work in the decompactification limit. If we begin with a string state in the zero winding sector and act on the state with a symmetry generator that acts locally, there is no way to obtain a state with non-zero winding.
Similarly, the scattering of two excitations without any winding will not result in non-trivial winding of the out-going states. In the zero-winding sector the~$\alg{u}(1)^4$ shift isometries of the~$\Torus^4$ are supplemented by an~$\alg{so}(4) $ symmetry, which we will discuss in the next subsection and will play an important role in~$\mathcal{A} $ when we will use it to constrain the S~matrix.

It is furthermore possible to check that, as long as we are in the decompactified theory with $P_-=\infty$, the light-cone Hamiltonian takes the same form in any sector with finite winding on~$\Torus^4 $. This indicates that the S~matrix that we will find by this treatment should be valid in any winding sector, and should not depend on the moduli of~$\Torus^4$. The dependence of the spectrum on winding numbers and torus moduli should then manifest itself only at the level of the Bethe-Yang equations, as it happens in the case of orbifolds, see \textit{e.g.}~\cite{Zoubos:2010kh,vanTongeren:2013gva} for a review.

\subsubsection{Gauge-fixed action with  \texorpdfstring{$\algSO(4)_1\oplus \algSO(4)_2$}{SO(4) + SO(4)} bispinor fermions}
\label{sec:gf-action-with-bisponors}

The fermions appearing in the action~\eqref{eq:lkingfix},~\eqref{eq:lwzgfix} are 32-component 9+1-dimensional spinors. However, these spinors satisfy a number of projections: the 9+1-dimensional Weyl projection, the kappa gauge condition~\eqref{eq:bmn-lc-kappa-gauge} as well as equation~\eqref{eq:theta-chi-eta-def-t}. Because of these, writing the fermions as 32 component spinors is rather redundant. In this sub-section we will write down the fully gauge-fixed action in terms of non-redundant physical spinors. 

As a result of the above projections, the physical spinors $\eta_I$ and $\chi_I$ are in fact bispinors of $\algSO(4)_1\oplus \algSO(4)_2\subset \algSO(8)$, with 
$\algSO(8)$ corresponding to rotations transverse to light-cone directions. The algebras $\algSO(4)_1$ and $\algSO(4)_2$ correspond to rotations along the non-light-cone 
$\AdS_3\times \Sphere^3$ and $\Torus^4$, directions, respectively.\footnote{In the next section we will write the $\Torus^4$ part of this algebra as $\algSO(4)_2 = \algSU(2)_{\suA} \oplus \algSU(2)_{\suB}$.} While the latter algebra remains unbroken by the background, $\algSO(4)_1$ is in fact broken to $\algSO(2)\oplus\algSO(2)$, as can be already seen in the plane-wave limit~\cite{Gava:2002xb,Gomis:2002qi}. We will see this breaking in the Lagrangian we write down in this subsection. Nevertheless, it is still convenient to express the fermionic fields that enter the Lagrangian as bispinors of $\algSO(4)_1\oplus \algSO(4)_2$. We will use the indices $\underline{a}$, $\dot{\underline{a}}$ (respectively, $a$, $\dot{a}$) to denote the positive and negative chirality $\algSO(4)_1$ ($\algSO(4)_2$) spinors. Further, we introduce gamma matrices, $\hat{\gamma}^{\underline{i}}$ with $\underline{i}=1,2,3,4$ and $\hat{\tau}^i$, $i=6,7,8,9$ for $\algSO(4)_1$ and $\algSO(4)_2$ .  We write these matrices as\footnote{%
  The matrices $\gamma^{\underline{i}}$ introduced here should not be confused with the three dimensional gamma matrices for $\AdS_3$ and $\Sphere^3$ used to express the Killing spinors in section~\ref{sec:Killing-spinors} and in appendix~\ref{sec:gamma-matrix-conventions} to construct the ten dimensional gamma matrices. Since the two types of matrices never appear in the same setting we hope that the meaning of $\gamma$ is clear from the context it appears in.%
}%
\begin{equation}
\label{eq:so(4)-gamma-matrices}
  (\hat{\gamma}^{\underline{i}})^{\underline{a\dot{a}}}{}_{\underline{b\dot{b}}} = \begin{pmatrix} 0 & (\gamma^{\underline{i}})^{\underline{a}}{}_{\underline{\dot{b}}} \\ (\tilde{\gamma}^{\underline{i}})^{\underline{\dot{a}}}{}_{\underline{b}} & 0 \end{pmatrix} ,
  \qquad
  (\hat{\tau}^i)^{a\dot{a}}{}_{b\dot{b}} = \begin{pmatrix} 0 & (\tau^i)^a{}_{\dot{b}} \\ (\tilde{\tau}^i)^{\dot{a}}{}_b & 0 \end{pmatrix} ,
\end{equation}
with the Clebsch-Gordan coefficients for the decomposition of two $\algSO(4)$ Weyl spinors of opposite chirality given by
\begin{equation}
  \begin{aligned}
  \gamma^1 &= + \sigma_3 , \quad &
  \gamma^2 &= - i \mathds{1} , \quad &
  \gamma^3 &= + \sigma_2 , \quad &
  \gamma^4 &= + \sigma_1 , \quad &
  \tilde{\gamma}^{\underline{i}} &= + ( \gamma^{\underline{i}} )^\dag ,
  \\
  \tau^6 &= + \sigma_1 , \quad &
  \tau^7 &= + \sigma_2 , \quad &
  \tau^8 &= + \sigma_3 , \quad &
  \tau^9 &= + i \mathds{1} , \quad &
  \tilde{\tau}^i &= - ( \tau^i )^\dag \,.
  \end{aligned}
\end{equation}
The notation introduced above is purposefully reminiscent of the light-cone gauge in flat space~\cite{Green:1982tk} but our exact conventions are slightly different to, for example, those in~\cite{Green:1982tc}. The matrices $\hat{\gamma}^{\underline{i}}$ and $\hat{\tau}^i$ satisfy the Clifford algebra relations
\begin{equation}
  \begin{aligned}
    \acomm{\hat{\gamma}^{\underline{i}}}{\hat{\gamma}^{\underline{j}}} &= + 2 \delta^{\underline{ij}} , \qquad &
    ( \hat{\gamma}^{\underline{i}} )^t &= + t \hat{\gamma}^{\underline{i}} t^{-1} ,
    \\
  \acomm{\hat{\tau}^i}{\hat{\tau}^j} &= - 2 \delta^{ij} , \qquad &
  ( \hat{\tau}^i )^t &= - s \hat{\tau}^i s^{-1} ,
  \end{aligned}
\end{equation}
where $t = s = \sigma_3 \otimes \sigma_2$.
We also introduce
\begin{equation}
  \begin{aligned}
    (\gamma^{\underline{ij}})^{\underline{a}}{}_{\underline{b}} &= \frac{1}{2} ( \gamma^{\underline{i}} \tilde{\gamma}^{\underline{j}} - \gamma^{\underline{j}} \tilde{\gamma}^{\underline{i}} )^{\underline{a}}{}_{\underline{b}} , \qquad &
    (\tau^{ij})^a{}_b &= \frac{1}{2} ( \tau^i \tilde{\tau}^j - \tau^j \tilde{\tau}^i )^a{}_b , \\
    (\tilde{\gamma}^{\underline{ij}})^{\underline{\dot{a}}}{}_{\underline{\dot{b}}} &= \frac{1}{2} ( \tilde{\gamma} \gamma^{\underline{j}} - \tilde{\gamma}^{\underline{j}} \gamma^{\underline{i}} )^{\underline{\dot{a}}}{}_{\underline{\dot{b}}} , \qquad &
    (\tilde{\tau}^{ij})^{\dot{a}}{}_{\dot{b}} &= \frac{1}{2} ( \tilde{\tau}^i \tau^j - \tilde{\tau}^j \tau^i )^{\dot{a}}{}_{\dot{b}} ,
  \end{aligned}
\end{equation}
so that the Lorentz generators take the form
\begin{equation}
  \hat{\gamma}^{\underline{ij}} = \begin{pmatrix} \gamma^{\underline{ij}} & 0 \\ 0 & \tilde{\gamma}^{\underline{ij}} \end{pmatrix} , \qquad
  \hat{\tau}^{ij} = \begin{pmatrix} \tau^{ij} & 0 \\ 0 & \tilde{\tau}^{ij} \end{pmatrix} .
\end{equation}
Some useful relations involving these gamma matrices are collected in appendix~\ref{sec:SO4-gamma-identities}.

In order to obtain compact expressions for the gauge-fixed action we find it necessary to perform a change of basis on the gamma matrices presented in appendix~\ref{sec:gamma-matrix-conventions}. These matrices are written as tensor products of five $2 \times 2$ matrices. Our change of basis takes the form
\begin{equation}
\label{eq:gamma-matrix-basis-change}
  m_1 \otimes  m_2 \otimes  m_3 \otimes  m_4 \otimes  m_5
  \to
  n_1 \otimes  n_2 \otimes  n_3 \otimes  n_4 \otimes  n_5
\end{equation}
with
\begin{equation}
 n_1= m_1 \,,\qquad
 n_2\otimes n_3=   P (m_3 \otimes  m_4) P^{-1}  \,,\qquad
 n_4\otimes n_5 =  m_2 \otimes  m_5 ,
\end{equation}
and 
\begin{equation}
  P = \begin{pmatrix} 0 & 1 & 0 & 0 \\ 0 & 0 & 1 & 0 \\ 0 & 0 & 0 & 1 \\ 1 & 0 & 0 & 0 \end{pmatrix} .
\end{equation}
With this change of basis,  $\algSO(4)_1$ and  $\algSO(4)_2$ act non-trivially only on $n_2\otimes n_3$ and $n_4\otimes n_5$, respectively, while the 9+1-dimensional Weyl projection acts only on $n_1$. The kappa gauge-fixed spinors satisfy
\begin{equation}
    \Gamma^{1234} \chi_I = +\chi_I \,,\qquad 
    \Gamma^{6789} \chi_I = +\chi_I \,,\qquad
    \Gamma^{1234} \eta_I = -\eta_I \,, \qquad
    \Gamma^{6789} \eta_I = -\eta_I \,.
\end{equation}
Since the action of $\Gamma^{1234}$ and $\Gamma^{6789}$ reduces to $\hat{\gamma}^{1234}$ and $\hat{\tau}^{6789}$ when acting on $\eta_I$ and $\chi_I$ we see that $\chi_I$ and $\eta_I$ carry indices
\begin{equation}
  ( \chi_I )^{\underline{a} b} , \qquad
  ( \eta_I )^{\underline{\dot{a}} \dot{b}} .
\end{equation}

Having introduced this notation we can now re-write $\lagr_{\text{\scriptsize kin}}$ in equation~\eqref{eq:lkingfix} as\footnote{In appendix~\ref{sec:relations-large-small-gamma-matrices} we summarise the relations between the $\Gamma^A$ and the $\hat{\gamma}^{\underline{i}}$ and $\hat{\tau}^i$ that are useful in obtaining the following expressions.
}
\begin{equation}
\label{eq:L-kin-bispinor}
  \begin{aligned}
    \lagr_{\text{\scriptsize kin}} = -2i\gamma^{\alpha\beta} \bigl[
    & \hat{E}_\alpha^+\bar{\eta}_1  \partial_\beta \eta_1 
    + \check{E}_\alpha^+\bar{\eta}_2 \partial_\beta \eta_2  
    +  \hat{E}_\alpha^+\bar{\chi}_1 \partial_\beta \chi_1 
    + \check{E}_\alpha^+  \bar{\chi}_2 \partial_\beta \chi_2 
\\
    &-  
\partial_\alpha x^+\bigl(\hat{E}_\beta^+ \bar{\eta}_1 \tilde{\gamma}^{34} \eta_1 
    + \check{E}_\beta^+ \bar{\eta}_2  \tilde{\gamma}^{34} \eta_2\bigr)  \\
    &-  \frac{1}{4} \bigl(  \sum_{A,B=0}^5 \hat{E}_\alpha^A \hat{E}_\beta^B \eta_{AB} + \dot{E}_\alpha^i \dot{E}_\beta^i  -4\hat{E}_\alpha^+ \partial_\beta x^-\bigr)\bar{\chi}_1 \gamma^{34} \chi_1 \\
    &+  \frac{1}{4} \bigl( \sum_{A,B=0}^5\check{E}_\alpha^A \check{E}_\beta^B \eta_{AB} +  \dot{E}_\alpha^i \dot{E}_\beta^i  -4 \check{E}_\alpha^+ \partial_\beta x^-\bigr)\bar{\chi}_2 \gamma^{34} \chi_2
    \bigr] .
  \end{aligned}
\end{equation}
Above, we have suppressed the spinor indices for compactness and defined
\begin{equation}
  \bar{\eta}_I \equiv (\eta_I)^{\underline{\dot{b}}\dot{b}}
  \epsilon_{\dot{\underline{b}}\dot{\underline{a}}} \epsilon_{\dot{b}\dot{a}} \,,
  \qquad
  \bar{\chi}_I \equiv (\chi_I)^{\underline{b}b}
  \epsilon_{\underline{ba}} \epsilon_{ba} \,.
\end{equation}
Re-writing $\lagr_{\text{WZ}}$ in equation~\eqref{eq:lwzgfix}  in terms of 
 $\algSO(4)_1$ and  $\algSO(4)_2$ bispinors one arrives at a longer expression which we have relegated to appendix~\ref{sec:Lwz-components}.

The above Lagrangian still depends on the worldsheet metric. As discussed above, one way to complete the light-cone gauge fixing is to go to first-order formalism and solve the Virasoro constraints. Alternatively we can impose the condition $p_-=2$ by solving for the worldsheet metric. Doing this we find that to the relevant order the metric is diagonal with components
\begin{equation}
  \begin{aligned}
    \gamma^{00} &= -1 + \frac{1}{2} ( z^2 - y^2 ) + \frac{1}{8} (z^2 + y^2) ( \dot{z}^2 + \pri{z}^2 + \dot{y}^2 + \pri{y}^2 - (z-y)^2 ) , \\
    \gamma^{11} &= +1 + \frac{1}{2} ( z^2 - y^2 ) + \frac{1}{8} (z^2 + y^2) ( \dot{z}^2 + \pri{z}^2 + \dot{y}^2 + \pri{y}^2 + (z-y)^2 ) .
  \end{aligned}
\end{equation}
The derivatives of the nondynamic field $x^-$ can then be found from the Virasoro constraints.

\subsection{Supercurrents}
\label{sec:charges}

In section~\ref{sec:action-before-lc} we wrote down an action which realised linearly all 16 supersymmetries of our background. However, half of the supervariations~\eqref{eq:susy-transformations} are incompatible with the BMN light-cone kappa gauge choice~\eqref{eq:bmn-lc-kappa-gauge}. This is a well known aspect of the light-cone gauge formalism~\cite{Green:1983wt}---it implies that such supervariations have to be combined with a compensating kappa transformation in order to preserve the gauge choice~\eqref{eq:bmn-lc-kappa-gauge}. The eight supercharges that commute with the Hamiltonian and form the fermionic part of $\mathcal{A}$ are associated with variations of precisely of this form.

Since kappa gauge transformations are known explicitly~\cite{Grisaru:1985fv}, it is in principal possible to find expression for such compensating kappa gauge transformations. The procedure is however computationally involved. To simplify matters, we will write down the supercurrents corresponding to the $\mathcal{A}$ supercharges to first order in fermions and third order in the transverse bosons. When computing the algebra $\mathcal{A}$ later in this section we will only need these expressions.

For notational convenience, we split the full supercurrent into parts involving only massless fields, only massive fields and a part involving a mix of massive and massless fields,
\begin{equation}\label{eq:total-current-tau}
  j^{\alpha}_I = j^{\alpha}_{I,\text{massless}} + j^\alpha_{I,\text{massive}} + j^{\alpha}_{I,\text{mixed}} , \qquad I = 1,2.
\end{equation}
The supercurrents are given by
\begingroup%
\abovedisplayshortskip=8pt plus 2pt minus 2pt%
\abovedisplayskip=8pt plus 2pt minus 2pt%
\belowdisplayshortskip=4pt plus 2pt minus 2pt%
\belowdisplayskip=4pt plus 2pt minus 2pt%
\begin{equation}\label{eq:massless-current-tau}
  \begin{aligned}
    j^{\tau}_{1,\text{massless}} =
    e^{+x^- \gamma^{34}} \Bigl( &
    \dot{x}^i \gamma^{34} \tilde{\tau}^i \chi_1 
    - \pri{x}^i \gamma^{34} \tilde{\tau}^i \chi_2 
    \Bigl) ,
    \\
    j^{\tau}_{2,\text{massless}} =
    e^{-x^- \gamma^{34}} \Bigl( &
    \dot{x}^i \gamma^{34} \tilde{\tau}^i \chi_2 
    - \pri{x}^i \gamma^{34} \tilde{\tau}^i \chi_1
    \Bigl) ,
  \end{aligned}
\end{equation}
\begin{equation}
  \begin{aligned}
    j^{\tau}_{1,\text{mixed}} =
    e^{+x^- \gamma^{34}} \Bigl( &
    -\tfrac{1}{2} ( z^2 - y^2 ) ( \dot{x}^i \gamma^{34} \tilde{\tau}^i \chi_1 + \pri{x}^i \gamma^{34} \tilde{\tau}^i \chi_2 )
    + z^{\underline{i}} y^{\underline{j}} \pri{x}^i  \gamma^{34} \gamma^{\underline{ij}}  \tilde{\tau}^i \chi_2
    \\
    & + \tfrac{1}{2} \dot{x} \cdot \pri{x} ( z^{\underline{i}} - y^{\underline{i}})  \gamma^{34} \gamma^{\underline{i}}  \eta_2
    + \tfrac{1}{4} ( \dot{x}^2 + \pri{x}^2 ) ( z^{\underline{i}} - y^{\underline{i}} )  \gamma^{34} \gamma^{\underline{i}}  \eta_1
    \Bigl) ,
    \\
    j^{\tau}_{2,\text{mixed}} =
    e^{-x^- \gamma^{34}} \Bigl( &
    -\tfrac{1}{2} ( z^2 - y^2 ) ( \dot{x}^i \gamma^{34} \tilde{\tau}^i \chi_2 + \pri{x}^i \gamma^{34} \tilde{\tau}^i \chi_1 )
    + z^{\underline{i}} y^{\underline{j}} \pri{x}^i  \gamma^{34} \gamma^{\underline{ij}}  \tilde{\tau}^i \chi_1
    \\
    & - \tfrac{1}{2} \dot{x} \cdot \pri{x} ( z^{\underline{i}} - y^{\underline{i}}) \gamma^{34} \gamma^{\underline{i}}  \eta_1
    - \tfrac{1}{4} ( \dot{x}^2 + \pri{x}^2 ) ( z^{\underline{i}} - y^{\underline{i}} )  \gamma^{34} \gamma^{\underline{i}}  \eta_2
    \Bigl) ,
  \end{aligned}
\end{equation}
\begin{equation}
  \begin{aligned}
    j^{\tau}_{1,\text{massive}} =
    e^{+x^- \gamma^{34}} \Bigl( &
    (\dot{z}^{\underline{i}} - \dot{y}^{\underline{i}}) \gamma^{\underline{i}} \eta_1 + (z^{\underline{i}} + y^{\underline{i}}) \gamma^{34} \gamma^{\underline{i}} \eta_1 - (\pri{z}^{\underline{i}} - \pri{y}^{\underline{i}}) \gamma^{\underline{i}} \eta_2
    \Bigr) ,
    \\
    j^{\tau}_{2,\text{massive}} =
    e^{-x^- \gamma^{34}} \Bigl( &
    (\dot{z}^{\underline{i}} - \dot{y}^{\underline{i}}) \gamma^{\underline{i}} \eta_2 - (z^{\underline{i}} + y^{\underline{i}}) \gamma^{34} \gamma^{\underline{i}} \eta_2 - (\pri{z}^{\underline{i}} - \pri{y}^{\underline{i}}) \gamma^{\underline{i}} \eta_1
    \Bigr) .
  \end{aligned}
\end{equation}
\begin{equation}\label{eq:massless-current-sigma}
  \begin{aligned}
    j^{\sigma}_{1,\text{massless}} =
    - e^{+x^- \gamma^{34}} \Bigl( &
    \pri{x}^i \gamma^{34} \tilde{\tau}^i \chi_1 - \dot{x}^i \gamma^{34} \tilde{\tau}^i \chi_2 
    \Bigl) ,
    \\
    j^{\sigma}_{2,\text{massless}} =
    - e^{-x^- \gamma^{34}} \Bigl( &
    \pri{x}^i \gamma^{34} \tilde{\tau}^i \chi_2 - \dot{x}^i \gamma^{34} \tilde{\tau}^i \chi_1
    \Bigl) ,
  \end{aligned}
\end{equation}
\begin{equation}
  \begin{aligned}
    j^{\sigma}_{1,\text{mixed}} =
    -e^{+x^- \gamma^{34}} \Bigl( &
    +\tfrac{1}{2} ( z^2 - y^2 ) ( \pri{x}^i \gamma^{34} \tilde{\tau}^i \chi_1 - \dot{x}^i \gamma^{34} \tilde{\tau}^i \chi_2 )
    + z^{\underline{i}} y^{\underline{j}} \dot{x}^i  \gamma^{34} \gamma^{\underline{ij}}  \tilde{\tau}^i \chi_2
    \\
    & + \tfrac{1}{2} \dot{x} \cdot \pri{x} ( z^{\underline{i}} - y^{\underline{i}})  \gamma^{34} \gamma^{\underline{i}} \eta_1
    + \tfrac{1}{4} ( \dot{x}^2 + \pri{x}^2 ) ( z^{\underline{i}} - y^{\underline{i}} )  \gamma^{34} \gamma^{\underline{i}}  \eta_2
    \Bigl) ,
    \\
    j^{\sigma}_{2,\text{mixed}} =
    -e^{-x^- \gamma^{34}} \Bigl( &
    +\tfrac{1}{2} ( z^2 - y^2 ) ( \pri{x}^i \gamma^{34} \tilde{\tau}^i \chi_2 - \dot{x}^i \gamma^{34} \tilde{\tau}^i \chi_1 )
    + z^{\underline{i}} y^{\underline{j}} \dot{x}^{\underline{i}}  \gamma^{34} \gamma^{\underline{ij}}  \tilde{\tau}^i \chi_1
    \\
    & - \tfrac{1}{2} \dot{x} \cdot \pri{x} ( z^{\underline{i}} - y^{\underline{i}})  \gamma^{34} \gamma^{\underline{i}}  \eta_2
    - \tfrac{1}{4} ( \dot{x}^2 + \pri{x}^2 ) ( z^{\underline{i}} - y^{\underline{i}} )  \gamma^{34} \gamma^{\underline{i}}  \eta_1
    \Bigl) ,
  \end{aligned}
\end{equation}
\begin{equation}
  \begin{aligned}
    j^{\sigma}_{1,\text{massive}} =
    -e^{+x^- \gamma^{34}} \Bigl( &
    (\pri{z}^{\underline{i}} - \pri{y}^{\underline{i}}) \gamma^{\underline{i}} \eta_1 - (\dot{z}^{\underline{i}} - \dot{y}^{\underline{i}}) \gamma^{\underline{i}} \eta_2 - (z^{\underline{i}} + y^{\underline{i}}) \gamma^{34} \gamma^{\underline{i}} \eta_2
    \Bigr) ,
    \\
    j^{\sigma}_{2,\text{massive}} =
    -e^{-x^- \gamma^{34}} \Bigl( &
    (\pri{z}^{\underline{i}} - \pri{y}^{\underline{i}}) \gamma^{\underline{i}} \eta_2 - (\dot{z}^{\underline{i}} - \dot{y}^{\underline{i}}) \gamma^{\underline{i}} \eta_1 + (z^{\underline{i}} + y^{\underline{i}}) \gamma^{34} \gamma^{\underline{i}} \eta_1
    \Bigr) .
  \end{aligned}
\end{equation}%
\endgroup%
\vspace{8pt}

\noindent Above, for the massive part of the supercurrents we have only written down the lowest-order-in-bosons expression since it will be all we need later on. For compactness we have also suppressed all spinor indices; re-instating these  we have, for example,
\begin{equation}
\gamma^{34}{\tilde\tau}^i\chi_1\equiv 
(\gamma^{34})^{\underline{a}}{}_{\underline{b}}({\tilde\tau}^i)^{\dot{a}}{}_b(\chi_1)^{\underline{b}b}\,.
\end{equation}
 Using the equations of motion, which are presented in appendix~\ref{sec:eoms}, we have checked that the currents satisfy the equation
\begin{equation}
\partial_\tau j_I^\tau+\partial_\sigma j_I^\sigma=0\,,
\end{equation}
and hence are conserved.

\subsection{The \texorpdfstring{$\mathcal{A}$}{A} algebra}
\label{sec:A-computation}

In this sub-section we will compute the algebra $\mathcal{A}$. Our computation will be done in a field expansion discussed below. In particular, we will work to leading order in fermions and sub-leading order in bosons. This is the same order to which the corresponding algebra was computed for Type IIB strings on $\AdS_5\times \Sphere^5$~\cite{Arutyunov:2006ak}. Before describing the details of the computations, let us pause briefly to make two general observations. 

Firstly, on-shell $\mathcal{A}$ reduces to $\algPSU(1|1)^4$ extended by the Hamiltonian and a central angular momentum and by the torus isometries. This is simply the part of superisometries of the classical string theory on $\AdS_3\times \Sphere^3\times \Torus^4$ that commutes with the Hamiltonian, and amounts to $\algPSU(1|1)^4\oplus\algU(1)^2\oplus\algSO(4)$. The important consistency check then is to see that when going off-shell, by relaxing the level-matching condition, the algebra becomes centrally extended in just the right way. In other words, the Poisson bracket between two different supercharges should result in an expression of the form
\begin{equation}
  \label{eq:central-extension-part-of-A}
  \acommPB{(Q_1)^{\underline{a}\dot{a}}}{(Q_2)^{\underline{b}\dot{b}}} = -i \, C^{\underline{ab},\dot{a}\dot{b}} ,
\end{equation}
where the matrix on the right-hand side can be decomposed into the two central charges extending the symmetry algebra.

Secondly, we note that the massive ($y_{\underline{i}}$, $z_{\underline{i}}$ and $\eta_I$) and the massless fields ($x_i$ and $\chi_I$) each form a consistent closed sector of the equations of motion of the theory. In the classical theory the massive sector is isomorphic to a closed sub-sector of the  Type IIB string on $\AdS_5\times \Sphere^5$.\footnote{\label{ft:massive-boring}This is a consequence of the fact that the kappa gauge-fixed massive sector of Type IIB strings on $\AdS_3\times \Sphere^3\times \Torus^4$ can be described in terms of a super-coset. In turn it is easy to see that that super-coset is a sub-super-coset of the kappa gauge-fixed Type IIB strings on $\AdS_5\times \Sphere^5$.} As a result, the off-shell computation of $\mathcal{A}$ in the massive sub-sector is identical to the computation performed in~\cite{Arutyunov:2006ak} and so we will not repeat it here. Instead, we will perform two types of computations that are new to Type IIB strings on $\AdS_3\times \Sphere^3\times \Torus^4$. In section~\ref{sec:massless-A-calculation} we restrict to the massless sector of the theory and compute the off-shell algebra $\mathcal{A}$; as anticipated in the previous paragraph, we explicitly see that on-shell the algebra does indeed reduce to $\algPSU(1|1)^4\oplus \algU(1)^2$. In section~\ref{sec:mixed-A-calculation}, we compute, off-shell, in the full massive and massless theory the relation~\eqref{eq:central-extension-part-of-A}.

We only determine the part of the central charges that depend on the bosonic fields. Since the central charges have to vanish for zero total worldsheet momentum this is enough to reconstruct the full charges.
As we will see, the momentum dependence of the central charges comes in through the nonlocal and nondynamic field $x^-$. To capture this dependence we employ a ``hybrid'' expansion similar to what was used in $\AdS_5 \times \Sphere^5$ in~\cite{Arutyunov:2006ak}. This means that we expand the action in the transverse fields to quadratic order in fermions and quartic order in the transverse bosons, but keep any explicit factors of $x^-$ unexpanded. This allows us to capture the full momentum dependence of the central charges. It is worth noting at this point that central extensions of the on-shell algebra of this kind had been studied for the plane-wave limit of $\AdS_3\times \Sphere^3\times \Torus^4$ in~\cite{Gava:2002xb}.

The fermions in the Lagrangian~\eqref{eq:L-kin-bispinor} and~\eqref{eq:L-wz-bispinor} do not have a canonical kinetic term and so will not have a conventional Poisson bracket. It is possible to further redefine the fermionic fields order by order in the field expansion to correct this. However, for our purpose it will be simpler to work with  the non-canonical Poisson bracket for the fermions that follows from the Lagrangian~\eqref{eq:L-kin-bispinor} and~\eqref{eq:L-wz-bispinor}. The Poisson bracket of the fermions $\eta_I$ and $\chi_I$ is presented below.

\subsubsection{The massless sub-sector}
\label{sec:massless-A-calculation}

As we have noted above, the equations of motion of our system are such that it is consistent to set the transverse massive excitations to zero. In this sub-section we focus on computing $\mathcal{A}$ in the purely massless sector. This sector turns out to have a number of simplifying features compared to the full theory and so serves as a good warm-up exercise. We will therefore repeat some of the steps discussed above in more detail. What is more, this sector is \emph{not} described by a semi-symmetric space coset and so understanding how it enters the integrable machinery is one of the central results of this paper. 
The massless part of the gauge fixed Lagrangian is given by
\begin{equation}
  \begin{gathered}
    \lagr_B^{(m)} = -\frac{1}{2} \gamma^{\alpha\beta} \Bigl(
    4 \partial_\alpha x^+ \partial_\beta x^- + \partial_\alpha x^i \partial_\beta x^i
    \Bigr)
    \\
    \lagr_{\text{\scriptsize kin}}^{(m)} = - 2i \gamma^{\alpha\beta} \Bigl(
    \bar{\chi}_I \partial_\alpha  \chi_I  \partial_\beta x^+
    - \frac{1}{4}\sigma^3_{IJ} \bar{\chi}_I  \gamma^{34} \chi_J  \partial_\alpha x^i \partial_\beta x^i
    \Bigr) ,
    \\
    \lagr_{\text{WZ}}^{(m)} = - 2i \epsilon^{\alpha\beta} \Bigl(\sigma^1_{IJ}
    \bar{\chi}_I  \partial_\alpha \chi_J  \partial_\beta x^+
    +\epsilon_{IJ} \frac{1}{4}  \bar{\chi}_I\gamma^{34}  \tau^{ij} \chi_J  \partial_\alpha x^i \partial_\beta x^j 
    \Bigr) .
  \end{gathered}
\end{equation}
The massless parts of the supercurrents were given in equations~\eqref{eq:massless-current-tau} and~\eqref{eq:massless-current-sigma}.
Notice that to this order in fermions the supercurrents do not contain a term cubic in the bosons, and are in fact the same as they would be in flat space.\footnote{%
 We expect this to be only true because we are working to lowest order in fermions. There does not appear to be an obvious reason for the term quartic in fermions in the action to vanish.  Such a term would likely lead to corrections to the supercurrents that are cubic in fermions. Such terms are however absent in the free theory.%
} %
The non-linear terms in the equations of motion for the fermions are exactly cancelled by the nonlocal exponential part of the supercurrents.

We now want to calculate the algebra $\mathcal{A}$ obtained by taking Poisson brackets between the supercharges obtained from the currents. To do this we write the action in first-order formalism. The conjugate momenta are of the bosonic fields are given by
\begin{equation}
  \begin{aligned}
    p_-
    = \frac{\delta\action}{\delta \dot{x}^-}
    &=
    - 2\gamma^{0\beta} \partial_\beta x^+,
    \\
    p_+
    = \frac{\delta\action}{\delta \dot{x}^+}
    &= - 2\gamma^{0\beta} \partial_\beta x^-
    - 2i \gamma^{0\beta} \bar{\chi}_I \partial_\beta \chi_I
    + 2i\sigma^1_{IJ}  \bar{\chi}_I  \pri{\chi}_J
    ,
    \\
    p_i
    = \frac{\delta\action}{\delta \dot{x}^i}
    &=
    - \gamma^{0\beta} \partial_\beta x^i
    + i \gamma^{0\beta} \sigma^3_{IJ} \bar{\chi}_I\gamma^{34} \chi_J
    \partial_\beta x^i
    - i \epsilon_{IJ} \bar{\chi}_I \gamma^{34} \tau^{ij} \chi_J
    \pri{x}^j .
  \end{aligned}
\end{equation}
Inserting this into the Lagrangian we get
\begin{equation}
  \lagr^{(m)} = p_- \dot{x}^- + p_i \dot{x}^i + i p_- 
  \bar{\chi}_I \dot{\chi}_I
  + \frac{\gamma^{01}}{\gamma^{00}} C_1 + \frac{1}{2\gamma^{00}} C_2 ,
\end{equation}
with the constraints given by
\begin{equation}
  \begin{gathered}
    C_1 = p_- \pri{x}^- + p_i \pri{x}^i + i p_- \bar{\chi}_I\pri{\chi}_I ,
    \\
    C_2 =
    p_+ p_- + p_i p_i + \pri{x}^2
    - 2 i p_ -  
    \sigma^1_{IJ}\bar{\chi}_I \pri{\chi}_J
    + i \sigma^3_{IJ}
    \bar{\chi}_I \gamma^{34}\chi_J
    ( p_i p_i - \pri{x}^2 )
    \\ \qquad\qquad
    + 2i \epsilon_{IJ}
    \bar{\chi}_I\gamma^{34} \tau^{ij}\chi_J
    p_i \pri{x}_j .
  \end{gathered}
\end{equation}
By solving the constraint $C_1 = 0$ we obtain
\begin{equation}
  \pri{x}^- = 
  -\frac{1}{p_-} p_i \pri{x}_i 
  - i\bar{\chi}_I\pri{\chi}_I
  \,.
\end{equation}
Up to quadratic order in the fermions the massless Hamiltonian density  $\mathcal{H}^{(m)} = -p_+$ is given by
\begin{equation}
  \begin{split}
    \mathcal{H}^{(m)} =
    \frac{1}{p_-} \bigl(&p_i p_i +  \pri{x}^2 
    - 2 i p_-\sigma^1_{IJ}\bar{\chi}_I\pri{\chi}_J
    \\
    &+ 
    i\sigma^3_{IJ} 
    \bar{\chi}_I \gamma^{34}\chi_J
    ( p_i p_i - \pri{X}^2 )
    + 
    2i\epsilon_{IJ}
    \bar{\chi}_I\gamma^{34} \tau^{ij}\chi_J
    p_i \pri{x}_j \bigr)\,.
  \end{split}
\end{equation}
From the kinetic term of the action we can read off the (canonical) Poisson brackets
\begin{equation}
  \begin{aligned}
    \commPB{x^i(\sigma)}{p_j(\sigma')} &= \delta^i_j \delta(\sigma - \sigma') , \\
    \acommPB{(\chi_I)^{\underline{a}a}(\sigma)}{(\chi_J)^{\underline{b}b}(\sigma')} &= -\frac{i}{2p_-} \epsilon^{\underline{ab}} \epsilon^{ab} \delta(\sigma - \sigma') \,.
  \end{aligned}
\end{equation}
Using results from appendix~\ref{app:mless-pbs} one can check that the supercharge densities
\begin{align}
  \mathcal{Q}_1 &= 
  e^{+\gamma^{34} x^-} \bigl( p_i \tilde{\tau}^i \chi_1 - \pri{x}^i \tilde{\tau}^i \chi_2 \bigr) \equiv 
  e^{+\gamma^{34}x^-}  \mathcal{Q}_1^{(2)}
\,,\\
  \mathcal{Q}_2 &= e^{-\gamma^{34} x^-} \bigl( \pri{x}^i \tilde{\tau}^i \chi_1 - p_i \tilde{\tau}^i \chi_2 \bigr)\equiv 
  e^{-\gamma^{34} x^-}  \mathcal{Q}_2^{(2)}
\end{align}
both lead to conserved charges
\begin{equation}
  \label{eq:conc-suchg-massless}
  \commPB{Q_1}{\ham^{(m)}} = \commPB{Q_2}{\ham^{(m)}} = 0\,.
\end{equation}
Next we note that up to integrating by parts\footnote{Here it is useful to note the identities~\eqref{eq:useful-taus}.}
\begin{equation}
  \acommPB{ (\mathcal{Q}_1^{(2)})^{\underline{a}\dot{a}} }{ (\mathcal{Q}_1^{(2)})^{\underline{b}\dot{b}} }
  =
  + \frac{i}{2 p_-} \epsilon^{\underline{ab}} \epsilon^{\dot{a}\dot{b}} \Bigl( p_i p_i + \pri{x}^2
  - 2 i p_- \sigma^1_{IJ}
  \bar{\chi}_I\pri{\chi}_J
  \Bigr) \,.
\end{equation}
The last term in the expression in the bracket is exactly the quadratic Hamiltonian in the massless sector. So up to quadratic order in excitations we find
\begin{equation}
  \begin{aligned}
    \acommPB{ (Q_1)_{\underline{a}\dot{a}} }{ (Q_1)_{\underline{b}\dot{b}} }
    &=
    + \frac{i}{2}
    \int_{-\infty}^{+\infty} d\sigma
    \bigl( e^{+\gamma^{34} x^-} \bigr)^{\underline{a}}{}_{\underline{c}} \bigl(e^{+\gamma^{34} x^-} \bigr)^{\underline{b}}{}_{\underline{d}}
    \epsilon^{\underline{cd}} \epsilon^{\dot{a}\dot{b}} \mathcal{H}^{(m)}
    \\
    &=
    + \frac{i}{2} \epsilon^{\underline{ab}} \epsilon^{\dot{a}\dot{b}} \ham^{(m)} \,.
  \end{aligned}
\end{equation}

Similarly we can calculate the commutator between two different supercharges
\begin{equation}
  \begin{aligned}
    \acommPB{ (\mathcal{Q}_1^{(2)})^{\underline{a}\dot{a}} }{ (\mathcal{Q}_2^{(2)})^{\underline{b}\dot{b}} }
    &=
    \biggl( \frac{i}{p_-} p_i \pri{x}^i 
    - \bar{\chi}_I \pri{\chi}_I 
    \biggr) \epsilon^{\underline{ab}} \epsilon^{\dot{a}\dot{b}}
    \\
    &= - i \pri{x}^- \epsilon^{\underline{ab}} \epsilon^{\dot{a}\dot{b}} .
  \end{aligned}
\end{equation}
At quadratic order we then find
\begin{equation}
  \begin{aligned}
    \acommPB{ (Q_1)^{\underline{a}\dot{a}} }{ (Q_2)^{\underline{b}\dot{b}} }
    &=
    -i \int_{-\infty}^{+\infty} d\sigma \,
    \bigl( e^{+\gamma^{34} x^-} \bigr)^{\underline{a}}{}_{\underline{c}} \bigl( e^{-\gamma^{34} x^-} \bigr)^{\underline{b}}{}_{\underline{d}}
    \epsilon^{cd} \epsilon^{\dot{a}\dot{b}} \, \pri{x}^- \\
    &=
    - \frac{i}{2}
    \int_{-\infty}^{+\infty} d\sigma \, \partial_{\sigma} \bigl( e^{+2 \gamma^{34} x^- } \gamma^{34} \epsilon \bigr)^{\underline{ab}} \, \epsilon^{\dot{a}\dot{b}}  \\
    &=
    - \frac{i}{2} 
    \bigl( e^{+2 \gamma^{34} x^-(+\infty) } - e^{+2 \gamma^{34} x^-(-\infty) } \bigr)^{\underline{a}}{}_{\underline{c}} ( \gamma^{34} \epsilon )^{\underline{cb}} \, \epsilon^{\dot{a}\dot{b}} \\
    &=
    - \frac{i}{2} \bigl( e^{+2 \gamma^{34} x^-(-\infty)} \bigr)^{\underline{a}}{}_{\underline{c}}
    \bigl( e^{+2 \gamma^{34} \Delta x^-} - 1 \bigr)^{\underline{c}}{}_{\underline{d}} ( \gamma^{34} \epsilon )^{\underline{db}} \, \epsilon^{\dot{a}\dot{b}} \\
    &=
    - \frac{i}{2} \bigl( e^{+2 \gamma^{34} x^-(-\infty)} \bigr)^{\underline{a}}{}_{\underline{c}} 
    \bigl( e^{+\frac{2}{p_-} \gamma^{34} p_{\text{ws}}} - 1 \bigr)^{\underline{c}}{}_{\underline{d}} ( \gamma^{34} \epsilon )^{\underline{db}} \, \epsilon^{\dot{a}\dot{b}} .
  \end{aligned}
\end{equation}
Hence, the central charge takes the form\footnote{The Poisson bracket between the charges $Q_1$ and $Q_2$ contains both the charge $C$ and its conjugate $\overline{C}$. In section~\ref{sec:algebra} we will write the full algebra $\mathcal{A}$ in a more convenient form. Here we report the expression corresponding to the charge $C$.} 
\begin{equation}
  C = \frac{i\zeta}{2} (e^{+i p_{\text{ws}}} - 1) ,
\end{equation}
with $\zeta = \exp(+2 i x^-(-\infty))$. This is exactly the form found in~\cite{Arutyunov:2006ak} for $\AdS_5\times\Sphere^5 $, which as we argued coincides with what we must have in our massive sector, as can be seen by an appropriate truncation of the supercoset~\cite{Sfondrini:2014via}. 

To summarize, in this sub-section we have worked in the massless subsector of the full string theory. We have constructed the supercharges and hamiltonian of the theory in the first-order formalism and have shown that they satisfy the commutation relations of $\mathcal{A}$. We have also found that in the off-shell theory the central extension $C$ takes precisely the form expected for $\mathcal{A}$.

\subsubsection{Fermionic Poisson brackets}
\label{sec:fermion-poisson}

Having found the central charges of $\mathcal{A}$ in the massless sector we now want to perform the same calculation again but now including both massive and massless fields.
To the order that we will be working in, we only need the dependence of the central charges on the bosonic fields. As explained below, we will in fact only need to consider terms up to quadratic order in both the massless and massive fields, so that the only bosonic quartic terms that we will be interested in contain fields of both masses. Each term in the supercharges contains at least one fermionic field. Since we are only interested in the bosonic field dependence of the central charge, we only need the contribution from the Poisson bracket of two supercharges that comes from the Poisson bracket between two fermions. Any other term will be higher order in fermions.

Since we only need to calculate Poisson brackets between fermionic fields, we do not need to introduce canonical momenta for the bosons. The kinetic terms for the fermions is quite complicated which leads to an involved Poisson structure. We relegate the calculation to appendix~\ref{sec:app-poisson-b-fermions} and simply state the non-zero Poisson brackets here
\begin{align}
  \label{eq:pb-between-etas-and-chis}
  \acommPB{\eta_1}{\eta_1} &= - \frac{i}{4} ( 1 + A_{11} ) \epsilon \epsilon \,,\quad &
  \acommPB{\eta_1}{\eta_2} &= - \frac{i}{4} A_{12}  \epsilon \epsilon \,,\quad &
  \acommPB{\eta_1}{\chi_2} &= - \frac{i}{4}  A_{14}  \epsilon \epsilon  \nonumber
  \\
  \acommPB{\eta_2}{\eta_2} &= - \frac{i}{4} ( 1 + A_{22} ) \epsilon \epsilon \,,\quad &
  \acommPB{\eta_2}{\chi_1} &= - \frac{i}{4} A_{23}  \epsilon \epsilon \,,\quad &
  \nonumber
  \\
  \acommPB{\chi_1}{\chi_1} &= - \frac{i}{4}  (1+A_{33})  \epsilon \epsilon \,,\quad &
  \acommPB{\chi_1}{\chi_2} &= - \frac{i}{4}  A_{34}  \epsilon \epsilon \,,\quad &
  \nonumber 
  \\ 
  \acommPB{\chi_2}{\chi_2} &= - \frac{i}{4}  (1+A_{44})  \epsilon \epsilon \,.
\end{align}
The \emph{bi-spinor valued} matrices $A_{ij}$ are given in equation~\eqref{eq:PB-A-coeffs} and we have suppressed the bispinor indices in the above so as not to over-clutter the notation.\footnote{The reader should note that the epsilon symbols appearing in the Poisson brackets carry different kinds of indices depending on which particular fermions' Poisson bracket one is computing.}%

\subsubsection{Computing the central charge \texorpdfstring{$C$}{C} in the full theory}
\label{sec:mixed-A-calculation}

To establish the off-shell symmetry algebra of Type IIB
string theory on $\AdS_3\times \Sphere^3\times \Torus^4$ we need to check whether the commutation relation~\eqref{eq:central-extension-part-of-A} holds \emph{in the full theory}.
Above, we have demonstrated such a relation in the massless sector of the theory (that is, when massive fields are turned off). Since, as we argued, a similar calculation for the massive sector follows directly from~\cite{Arutyunov:2006ak}, all that we need to worry about now  are the mixed-mass terms.
In this sub-section, we will indeed establish~\eqref{eq:central-extension-part-of-A} by taking into account mixed-mass terms the supercharges expanded to linear order in fermions and cubic order in bosons. To this order we will be showing that such a relation holds with the central charge $C$ taken to zeroth order in fermions and quartic order in bosons.


Using the Poisson brackets given in equation~\eqref{eq:pb-between-etas-and-chis}, we find the Poisson bracket between two supercharges
\begin{equation}
\label{eq:pb-of-sucharges-for-centr-ext}
  \begin{aligned}
    \int d\sigma \, d\sigma' \, \acommPB{j_1^\tau(\sigma)}{j_2^\tau(\sigma')}
    =&
    -\frac{i}{p_-} \int d\sigma \, e^{+2\gamma^{34} x^-} \bigl(
    ( \dot{z} \cdot \pri{z} + \dot{y} \cdot \pri{y} + \dot{x} \cdot \pri{x} ) \epsilon \epsilon
    \\ & \phantom{{}-\frac{i}{p_-} \int d\sigma \, e^{+2\gamma^{34} x^-} \bigl( } 
    \mathllap{+} ( z \cdot \pri{z} - y \cdot \pri{y} ) \gamma^{34} \epsilon \epsilon
    \bigr)
    \\ &
    + \frac{i}{2} \int d\sigma \, ( z^{\underline{i}} \pri{y}^{\underline{j}} + \pri{z}^{\underline{i}} y^{\underline{j}} ) \gamma^{34} \gamma^{\underline{ij}} \epsilon\epsilon\, .
  \end{aligned}
\end{equation}
The details of the calculation are given in appendix~\ref{app:der-of-centr-ext-eq}. The last line above is a total derivative and can be dropped with a suitable choice of boundary conditions. By partially integrating the second line and again dropping the total derivative we find that the remaining integrand takes the form
\begin{equation}
  \dot{z} \cdot \pri{z} + \dot{y} \cdot \pri{y} + \dot{x} \cdot \pri{x} + (z^2 - y^2) \pri{x}^-
  =
  \dot{z} \cdot \pri{z} + \dot{y} \cdot \pri{y} + \dot{x} \cdot \pri{x} - \frac{1}{2} (z^2 - y^2) ( \dot{x} \cdot \pri{x} ) ,
\end{equation}
up to terms that are quartic in the massive fields. The last expression is equal to $-p_-\pri{x}^-$ and so we may write
\begin{equation}
  \begin{aligned}
    \int d\sigma \, d\sigma' \, \acommPB{j_1^\tau(\sigma)}{j_2^\tau(\sigma')}
    =&
    +i \int d\sigma \, e^{+2\gamma^{34} x^-} \, \pri{x}^- \epsilon\epsilon
    \\
    =&
    -\frac{i}{2} \int d\sigma \, \partial_\sigma \bigl( e^{+2\gamma^{34} x^-} \bigr) \gamma^{34} \epsilon\epsilon
    \\
    =&
    -\frac{i}{2} \bigl( e^{+2\gamma^{34} x^-(+\infty)} - e^{+2\gamma^{34} x^-(-\infty)} \bigr) \gamma^{34} \epsilon\epsilon
    \\
    =&
    -\frac{i}{2} e^{+2\gamma^{34} x^-(-\infty)} \bigl( e^{+2\gamma^{34} \Delta x^-} - 1 \bigr) \gamma^{34} \epsilon\epsilon
    \\
    =&
    -\frac{i}{2} e^{+2\gamma^{34} x^-(-\infty)} \bigl( e^{+\gamma^{34} p_{\text{ws}}} - 1 \bigr) \gamma^{34} \epsilon\epsilon\,.
  \end{aligned}
\end{equation}
Hence, the central charge takes the form
\begin{equation}
\label{eq:central-charge-ST}
  C = \frac{i\zeta}{2} (e^{+i p_{\text{ws}}} - 1) ,
\end{equation}
with $\zeta = \exp(+2 i x^-(-\infty))$, in agreement with the expression found the previous sub-section.

\section{Symmetry algebra}
\label{sec:algebra}
In the previous section we have found the off-shell symmetry algebra ${\cal A}$ for type IIB superstrings on $\AdS_3\times\Sphere^3\times\Torus^4$ . We showed that~$\mathcal{A}$ is given by a central extension of $\alg{psu}(1|1)^4\oplus\alg{so}(4)_{2}$, where $\alg{so}(4)_{2}$ comes from the torus coordinates.\footnote{%
As discussed at the end of section~\ref{sec:first-order-uniform-lc}, such~$\alg{so}(4)_{2}$ is unbroken as long as we are in the decompactification limit~$P_-=\infty$.
} 

In this section we will first review how this algebra can be constructed by tensoring two copies of~$\alg{su}(1|1)^2_{\ce}$. Then, in subsection~\ref{sec:near-BMN-representations} we will investigate the representations of ${\cal A}$ in the near-plane wave limit. This can be read-off from the supercurrents obtained in section~\ref{sec:charges} but we collect the results here to set up the notation and conventions that we will use in later sections.

As a preliminary step, it is convenient to rewrite~$\mathcal{A}$ in components, using the notation introduced in appendix~\ref{app:algebra}. We then find the anti-commutation relations take the form
\begin{equation}
\label{eq:cealgebra}
\begin{aligned}
&\{\gen{Q}_{\smallL}^{\ \dot{a}},\overline{\gen{Q}}{}_{\smallL \dot{b}}\} =\frac{1}{2}\delta^{\dot{a}}_{\ \dot{b}}\,(\gen{H}+\gen{M}),
&\qquad &\{\gen{Q}_{\smallL}^{\ \dot{a}},{\gen{Q}}{}_{\smallR \dot{b}}\} =\delta^{\dot{a}}_{\ \dot{b}}\,\gen{C},
\\
&\{\gen{Q}_{\smallR  \dot{a}},\overline{\gen{Q}}{}_{\smallR}^{\ \dot{b}}\} =\frac{1}{2}\delta^{\ \dot{b}}_{\dot{a}}\,(\gen{H}-\gen{M}),
&\qquad &\{\overline{\gen{Q}}{}_{\smallL \dot{a}},\overline{\gen{Q}}{}_{\smallR}^{\ \dot{b}}\} =\delta^{\ \dot{b}}_{\dot{a}}\,\overline{\gen{C}},
\end{aligned}
\end{equation}
where we introduced labels~``L'' and~``R'' (left and right) for the supercharges in~$\alg{psu}(1|1)^4$. These are inherited from the superisometry algebra $\alg{su}(1,1|2)_{\sL}\oplus\alg{su}(1,1|2)_{\sR} $, where they refer to the chirality in the dual~$\CFT_2 $.
Note that in the leading-order expansion of appendix~\ref{app:algebra}, the central charges~$\gen{C},\overline{\gen{C}}$ were linear functions of the worldsheet momentum, $\gen{C}=\overline{\gen{C}}=-\frac{1}{2}\gen{P}$. This is indeed the leading order term in the expansion of the non-linear relation
\begin{equation}
\label{eq:allloop-centralcharges}
\gen{C}=+\frac{i\zeta}{2}(e^{+i\gen{P}}-1),
\qquad\qquad
\overline{\gen{C}}=-\frac{i\bar{\zeta}}{2}(e^{-i\gen{P}}-1),
\end{equation}
found in the previous section, \cf~\eqref{eq:central-charge-ST}. The $\alg{so}(4)_2$ subalgebra arising from the torus directions can be decomposed into~$\alg{su}(2)_{\bullet}\oplus\alg{su}(2)_{\circ}$, satisfying
\begin{equation}
\comm{{\gen{J}_{\bullet \dot{a}}}^{\dot{b}}
}{
{\gen{J}_{\bullet \dot{c}}}^{\dot{d}}}
 = 
\delta^{\dot{b}}_{\ \dot{c}}\, {\gen{J}_{\bullet \dot{a}}}^{\dot{d}}
-
\delta^{\dot{d}}_{\ \dot{a}}\, {\gen{J}_{\bullet \dot{c}}}^{\dot{b}},
\qquad
\comm{{\gen{J}_{\circ a}}^b}{{\gen{J}_{\circ c}}^d} = \delta^b_{\ c}\, {\gen{J}_{\circ a}}^d
-
\delta^d_{\ a}\, {\gen{J}_{\circ c}}^b, 
\end{equation}
The supercharges $\gen{Q}_{j\,\smallL,\smallR}$ are in the fundamental representation of $\alg{su}(2)_{\bullet}$. Indices are therefore raised and lowered by the antisymmetric tensor~$\epsilon^{\dot{a}\dot{b}}$ and its inverse, so that charges with upper indices transform in the anti-fundamental representation. We then have
\begin{equation}
\comm{{\gen{J}_{\bullet {\dot{a}}}}^{\dot{b}}}{\gen{Q}_{\dot{c}}} = \delta^{\dot{b}}_{\ {\dot{c}}} \gen{Q}_{\dot{a}} - \frac{1}{2} \delta^{\ {\dot{b}}}_{\dot{a}} \gen{Q}_{\dot{c}},
\qquad
\comm{{\gen{J}_{\bullet {\dot{a}}}}^{\dot{b}}}{\gen{Q}^{\dot{c}}} = -\delta^{\ {\dot{c}}}_{\dot{a}} \gen{Q}^{\dot{b}} + \frac{1}{2} \delta^{\ {\dot{b}}}_{\dot{a}} \gen{Q}^{\dot{c}}, 
\end{equation}
where $\gen{Q}$ is any supercharge in the appropriate representation.
All of the generators of the centrally extended~$\alg{psu}(1|1)^4$ commute with $\alg{su}(2)_{\circ}$.
The $\alg{u}(1)$ charges of ${\cal A}$ are therefore given by the Hamiltonian~$\gen{H}$, the angular momentum~$\gen{M}$, two  Cartan elements coming from the two~$\alg{su}(2)$'s and the central elements $\gen{C},\overline{\gen{C}}$.

The centrally extended $\alg{psu}(1|1)^4$ superalgebra appeared already in the study of the massive sector of the theory in~\cite{Borsato:2013qpa}, and as discussed there it could be obtained from two copies of the centrally extended $\alg{su}(1|1)^2$. In the next subsection we briefly review that construction.

\subsection{From \texorpdfstring{$\alg{su}(1|1)^2_{\text{c.e.}}$}{su(1|1)**2 c.e.} to \texorpdfstring{$\alg{psu}(1|1)^4_{\text{c.e.}}$}{psu(1|1)**4 c.e.}}
Let us consider $\alg{su}(1|1)_{\sL}\oplus\alg{su}(1|1)_{\sR}$, given by
\begin{equation}
\acomm{\gen{Q}_{\smallL}}{\overline{\gen{Q}}{}_{\smallL}} = \gen{H}_{\smallL},
\qquad
\acomm{\gen{Q}_{\smallR}}{\overline{\gen{Q}}{}_{\smallR}} = \gen{H}_{\smallR}.
\end{equation}
Physically, if we intend to couple these two systems, it is natural to define the positive-define combination of the two central charges to be the Hamiltonian, while the other one will be an angular momentum:
\begin{equation}
\gen{H}=\gen{H}_{\smallL}+\gen{H}_{\smallR},\qquad
\gen{M}=\gen{H}_{\smallL}-\gen{H}_{\smallR}\,.
\end{equation}
Let us consider a central extension of $\alg{su}(1|1)^2$ by setting
\begin{equation}
\acomm{\gen{Q}_{\smallL}}{\gen{Q}_{\smallR}} = \gen{C}\,,
\qquad
\acomm{\overline{\gen{Q}}{}_{\smallL}}{\overline{\gen{Q}}{}_{\smallR}} = \overline{\gen{C}}\,.
\end{equation}

If we now consider a tensor product of two copies of the above  algebra, we have
\begin{equation}\label{eq:supercharges-tensor-product}
  \begin{aligned}
    \gen{Q}_{\smallL}^{\ 1} = \gen{Q}_{\smallL} \otimes \1 ,\, \qquad
    \overline{\gen{Q}}{}_{\smallL 1} = \overline{\gen{Q}}{}_{\smallL} \otimes \1 ,\, \qquad
    \gen{Q}_{\smallL}^{\ 2} = \1 \otimes \gen{Q}_{\smallL} ,\, \qquad
    \overline{\gen{Q}}{}_{\smallL 2} = \1 \otimes \overline{\gen{Q}}{}_{\smallL} , \\
    \gen{Q}_{\smallR 1} = \gen{Q}_{\smallR} \otimes \1 , \qquad
    \overline{\gen{Q}}{}_{\smallR}^{\ 1} = \overline{\gen{Q}}{}_{\smallR} \otimes \1 , \qquad
    \gen{Q}_{\smallR 2} = \1 \otimes \gen{Q}_{\smallR} , \qquad
    \overline{\gen{Q}}{}_{\smallR}^{\ 2} = \1 \otimes \overline{\gen{Q}}{}_{\smallR} ,
  \end{aligned}
\end{equation}
and for the central elements
\begin{equation}\label{eq:supercharges-tensor-product-C}
  \begin{aligned}
    &\gen{H}_{\smallL}^{\ 1} = \gen{H}_{\smallL} \otimes \1 , \qquad &&
    \gen{H}_{\smallL}^{\ 2} = \1 \otimes \gen{H}{\smallL} , \qquad &&
    \gen{C}^{1} = \gen{C} \otimes \1 ,
\qquad &&
    \gen{C}^{2} = \1 \otimes \gen{C} , \\
    &\gen{H}_{\smallR}^{\ 1} = \gen{H}_{\smallR} \otimes \1 , \qquad &&
    \gen{H}_{\smallR}^{\ 2} = \1 \otimes \gen{H}_{\smallR} , \qquad &&
    \overline{\gen{C}}{}^{1} = \overline{\gen{C}} \otimes \1 ,
\qquad &&
    \overline{\gen{C}}{}^{2} = \1 \otimes \overline{\gen{C}} . 
  \end{aligned}
\end{equation}
If we now identify the central charges as
\begin{equation}
\gen{H}_{\smallL}^{\ 1}=\gen{H}_{\smallL}^{\ 2},
\quad
\gen{H}_{\smallR}^{\ 1}=\gen{H}_{\smallR}^{\ 2},
\qquad
\gen{C}^1=\gen{C}^2,
\quad
\overline{\gen{C}}{}^1=\overline{\gen{C}}{}^2,
\end{equation}
and consequently drop the indices $1,2$, we are left precisely with~\eqref{eq:cealgebra}. 
Constructing $\alg{psu}(1|1)^4_{\text{c.e.}}$ as a tensor product of two~$\alg{su}(1|1)^2_{\text{c.e.}}$ as described above will be particularly useful in the study of its representations below~\cite{Borsato:2013qpa,Sfondrini:2014via}.

\subsection{Representations in the near-plane-wave limit}
\label{sec:near-BMN-representations}
We expect the fundamental excitations of the GS string to  transform in \emph{two} distinct (not necessarily irreducible) representations of~$\mathcal{A}$, for massive and massless particles\footnote{%
This follows from the fact that the Hamiltonian~$\gen{H}$ takes different values on massless and massive excitations.
}, with the corresponding modules having the same dimension.
This constrains the dimension of the representations of $\alg{psu}(1|1)^4_{\ce}\subset\mathcal{A}$ that may appear. Since long representations are at least sixteen-dimensional, our representations must instead be \emph{short}, \ie they must satisfy the shortening condition~\cite{Borsato:2012ud,Borsato:2013qpa}
\begin{equation}
\gen{H}_{\sL}\,\gen{H}_{\sR}=\gen{C}\,\overline{\gen{C}}\,,
\end{equation}
which can be recast in the form of a dispersion relation
\begin{equation}
\label{eq:shortening}
\gen{H}^2=\gen{M}^2+4\,\gen{C}\,\overline{\gen{C}}\,,
\end{equation}
in which the eigenvalues of~$\gen{M}$ play the role of a mass term\footnote{%
Recall that the central charges~$\gen{C},\overline{\gen{C}}$ are functions of the momentum $p$ and vanish at $p=0$.
}.

In section~\ref{sec:charges}  we obtained the representation of~$\mathcal{A}$ in terms of the fields. In order study it more easily, it is useful to go to momentum-space and introduce oscillators. It will be enough to consider the leading-order in the field expansion of the supercharges, which  coincides with the leading order in a near-plane-wave~\cite{Gava:2002xb} or BMN expansion~\cite{Berenstein:2002jq}.

Let us introduce bosonic creation and annihilation operators, that schematically take the form
\begin{equation}
\begin{aligned}
a^{\dagger}(p)\approx \int\frac{\de\sigma}{\sqrt{\omega(p,m)}} \big(\omega(p,m)\,X-i P\big) e^{+ip\sigma},\\
a(p)\approx \int\frac{\de\sigma}{\sqrt{\omega(p,m)}} \big(\omega(p,m)\,X+i P\big) e^{-ip\sigma},
\end{aligned}
\end{equation}
where~$\omega(p,m)$ is the dispersion, and fermionic ones
\begin{equation}
\begin{aligned}
d^{\dagger}(p)\approx \int\frac{\de\sigma}{\sqrt{\omega(p,m)}} \big(f(p,m)\,\eta-i g(p,m) \bar{\eta}\big) e^{+ip\sigma},\\
d(p)\approx \int\frac{\de\sigma}{\sqrt{\omega(p,m)}} \big(f(p,m)\,\eta+i g(p,m) \bar{\eta}\big) e^{-ip\sigma},
\end{aligned}
\end{equation}
where $f(p,m),$ $g(p,m)$ are wavefunction parameters. We will have eight such pairs of operators for bosons and eight for fermions, whose precise form is given in appendix~\ref{app:algebra}. 
We can use them to construct the module of the representation, which is then given by the eight massive states
\begin{equation}
\ket{Z^{\sL,\sR}}=a^\dagger_{\sL,\sR\, z}\ket{0},\quad
\ket{Y^{\sL,\sR}}=a^\dagger_{\sL,\sR\, y}\ket{0},\quad
\ket{\eta^{\sL \dot{a}}}=d^{\ \dot{a} \dagger}_{\sL}\ket{0},\quad
\ket{\eta^{\sR}_{\ \dot{a}}}=d^{\dagger}_{\sR \dot{a}}\ket{0},
\end{equation}
and the eight massless ones
\begin{equation}
\ket{T^{\dot{a}a}}=a^{\dot{a} a\dagger}\ket{0},\qquad
\ket{\chi^{a}}=d^{a\,\dagger}\ket{0},\qquad
\ket{\widetilde{\chi}^{a}}=\tilde{d}^{a\,\dagger}\ket{0}.
\end{equation}
Since at this order all excitations are relativistic, we have
\begin{equation}
\omega(p,m)=\sqrt{m^2+p^2},\qquad
f(p,m)=\sqrt{\frac{\omega(p,m)+|m|}{2}},\qquad
g(p,m)=\frac{-p}{2f(p,m)}\,,
\end{equation}
see also equations~(\ref{eq:fermion-par-massive}--\ref{eq:fermion-par-massless}). Note that formul\ae{} depend on the eigenvalue $m$ of $\gen{M}$, which can take value~$\pm 1$ for massive excitations and $0$ for massless ones. It will be convenient to denote~$\omega_p=\omega(p,\pm1)$, $\tilde{\omega}_p=\omega(p,0)$, and similarly for $f$ and $g$.

In terms of the ladder operators the supercharges take a very transparent form, whence their action can be easily read off:
\begin{equation}
\label{eq:supercharges-leading-order-rep}
\begin{aligned}
&\gen{Q}_{\smallL}^{\ {\dot{a}}}= \int \de p \ \Bigg[
 (d_{\sL}^{\ {\dot{a}}\,\dagger} a_{\sL y} + \epsilon^{{\dot{a}\dot{b}}}\, a_{\sL
z}^\dagger  d_{\sL {\dot{b}}})f_p
+ (a_{\sR y}^\dagger  d_{\sR}^{\ {\dot{a}}} +\epsilon^{{\dot{a}\dot{b}}}\, d_{\sR
{\dot{b}}}^\dagger  a_{\sR z})g_p \\
&\qquad\qquad\qquad\qquad\qquad\qquad\qquad\qquad\qquad\qquad
+ \left( \epsilon^{{\dot{a}\dot{b}}}\, \tilde{d}^{{a}\,\dagger}a_{{\dot{b}a}}+
a^{{\dot{a}a}\,\dagger}d_{{a}}\right)\tilde{f}_p\Bigg],\\
&\gen{Q}_{\smallR {\dot{a}}}=\int \de p \ \Bigg[
 (d_{\sR {\dot{a}}}^\dagger  a_{\sR y} -\epsilon_{{\dot{a}\dot{b}}}\, a_{\sR z}^\dagger
d_{\sR}^{\ {\dot{b}}})f_p
+ (a_{\sL y}^\dagger  d_{\sL {\dot{a}}} -\epsilon_{{\dot{a}\dot{b}}}\,  d_{\sL}^{\
{\dot{b}}\,\dagger} a_{\sL z})g_p\\
&\qquad\qquad\qquad\qquad\qquad\qquad\qquad\qquad\qquad\qquad
 + \left( d^{{a}\,\dagger}a_{{\dot{a}a}}-
\epsilon_{{\dot{a}\dot{b}}}\, a^{{\dot{b}a}\,\dagger}\tilde{d}_{{a}}\right)\tilde{g}_p\Bigg],
\end{aligned}
\end{equation}
where we suppressed the dependence of $a^{\dagger},a$ and $d^\dagger,d$ on the momentum~$p$.
Similarly, the Hamiltonian~$\gen{H}$ and the angular momentum~$\gen{M}$ read
\begin{equation}
\begin{aligned}
\gen{H}=&\!\!\int \de p \big[
(a^{\dagger}_{\sL z}a_{\sL z}+a^{\dagger}_{\sL y}a_{\sL y}+d^{\ \dot{a}\dagger}_{\sL}d_{\sL \dot{a}}
+a^{\dagger}_{\sR z}a_{\sR z}+a^{\dagger}_{\sR y}a_{\sR y}+d^{\ \dot{a}\dagger}_{\sR}d_{\sR \dot{a}})\,\omega_p\\
&\qquad\qquad\qquad\qquad\qquad\qquad\qquad\qquad +(a^{\dagger}_{\dot{a}a}a^{\dot{a}a} +d^{a\dagger}d_{a} +\tilde{d}^{a\dagger}\tilde{d}_{a})\,\tilde{\omega}_p\big],\\
\gen{M}=&\!\!\int \de p \big[
(a^{\dagger}_{\sL z}a_{\sL z}+a^{\dagger}_{\sL y}a_{\sL y}+d^{\ \dot{a}\dagger}_{\sL}d_{\sL \dot{a}})-(a^{\dagger}_{\sR z}a_{\sR z}+a^{\dagger}_{\sR y}a_{\sR y}+d^{\ \dot{a}\dagger}_{\sR}d_{\sR \dot{a}})\big].
\end{aligned}
\end{equation}

This sixteen-dimensional module will split in several irreducible ones, which we will describe below one by one. We can label them by the eigenvalue of the angular momentum~$\gen{M}$.

\begin{figure}[t]
  \centering
  \begin{tikzpicture}[%
    box/.style={outer sep=1pt},
    Q node/.style={inner sep=1pt,outer sep=0pt},
    arrow/.style={-latex}
    ]%

    \node [box] (PhiM) at ( 0  , 2cm) {\small $\ket{Y^{\sL}}$};
    \node [box] (PsiP) at (-2cm, 0cm) {\small $\ket{\eta^{\sL 1}}$};
    \node [box] (PsiM) at (+2cm, 0cm) {\small $\ket{\eta^{\sL 2}}$};
    \node [box] (PhiP) at ( 0  ,-2cm) {\small $\ket{Z^{\sL}}$};

    \newcommand{\horshift}{0.09cm,0cm}
    \newcommand{\vershift}{0cm,0.10cm}
 
    \draw [arrow] ($(PhiM.west) +(\vershift)$) -- ($(PsiP.north)-(\horshift)$) node [pos=0.5,anchor=south east,Q node] {\scriptsize $\gen{Q}^{\ 1}_{\sL},\overline{\gen{Q}}{}^{\ 1}_{\sR}$};
    \draw [arrow] ($(PsiP.north)+(\horshift)$) -- ($(PhiM.west) -(\vershift)$) node [pos=0.5,anchor=north west,Q node] {};

    \draw [arrow] ($(PsiM.south)-(\horshift)$) -- ($(PhiP.east) +(\vershift)$) node [pos=0.5,anchor=south east,Q node] {};
    \draw [arrow] ($(PhiP.east) -(\vershift)$) -- ($(PsiM.south)+(\horshift)$) node [pos=0.5,anchor=north west,Q node] {\scriptsize $\overline{\gen{Q}}{}_{\sL 1},\gen{Q}_{\sR 1}$};

    \draw [arrow] ($(PhiM.east) -(\vershift)$) -- ($(PsiM.north)-(\horshift)$) node [pos=0.5,anchor=north east,Q node] {};
    \draw [arrow] ($(PsiM.north)+(\horshift)$) -- ($(PhiM.east) +(\vershift)$) node [pos=0.5,anchor=south west,Q node] {\scriptsize $\overline{\gen{Q}}{}_{\sL 2},{\gen{Q}}{}_{\sR 2}$};

    \draw [arrow] ($(PsiP.south)-(\horshift)$) -- ($(PhiP.west) -(\vershift)$) node [pos=0.5,anchor=north east,Q node] {\scriptsize $
    \gen{Q}^{\ 2}_{\sL},\overline{\gen{Q}}{}^{\ 2}_{\sR}$};
    \draw [arrow] ($(PhiP.west) +(\vershift)$) -- ($(PsiP.south)+(\horshift)$) node [pos=0.5,anchor=south west,Q node] {};

    \draw [arrow] (PsiM) -- (PsiP) node [pos=0.6,anchor=south west,Q node] {\scriptsize $\gen{J}^{\ a}_{\suA}$};
    \draw [arrow] (PsiP) -- (PsiM);

\draw[rounded corners=5mm] (-2.8cm,-2.5cm)rectangle (2.8cm,2.5cm);
  \end{tikzpicture}
\hspace{2cm}
  \begin{tikzpicture}[%
    box/.style={outer sep=1pt},
    Q node/.style={inner sep=1pt,outer sep=0pt},
    arrow/.style={-latex}
    ]%

    \node [box] (PhiM) at ( 0  , 2cm) {\small $\ket{Z^{\sR}}$};
    \node [box] (PsiP) at (-2cm, 0cm) {\small $\ket{\eta^{\sR}_{\  1}}$};
    \node [box] (PsiM) at (+2cm, 0cm) {\small $\ket{\eta^{\sR}_{\  2}}$};
    \node [box] (PhiP) at ( 0  ,-2cm) {\small $\ket{Y^{\sR}}$};

    \newcommand{\horshift}{0.09cm,0cm}
    \newcommand{\vershift}{0cm,0.10cm}
 
    \draw [arrow] ($(PsiP.north)-(\horshift)$) -- ($(PhiM.west) +(\vershift)$) node [pos=0.5,anchor=south east,Q node] {\scriptsize $\gen{Q}_{\sR 2},\overline{\gen{Q}}{}_{\sL 2}$};
    \draw [arrow] ($(PhiM.west) -(\vershift)$) -- ($(PsiP.north)+(\horshift)$) node [pos=0.5,anchor=north west,Q node] {};

    \draw [arrow] ($(PhiP.east) +(\vershift)$) -- ($(PsiM.south)-(\horshift)$) node [pos=0.5,anchor=south east,Q node] {};
    \draw [arrow] ($(PsiM.south)+(\horshift)$) -- ($(PhiP.east) -(\vershift)$) node [pos=0.5,anchor=north west,Q node] {\scriptsize $\overline{\gen{Q}}{}^{\ 2}_{\sR},\gen{Q}_{\sL}^{\ 2}$};

    \draw [arrow] ($(PsiM.north)-(\horshift)$) -- ($(PhiM.east) -(\vershift)$) node [pos=0.5,anchor=north east,Q node] {};
    \draw [arrow] ($(PhiM.east) +(\vershift)$) -- ($(PsiM.north)+(\horshift)$) node [pos=0.5,anchor=south west,Q node] {\scriptsize $\overline{\gen{Q}}{}^{\ 1}_{\sR}, \gen{Q}_{\sL}^{\ 1}$};

    \draw [arrow] ($(PhiP.west) -(\vershift)$) -- ($(PsiP.south)-(\horshift)$) node [pos=0.5,anchor=north east,Q node] {\scriptsize $\gen{Q}_{\sR 1}, \overline{\gen{Q}}{}_{\sL 1}$};
    \draw [arrow] ($(PsiP.south)+(\horshift)$) -- ($(PhiP.west) +(\vershift)$) node [pos=0.5,anchor=south west,Q node] {};

    \draw [arrow] (PsiM) -- (PsiP) node [pos=0.6,anchor=south west,Q node] {\scriptsize $\gen{J}^{\ a}_{\suA}$};
    \draw [arrow] (PsiP) -- (PsiM);
    
\draw[rounded corners=5mm] (-2.8cm,-2.5cm)rectangle (2.8cm,2.5cm);
  \end{tikzpicture}
  \caption{The left and right $\psu(1|1)^4_{\ce}$ multiplets consists of two bosons~$Y^{\sL,\sR}$, $Z^{\sL,\sR}$ and of two fermions~$\eta_{\ \dot{a}}^{\sL,\sR}$, corresponding to transverse directions on~$\AdS_3\times\Sphere^3$. The fermions carry an index~$\dot{a}$ of the fundamental representation of~$\su(2)_{\suA}$.
  Note that off-shell any excitation is charged under all supercharges, whereas on-shell left (respectively right) excitations are charged only under left (respectively right) supercharges. We indicate the supercharges whose action corresponds to the outermost arrows of the diagram. The innermost ones follow by Hermitian conjugation.}
  \label{fig:massive}
\end{figure}
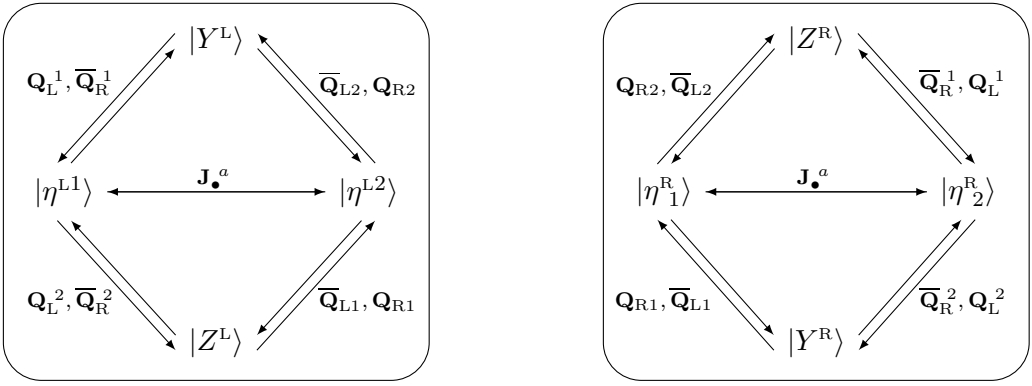

\subsubsection{Left representation}
One representation of dimension four (two bosons and two fermions) has eigenvalue~$+1$ under~$\gen{M}$, and consists only of excitations labelled~``L''. These correspond to half of the transverse modes on~$\AdS_3\times\Sphere^3$, and it can be represented as in the left panel of figure~\ref{fig:massive}. This is a bi-fundamental representation of~$\alg{psu}(1|1)^4_{\ce}$, supplemented by the action of~$\su(2)_{\bullet}$ on the fermions. In particular, the fermions are in the fundamental representation
\begin{equation}
\label{eq:su2-repr}
\gen{J}_{\bullet \dot{a}}^{\ \ \dot{b}} \ket{\eta^{\sL \dot{c}}} = -\delta_a^{\ \dot{c}} \ket{\eta^{\sL \dot{b}}} + \frac{1}{2} \delta^{\ \dot{b}}_{\dot{a}} \ket{\eta^{\sL \dot{c}}}.
\end{equation}

It is interesting to see what happens if we consider the representation on-shell, \ie (since we are dealing with a single-particle representation) at zero momentum. Observing that one of the fermion wave-function parameters vanishes then, $g_{p=0}=0$, greatly simplifies the action of the supercharges~\eqref{eq:supercharges-leading-order-rep}. In fact, only left supercharges act non-trivially. For this reason we call this representation \emph{left}, which explains the name of the corresponding excitations, which then can be thought of as left-movers in the dual $\CFT_2 $. Of course this is only true for an on-shell one-particle state---generally, all states are charged under the whole~$\alg{psu}(1|1)^4_{\ce}$.

\subsubsection{Right representation}
We have a similar representation with eigenvalue~$-1$ under~$\gen{M}$, which is depicted in the right panel of figure~\ref{fig:massive}, and consists of ``R'' excitations. This is again a bi-fundamental representation, closely resembling the left one. We see however that if we take the~$\Sphere^3$ excitation~$\ket{Y^{\sL}}$ to be the highest weight state in the left-representation, we must take the~$\AdS_3$ excitation~$\ket{Z^{\sR}}$ to be the highest weight state here. The reason is that we cannot take e.g.~$\gen{Q}^{\ 1}_{\sL}$ and~$\gen{Q}_{\sR 1}$ to be both lowering operators, if  we want the central charge~$\gen{C}$ to be non-vanishing. Instead, we should take e.g.~$\gen{Q}^{\ 1}_{\sL}$ and~$\overline{\gen{Q}}{}^{\ 1}_{\sR}$. This  in turn forces the choice of different highest weight states in the two representations.
The fermions $\eta^{\sR}_{\ \dot{a}}$ are in the anti-fundamental of $\su(2)_{\bullet}$.

In the same way as earlier, taking the representation to be on-shell makes it charged under the right supercharges only.

\subsubsection{Massless representation}
We expect the remaining eight particles (four bosons and four fermions) to be all massless, at least in this semi-classical analysis. This is indeed the case, and we can check that the massless particles arrange themselves into two irreducible representations of~$\alg{psu}(1|1)^4_{\ce}$, both with fermionic highest weight states, see figure~\ref{fig:massless}. Additionally, note that these representations seem to be left by the argument above. This should be taken with a pinch of salt, because we are considering the situation when the eigenvalue of~$\gen{M}$ precisely vanishes, which as we will see makes the left and right representations practically equivalent.

We have to take the action of~$\alg{so}(4)_2$ into account. Under this, all of the torus bosons are obviously charged. In our~$\su(2)$ decomposition we see that the two $\alg{psu}(1|1)^4_{\ce}$ modules are rotated one into  another by the action of~$\su(2)_{\circ}$. We then conclude that the massless particles transform into \emph{a single irreducible representation} of the symmetry algebra~$\mathcal{A}$.

\begin{figure}[t]
  \centering
  \begin{tikzpicture}[%
    box/.style={outer sep=1pt},
    Q node/.style={inner sep=1pt,outer sep=0pt},
    arrow/.style={-latex}
    ]%
\begin{scope}[xshift=-4cm]
    \node [box] (PhiM) at ( 0  , 2cm) {\small $\ket{\chi^{1}}$};
    \node [box] (PsiP) at (-2cm, 0cm) {\small $\ket{T^{11}}$};
    \node [box] (PsiM) at (+2cm, 0cm) {\small $\ket{T^{21}}$};
    \node [box] (PhiP) at ( 0  ,-2cm) {\small $\ket{\widetilde{\chi}^{1}}$};

    \newcommand{\horshift}{0.09cm,0cm}
    \newcommand{\vershift}{0cm,0.10cm}
 
    \draw [arrow] ($(PhiM.west) +(\vershift)$) -- ($(PsiP.north)-(\horshift)$) node [pos=0.5,anchor=south east,Q node] {\scriptsize $\gen{Q}^{\ 1}_{\sL},\overline{\gen{Q}}{}_{\sR}^{\ 1}$};
    \draw [arrow] ($(PsiP.north)+(\horshift)$) -- ($(PhiM.west) -(\vershift)$) node [pos=0.5,anchor=north west,Q node] {};

    \draw [arrow] ($(PsiM.south)-(\horshift)$) -- ($(PhiP.east) +(\vershift)$) node [pos=0.5,anchor=south east,Q node] {};
    \draw [arrow] ($(PhiP.east) -(\vershift)$) -- ($(PsiM.south)+(\horshift)$) node [pos=0.5,anchor=north west,Q node] {\scriptsize $\overline{\gen{Q}}{}_{\sL 1}, \gen{Q}_{\sR 1}$};

    \draw [arrow] ($(PhiM.east) -(\vershift)$) -- ($(PsiM.north)-(\horshift)$) node [pos=0.5,anchor=north east,Q node] {};
    \draw [arrow] ($(PsiM.north)+(\horshift)$) -- ($(PhiM.east) +(\vershift)$) node [pos=0.5,anchor=south west,Q node] {\scriptsize $\overline{\gen{Q}}{}_{\sL 2}, \gen{Q}_{\sR 2}$};

    \draw [arrow] ($(PsiP.south)-(\horshift)$) -- ($(PhiP.west) -(\vershift)$) node [pos=0.5,anchor=north east,Q node] {\scriptsize $\gen{Q}^{\ 2}_{\sL},\overline{\gen{Q}}{}_{\sR}^{\ 2}$};
    \draw [arrow] ($(PhiP.west) +(\vershift)$) -- ($(PsiP.south)+(\horshift)$) node [pos=0.5,anchor=south west,Q node] {};

    \draw [arrow] (PsiM) -- (PsiP) node [pos=0.65,anchor=south west,Q node] {\scriptsize $\gen{J}^{\ a}_{\suA}$};
    \draw [arrow] (PsiP) -- (PsiM);

\draw[rounded corners=5mm] (-2.8cm,-2.5cm)rectangle (2.8cm,2.5cm);
\end{scope}
\begin{scope}[xshift=0cm]
    \draw [arrow] (-1cm,0cm) -- (1cm,0cm) node [Q node] at (0cm,0.15cm) {\scriptsize $\gen{J}^{\ \alpha}_{\suB}$};
    \draw [arrow] (1cm,0cm) -- (-1cm,0cm);
    %
\end{scope}
\begin{scope}[xshift=4cm]

    \node [box] (PhiM) at ( 0  , 2cm) {\small $\ket{\chi^{2}}$};
    \node [box] (PsiP) at (-2cm, 0cm) {\small $\ket{T^{12}}$};
    \node [box] (PsiM) at (+2cm, 0cm) {\small $\ket{T^{22}}$};
    \node [box] (PhiP) at ( 0  ,-2cm) {\small $\ket{\widetilde{\chi}^{2}}$};

    \newcommand{\horshift}{0.09cm,0cm}
    \newcommand{\vershift}{0cm,0.10cm}
 
    \draw [arrow] ($(PhiM.west) +(\vershift)$) -- ($(PsiP.north)-(\horshift)$) node [pos=0.5,anchor=south east,Q node] {\scriptsize $\gen{Q}^{\ 1}_{\sL},\overline{\gen{Q}}{}_{\sR}^{\ 1}$};
    \draw [arrow] ($(PsiP.north)+(\horshift)$) -- ($(PhiM.west) -(\vershift)$) node [pos=0.5,anchor=north west,Q node] {};

    \draw [arrow] ($(PsiM.south)-(\horshift)$) -- ($(PhiP.east) +(\vershift)$) node [pos=0.5,anchor=south east,Q node] {};
    \draw [arrow] ($(PhiP.east) -(\vershift)$) -- ($(PsiM.south)+(\horshift)$) node [pos=0.5,anchor=north west,Q node] {\scriptsize $\overline{\gen{Q}}{}_{\sL 1}, \gen{Q}_{\sR 1}$};

    \draw [arrow] ($(PhiM.east) -(\vershift)$) -- ($(PsiM.north)-(\horshift)$) node [pos=0.5,anchor=north east,Q node] {};
    \draw [arrow] ($(PsiM.north)+(\horshift)$) -- ($(PhiM.east) +(\vershift)$) node [pos=0.5,anchor=south west,Q node] {\scriptsize $\overline{\gen{Q}}{}_{\sL 2}, \gen{Q}_{\sR 2}$};

    \draw [arrow] ($(PsiP.south)-(\horshift)$) -- ($(PhiP.west) -(\vershift)$) node [pos=0.5,anchor=north east,Q node] {\scriptsize $\gen{Q}^{\ 2}_{\sL}, \overline{\gen{Q}}{}_{\sR}^{\ 2}$};
    \draw [arrow] ($(PhiP.west) +(\vershift)$) -- ($(PsiP.south)+(\horshift)$) node [pos=0.5,anchor=south west,Q node] {};

    \draw [arrow] (PsiM) -- (PsiP) node [pos=0.65,anchor=south west,Q node] {\scriptsize $\gen{J}^{\ a}_{\suA}$};
    \draw [arrow] (PsiP) -- (PsiM);

\draw[rounded corners=5mm] (-2.8cm,-2.5cm)rectangle (2.8cm,2.5cm);
\end{scope}
\draw[rounded corners=5mm] (-7.5cm,-3cm)rectangle (7.5cm,3cm);
  \end{tikzpicture}
  \caption{The eight massless excitations form two $\psu(1|1)^4_{\ce}$ multiplets. The four bosons $T^{\dot{a}a}$ are charged both under $\su(2)_{\suA}$ and $\su(2)_{\suB}$, while the four fermions~$\chi^{a}, \widetilde{\chi}^{a}$ are in the fundamental representation of $\su(2)_{\suB}$ only.  Again we indicate the charges whose action corresponds to the outermost arrows of the diagram, while  the innermost ones follow by Hermitian conjugation. Note how~$\su(2)_{\suB}$ relates the two $\psu(1|1)^4_{\ce}$~modules, yielding a single irreducible representation of~$\mathcal{A}$, denoted by a box.}
  \label{fig:massless}
\end{figure}

\section{Exact representations}
\label{sec:representations-allloop}
In the previous section we constructed representations of~$\mathcal{A} $ that are valid in the near-plane-wave limit, when~$\gen{C}=\overline{\gen{C}}=-\frac{1}{2}\gen{P}$. On the other hand, we know that instead $\gen{C}\neq\overline{\gen{C}}$ should be non-linear functions of the world-sheet momentum~$\gen{P} $ given by equation~\eqref{eq:allloop-centralcharges}. In this section, we will show that the representations of section~\ref{sec:near-BMN-representations} can be deformed in such a way so as to satisfy~\eqref{eq:allloop-centralcharges} together with the shortening condition~\eqref{eq:shortening} without resorting to any perturbative expansions.

To do this, it will be sufficient to suitably deform the representation parameters~$\omega_p,f_p$ and~$g_p$. In fact, since the Hamiltonian follows from the central charges,~$\omega_p$ will be fixed in terms of the fermion wave-function parameters. In order to find the central charges~$\gen{C}\neq\overline{\gen{C}}$, which will now be complex and conjugate to each other, we must introduce complex representation coefficients. Hence, $f_p$ will be replaced by~$a_p$ or its conjugate~$\bar{a}_p$, and~$g_p$ by~$b_p$ or~$\bar{b}_p$. These will also suitably depend on the mass~$m$. When expanded in the near-BMN limit, $a_p$ and $\bar{a}_p$ reduce to~$f_p$ (or $\tilde{f}_p$ if $m=0$) while $b_p$ and $\bar{b}_p$ to~$g_p$ (or $\tilde{g}_p$).

We will proceed as follows. Since the representations of~$\mathcal{A}$ discussed above must in particular be bi-fundamental representations of~$\alg{psu}(1|1)^4_{\ce}$, and these can be obtained~\cite{Borsato:2012ud,Sfondrini:2014via} from fundamental representations of~$\alg{su}(1|1)^2_{\ce}$, we will focus on these first. In subsection~\ref{sec:rep-su11} we will construct the most general $\alg{su}(1|1)^2_{\ce}$ short fundamental representations. In subsections~\ref{sec:repr-massive} and~\ref{sec:repr-massless} we will show how from these we can find massive and massless~$\alg{psu}(1|1)^4_{\ce}$ representations, respectively, and comment on their properties. Then in subsection~\ref{sec:par-abcd} we write down the deformed representation parameters~$a_p,b_p$ for the massive and massless case, and finally in subsection~\ref{sec:correction-dispersion} we will rule out quantum corrections to the massless dispersion relation by a symmetry argument.

\subsection{Short representations of \texorpdfstring{$\alg{su}(1|1)^2$}{psu(1|1)**2} c.e.}
\label{sec:rep-su11}

In~\cite{Borsato:2012ud} short representations of the centrally extended $\alg{psu}(1|1)_{\sL} \oplus \alg{psu}(1|1)_{\sR} $ algebra were considered. Denoting a boson and  fermion excitation of definite momentum~$p$ by $\ket{\phi^{\sL}_p},\ket{\psi^{\sL}_p}$ respectively, the fundamental representation $\varrho_{\sL}$ is given by
\begin{equation}\label{eq:su(1|1)2-repr1}
\boxed{\varrho_{\sL}:} \qquad\qquad
  \begin{aligned}
    \gen{Q}_{\smallL} \ket{\phi^{\sL}_p} &= a_p \ket{\psi^{\sL}_p} , \qquad &
    \gen{Q}_{\smallL} \ket{\psi^{\sL}_p} &= 0 , \\
    \overline{\gen{Q}}{}_{\smallL} \ket{\phi^{\sL}_p} &= 0 , \qquad &
    \overline{\gen{Q}}{}_{\smallL} \ket{\psi^{\sL}_p} &= \bar{a}_p \ket{\phi^{\sL}_p} , \\
    \gen{Q}_{\smallR} \ket{\phi^{\sL}_p} &= 0 , \qquad &
    \gen{Q}_{\smallR} \ket{\psi^{\sL}_p} &= b_p \ket{\phi^{\sL}_p} , \\
    \overline{\gen{Q}}{}_{\smallR} \ket{\phi^{\sL}_p} &= \bar{b}_p \ket{\psi^{\sL}_p} , \qquad &
    \overline{\gen{Q}}{}_{\smallR} \ket{\psi^{\sL}_p} &= 0 .
  \end{aligned}
\end{equation}
Our choice of the representation coefficients ensures that the left- and right-Hamiltonians are positive definite. Furthermore, when we reduce to an one-particle \emph{on-shell} representation, it must be $b_{p=0}=\bar{b}_{p=0}=0$. In this sense, this representation is a left one.

We can consider a right representation~$\varrho_{\sR} $. The module consists of two excitations $\ket{\phi^{\sR}_p}$ and $\ket{\psi^{\sR}_p}$ transforming as
\begin{equation}\label{eq:su(1|1)2-reprR}
\boxed{\varrho_{\sR}:} \qquad\qquad
  \begin{aligned}
    \gen{Q}_{\smallR} \ket{\phi_p^{\sR}} &= a_p \ket{\psi_p^{\sR}} , \qquad &
    \gen{Q}_{\smallR} \ket{\psi_p^{\sR}} &= 0 , \\
    \overline{\gen{Q}}{}_{\smallR} \ket{\phi_p^{\sR}} &= 0 , \qquad &
    \overline{\gen{Q}}{}_{\smallR} \ket{\psi_p^{\sR}} &= \bar{a}_p \ket{\phi_p^{\sR}} , \\
    \gen{Q}_{\smallL} \ket{\phi_p^{\sR}} &= 0 , \qquad &
    \gen{Q}_{\smallL} \ket{\psi_p^{\sR}} &= b_p \ket{\phi_p^{\sR}} , \\
    \overline{\gen{Q}}{}_{\smallL} \ket{\phi_p^{\sR}} &= \bar{b}_p \ket{\psi_p^{\sR}} , \qquad &
    \overline{\gen{Q}}{}_{\smallL} \ket{\bar{\psi}_p} &= 0 .
  \end{aligned}
\end{equation}
Comparing with~\eqref{eq:su(1|1)2-repr1}, we see that the two representations are related by exchanging the labels~L$\leftrightarrow$R.

It we will also be useful to consider a representation $\widetilde{\varrho}_{\sL}$,
\begin{equation}\label{eq:su(1|1)2-repr2}
\boxed{\widetilde{\varrho}_{\sL}:} \qquad\qquad
  \begin{aligned}
    \gen{Q}_{\smallL} \ket{\widetilde{\psi}_p^{\sL}} &= a_p \ket{\widetilde{\phi}_p^{\sL}} , \qquad &
    \gen{Q}_{\smallL} \ket{\widetilde{\phi}_p^{\sL}} &= 0 , \\
    \overline{\gen{Q}}{}_{\smallL} \ket{\widetilde{\psi}_p^{\sL}} &= 0 , \qquad &
    \overline{\gen{Q}}{}_{\smallL} \ket{\widetilde{\phi}_p^{\sL}} &= \bar{a}_p \ket{\widetilde{\psi}_p^{\sL}} , \\
    \gen{Q}_{\smallR} \ket{\widetilde{\psi}_p^{\sL}} &= 0 , \qquad &
    \gen{Q}_{\smallR} \ket{\widetilde{\phi}_p^{\sL}} &= b_p \ket{\widetilde{\psi}_p^{\sL}} , \\
    \overline{\gen{Q}}{}_{\smallR} \ket{\widetilde{\psi}_p^{\sL}} &= \bar{b}_p \ket{\widetilde{\phi}_p^{\sL}} , \qquad &
    \overline{\gen{Q}}{}_{\smallR} \ket{\widetilde{\phi}_p^{\sL}} &= 0 .
  \end{aligned}
\end{equation}
This representation differs from~$\varrho_{\sL}$ by the choice of the highest weight state, which is fermionic here rather than bosonic. In a similar way, one can construct a~$\widetilde{\varrho}_{\sR}$ representation. 
It is easy to check that any short representation of~$\alg{psu}(1|1)^{4}_{\ce}$ for which $\gen{H}$ has real eigenvalues takes the form of one of~$\varrho_{\sL}$,~$\varrho_{\sR}$,~$\widetilde{\varrho}_{\sL}$,
~$\widetilde{\varrho}_{\sR}$. We will use graded tensor products of these representations to construct  short representations of~$\alg{psu}(1|1)^{4}_{\ce}$.

\subsection{\texorpdfstring{$\alg{psu}(1|1)^4_{\ce}$}{psu(1|1)**4 c.e.} representations for massive excitations}
\label{sec:repr-massive}
We now want to take the massive near-plane-wave representations discussed in section~\ref{sec:near-BMN-representations} and deform them in such a way as to reproduce the non-linear relation~\eqref{eq:allloop-centralcharges}. We will then see how these can indeed  be thought of as coming from the tensor product of suitable pairs of the~$\alg{su}(1|1)^2_{\ce}$ representations that we just constructed.

If we restrict to on-shell one-particle states, only left (respectively right) supercharges have a non-trivial action on left (respectively right) excitations, of the form
\begin{equation}\label{eq:repr-massive}
\boxed{\varrho_{\sL}\otimes\varrho_{\sL}:} \qquad
  \begin{aligned}
    \gen{Q}_{\sL}^{\ \dot{a}} \ket{Y_p^{\sL}} &= a_p \ket{\eta^{\sL \dot{a}}_p},
    \qquad
    &\gen{Q}_{\sL}^{\ \dot{a}} \ket{\eta^{\sL \dot{b}}_p} &= \epsilon^{\dot{a}\dot{b}} \, a_p \ket{Z_p^{\sL}}, \\
    \overline{\gen{Q}}{}_{\sL \dot{a}} \ket{Z_p^{\sL}} &=  - \epsilon_{\dot{a}\dot{b}}  \, \bar{a}_p \ket{\eta^{\sL \dot{b}}_p},
    \qquad
    &\overline{\gen{Q}}{}_{\sL \dot{a}} \ket{\eta^{\sL \dot{b}}_p}& =  \delta_{\dot{a}}^{\ \dot{b}}  \, \bar{a}_p \ket{Y_p^{\sL}}, \\[8pt]
    \gen{Q}_{\sR \dot{a}} \ket{Y_p^{\sR}} &=  \epsilon_{\dot{a}\dot{b}} \,  a_p \ket{\eta^{\sR \dot{b}}_p},
    \qquad
    &\gen{Q}_{\sR \dot{a}} \ket{\eta^{\sR \dot{b}}_p} &= \delta_{\dot{a}}^{\ \dot{b}} \,  a_p \ket{Z_p^{\sR}}, \\
    \overline{\gen{Q}}{}_{\sR}^{\ \dot{a}} \ket{Z_p^{\sR}} &= \bar{a}_p \ket{\eta^{\sR \dot{a}}_p},
    \qquad
    &\overline{\gen{Q}}{}_{\sR}^{\ \dot{a}} \ket{\eta^{\sR \dot{b}}_p} &= - \epsilon^{\dot{a}\dot{b}}  \, \bar{a}_p \ket{Y_p^{\sR}},
  \end{aligned}
\end{equation}
where we anticipated the tensor-product structure in the formula label.
Off shell, by virtue of the central extension this is supplemented by the action of left supercharges on right-moving states and vice-versa:
\begin{equation}\label{eq:repr-massive-central-extension}
\boxed{\varrho_{\sR}\otimes\varrho_{\sR}:} \qquad
  \begin{aligned}
    &\gen{Q}_{\sL}^{\ \dot{a}} \ket{Z_p^{\sR}} =  b_p \ket{\eta^{\sR \dot{a}}_p},
    \qquad
    &\gen{Q}_{\sL}^{\ \dot{a}} \ket{\eta^{\sR \dot{b}}_p} &=- \epsilon^{\dot{a}\dot{b}} \,  b_p \ket{Y_p^{\sR}},\\
    &\overline{\gen{Q}}{}_{\sL \dot{a}} \ket{Y_p^{\sR}} = \epsilon_{\dot{a}\dot{b}} \,  \bar{b}_p \ket{\eta^{\sR \dot{b}}_p},
    \qquad
    &\overline{\gen{Q}}{}_{\sL \dot{a}} \ket{\eta^{\sR \dot{b}}_p} &= \delta_{\dot{a}}^{\ \dot{b}} \,  \bar{b}_p \ket{Z_p^{\sR}}, \\[8pt]
    &\gen{Q}_{\sR \dot{a}} \ket{Z^{\sL}_p} = - \epsilon_{\dot{a}\dot{b}} \,  b_p \ket{\eta^{\sL \dot{b}}_p},
    \qquad
    &\gen{Q}_{\sR \dot{a}} \ket{\eta^{\sL \dot{b}}_p} &= \delta_{\dot{a}}^{\ \dot{b}} \, b_p \ket{Y^{\sL}_p},\\
    &\overline{\gen{Q}}{}_{\sR}^{\ \dot{a}} \ket{Y^{\sL}_p} = \bar{b}_p \ket{\eta^{\sL \dot{a}}_p},
    \qquad
    &\overline{\gen{Q}}{}_{\sR}^{\ \dot{a}} \ket{\eta^{\sL \dot{b}}_p} &= \epsilon^{\dot{a}\dot{b}} \,  \bar{b}_p \ket{Z^{\sL}_p}.
  \end{aligned}
\end{equation}

\subsubsection{Bi-fundamental structure}
The above left and right $\alg{psu}(1|1)^4$ representations can  be constructed by tensoring the fundamental representation of (\ref{eq:su(1|1)2-repr1},\ref{eq:su(1|1)2-reprR}). To this end, it is sufficient to identify the excitations as
\begin{equation}
\label{eq:mv-tensor}
  \begin{aligned}
    Y^{\sL} = \phi^{\sL} \otimes \phi^{\sL} , \qquad
    \eta^{\sL 1} = \psi^{\sL} \otimes \phi^{\sL} ,\;  \qquad
    \eta^{\sL 2} = \phi^{\sL} \otimes \psi^{\sL} ,\;  \qquad
    Z^{\sL} = \psi^{\sL} \otimes \psi^{\sL} , \\
    Y^{\sR} = {\phi}^{\sR} \otimes {\phi}^{\sR} , \qquad
    \eta^{\sR}_{\ 1} = {\psi}^{\sR} \otimes {\phi}^{\sR} , \qquad
    \eta^{\sR}_{\ 2} = {\phi}^{\sR} \otimes {\psi}^{\sR} , \qquad
    Z^{\sR} = {\psi}^{\sR} \otimes {\psi}^{\sR} ,
  \end{aligned}
\end{equation}
where indices have been appropriately raised and lowered. The left $\alg{psu}(1|1)^4$ module is then isomorphic to~$\varrho_{\sL}\otimes\varrho_{\sL}$, while the right one to~$\varrho_{\sR}\otimes\varrho_{\sR}$.

\subsubsection{Left-right symmetry}
It is clear that consistency with the string theory picture requires considering both the left and right representations\footnote{%
As we will see, this is also required if we want the worlsheet theory to be crossing-invariant.
}.
On the other hand, ``left'' and ``right'' are just labels that can be swapped without affecting the description. This results in a $\mathbb{Z}_2$ \emph{left-right symmetry} (LR-symmetry)~\cite{Borsato:2012ud,Borsato:2013qpa,Sfondrini:2014via}. 
In particular, this relates  left and right massive excitations as
\begin{equation}\label{eq:LR-massive}
Y^{\sL} \longleftrightarrow {Y}^{\sR}, \qquad Z^{\sL} \longleftrightarrow {Z}^{\sR}, \qquad \eta^{\sL \dot{a}} \longleftrightarrow \eta^{\sR}_{\ \dot{a}},
\end{equation}
which is compatible with~\eqref{eq:LR-oscillator-massive}.


\subsection{\texorpdfstring{$\alg{psu}(1|1)^4_{\ce}$}{psu(1|1)**4 c.e.} representations for massless excitations}
\label{sec:repr-massless}
Similarly to the previous subsection, we now deform the massless near-plane-wave representations and show how they also enjoy a tensor-product structure.
We know that massless bosons transform in the fundamental representation of both $\su(2)_{\bullet}$  and  $\su(2)_{\circ}$, while fermions are singlets of~$\su(2)_{\bullet}$ and are in the fundamental of~$\su(2)_{\circ}$. All these indices correspond to the fundamental representation of equation~\eqref{eq:su2-repr}
The action of the supercharges on the massless excitations is
\begin{equation}\label{eq:repr-massless}
\boxed{(\varrho_{\sL}\otimes\widetilde{\varrho}_{\sL})^{\oplus 2}}: \qquad
  \begin{aligned}
    \gen{Q}_{\sL}^{\ \dot{a}} \ket{T^{\dot{b}a}_p}& = \epsilon^{\dot{a}\dot{b}} a_p \ket{\widetilde{\chi}^a_p},
    \qquad
    &\gen{Q}_{\sL}^{\ \dot{a}} \ket{\chi^{a}_p} \;&=  a_p \ket{T^{\dot{a}a}_p}, \\
    \overline{\gen{Q}}{}_{\sL \dot{a}} \ket{\widetilde{\chi}^{a}_p}\;& = -\epsilon_{\dot{a}\dot{b}} \bar{a}_p \ket{T^{\dot{b}a}_p},
    \qquad
    &\overline{\gen{Q}}{}_{\sL \dot{a}} \ket{T^{\dot{b}a}_p} &= \delta_{\dot{a}}^{\ \dot{b}} \bar{a}_p \ket{\chi^a_p}, \\[8pt]
    \gen{Q}_{\sR \dot{a}} \ket{T^{\dot{b}a}_p} &= \delta_{\dot{a}}^{\ \dot{b}} b_p \ket{\chi^a_p},
    \qquad
    &\gen{Q}_{\sR \dot{a}} \ket{\widetilde{\chi}^a_p} \;&= -\epsilon_{\dot{a}\dot{b}} b_p \ket{T^{\dot{b}a}_p}, \\
    \overline{\gen{Q}}{}_{\sR}^{\ \dot{a}} \ket{\chi^a_p}\;& = \bar{b}_p \ket{T^{\dot{a}a}_p},
    \qquad
    &\overline{\gen{Q}}{}_{\sR}^{\ \dot{a}} \ket{T^{\dot{b}a}_p} &= \epsilon^{\dot{a}\dot{b}} \bar{b}_p \ket{\widetilde{\chi}^a_p}.
  \end{aligned}
\end{equation}
Masslessness of the excitations is encoded in the fact that they are  annihilated by~$\gen{M}$, which due to the shortening condition~\eqref{eq:shortening} plays the role of mass. This results in a constraint on the representation coefficients
\begin{equation}
\label{eq:masslessness}
|a_p|^2 = |b_p|^2.
\end{equation}

\subsubsection{Bi-fundamental structure}
The above representation of~$\mathcal{A}$ can be constructed out of two bi-fundamental $\alg{psu}(1|1)^4_{\ce}$ representations. To see this we note that the massless excitations can be re-written as 
\begin{equation}\label{eq:ml-tensor}
  \begin{aligned}
    T^{1a} = \big(\psi^{\sL} \otimes \tilde{\psi}^{\sL}\big)^{a} ,\quad
    \widetilde{\chi}^{a} = \big(\psi^{\sL} \otimes \tilde{\phi}^{\sL}\big)^{a} , \quad
    \chi^{a} = \big(\phi^{\sL} \otimes \tilde{\psi}^{\sL}\big)^{a} , \quad
     T^{2a} = \big(\phi \otimes \tilde{\phi}^{\sL}\big)^{a} , 
  \end{aligned}
\end{equation}
where the two copies are labelled by an $\alg{su}(2)_{\circ}$ index $a=1,2$.
Note that we used two modules of the form $(\varrho_{\sL}\otimes\widetilde{\varrho}_{\sL}) \oplus(\varrho_{\sL}\otimes\widetilde{\varrho}_{\sL})$,  in agreement with the fact that for massless representations the highest weight state is fermionic.

\subsubsection{Equivalent descriptions}
It may appear strange that massless excitations come from left representations only.
Actually, there are several ways to obtain the massless excitations out of tensor product of $\alg{su}(1|1)^2$ modules. Let us introduce rescaled excitations\footnote{%
As we will see in the next subsection, the ratio $\tfrac{a_p}{b_p}$ appearing here is essentially the sign of the momentum, due to the massless condition~\eqref{eq:masslessness}.
}
\begin{equation}
  \label{eq:mless-rescaling}
  \ket{ \cbchi_p^a } = -\frac{a_p}{b_p} \ket{ \widetilde{\chi}_p^a }, \qquad \ket{ \cchi_p^a } = \frac{b_p}{a_p} \ket{ \chi_p^a },
\end{equation}
which gives the following action of the supercharges
\begin{equation}\label{eq:repr-massless2}
\boxed{(\varrho_{\sR}\otimes\widetilde{\varrho}_{\sR})^{\oplus 2}}: \qquad
  \begin{aligned}
    \gen{Q}_{\sL}^{\ \dot{a}} \ket{T^{\dot{b}a}_p} &= -\epsilon^{\dot{a}\dot{b}} b_p \ket{\cbchi^\alpha_p},
    \qquad &
    \gen{Q}_{\sL}^{\ \dot{a}} \ket{\cchi^{a}_p}\ &=  b_p \ket{T^{\dot{a}a}_p}, \\
    \overline{\gen{Q}}{}_{\sL \dot{a}} \ket{\cbchi^{a}_p}\ &= \epsilon_{\dot{a}\dot{b}} \bar{b}_p \ket{T^{\dot{b}a}_p},
    \qquad &
    \overline{\gen{Q}}{}_{\sL \dot{a}} \ket{T^{\dot{b}a}_p} &= \delta_{\dot{a}}^{\ \dot{b}} \bar{b}_p \ket{\cchi^a_p}, \\[8pt]
    \gen{Q}_{\sR \dot{a}} \ket{T^{\dot{b}a}_p} &= \delta_{\dot{a}}^{\ \dot{b}} a_p \ket{\cchi^a_p},
    \qquad &
    \gen{Q}_{\sR \dot{a}} \ket{\cbchi^a_p}\ &= \epsilon_{\dot{a}\dot{b}} a_p \ket{T^{\dot{b}a}_p}, \\
    \overline{\gen{Q}}{}_{\sR}^{\ \dot{a}} \ket{\cchi^a_p}\ &= \bar{a}_p \ket{T^{\dot{a}a}_p},  
    \qquad &
    \overline{\gen{Q}}{}_{\sR}^{\ \dot{a}} \ket{T^{\dot{b}a}_p} &= -\epsilon^{\dot{a}\dot{b}} \bar{a}_p \ket{\cbchi^a_p},
  \end{aligned}
\end{equation}
where we used~\eqref{eq:masslessness}. Thanks to this change of basis, we can identify a \emph{different} tensor product structure, given by
\begin{equation}\label{eq:ml-tensor-R}
  \begin{aligned}
    T_{1a} = \big({\psi}^{\sR} \otimes \widetilde{\psi}^{\sR}\big)_{a} , \quad
    \cbchi_{a} = \big({\phi}^{\sR} \otimes \widetilde{\psi}^{\sR}\big)_{a} , \quad
    \cchi_{a} = \big({\psi}^{\sR} \otimes \widetilde{\phi}^{\sR}\big)_{a} , \quad
    T_{2a} = \big({\phi}^{\sR} \otimes \widetilde{\phi}^{\sR}\big)_{a} .
  \end{aligned}
\end{equation}
This amounts to considering two modules of the form $({\varrho}_{\sR}\otimes\widetilde{\varrho}_{\sR}) \oplus({\varrho}_{\sR}\otimes\widetilde{\varrho}_{\sR})$, \ie to obtain the massless representation out of the massless limit of two right modules, rather than of two left ones.
Yet another possibility is to perform the rescaling~\eqref{eq:mless-rescaling} only on one of the~$\alg{psu}(1|1)^4_{\ce}$ representations, for instance setting
\begin{equation}\label{eq:ml-tensor-LR}
\ket{ \cbchi_p^1 } = \ket{ \widetilde{\chi}_p^1 }, \quad \ket{ \cchi_p^1 } = \ket{ \chi_p^1 },
\qquad
\ket{ \cbchi_p^2 } = -\frac{a_p}{b_p} \ket{ \widetilde{\chi}_p^2 }, \quad \ket{ \cchi_p^2 } = \frac{b_p}{a_p} \ket{ \chi_p^2 },
\end{equation}
on the fermions. In this way, the massless modules have the form $({\varrho}_{\sL}\otimes\widetilde{\varrho}_{\sL}) \oplus({\varrho}_{\sR}\otimes\widetilde{\varrho}_{\sR})$, where the former term in the direct sum corresponds to $a=1$ and the latter to $a=2$.

The conclusion is that any of the two $\alg{psu}(1|1)^4$ modules describing the massless excitations can be equivalently taken to be right or left, up to an inessential change of basis.\footnote{This is true only when~\eqref{eq:masslessness} holds, which is not surprising: if that is not the case, left and right representations have opposite, non-vanishing charge under~$\gen{M}$ and therefore cannot be equivalent.}
Therefore, we can think of our representation as coming from any of these choices depending on what is most convenient, as we will discuss in the next subsection.

\subsubsection{Left-right symmetry}
We have seen that the massive sector should be invariant under exchanging the ``left'' and ``right'' labels. This imposes a discrete symmetry on the S-matrix in that sector~\cite{Borsato:2012ud,Borsato:2013qpa}. When we include massless excitations and consider two-particle states as it is necessary for the S-matrix, a consistency condition arises. One should give a prescription on how the massless excitations transform  when exchanging L$\leftrightarrow$R in the  massive sector. 

Let us start from the case where the massless excitations are in the representation $({\varrho}_{\sL}\otimes\widetilde{\varrho}_{\sL}) \oplus({\varrho}_{\sR}\otimes\widetilde{\varrho}_{\sR})$. Then, a natural extension of LR symmetry is exchanging the L and  R labels in the massless sector too. Following the discussion of the previous subsection, it is clear that those labels are somewhat artificial, because the left and right massless representation are in fact equivalent. 

In what follows we will find it more convenient to describe massless excitations using the $({\varrho}_{\sL}\otimes\widetilde{\varrho}_{\sL}) \oplus({\varrho}_{\sL}\otimes\widetilde{\varrho}_{\sL})$ representation. This is because the action of the $\alg{su}(2)_{\circ}$ raising and lowering operators is simpler in this case (in the mixed case, one finds additional factors of~${a_p}/{b_p}$). 
Even in this case, we can consistently implement LR symmetry, by combining the previous prescription with the change of basis~\eqref{eq:ml-tensor-LR}, see also equation~\eqref{eq:LR-oscillator-massless}. This gives the prescription
\begin{equation}
\label{eq:LR-massless}
\ket{T^{\dot{a}a}} \longleftrightarrow \ket{T_{\dot{a} a}}, \qquad \ket{\widetilde{\chi}^a} \longleftrightarrow +\frac{b_p}{a_p} \ket{\chi_a}, \qquad \ket{\chi^a} \longleftrightarrow -\frac{a_p}{b_p} \ket{\widetilde{\chi}_a}, 
\end{equation}
where we suppressed the momentum label on the excitations for clarity.
Note that the symmetry now depends on the momentum of the particles through the combination $\tfrac{a_p}{b_p}$. We will comment on the interpretation of these ubiquitous coefficients in the next section.

\subsection{Representation coefficients}\label{sec:par-abcd}
The representations we constructed can be labelled by the values of the central charges, in terms of which the representation coefficients~$a_p,b_p$ and their conjugates must be determined. There are essentially two independent charges: the angular momentum~$\gen{M}$, whose absolute value plays the role of the mass, and the central charge~$\gen{C}$ (together with its conjugate~$\overline{\gen{C}}$).
Their eigenvalues on a one-particle off shell representation are
\begin{equation}\label{eq:rep-evalues}
\begin{aligned}
\gen{M} \ket{\mathcal{X}_p} &= m \ket{\mathcal{X}_p}, \qquad \gen{C} \ket{\mathcal{X}_p} &= C_p \ket{\mathcal{X}_p}, \qquad \overline{\gen{C}} \ket{\mathcal{X}_p} &= \bar{C}_p \ket{\mathcal{X}_p}, 
\end{aligned}
\end{equation}
In section~\ref{sec:off-shell-algebra} we have found their expressions in terms of the momentum~$p$ and the coupling constant~$h$ to be
\begin{equation}
m=\begin{cases}
\pm 1&\text{massive}\\
\phantom{+}0&\text{massless}\\
\end{cases}\ ,
\qquad C_p = h \frac{i}{2}(e^{ip}-1)\,\zeta,
\end{equation}
where we explicitly extracted the dependence on the coupling constant~$h$, and $\zeta=e^{2i\xi}$ characterizes the representation. From the shortening condition~\eqref{eq:shortening} we immediately find the dispersion relation
\begin{equation}
E_p = \sqrt{ m^2 + 4 h^2 \sin^2 \frac{p}{2}}\,.
\end{equation}
For $m^2=1$, this dispersion relation, strongly reminiscent of the~$\AdS_5\times\Sphere^5$ case~\cite{Hofman:2006xt}, was first found in this context from the study of giant magnons on $\AdS_3\times \Sphere^3$~\cite{David:2008yk}.
The parametrisation
\begin{equation}
\begin{aligned}
a_p &= \eta_p e^{i\xi}, \qquad \bar{a}_p &= \eta_p e^{-ip/2} e^{-i\xi}, \qquad b_p &= -\frac{\eta_p}{x^-_p} e^{-ip/2} e^{i\xi}, \qquad \bar{b}_p &= -\frac{\eta_p}{x^+_p} e^{-i\xi},
\end{aligned}
\end{equation}
with
\begin{equation}
\eta_p = e^{ip/4}  \sqrt{\frac{ih}{2}(x^-_p - x^+_p)}\,,
\end{equation}
satisfies equation~\eqref{eq:rep-evalues} if the Zhukovski parameters $x^\pm$ satisfy
\begin{equation}
\label{eq:zhukovski}
x^+_p +\frac{1}{x^+_p} -x^-_p -\frac{1}{x^-_p} = \frac{2i \, |m|}{h}, \qquad \frac{x^+_p}{x^-_p}=e^{ip}.
\end{equation}
As for the value of $\xi$, as discussed at length in~\cite{Arutyunov:2006yd} we can take it to vanish on the one-particle representation, but it plays an important role in the multi-particle ones. When constructing the two-particle representation out of the tensor product of two one-particle ones, we see that to reproduce the correct value of the central charges we cannot take $\xi_1=\xi_2=0$ for both of the constituent particles. In fact, imposing that on a two-particle state $\gen{C}$ gives
\begin{equation}
\gen{C} \ket{\mathcal{X}_1 \mathcal{X}_2 } = h \frac{i}{2} (e^{i(p_1+p_2)}-1)\,\ket{\mathcal{X}_1 \mathcal{X}_2 }.
\end{equation}
we see that the parameter~$\xi$ must be non-vanishing in one of the representations, resulting in a non-trivial coproduct.~\cite{Plefka:2006ze,Torrielli:2011zz,Torrielli:2011gg} We solve this condition by setting
\begin{equation}
\xi_1 = 0\,, \qquad \xi_2 = \frac{p_1}{2}\,.
\end{equation}

We see that the case~$m=1$ does not introduce any new feature with respect to the treatment of~\cite{Borsato:2013qpa}. However, at $m=0$ something new happens already at the level of the representation parameters, in terms of the non-analyticities
\begin{equation}
E_p = 2\, h \left|\sin \frac{p}{2}\right|\,,
\qquad
x^{\pm}_p=e^{\pm \frac{i}{2}p}\,\text{sign} \big(\sin \frac{p}{2}\big)\,.
\end{equation}
This is typical in a theory featuring massless excitations, and naturally leads to distinguish left- and right-movers \emph{on the worldsheet}. Indeed the combination
\begin{equation}
\frac{a_p}{b_p}=-\text{sign} \big(\sin \frac{p}{2}\big)\,,
\end{equation}
which appears in the formulation of (target-space) left-right symmetry or in the action of the $\alg{su}(2)_{\circ}$ raising and lowering operators, shows that excitations have different transformation properties depending on the sign of their momentum. A way to consistently get rid of the non-analyticities is to treat worldsheet left- and right-movers as two genuinely different species of particles, with dispersion
\begin{equation}
\label{eq:lr-dispersion}
E_p=\begin{cases}
E_{\text{left}}\,\ =+ 2\, h \sin \frac{p}{2} \qquad & \phantom{-}0< p \leq \tfrac{\pi}{2}\,,\\
E_{\text{right}}=- 2\, h \sin \frac{p}{2} & -\tfrac{\pi}{2}\leq p < 0\,.
\end{cases}
\end{equation}
If we take both massless~$\alg{psu}(1|1)^4_{\ce}$ modules to be in the representation~$\varrho_{\sL}\otimes\widetilde{\varrho}_{\sL}$, we find that worldsheet left-movers (respectively right-movers) transform in a definite representation of~$\alg{su}(2)_{\circ}$. On the other hand, by virtue of~\eqref{eq:LR-massless}, the LR symmetry transformation also swaps worldsheet left- and right-moving fermions. In the representation~$\varrho_{\sL}\otimes\widetilde{\varrho}_{\sR}$, instead, the situation is reversed as can be seen from~\eqref{eq:mless-rescaling}. LR symmetry preserves the worldsheet chirality, while~$\alg{su}(2)_{\circ}$ rotates worldsheet left- and right-moving fermions into each other.

\subsection{Corrections to the massless dispersion relation}
\label{sec:correction-dispersion}

As we have seen in the previous section (see for example equation~\eqref{eq:lr-dispersion}), the massless modes have rather different properties to the massive ones.
One may wonder whether the massless modes could receive corrections which would give them masses in the quantum theory.
 Even in integrable theories, it may happen that quantum corrections dress the particle with a mass,\footnote{%
A prototypical example being the Gross-Neveu model~\cite{Gross:1974jv}.}
and it is interesting to investigate whether this can be the case here. As we are about to show, this is impossible unless part of the~$\alg{so}(4)_2$ invariance is also broken.

We have seen that the massless module is constructed out of two bi-fundamental $\alg{psu}(1|1)^4_{\ce}$ ones. Each of them can be equivalently left (L) or right (R), giving the possibilities
\begin{equation}\begin{aligned}
\text{LL:}\ (\varrho_{\sL}\otimes\widetilde{\varrho}_{\sL}) \oplus(\varrho_{\sL}\otimes\widetilde{\varrho}_{\sL}),
\qquad
\text{RR:}\ (\varrho_{\sR}\otimes\widetilde{\varrho}_{\sR}) \oplus(\varrho_{\sR}\otimes\widetilde{\varrho}_{\sR}),\\
\text{LR:}\ (\varrho_{\sL}\otimes\widetilde{\varrho}_{\sL}) \oplus(\varrho_{\sR}\otimes\widetilde{\varrho}_{\sR}),
\qquad
\text{RL:}\ (\varrho_{\sR}\otimes\widetilde{\varrho}_{\sR}) \oplus(\varrho_{\sL}\otimes\widetilde{\varrho}_{\sL}).
\end{aligned}
\end{equation}
However, only when the representation coefficients satisfy~$|a_p^2|=|b_p|^2$, all of these cases are equivalent, up to a rescaling of the basis vectors. If we think of deforming the representations to switch on a mass $\varepsilon>0$ (coming from e.g. quantum corrections), these four choices will no longer be equivalent, and will lead to different deformations, which we will consider separately.
The non-vanishing masses will appear in particular in the matrix representation of~$\gen{M}$. This will read either
\begin{equation}
\label{eq:masscorrLLRR}
\text{LL/RR}:\ \gen{M}= 
\left(\begin{array}{cccc}
+ \mathbf{1}_4 & 0 & 0 & 0 \\
0 & - \mathbf{1}_4 & 0 & 0 \\
0 &0 & \pm \varepsilon \mathbf{1}_4 & 0  \\
0 &0 &0 & \pm \varepsilon \mathbf{1}_4   
\end{array}\right),
\end{equation}
where we used a block form and the $\pm$ signs correspond to LL/RR, or
\begin{equation}
\label{eq:masscorrLRRL}
\text{LR/RL}:\ \gen{M}= 
\left(\begin{array}{cccc}
+ \mathbf{1}_4 & 0 & 0 & 0 \\
0 & - \mathbf{1}_4 & 0 & 0 \\
0 &0 & \pm \varepsilon \mathbf{1}_4 & 0  \\
0 &0 &0 & \mp \varepsilon \mathbf{1}_4   
\end{array}\right).
\end{equation}
In this notation each block corresponds to a bi-fundamental representation, with the two upper blocks being the massive left and right representations of~\cite{Borsato:2013qpa}.

We can immediately establish that the choice of~\eqref{eq:masscorrLLRR} is incompatible with crossing invariance. In fact, charge conjugation will flip the sign of all the~$\alg{u}(1)$ charges, see also equation~\eqref{eq:crossign-u1-charges} below. Therefore, a necessary condition for crossing invariance is that each representation appears together with its conjugate, which is indeed the case in the massive sector, as one can easily read-off from the first two diagonal entries of $\gen{M}$ above. This is not the case when we choose to deform the LL (or RR) representation for $\varepsilon>0$: both representations will have positive (or negative) eigenvalues of modulus $\varepsilon$.
On the other hand, the choice of equation~\eqref{eq:masscorrLRRL} is compatible with crossing symmetry, but does not respect the~$\alg{so}(4)$ invariance. In fact, all of the supercharges should commute with~$\alg{su}(2)_{\circ}$,  and therefore so should~$\gen{M}$. But $\alg{su}(2)_{\circ}$ rotates the two mass-$\varepsilon$ representations into each other, so that its raising and lowering operators cannot commute with $\gen{M}$ when $\varepsilon\neq0$. This also clearly shows how both obstructions disappear at~$\varepsilon=0$.

\section{S~matrix}
\label{sec:smat}
In the previous section we have constructed the representations of~$\mathcal{A} $ valid for arbitrary values of the momentum and of the coupling constant~$h$, as long as we are in the decompactification limit. The mere fact that the two-particle S~matrix should commute with the off-shell symmetry algebra will allow us to fix it almost completely, up to a small number of functions---the dressing factors. 

In subsection~\ref{sec:Smat-su11} we recall the construction of the~$\alg{su}(1|1)^2_{\ce}$ invariant S~matrices of~\cite{Borsato:2012ud}. By tensoring these in a suitable way, in subsection~\ref{sec:Smat-tensor} we obtain the $\mathcal{A}$-invariant S~matrix up to the dressing factors. In subsection~\ref{sec:unitarity} we show that these dressing factors can in turn be constrained by unitarity. In subsection~\ref{sec:ybe} we show that the S~matrix satisfies the Yang-Baxter equation, and therefore can be used to define an integrable theory. Finally, in subsection~\ref{sec:crossing}, we constrain the dressing factors by crossing symmetry.

The S-matrix scattering two fundamental particles is an operator---in fact, a finite-dimensional matrix---that relates in- and out-states. Schematically,
\begin{equation}
\mathcal{S} \ket{\mathcal{X}^{(in)}_p \mathcal{Y}^{(in)}_q} = \ket{\mathcal{Y}^{(out)}_q \mathcal{X}^{(out)}_p}
\end{equation}
where outgoing momenta are permuted.
Due to the presence of massless excitations, defining a scattering matrix may appear problematic. In the familiar relativistic case, the equivalent of the dispersion relation~\eqref{eq:lr-dispersion} is linear in~$p$. 	Therefore, the group velocity of a wave-packet is
\begin{equation}
v_{\text{rel}}=\frac{\partial E_p}{\partial p}=\pm \text{const}\,,
\end{equation}
\ie, massless relativistic particles move at a constant velocity---the speed of light. Particles with the same worldsheet chirality then cannot scatter, regardless of the value of their  momentum.\footnote{Nevertheless, in the relativistic case, a formal treatment of factorised scattering matrices is still possible~\cite{Zamolodchikov:1992zr,Fendley:1993wq,Fendley:1993xa, Fendley:1993jh}. To this end, it is necessary to introduce appropriate rapidity variables and take suitable limits of them.}
In the case of interest to the present paper however, factorised massless scattering appears to be \emph{simpler} than in the relativistic case. In fact, from equation~\eqref{eq:lr-dispersion} we find the group velocity
\begin{equation}
  v_{\text{non-rel}}=\pm\, h\,\cos\big(\frac{p}{2}\big).
\end{equation}
We see then that massless excitations of different momenta have different velocity, so that we can expect them to scatter in the usual way.
It could be interesting to investigate the near-plane-wave limit of our S~matrix, in which the theory becomes approximately relativistic.

The two-particle S-matrix has to satisfy a number of consistency conditions.
The first requirement, is that $\mathcal{S}$ commutes with all the generators of~${\cal A} $ acting on two-particle excitations
\begin{equation}
\label{eq:smat-commutation}
\mathcal{S}_{(12)}(p,q) \, \gen{Q}_{(12)}(p,q) =\gen{Q}_{(12)}(q,p) \, \mathcal{S}_{(12)}(p,q) .
\end{equation}
Additionally, \emph{braiding unitarity} is the requirement that acting twice with the S-matrix is equivalent to acting with the identity operator
\begin{equation}
\mathcal{S}_{(12)}(q,p)\, \mathcal{S}_{(12)}(p,q) =\mathbf{1}
\end{equation}
and \emph{physical unitarity} demands that $\mathcal{S}$ is unitary as a matrix
\begin{equation}
\mathcal{S}_{(12)}(p,q) \; \big(\mathcal{S}_{(12)}(p,q)\big)^\dagger =  \mathbf{1}.
\end{equation}
Furthermore, as we will see our S-matrix will satisfy the Yang-Baxter equation, meaning that it can consistently be used to define multi-particle scattering. In what follows, it will be useful to define, by means of the permutation matrix~$\Pi$,
\begin{equation}\label{eq:s-versus-s}
\mathbf{S}=\Pi\, \mathcal{S}\,,
\end{equation}
which then satisfies
\begin{equation}
\mathbf{S}_{(12)}(p,q) \, \gen{Q}_{(12)}(p,q) =\gen{Q}_{(21)}(q,p) \, \mathbf{S}_{(12)}(p,q) ,
\qquad
\mathbf{S}_{(21)}(q,p)\, \mathbf{S}_{(12)}(p,q) =\mathbf{1}.
\end{equation}

Conservation of energy and of the other central charges implies that the scattering can be broken down in several sectors, consistently with our discussion of the representations. We naturally find  purely \emph{massive} scattering, which is the one that was already discussed in~\cite{Borsato:2013qpa}, purely \emph{massless} scattering, and finally  \emph{mixed} massive-massless scattering.
In each of these parts we will construct the S~matrix as one that describes the scattering of suitable $\alg{su}(1|1)^4_{\ce}$ modules. In fact, in view of our discussion of these modules, it is convenient to first consider suitable $\alg{su}(1|1)^2_{\ce}$ invariant S~matrices that will serve as building blocks of the full S~matrix.

\subsection{The \texorpdfstring{$\alg{su}(1|1)^2_{\ce}$}{su(1|1)**2 c.e.} invariant S~matrices}
\label{sec:Smat-su11}
In section \ref{sec:rep-su11} we constructed the short representations of $\alg{psu}(1|1)^2$, namely $\varrho_{\sL},\varrho_{\sR},\widetilde{\varrho}_{\sL},\widetilde{\varrho}_{\sR}$.
The scattering of these leads to distinct S~matrices. We start by discussing the case in which both excitations are left ones, which allows for four different S~matrices:
\begin{equation}\label{eq:su(1|1)2-Smat-grad1}
\begin{aligned}
\mathcal{S}^{\sL\sL} \ket{\phi_p^{\sL} \phi_q^{\sL}} &= \phantom{+} A_{pq}^{\sL\sL} \ket{\phi_q^{\sL} \phi_p^{\sL}},
\qquad
&\mathcal{S}^{\sL\sL} \ket{\phi_p^{\sL} \psi_q^{\sL}} &= B_{pq}^{\sL\sL} \ket{\psi_q^{\sL} \phi_p^{\sL}} +  C_{pq}^{\sL\sL} \ket{\phi_q^{\sL} \psi_p^{\sL}}, \\
\mathcal{S}^{\sL\sL} \ket{\psi_p^{\sL} \psi_q^{\sL}} &= \phantom{+} F_{pq}^{\sL\sL} \ket{\psi_q^{\sL} \psi_p^{\sL}},\qquad
&\mathcal{S}^{\sL\sL} \ket{\psi_p^{\sL} \phi_q^{\sL}} &= D_{pq}^{\sL\sL} \ket{\phi_q^{\sL} \psi_p^{\sL}} +  E_{pq}^{\sL\sL} \ket{\psi_q^{\sL} \phi_p^{\sL}},
\end{aligned}
\end{equation}

\begin{equation}\label{eq:su(1|1)2-Smat-grad2}
\begin{aligned}
\mathcal{S}^{\tilde{\sL}\tilde{\sL}} \ket{\tilde{\phi}^{\sL}_p \tilde{\phi}^{\sL}_q} &= -F_{pq}^{\sL\sL} \ket{\tilde{\phi}^{\sL}_q \tilde{\phi}^{\sL}_p}, 
\qquad
&\mathcal{S}^{\tilde{\sL}\tilde{\sL}} \ket{\tilde{\phi}^{\sL}_p \tilde{\psi}^{\sL}_q} &= D_{pq}^{\sL\sL} \ket{\tilde{\psi}^{\sL}_q \tilde{\phi}^{\sL}_p}  -E_{pq}^{\sL\sL} \ket{\tilde{\phi}^{\sL}_q \tilde{\psi}^{\sL}_p}, \\
\mathcal{S}^{\tilde{\sL}\tilde{\sL}} \ket{\tilde{\psi}^{\sL}_p \tilde{\psi}^{\sL}_q} &= -A_{pq}^{\sL\sL} \ket{\tilde{\psi}^{\sL}_q \tilde{\psi}^{\sL}_p}, 
\qquad
&\mathcal{S}^{\tilde{\sL}\tilde{\sL}} \ket{\tilde{\psi}^{\sL}_p \tilde{\phi}^{\sL}_q} &= B_{pq}^{\sL\sL} \ket{\tilde{\phi}^{\sL}_q \tilde{\psi}^{\sL}_p}  -C_{pq}^{\sL\sL} \ket{\tilde{\psi}^{\sL}_q \tilde{\phi}^{\sL}_p},
\end{aligned}
\end{equation}

\begin{equation}\label{eq:su(1|1)2-Smat-grad3}
\begin{aligned}
\mathcal{S}^{{\sL}\tilde{\sL}} \ket{\phi^{\sL}_p \tilde{\phi}^{\sL}_q} &= \phantom{+}B_{pq}^{\sL\sL} \ket{\tilde{\phi}^{\sL}_q \phi^{\sL}_p} -C_{pq}^{\sL\sL} \ket{\tilde{\psi}^{\sL}_q \psi^{\sL}_p},
\qquad
&\mathcal{S}^{{\sL}\tilde{\sL}} \ket{\phi^{\sL}_p \tilde{\psi}^{\sL}_q} &= \phantom{+}A_{pq}^{\sL\sL} \ket{\tilde{\psi}^{\sL}_q \phi^{\sL}_p} , \\
\mathcal{S}^{{\sL}\tilde{\sL}} \ket{\psi^{\sL}_p \tilde{\psi}^{\sL}_q} &= -D_{pq}^{\sL\sL} \ket{\tilde{\psi}^{\sL}_q \psi^{\sL}_p}+E_{pq}^{\sL\sL} \ket{\tilde{\phi}^{\sL}_q \phi^{\sL}_p} ,
\qquad
&\mathcal{S}^{{\sL}\tilde{\sL}} \ket{\psi^{\sL}_p \tilde{\phi}^{\sL}_q} &= -F_{pq}^{\sL\sL} \ket{\tilde{\phi}^{\sL}_q \psi^{\sL}_p} ,
\end{aligned}
\end{equation}

\begin{equation}\label{eq:su(1|1)2-Smat-grad4}
\begin{aligned}
\mathcal{S}^{\tilde{\sL}\sL} \ket{\tilde{\phi}^{\sL}_p \phi^{\sL}_q} &= \phantom{+}D_{pq}^{\sL\sL} \ket{\phi^{\sL}_q \tilde{\phi}^{\sL}_p} + E_{pq}^{\sL\sL} \ket{\psi^{\sL}_q \tilde{\psi}^{\sL}_p},
\qquad
&\mathcal{S}^{\tilde{\sL}\sL} \ket{\tilde{\phi}^{\sL}_p \psi^{\sL}_q} &= -F_{pq}^{\sL\sL} \ket{\psi^{\sL}_q \tilde{\phi}^{\sL}_p} , \\
\mathcal{S}^{\tilde{\sL}\sL} \ket{\tilde{\psi}^{\sL}_p \psi^{\sL}_q} &= -B_{pq}^{\sL\sL} \ket{\psi^{\sL}_q \tilde{\psi}^{\sL}_p}-C_{pq}^{\sL\sL} \ket{\phi^{\sL}_q \tilde{\phi}^{\sL}_p} ,
\qquad
&\mathcal{S}^{\tilde{\sL}\sL} \ket{\tilde{\psi}^{\sL}_p \phi^{\sL}_q} &= \phantom{+}A_{pq}^{\sL\sL} \ket{\phi^{\sL}_q \tilde{\psi}^{\sL}_p} .
\end{aligned}
\end{equation}
As indicated by their labels, the S~matrices above scatter particles in the representations  ${\varrho}_{\sL} \otimes \varrho_{\sL}$, $\widetilde{\varrho}_{\sL} \otimes \widetilde{\varrho}_{\sL}$, $\varrho_{\sL} \otimes \widetilde{\varrho}_{\sL}$ and $\widetilde{\varrho}_{\sL} \otimes \varrho_{\sL}$. 
The four S~matrices are related by simple changes of bases. Their different structures account for the fact that a pair of highest- or lowest-weight states should scatter diagonally.

The ratios $B^{\sL\sL}_{pq}/A^{\sL\sL}_{pq}$,  $C^{\sL\sL}_{pq}/A^{\sL\sL}_{pq}$, $\dotsc$,  $F^{\sL\sL}_{pq}/A^{\sL\sL}_{pq}$ are fixed by \eqref{eq:smat-commutation}. However, that linear relation allows for an arbitrary prefactor in each S~matrix. Such prefactors, until we take non-linear constraints such as unitarity and crossing symmetry into account, are merely a matter of convention. 
The explicit parametrisation of the S-matrix element is given in appendix~\ref{app:smat-param}, with a convention that slightly differs from the original one of~\cite{Borsato:2012ud}.

When the two excitations do not have the same LR flavor we find  S-matrices such as 
\begin{equation}\label{eq:su(1|1)2-Smat-LRgrad1}
\begin{aligned}
\mathcal{S}^{\sL\sR} \ket{\phi^{\sL}_p \phi^{\sR}_q} &= A^{\sL\sR}_{pq} \ket{\phi^{\sR}_q \phi^{\sL}_p} + B^{\sL\sR}_{pq} \ket{\psi^{\sR}_q \psi^{\sL}_p}, \qquad 
&\mathcal{S}^{\sL\sR} \ket{\phi^{\sL}_p \psi^{\sR}_q} &= C^{\sL\sR}_{pq} \ket{\psi^{\sR}_q \phi^{\sL}_p} , \\
\mathcal{S}^{\sL\sR} \ket{\psi^{\sL}_p \psi^{\sR}_q} &= E^{\sL\sR}_{pq} \ket{\psi^{\sR}_q \psi^{\sL}_p}+F^{\sL\sR}_{pq} \ket{\phi^{\sR}_q \phi^{\sL}_p} ,  \qquad 
& \mathcal{S}^{\sL\sR} \ket{\psi^{\sL}_p \phi^{\sR}_q} &= D^{\sL\sR}_{pq} \ket{\phi^{\sR}_q \psi^{\sL}_p} ,
\end{aligned}
\end{equation}

\begin{equation}\label{eq:su(1|1)2-Smat-RLgrad1}
\begin{aligned}
\mathcal{S}^{\sR\sL} \ket{\phi^{\sR}_p \phi^{\sL}_q} &= A^{\sR\sL}_{pq} \ket{\phi^{\sL}_q \phi^{\sR}_p} + B^{\sL\sR}_{pq} \ket{\psi^{\sL}_q \psi^{\sR}_p},  \qquad 
& \mathcal{S}^{\sR\sL} \ket{\phi^{\sR}_p \psi^{\sL}_q} &= C^{\sR\sL}_{pq} \ket{\psi^{\sL}_q \phi^{\sR}_p} , \\
\mathcal{S}^{\sR\sL} \ket{\psi^{\sR}_p \psi^{\sL}_q} &= E^{\sR\sL}_{pq} \ket{\psi^{\sL}_q \psi^{\sR}_p}+F^{\sR\sL}_{pq} \ket{\phi^{\sL}_q \phi^{\sR}_p} ,  \qquad 
& \mathcal{S}^{\sR\sL} \ket{\psi^{\sR}_p \phi^{\sL}_q} &= D^{\sR\sL}_{pq} \ket{\phi^{\sL}_q \psi^{\sR}_p} ,
\end{aligned}
\end{equation}

\begin{equation}\label{eq:su(1|1)2-Smat-LRgrad2}
\begin{aligned}
\mathcal{S}^{\tilde{\sL}\sR} \ket{\tilde{\phi}^{\sL}_p \phi^{\sR}_q} &= +D^{\sL\sR}_{pq} \ket{\phi^{\sR}_q \tilde{\phi}^{\sL}_p},  \qquad 
& \mathcal{S}^{\tilde{\sL}\sR} \ket{\tilde{\phi}^{\sL}_p \psi^{\sR}_q} &= -E^{\sL\sR}_{pq} \ket{\psi^{\sR}_q \tilde{\phi}^{\sL}_p} -F^{\sL\sR}_{pq} \ket{\phi^{\sR}_q \tilde{\psi}^{\sL}_p}, \\
\mathcal{S}^{\tilde{\sL}\sR} \ket{\tilde{\psi}^{\sL}_p \psi^{\sR}_q} &= -C^{\sL\sR}_{pq} \ket{\psi^{\sR}_q \tilde{\psi}^{\sL}_p},  \qquad 
& \mathcal{S}^{\tilde{\sL}\sR} \ket{\tilde{\psi}^{\sL}_p \phi^{\sR}_q} &= +A^{\sL\sR}_{pq} \ket{\phi^{\sR}_q \tilde{\psi}^{\sL}_p} -B^{\sL\sR}_{pq} \ket{\psi^{\sR}_q \tilde{\phi}^{\sL}_p},
\end{aligned}
\end{equation}

\begin{equation}\label{eq:su(1|1)2-Smat-RLgrad2}
\begin{aligned}
\mathcal{S}^{\sR\tilde{\sL}} \ket{\phi^{\sR}_p \tilde{\phi}^{\sL}_q} &= +C^{\sR\sL}_{pq} \ket{\tilde{\phi}^{\sL}_q \phi^{\sR}_p},  \qquad 
& \mathcal{S}^{\sR\tilde{\sL}} \ket{\phi^{\sR}_p \tilde{\psi}^{\sL}_q} &= +A^{\sR\sL}_{pq} \ket{\tilde{\psi}^{\sL}_q \phi^{\sR}_p} - B^{\sR\sL}_{pq} \ket{\tilde{\phi}^{\sL}_q \psi^{\sR}_p}, \\
\mathcal{S}^{\sR\tilde{\sL}} \ket{\psi^{\sR}_p \tilde{\psi}^{\sL}_q} &= -D^{\sR\sL}_{pq} \ket{\tilde{\psi}^{\sL}_q \psi^{\sR}_p},  \qquad 
& \mathcal{S}^{\sR\tilde{\sL}} \ket{\psi^{\sR}_p \tilde{\phi}^{\sL}_q} &= -E^{\sR\sL}_{pq} \ket{\tilde{\phi}^{\sL}_q \psi^{\sR}_p} +  F^{\sR\sL}_{pq} \ket{\tilde{\psi}^{\sL}_q \phi^{\sR}_p}.
\end{aligned}
\end{equation}
The S~matrices above scatter particles in the representations  ${\varrho}_{\sL} \otimes \varrho_{\sR}$, ${\varrho}_{\sR} \otimes \varrho_{\sL}$ and $\widetilde{\varrho}_{\sL} \otimes \varrho_{\sR}$, $\varrho_{\sR} \otimes \widetilde{\varrho}_{\sL}$. The former pair was one of the main results of~\cite{Borsato:2012ud}, while the latter corresponds to the so-called second central extension in appendix D in the same reference.
Note that we could write down four more S~matrices corresponding to scattering processes where one of the excitations is in the representation $\widetilde{\varrho}_{\sR}$. We will not be needing their explicit form, which in any case follows from similar changes of bases.
The S-matrix elements obey a constraint due to left-right symmetry, which simply reads
\begin{equation}\label{eq:rel-s-mat-LR-el-gr}
\begin{aligned}
& A^{\sR\sL}=A^{\sL\sR}, \qquad & B^{\sR\sL}=B^{\sL\sR}, \qquad & C^{\sR\sL}=C^{\sL\sR}, \\
& D^{\sR\sL}=D^{\sL\sR}, \qquad & E^{\sR\sL}=E^{\sL\sR}, \qquad & F^{\sR\sL}=F^{\sL\sR}.
\end{aligned}
\end{equation}
The explicit expression of these S-matrix elements is also given in appendix~\ref{app:smat-param}. The case where we scatter particles in the representations ${\varrho}_{\sR} \otimes \varrho_{\sR}$, $\widetilde{\varrho}_{\sR} \otimes \varrho_{\sR}$, {\it etc.} follows from~equations~\eqref{eq:su(1|1)2-Smat-grad1}-\eqref{eq:su(1|1)2-Smat-grad4} by LR symmetry.

\subsection{The S-matrix from a tensor product}
\label{sec:Smat-tensor}
As discussed in section~\ref{sec:representations-allloop}, the excitations of the $\AdS_3\times\Sphere^3\times\Torus^4$ superstring transform in four bi-fundamental representations of $\alg{psu}(1|1)^4_{\ce}$, two massive (${\varrho}_{\sL} \otimes \varrho_{\sL}$ and ${\varrho}_{\sR} \otimes \varrho_{\sR}$) and two massless ones
(both of the form ${\varrho}_{\sL} \otimes \widetilde{\varrho}_{\sL}$). The corresponding S~matrices can be obtained from graded tensor products of the ones introduced in the previous section, which takes the general form
\begin{equation}
\mathbf{S}_{\alg{psu}(1|1)^4}\approx\mathbf{S}_{\alg{su}(1|1)^2} \;\hat{\otimes}\; \mathbf{S}_{\alg{su}(1|1)^2},
\end{equation}
up to a prefactor.
The tensor product~$\hat{\otimes}$ is graded, \ie it takes into account the signs arising from fermion permutations
\begin{equation}
\label{eq:gradedtensorpr}
\left( \mathbf{A}\,\hat{\otimes}\,\mathbf{B} \right)_{MM',NN'}^{KK',LL'} = (-1)^{\epsilon_{M'}\epsilon_{N}+\epsilon_{L'}\epsilon_{K}} \ \mathbf{A}_{MN}^{KL} \  \mathbf{B}_{M'N'}^{K'L'}\,,
\end{equation}
where the symbol~$\epsilon$ is one for fermions and zero for bosons. We collect explicit expressions for the graded tensor products used in the construction of the S-matrix blocks in appendix~\ref{app:explicitSmatelements}. Each of the $\alg{psu}(1|1)^4_{\ce}$-invariant blocks that we obtain in this way should be multiplied by a (dressing) scalar factor. A priori these may all differ, but as we will see some of those are in fact related by additional symmetries, namely left-right symmetry and~$\alg{su}(2)_{\circ}$.
To describe these blocks it is convenient to split the S-matrix in three sectors: massive, massless, and mixed-mass. Schematically
\begin{equation}
\mathbf{S}=
\left(\begin{array}{cc}
\mathbf{S}^{\bullet\bullet} & \mathbf{S}^{\circ\bullet}\\
\mathbf{S}^{\bullet\circ} & \mathbf{S}^{\circ\circ}
\end{array}\right)\,,
\end{equation}
where $\mathbf{S}^{\bullet\bullet}$ scatters two massive particles, $\mathbf{S}^{\circ\circ}$ scatters two massless ones, and the remaining blocks describe mixed-mass scattering.\footnote{%
The use of the $\bullet$ and $\circ$ symbols here to denote massive and massless excitations respectively is reminiscent of the notation introduced for the algebras $\alg{su}(2)_{\bullet}$ and $\alg{su}(2)_{\circ}$. In fact, only massive fermions are charged under $\alg{su}(2)_{\bullet}$, and all  massless excitations are charged under $\alg{su}(2)_{\circ}$.
}

Below, we will construct the S~matrix block by block, having particular care to keep track of the number of independent scalar factor that we should allow for. We will come back to those factors at the end of the discussion.

\subsubsection{Massive sector \texorpdfstring{($\bullet\bullet$)}{}}
In the massive sector, the  $\alg{psu}(1|1)^4_{\ce}$ modules are~$\varrho_{\sL}{\otimes} \varrho_{\sL}$ and~$\varrho_{\sR}{\otimes} \varrho_{\sR}$, which we  labelled ``left'' and ``right'' depending on their eigenvalues under~$\gen{M}$.  The S~matrix that scatters two left-modules is then
\begin{equation}
\text{left - left:}\quad
\mathbf{S}^{\sL\sL}\,\hat{\otimes}\,\mathbf{S}^{\sL\sL}\,.
\end{equation}
In a similar way, the matrix scattering a left excitation with a right one is 
\begin{equation}
\text{left - right:}\quad
\mathbf{S}^{\sL\sR}\,\hat{\otimes}\,\mathbf{S}^{\sL\sR}\,.
\end{equation}
These two matrices can in principle be multiplied by two arbitrary scalar factors containing the dressing factors, which we call~$\sigma^{\bullet\bullet}$ and~$\widetilde{\sigma}^{\bullet\bullet}$ respectively. The two remaining blocks (RR and RL) can be obtained in a similar way, and moreover are related to the LL and LR blocks by left-right symmetry. In particular, this symmetry constrains the RR and RL scalar factors in terms of~$\sigma^{\bullet\bullet}$ and~$\widetilde{\sigma}^{\bullet\bullet}$.

The massive fermions are also charged under~$\alg{su}(2)_{\bullet}$. It is easy to see that invariance under such transformations is guaranteed by the tensor product structure, so that no additional requirement should be imposed on~$\mathbf{S}^{\bullet\bullet}$. Therefore, the massive sector of our S-matrix is given precisely by the matrix proposed in~\cite{Borsato:2013qpa}, and schematically reads
\begin{equation}
  \mathbf{S}^{\bullet\bullet} = \!\left(\!
    \begin{array}{ccc}
      \sigma^{\bullet\bullet}\; \mathbf{S}^{\sL\sL} \hat{\otimes} \mathbf{S}^{\sL\sL} & & \widetilde{\sigma}^{\bullet\bullet}\; \mathbf{S}^{\sR\sL} \hat{\otimes} \mathbf{S}^{\sR\sL} \\[4pt]
      \widetilde{\sigma}^{\bullet\bullet}\; \mathbf{S}^{\sL\sR} \hat{\otimes} \mathbf{S}^{\sL\sR} & & \sigma^{\bullet\bullet}\; \mathbf{S}^{\sR\sR} \hat{\otimes} \Smat^{\sR\sR}
    \end{array}\!\right), 
\end{equation}

\subsubsection{Mixed-mass sector \texorpdfstring{($\bullet\circ$ and $\circ\bullet$)}{}}
In the mixed-mass sector we scatter one massive particle with one massless one, or vice versa. Let us focus on the former possibility. Massive excitations are given by $\varrho_{\sL}{\otimes} \varrho_{\sL}$ or $\varrho_{\sR}{\otimes} \varrho_{\sR}$, while massless ones consist of two identical modules\footnote{%
Recall that we could equivalently (up to a change of basis) obtain either or both of these two modules from the massless limit of~$\varrho_{\sL}{\otimes} \widetilde{\varrho}_{\sL}$ or~$\varrho_{\sR}{\otimes} \widetilde{\varrho}_{\sR}$. For definiteness, we take the former representation for all massless particles.
}
 of the form~$\varrho_{\sL}{\otimes} \widetilde{\varrho}_{\sL}$, which together form a doublet of $\alg{su}(2)_{\circ}$. Let us consider first the case where the massive particles have left flavor. Then we find two blocks (one for each $\varrho_{\sL}{\otimes} \widetilde{\varrho}_{\sL}$ module), each of the form
\begin{equation}
\label{eq:mssive-mless1}
 \text{massive (left) -  massless:}\quad
\mathbf{S}^{\sL\sL}\,\hat{\otimes}\,\mathbf{S}^{\sL\tilde{\sL}}\,.
\end{equation}
The relative coefficient between the blocks is fixed by $\alg{su}(2)_{\circ}$ action.
 
If instead we started from a right massive particle, we would have found two blocks of the form
\begin{equation}
\label{eq:mssive-mless2}
 \text{massive (right) -  massless:}\quad
\mathbf{S}^{\sR\sL}\,\hat{\otimes}\,\mathbf{S}^{\sR\tilde{\sL}}\,,
\end{equation}
which again form a doublet of $\alg{su}(2)_{\circ}$.

The two pairs of blocks of \eqref{eq:mssive-mless1} and \eqref{eq:mssive-mless2} are related to each other by left-right symmetry. Therefore,  we are left only with a single undetermined scalar factor for the scattering of a massive particle with a massless one, \ie for the whole~$\mathbf{S}^{\bullet\circ}$. We denote the corresponding dressing phase as~$\sigma^{\bullet\circ}$, and we have
\begin{equation}
\mathbf{S}^{\bullet\circ} = \sigma^{\bullet\circ} \; \left[ \left( \mathbf{S}^{\sL\sL} \hat{\otimes} \mathbf{S}^{\sL\stL} \right) \oplus \left( \mathbf{S}^{\sR\sL}  \hat{\otimes} \mathbf{S}^{\sR\stL} \right) \right]^{\oplus 2}.
\end{equation}

If we now consider the scattering of a massless particle with a massive one, analogous considerations yield the~$\mathbf{S}^{\circ\bullet}$ up to a dressing factor~$\sigma^{\circ\bullet}$,
\begin{equation}
\mathbf{S}^{\circ\bullet} = \sigma^{\circ\bullet} \; \left[ \left( \mathbf{S}^{\sL\sL} \hat{\otimes} \mathbf{S}^{\stL\sL} \right) \oplus \left( \mathbf{S}^{\sL\sR} \hat{\otimes} \mathbf{S}^{\stL\sR} \right) \right]^{\oplus 2}\,.
\end{equation} 

\subsubsection{Massless sector \texorpdfstring{($\circ\circ$)}{}}
We are left with the scattering of massless particles, each transforming in two copies of  $\varrho_{\sL}{\otimes} \widetilde{\varrho}_{\sL}$. Constructing the S-matrix for these
$\alg{psu}(1|1)^4_{\ce}$ modules would lead to~16 seemingly unrelated blocks. However, each of the modules is part of a  $\alg{su}(2)_{\circ}$ doublet. As a result, the blocks must arrange themselves in an $\alg{su}(2)_{\circ}$ covariant expression. In other words we may decompose $\mathbf{S}^{\circ\circ}$ as
\begin{equation}
\mathbf{S}_{\alg{su}(2)}\;{\otimes}\;\big(\mathbf{S}^{\sL\sL}\;\hat{\otimes}\;\mathbf{S}^{\tilde{\sL}\tilde{\sL}}\big)\,,
\end{equation}
and the $\alg{su}(2)$ invariant S-matrix takes the familiar form\footnote{%
Recall that in our notation the matrices~$\mathbf{S}$ do not permute the excitations, {\it cf.} equation~\eqref{eq:s-versus-s}.
}
\begin{equation}
\label{eq:su2smat}
\mathbf{S}_{\alg{su}(2)}(p,q)=\frac{1}{1+\varsigma_{pq}}\big(\Pi+\varsigma_{pq} \mathbf{1}\big)\,,
\end{equation}
in terms of the permutation operator~$\Pi$ and of an undetermined function~$\varsigma_{pq}$ which we will constrain later. The action of $\mathbf{S}_{\alg{su}(2)}$ can be represented in a block matrix form, whereby the whole $\mathbf{S}^{\circ\circ}$ takes the form
\begin{equation}
\left(\begin{array}{cccc}
\mathbf{S}^{\sL\sL}\hat{\otimes}\mathbf{S}^{\tilde{\sL}\tilde{\sL}} &	0		&	0		&	0		\\
0		&\frac{\varsigma_{pq}}{1+\varsigma_{pq}}\mathbf{S}^{\sL\sL}\hat{\otimes}\mathbf{S}^{\tilde{\sL}\tilde{\sL}} 	&\frac{1}{1+\varsigma_{pq}}\mathbf{S}^{\sL\sL}\hat{\otimes}\mathbf{S}^{\tilde{\sL}\tilde{\sL}} &	0		\\
0		&\frac{1}{1+\varsigma_{pq}}\mathbf{S}^{\sL\sL}\hat{\otimes}\mathbf{S}^{\tilde{\sL}\tilde{\sL}} &\frac{\varsigma_{pq}}{1+\varsigma_{pq}}\mathbf{S}^{\sL\sL}\hat{\otimes}\mathbf{S}^{\tilde{\sL}\tilde{\sL}} &	0		\\
0		&	0		&	0		&	\mathbf{S}^{\sL\sL}\hat{\otimes}\mathbf{S}^{\tilde{\sL}\tilde{\sL}}		\\
\end{array}\right)
\end{equation}
up to an overall coefficient containing the dressing factor~$\sigma^{\circ\circ}$.

\subsubsection{Normalisation of the sectors}
\label{sec:normalisation}
Before moving on to discuss the non-linear constraints on the S~matrix it is convenient to fix its normalisation. There is some arbitrariness in doing so because, as we mentioned, any of the blocks discussed above can be multiplied by an arbitrary prefactor.

Here we chose these prefactors in such a way as to reproduce scattering elements that are compatible with perturbative results (in the massive sector, where there exists a proposal for the dressing factors~\cite{Borsato:2013hoa}) and make their symmetry properties as manifest as possible. 
In particular, we dictate the form of the following boson-boson processes
\begin{equation}
\begin{aligned}
\bra{Y^{\sL}_q \, Y^{\sL}_p} \mathcal{S} \ket{Y^{\sL}_p \, Y^{\sL}_q} & = \frac{x^+_p}{x^-_p} \, \frac{x^-_q}{x^+_q} \, \frac{x^-_p - x^+_q}{x^+_p - x^-_q} \, \frac{1-\frac{1}{x^-_p x^+_q}}{1-\frac{1}{x^+_p x^-_q}} \, \frac{1}{\left(\sigma^{\bullet\bullet}_{pq} \right)^2 }, \\
\bra{Y^{\sR}_q \, Y^{\sL}_p} \mathcal{S} \ket{Y^{\sL}_p \, Y^{\sR}_q} & = \frac{x^+_p}{x^-_p} \, \frac{x^-_q}{x^+_q} \, \frac{1-\frac{1}{x^+_p x^-_q}}{1-\frac{1}{x^+_p x^+_q}}  \, \frac{1-\frac{1}{x^-_p x^+_q}}{1-\frac{1}{x^-_p x^-_q}} \, \frac{1}{\left(\tilde{\sigma}^{\bullet\bullet}_{pq} \right)^2 }, \end{aligned}
\end{equation}
\begin{equation}
\begin{aligned}
\bra{T^{\dot{a}a}_q \, Y^{\sL}_p} \mathcal{S} \ket{Y^{\sL}_p \, T^{\dot{a}a}_q} 
&=
 \left( \frac{1-\frac{1}{x^+_p x^-_q}}{1-\frac{1}{x^+_p x^+_q}}  \, \frac{1-\frac{1}{x^-_p x^+_q}}{1-\frac{1}{x^-_p x^-_q}} \right)^{1/2} \, \frac{1}{\left(\sigma^{\bullet\circ}_{pq} \right)^2 }, \\
\bra{Y^{\sL}_q \, T^{\dot{a}a}_p} \mathcal{S} \ket{T^{\dot{a}a}_p \, Y^{\sL}_q} 
&=
 \left( \frac{1-\frac{1}{x^+_p x^-_q}}{1-\frac{1}{x^+_p x^+_q}}  \, \frac{1-\frac{1}{x^-_p x^+_q}}{1-\frac{1}{x^-_p x^-_q}} \right)^{1/2} \, \frac{1}{\left(\sigma^{\circ\bullet}_{pq} \right)^2 },
\end{aligned}
\end{equation}
\begin{equation}
\begin{aligned}
\label{eq:unitarity-varsigma}
\bra{T^{\dot{a}a}_q \, T^{\dot{a}a}_p} \mathcal{S} \ket{T^{\dot{a}a}_p \, T^{\dot{a}a}_q} & =  \frac{1}{\left(\sigma^{\circ\circ}_{pq} \right)^2 }.
\end{aligned}
\end{equation}
In the massive sector, this is the same normalisation used in~\cite{Borsato:2013qpa,Borsato:2013hoa}. The necessary prefactors by which the formulae of appendix~\ref{app:smat-param} must be multiplied are given in appendix~\ref{app:dressingfactors}.

\subsection{Physical and braiding unitarity}
\label{sec:unitarity}
We have already mentioned that braiding and physical unitarity are necessary properties for the self-consistency of our construction. It is easy to see that they result in (mild) constrains on the five scalar factors, \ie
\begin{equation}
\begin{aligned}
\sigma^{\bullet\bullet}_{qp}=\big(\sigma^{\bullet\bullet}_{pq}\big)^*=\frac{1}{\sigma^{\bullet\bullet}_{pq}}\,,
\qquad
\tilde{\sigma}^{\bullet\bullet}_{qp}=\big(\tilde{\sigma}^{\bullet\bullet}_{pq}\big)^*=\frac{1}{\tilde{\sigma}^{\bullet\bullet}_{pq}}\,,
\qquad
\sigma^{\circ\circ}_{qp}=\big(\sigma^{\circ\circ}_{pq}\big)^*=\frac{1}{\sigma^{\circ\circ}_{pq}}\,,\\
\qquad
\sigma^{\bullet\circ}_{qp}=\big(\sigma^{\bullet\circ}_{pq}\big)^*=\frac{1}{\sigma^{\circ\bullet}_{pq}}\,,
\qquad\qquad
\sigma^{\circ\bullet}_{qp}=\big(\sigma^{\circ\bullet}_{pq}\big)^*=\frac{1}{\sigma^{\bullet\circ}_{pq}}\,,
\qquad\qquad
\end{aligned}
\end{equation}
together with a constrain on the undetermined function $\varsigma_{pq}$ appearing in the $\alg{su}(2)$ S-matrix,
\begin{equation}
\varsigma_{qp}=\big(\varsigma_{pq}\big)^*=-\varsigma_{pq}\,,
\end{equation}
where ${}^*$ denotes complex conjugation.

\subsection{The Yang-Baxter equation}
\label{sec:ybe}
A necessary condition for our S-matrix to describe an integrable system is that the Yang-Baxter equation (YBE) holds. In a matrix language, this is a cubic equation on a three-particle vector space
\begin{equation}
\mathbf{1}\otimes \mathcal{S}(p,q) \cdot
\mathcal{S}(p,r)\otimes \mathbf{1} \cdot
 \mathbf{1}\otimes\mathcal{S}(q,r)
=
\mathcal{S}(q,r)\otimes \mathbf{1} \cdot
\mathbf{1}\otimes \mathcal{S}(p,r) \cdot
\mathcal{S}(p,q)\otimes \mathbf{1}\,.
\end{equation}

Since our S-matrix decomposes into a tensor-product structure, so does the Yang-Baxter equation, which therefore can be checked directly for the fundamental S-matrices of section~\ref{sec:Smat-su11} together with the~$\alg{su}(2)$ S-matrix of section~\ref{sec:Smat-tensor}. While the former automatically satisfy the YBE, the latter in general does not. In fact, it is well known from the study of the Heisenberg model that, for the $\alg{su}(2)$-invariant S-matrix~\eqref{eq:su2smat} to be integrable, the relative coefficient between between the identity and the permutation operator cannot be arbitrary. In our normalisation, the YBE imposes
\begin{equation}
\varsigma(p,q)-\varsigma(p,r)+\varsigma(q,r)=0\,.
\end{equation}
Therefore, $\varsigma_{pq}$ must be the difference of two appropriately defined rapidities,
\begin{equation}
\label{eq:varsigma-difference}
\varsigma(p,q)=i\big(w_p-w_q\big),
\end{equation}
which together with equation~\eqref{eq:unitarity-varsigma} implies that~$w(p)$ is real.

\subsection{Crossing invariance}
\label{sec:crossing}
Another natural requirement on our S-martrix is crossing invariance~\cite{Janik:2006dc}.
The crossing transformation involves analytic continuation of the S-matrix to an unphysical channel, so that momentum and energy flip signs
\begin{equation}
p\to -p,\qquad \omega_p\to -\omega_p\,.
\end{equation}
In string-theory discussions of crossing it is useful to introduce a complex variable $z$ that takes values in a ``rapidity torus'',\footnote{In this sub-section we use $z$ to denote this rapidity. We trust this causes no confusion with $z_{\underline{i
}}$ used to denote the tranverse $\AdS_3$ massive boson fields used in other sections of this paper.} so that the dispersion relations are uniformised~\cite{Janik:2006dc,Arutyunov:2009kf, Borsato:2013hoa,Sfondrini:2014via}.
The analytic continuation amounts to $z \to z+ \omega_2$, where $\omega_2$ is half of the imaginary period of the torus. In terms of the Zhukovski parameters $x^\pm$ and of the function $\eta$ this gives
\begin{equation}
x^\pm(z+\omega_2) = \frac{1}{x^\pm(z)}, \qquad \eta(z+\omega_2) = \frac{i}{x^+(z)} \eta(z).
\end{equation}

In order to impose crossing symmetry on the S-matrix, one needs to find the matrix~$\mathscr{C}_p$ that implements the crossing transformation on the one-particle states. In general~$\mathscr{C}_p$  is momentum dependent and it turns out that we will need this dependence for massless fermions. The charge conjugation matrix~$\mathscr{C}_p$ acts on the $\alg{su}(2)$ charges as
\begin{equation}
{\gen{J}_{\bullet \dot{b}}}^{\dot{a}}=-\mathscr{C}_p \, {\gen{J}_{\bullet \dot{a}}}^{\dot{b}}\,\mathscr{C}_p^{-1},
\qquad
{\gen{J}_{\circ b}}^a=- \mathscr{C}_p \, {\gen{J}_{\circ a}}^b\,\mathscr{C}_p^{-1} ,
\end{equation}
and flips the sign of the central charges, giving in particular
\begin{equation}
\label{eq:crossign-u1-charges}
\gen{H} = - \mathscr{C}_p\gen{H}\,\mathscr{C}_p^{-1},
\qquad
\gen{M} = - \mathscr{C}_p\gen{M}\,\mathscr{C}_p^{-1}.
\end{equation}
while the supercharges one has
\begin{equation}
\begin{aligned}
\gen{Q}^{\ \dot{a}}_{\sL}(z+\omega_2)^{\st} &= - e^{-\frac{i}{2}\, p}\mathscr{C}(z)\, \gen{Q}^{\ \dot{a}}_{\sL}(z)\, \mathscr{C}^{-1}(z), \\
\gen{Q}_{\sR \dot{a}}(z+\omega_2)^{\st} &= - e^{-\frac{i}{2}\, p}\mathscr{C}(z)\, \gen{Q}_{\sR \dot{a}}(z)\, \mathscr{C}^{-1}(z), \\
\overline{\gen{Q}}{}_{\sL  \dot{a}}(z+\omega_2)^{\st} &= -e^{+\frac{i}{2} p}\, \mathscr{C}(z)\, \overline{\gen{Q}}{}_{\sL \dot{a}}(z)\, \mathscr{C}^{-1}(z),\\
\overline{\gen{Q}}{}_{\sR}^{\  \dot{a}}(z+\omega_2)^{\st} &= -e^{+\frac{i}{2} p}\, \mathscr{C}(z)\, \overline{\gen{Q}}{}_{\sR}^{\  \dot{a}}(z)\, \mathscr{C}^{-1}(z).
\end{aligned}
\end{equation}
Here ${}^\st$ denotes supertransposition, defined as $\gen{Q}^{\st} = \gen{Q}^{\text{t}} \, \Sigma$. The diagonal matrix $\Sigma$ is the fermion-sign matrix, taking values $+1, -1$ on bosons and fermions respectively.

If we work in the basis 
\begin{equation}
\{ Y^{\sL}, \eta^{\sL 1}, \eta^{\sL 2}, Z^{\sL} \} \oplus \{ Y^{\sR}, \eta^{\sR 1}, \eta^{\sR 2}, Z^{\sR} \} \oplus \{ T^{11}, T^{21}, T^{12}, T^{22} \} \oplus \{ \widetilde{\chi}^1, \chi^1, \widetilde{\chi}^2, \chi^2 \} ,
\end{equation}
then the matrix for the crossing transformation can be written as\footnote{The solution for $\mathscr{C}_p$ is not unique, due to the fact that we are dealing with several irreducible representations of the symmetry algebra. Nevertheless the crossing equations that we will derive do not depend on this ambiguity.}
\begin{equation}
\newcommand{\0}{\color{black!40}0}
  \renewcommand{\arraystretch}{1.1}
  \setlength{\arraycolsep}{3pt}
  \mathscr{C}_p=\!\left(\!
    \mbox{\footnotesize$
      \begin{array}{cccc|cccc}
        \0 & \0 & \0 & \0 & 1 & \0 & \0 & \0 \\
        \0 & \0 & \0 & \0 & \0 & \0 & -i & \0 \\
        \0 & \0 & \0 & \0 & \0 & i & \0 & \0 \\
        \0 & \0 & \0 & \0 & \0 & \0 & \0 & 1 \\
        \hline
        1 & \0 & \0 & \0 & \0 & \0 & \0 & \0 \\
        \0 & \0 & i & \0 & \0 & \0 & \0 & \0 \\
        \0 & -i & \0 & \0 & \0 & \0 & \0 & \0 \\
        \0 & \0 & \0 & 1 & \0 & \0 & \0 & \0 \\
      \end{array}$}\!
  \right) 
	\oplus
	\!\left(\!
    \mbox{\footnotesize$
      \begin{array}{cccc|cccc}
        \0 & \0 & \0 & 1 & \0 & \0 & \0 & \0 \\
        \0 & \0 & -1 & \0 & \0 & \0 & \0 & \0 \\
        \0 & -1 & \0 & \0 & \0 & \0 & \0 & \0 \\
        1 & \0 & \0 & \0 & \0 & \0 & \0 & \0 \\
        \hline
        \0 & \0 & \0 & \0 & \0 & \0 & \0 & -i \frac{a_p}{b_p} \\
        \0 & \0 & \0 & \0 & \0 & \0 & i \frac{b_p}{a_p} & \0 \\
        \0 & \0 & \0 & \0 & \0 & i \frac{a_p}{b_p} & \0 & \0 \\
        \0 & \0 & \0 & \0 & -i \frac{b_p}{a_p} & \0 & \0 & \0 \\
      \end{array}$}\!
  \right).
\end{equation}

The crossing equations can be derived in a standard way~\cite{Arutyunov:2009ga,Sfondrini:2014via} and are most simply expressed in terms of the matrix~$\mathbf{S}$, and read
\begin{equation}\label{eq:cr-matrix-form}
\begin{aligned}
\mathscr{C}(z_p) \otimes \mathbf{1} \cdot \mathbf{S}^{\text{t}_1}(z_p+\omega_2,z_q) \cdot \mathscr{C}^{-1}(z_p) \otimes \mathbf{1} \cdot \mathbf{S}(z_p,z_q) &= \mathbf{1} \otimes \mathbf{1},\\
\mathbf{1} \otimes \mathscr{C}^{-1}(z_q) \cdot \mathbf{S}^{\text{t}_2}(z_p,z_q-\omega_2) \cdot  \mathbf{1} \otimes \mathscr{C}(z_q) \cdot \mathbf{S}(z_p,z_q) &= \mathbf{1} \otimes \mathbf{1}.
\end{aligned}
\end{equation}
In fact, taking the symmetry properties of the scalar factors into account, it will be sufficient to consider the crossing equation in either variable, e.g. the first.
Such a matrix equation automatically yields a left-hand side which is proportional to the identity matrix, and constrains the normalisation of certain products of S-matrix elements to be one. In appendix~\ref{app:crossing} we write down such constraints in components. They are equivalent to the following equations for the scalar factors
\begin{equation}\label{eq:cr-massive}
\begin{aligned}
\left(\sigma^{\bullet\bullet}_{pq}\right)^2 \ \left(\tilde{\sigma}^{\bullet\bullet}_{\bar{p}q}\right)^2 &= \left( \frac{x^-_q}{x^+_q} \right)^2 \frac{(x^-_p-x^+_q)^2}{(x^-_p-x^-_q)(x^+_p-x^+_q)} \frac{1-\frac{1}{x^-_px^+_q}}{1-\frac{1}{x^+_px^-_q}}, \\
\left(\sigma^{\bullet\bullet}_{\bar{p}q}\right)^2 \ \left(\tilde{\sigma}^{\bullet\bullet}_{pq}\right)^2 &= \left( \frac{x^-_q}{x^+_q} \right)^2 \frac{\left(1-\frac{1}{x^+_px^+_q}\right)\left(1-\frac{1}{x^-_px^-_q}\right)}{\left(1-\frac{1}{x^+_px^-_q}\right)^2} \frac{x^-_p-x^+_q}{x^+_p-x^-_q},
\end{aligned}
\end{equation}
\begin{equation}\label{eq:cr-mixed}
\begin{aligned}
\left(\sigma^{\bullet \circ}_{\bar{p}q} \right)^2 \ \left( \sigma^{\bullet \circ}_{pq} \right)^2 &= \frac{x^+_p}{x^-_p} \frac{x^-_p-x^+_q}{x^+_p-x^+_q} \frac{1-\frac{1}{x^+_px^+_q}}{1-\frac{1}{x^-_px^+_q}}, \\
\left(\sigma^{\circ \bullet}_{\bar{p}q} \right)^2 \ \left( \sigma^{\circ \bullet}_{pq} \right)^2 &= \frac{x^+_q}{x^-_q} \frac{x^+_p-x^-_q}{x^+_p-x^+_q} \frac{1-\frac{1}{x^+_px^+_q}}{1-\frac{1}{x^+_px^-_q}}
\end{aligned}
\end{equation}
\begin{equation}\label{eq:cr-massless}
\begin{aligned}
\left(\sigma^{\circ\circ}_{\bar{p}q} \right)^2 \ \left(\sigma^{\circ\circ}_{pq} \right)^2 &= \frac{\varsigma_{pq}-1}{\varsigma_{pq}} \, \frac{1-\frac{1}{x^+_p x^+_q}}{1-\frac{1}{x^+_p x^-_q}} \, \frac{1-\frac{1}{x^-_p x^-_q}}{1-\frac{1}{x^-_p x^+_q}},
\end{aligned}
\end{equation}
\begin{equation}\label{eq:cr-varsigma}
\begin{aligned}
\varsigma_{\bar{p}q} &= \varsigma_{pq}-1,
\end{aligned}
\end{equation}
where we indicate the crossed momenta by a bar,
\begin{equation}
\bar{p}=p(z+\omega_2).
\end{equation}
The equations in (\ref{eq:cr-massive}) are the crossing equations for the scalar factors in the massive sector. They were already derived in~\cite{Borsato:2013qpa} and a solution to them was proposed in~\cite{Borsato:2013hoa}.
Equation (\ref{eq:cr-massless}) contraints the scalar factor of the massless sector, while (\ref{eq:cr-mixed}) gives the crossing equations for massive-massless and massless-massive scalar factors.
Finally, if we use the fact that the scalar factor~$\varsigma_{pq}$ is given by the difference of two rapidities as in equation~\eqref{eq:varsigma-difference}, we have that its crossing symmetry simply amounts to the well-known equation
\begin{equation}
w(\bar{p})=w(p)+i\,.
\end{equation}

\section{Discussion and outlook}
\label{sec:conclusion}
In this paper we have given a detailed exposition of the results announced  in~\cite{Borsato:2014exa}. We have found the symmetries of type IIB~$\AdS_3\times\Sphere^3\times\Torus^4 $ superstrings with R-R flux from the light-cone gauge-fixed action, and illustrated how they can be used to fix the exact non-perturbative worldsheet S~matrix up to five crossing-symmetric dressing factors. We have also written down the crossing relations that these dressing factors satisfy.

Our results provide a comprehensive framework for investigating the  $\AdS_3/\CFT_2 $
correspondence using integrability tools. In particular, we have shown how to include the massless modes into the non-perturbative worldsheet S~matrix of the theory, thus solving this long-standing obstacle. 

The next natural step in this investigation is to determine the form of such factors: so far, a proposal~\cite{Borsato:2013hoa} exists only for the ones related to scattering processes in the massive sector, \ie $\sigma^{\bullet\bullet}$ and~$\widetilde{\sigma}^{\bullet\bullet}$. Finding the remaining factors will likely require new insights into the analytic structure of the rapidity curve for massless excitations, as well as  guidance from perturbative calculations~\cite{Sundin:2013ypa}. 

Another interesting direction would be to write down the Bethe-Yang equations describing the asymptotic spectrum and thereby extending the results presented for the massive modes in~\cite{Borsato:2013qpa} to the complete theory. While diagonalising the S~matrix is a relatively straightforward task, it would be very interesting to see how the asymptotic~$\mathcal{N}=(4,4)$ symmetry is realised on the spectrum. 
It will be particularly interesting to see how the AFS phase~\cite{Arutyunov:2004vx}  generalises to the massless modes setting and how the semi-classical features such as finite gap equations~\cite{Kazakov:2004qf,Beisert:2005bm} or Landau-Lifshitz equations~\cite{Kruczenski:2003gt,Kruczenski:2004kw,Hernandez:2004uw,Stefanski:2004cw,Stefanski:2005tr,Stefanski:2007dp} emerge in this setting.  The former would then need to be compared to the recent proposal for the finite-gap equations of~\cite{Lloyd:2013wza}. 

Since the massless dispersion relation is reminiscent of the ones of giant magnons~\cite{Hofman:2006xt}, it is natural to look for similar classical solitonic solution in the massless sector. It appears however that such solutions cannot be straightforwardly constructed only out of bosonic fields, as the related equations of motions are essentially free.

Additionally, note that our description is valid even in the presence of non-trivial winding on the torus because we are in the strict decompactification limit. At the level of the Bethe-Yang equations we should instead be able to distinguish the winding sectors and torus moduli. It would also be very interesting if a similar set of Bethe ansatz equations could be extracted from the dual CFT, perhaps by techniques similar to the ones described in~\cite{Pakman:2009mi}. Further, given the advances in the thermodynamical Bethe Ansatz/quantum spectral curve program~\cite{Ambjorn:2005wa,Arutyunov:2007tc, Arutyunov:2009zu, Gromov:2009tv,Bombardelli:2009ns,Arutyunov:2009ur,Cavaglia:2010nm, Gromov:2014caa}, it would be interesting to see how massless modes will appear in that setting.

Let us remark that the methods presented here should be applicable to more general cases. One is the $\AdS_3\times\Sphere^3\times\Sphere^3\times\Sphere^1 $ background, which is also classically integrable~\cite{Babichenko:2009dk} and whose massive-sector S~matrix and Bethe-Yang equations were found in refs.~\cite{Borsato:2012ud,Borsato:2012ss}. While this case is somewhat more complicated that the $\AdS_3\times\Sphere^3\times\Torus^4$ one---the dual CFT is still to be precisely identified~\cite{Gukov:2004ym,Tong:2014yna}---the presence of so-called {large} $\mathcal{N}=(4,4)$ symmetry yields a rich algebraic structure, which was fruitfully employed in the study of higher spin theories~\cite{Gaberdiel:2013vva}.

It would also be interesting to consider the case where~$\AdS_3$ backgrounds are supported by a mixture of R-R and NS-NS fluxes. These are also classically integrable~\cite{Cagnazzo:2012se}, and interpolate between the pure R-R superstrings described here and supersymmetric WZW models~\cite{Maldacena:2000hw, Maldacena:2000kv, Maldacena:2001km}. Recently, considerable effort was put into studying the massive sector of the mixed-flux $\AdS_3\times\Sphere^3\times\Torus^4$ background, both in terms of their world-sheet S~matrix~\cite{Hoare:2013pma,Hoare:2013ida,Bianchi:2014rfa} and by semi-classical techniques~\cite{Hoare:2013lja,Babichenko:2014yaa}. The study of their off-shell symmetries and representations could shed new light on the structure of the massive sector and of the massless one, which at this time remains quite obscure.

Finally, the prospect of studying deformations and orbifolds of these backgrounds---as it was done for $\AdS_5\times\Sphere^5 $, see \textit{e.g.}~\cite{Zoubos:2010kh,vanTongeren:2013gva} for a review---is extremely appealing, as they include BTZ-like black-hole backgrounds~\cite{Banados:1992wn}, and also appear to be classically integrable~\cite{David:2011iy,David:2012aq}. This may allow to put, for the first time, an integrability handle on string theory in  black-hole backgrounds.

We are confident that we will witness significant progress in these directions in the near future, and we hope to report on some of these topics soon.

\section*{Acknowledgments}
We would like to thank Gleb Arutyunov, Sergey Frolov, Ben Hoare, Alessandro Torrielli, Arkady Tseytlin, Linus Wulff and Kostya Zarembo for discussions. We are particularly grateful to Gleb Arutyunov, Arkady Tseytlin and Kostya Zarembo for their comments on the manuscript.\\
Part of A.S.'s work leading to these results was done during his employment at the Institute for Theoretical Physics of Utrecht University.\\
R.B.\@ acknowledges support by the Netherlands Organization for Scientific Research (NWO) under the VICI grant 680-47-602. 
His work is also part of the ERC Advanced grant research programme No.\ 246974,
 ``Supersymmetry: a window to non-perturbative physics'',
and of the D-ITP consortium, a program of the NWO that is funded by the Dutch Ministry of Education, Culture and Science (OCW).
O.O.S.'s  work was supported by the ERC Advanced grant No.\ 290456,
 ``Gauge theory -- string theory duality''.
A.S.'s work is funded by the People Programme (Marie Curie Actions) of the European Union, Grant Agreement No. 317089 (GATIS).
B.S.\@ acknowledges funding support from an STFC Consolidated Grant  ``Theoretical Physics at City University''
ST/J00037X/1.

\appendix

\section{Index conventions}
\label{app:index-conventions}

In this appendix we collect our index conventions. Indices $\alpha,\beta,\dotsc=\tau,\sigma$ are used for worldsheet coordinates. Indices $m,n,\dotsc=0,\dotsc,9$ are used for spacetime coordinates; the coordinates $m=0,5$ will form the light-cone directions and are denoted as $t$ and $\phi$, respectively. Indices $A,B,\dotsc=0,\dotsc,9$ are used for $\algSO(1,9)$ tangent indices. Indices $I,J,\dotsc=1,2$ denote the two sets of spacetime spinors.

We will often write expressions in $\algSO(4)_1\times \algSO(4)_2$ notation. $\algSO(4)_1$ corresponds to rotations along the $\AdS_3\times\Sphere^3$ directions transverse to the light-cone directions  $t$ and $\phi$. This is not a symmetry of the theory; nevertheless it will be useful to write the theory in terms of this algebra.  $\algSO(4)_2$ corresponds to rotations along $\Torus^4$. Underlined indices will always refer to $\algSO(4)_1$. Indices $\underline{a},\underline{b},\dotsc=1,2$ and $\underline{\dot{a}},\underline{\dot{b}},\dotsc=1,2$ are used for the two Weyl spinors of $\algSO(4)_1$, while $\underline{i},\underline{j},\dotsc=1,\dotsc,4$ are used for the vector of  $\algSO(4)_1$. Further, throughout the paper $\underline{i},\underline{j},\dotsc=1,2$ will denote the two directions of  $\AdS_3$ transverse to $t$; the corresponding coordinates will be denoted as $z_{\underline{i}}$, with the understanding that $z_3\equiv z_4\equiv 0$. Similarly, $\underline{i},\underline{j},\dotsc=3,4$ will denote the two directions of  $\Sphere_3$ transverse to $\phi$;  the corresponding coordinates will be denoted as $y_{\underline{i}}$, with $y_1\equiv y_2\equiv 0$. By a slight abuse of notation we will sometimes write expressions like $\epsilon^{\underline{ij}}z_{\underline{i}}\partial_\alpha z_{\underline{j}}$ or $\epsilon^{\underline{ij}}y_{\underline{i}}\partial_\alpha y_{\underline{j}}$ with the understanding that
\begin{equation}
\epsilon^{\underline{ij}}z_{\underline{i}}\partial_\alpha z_{\underline{j}}\equiv z_1\partial_\alpha z_2-z_2\partial_\alpha z_1\,,
\qquad
\epsilon^{\underline{ij}}y_{\underline{i}}\partial_\alpha y_{\underline{j}}\equiv y_3\partial_\alpha y_4-y_4\partial_\alpha y_3\,.
\end{equation}
Indices $a,b,\dotsc=1,2$ and $\dot{a},\dot{b},\dotsc=1,2$ are used for the two Weyl spinors of $\algSO(4)_2$ while $i,j,\dotsc=1,\dotsc,4$ are used for the vector of  $\algSO(4)_2$. We raise and lower the spinor indices using epsilon symbols which we normalize by
\begin{equation}
  \epsilon^{12} = - \epsilon_{12} = +1.
\end{equation}

\section{Spinor and Gamma matrix conventions}
\label{sec:gamma-matrix-conventions}

For $\AdS_3$ and $\Sphere^3$ we consider the three-dimensional gamma matrices\footnote{%
  Our conventions are the same as those of~\cite{Babichenko:2009dk}, except for the definition of $\gamma^0$ and $\gamma^2$.
}%
\begin{equation}
  \gamma^0 = -i\sigma_3 , \quad
  \gamma^1 = \sigma_1 , \quad
  \gamma^2 = \sigma_2 ,\quad
  \gamma^3 = \sigma_1 , \quad
  \gamma^4 = \sigma_2 , \quad
  \gamma^5 = \sigma_3 .
\end{equation}
We further define
\begin{equation}
  \gamma^6 = \sigma_1 , \quad
  \gamma^7 = \sigma_2 , \quad
  \gamma^8 = \sigma_3 .
\end{equation}
The ten-dimensional gamma matrices are then given by
\begin{equation}
    \newcommand{\I}{\mathrlap{\,\mathds{1}}\hphantom{\sigma_a}}
    \newcommand{\II}{\mathrlap{\,\mathds{1}}\hphantom{\gamma^A}}
  \begin{aligned}
    \Gamma^A &= +\sigma_1 \otimes \sigma_2 \otimes \gamma^A \otimes \II \otimes \II , & A&= 0,1,2, \\
    \Gamma^A &= +\sigma_1 \otimes \sigma_1 \otimes \II \otimes \gamma^A \otimes \II , & A&= 3,4,5, \\
    \Gamma^A &= +\sigma_1 \otimes \sigma_3 \otimes \II \otimes \II \otimes \gamma^A , & A&= 6,7,8, \\
    \Gamma^9 &= -\sigma_2 \otimes \I \otimes \II \otimes \II \otimes \II .
  \end{aligned}
\end{equation}
We then have
\begin{equation}
    \newcommand{\I}{\mathrlap{\,\mathds{1}}\hphantom{\sigma_a}}
  \begin{aligned}
    \Gamma^{05} &= \phantom{i} {-}\I \otimes \sigma_3  \otimes \sigma_3 \otimes \sigma_3 \otimes \I , \\
    \Gamma^{012} &= \phantom{i} {+} \sigma_1 \otimes \sigma_2  \otimes \I \otimes \I \otimes \I , \\
    \Gamma^{345} &= +i \sigma_1 \otimes \sigma_1  \otimes \I \otimes \I \otimes \I , \\
    \Gamma^{012345} &= +\phantom{i} \I \otimes \sigma_3  \otimes \I \otimes \I \otimes \I , \\
    \Gamma^{1234} &= -\phantom{i} \I \otimes  \I \otimes  \sigma_3 \otimes  \sigma_3 \otimes  \I , \\
    \Gamma^{6789} &= +\phantom{i} \sigma_3 \otimes \sigma_3 \otimes \I \otimes \I \otimes \I , \\
    \Gamma = \Gamma^{0123456789} &= \phantom{i} {+} \sigma_3 \otimes \I \otimes \I \otimes \I \otimes \I .
  \end{aligned}
\end{equation}
The gamma matrices satisfy
\begin{equation}
  (\Gamma^A)^t = - T \Gamma^A T^{-1} , \qquad
  (\Gamma^A)^\dag = - C \Gamma^A C^{-1} , \qquad
  (\Gamma^A)^* = + B \Gamma^A B^{-1} , \qquad
\end{equation}
where
\begin{equation}
  T = -i\sigma_2 \otimes \sigma_2 \otimes \sigma_2 \otimes \sigma_2 \otimes \sigma_2 , \qquad
  C = \Gamma^0 , \qquad
  B = -\Gamma^0 \, T .
\end{equation}
It is useful to note the relations
\begin{equation}
  \begin{gathered}
    T^\dag T = C^\dag C = B^\dag B = 1 , \qquad
    B^t = T C^\dag , \\
    T^\dag = - T = + T^t , \qquad
    C^\dag = - C = + C^t , \qquad
    B^\dag = + B = + B^t , \\
    T = - \Gamma^{01479} , \qquad
    C = -i \sigma_1 \otimes \sigma_2 \otimes \sigma_3 \otimes \mathds{1} \otimes \mathds{1} , \\
    B = + \sigma_3 \otimes \mathds{1} \otimes \sigma_1 \otimes \sigma_2 \otimes \sigma_2 = - \Gamma^{1479} , \\
    B \Gamma B^\dag = \Gamma^* .
  \end{gathered}
\end{equation}
The Majorana spinors satisfy the conditions
\begin{equation}
  \theta^* = B \theta , \qquad
  \bar{\theta} = \theta^\dag C = \theta^t T .
\end{equation}

\section{Killing spinors and a preferred choice of vielbeins}
\label{app:Killing-spinors}

In this appendix we collect some of the computational details that are useful for the calculation of Killing spinors done in section~\ref{sec:Killing-spinors}. We begin by 
presenting solutions of the Killing spinor equation on $\Sphere^3$ and $\AdS_3$ and, using these, we construct Killing spinors in the full $\AdS_3 \times \Sphere^3 \times \Torus^4$ geometry. In parallel with this construction, we also introduce a particular choice of vielbeins for the geometries in question. Such a choice is of course in some sense arbitrary and can be gauged away. However, many of the detailed computations performed in this paper simplify significantly in this frame.

\paragraph{Killing spinors on $\Sphere^3$.}

The $\Sphere^3$ metric~\eqref{eq:s3metric}
can be written in terms of a diagonal dreibein\footnote{%
  We denote the tangent space directions for $\Sphere^3$ with $A=3,4,5$.
}%
\begin{equation}
\label{eq:diag-s3-vielbein}
  E_m{}^A =
  \begin{pmatrix}
    \frac{1}{1 + \frac{y_3^2 + y_4^2}{4}} & 0 & 0 \\
    0 & \frac{1}{1 + \frac{y_3^2 + y_4^2}{4}} & 0 \\
    0 & 0 & \frac{1 - \frac{y_3^2 + y_4^2}{4}}{1 + \frac{y_3^2 + y_4^2}{4}}
  \end{pmatrix}
\end{equation}
and the spin connection
\begin{equation}
  \begin{aligned}
    \omega_{y_3 \, AB} &=
    \frac{1}{2}
    \begin{pmatrix}
      0 & - \frac{y_4}{1 + \frac{y_3^2 + y_4^2}{4}} & 0 \\
      + \frac{y_4}{1 + \frac{y_3^2 + y_4^2}{4}} & 0 & 0 \\
      0 & 0 & 0
    \end{pmatrix} ,
    \\
    \omega_{y_4 \, AB} &=
    \frac{1}{2}
    \begin{pmatrix}
      0 & + \frac{y_3}{1 + \frac{y_3^2 + y_4^2}{4}} & 0 \\
      - \frac{y_3}{1 + \frac{y_3^2 + y_4^2}{4}} & 0 & 0 \\
      0 & 0 & 0
    \end{pmatrix} ,
    \\
    \omega_{\phi \, AB} &=
    \begin{pmatrix}
      0 & 0 & + \frac{y_3}{1 + \frac{y_3^2 + y_4^2}{4}} \\
      0 & 0 & + \frac{y_4}{1 + \frac{y_3^2 + y_4^2}{4}} \\
      - \frac{y_3}{1 + \frac{y_3^2 + y_4^2}{4}} & - \frac{y_4}{1 + \frac{y_3^2 + y_4^2}{4}} & 0
    \end{pmatrix} .
  \end{aligned}
\end{equation}
The $\Sphere^3$ Killing spinors satisfy~\cite{Lu:1998nu}
\begin{equation}
  \label{eq:S3-Killing-spinor-eq}
  \partial_m \eta_{\Sphere^3}^I + \frac{1}{4} \omega_m^{AB} \gamma_{AB} \eta_{\Sphere^3}^I + \frac{i}{2} E_m{}^A \gamma_A \sigma_3^{IJ} \eta_{\Sphere^3}^J = 0.
\end{equation}
These equations are solved by
\begin{equation}
  \begin{aligned}
    \tilde{\eta}_{\Sphere^3}^1 &=
    \frac{1}{\sqrt{1 + \frac{y_3^2 + y_4^2}{4}}} \Bigl( 1 - \frac{i y_3}{2} \gamma^3 - \frac{i y_4}{2} \gamma^4 \Bigr) \, e^{-\frac{i\phi}{2} \gamma^5} \eta_0^1
    \equiv \hat{M}_{\Sphere^3} \eta_0^1 ,
    \\
    \tilde{\eta}_{\Sphere^3}^2 &=
    \frac{1}{\sqrt{1 + \frac{y_3^2 + y_4^2}{4}}} \Bigl( 1 + \frac{i y_3}{2} \gamma^3 + \frac{i y_4}{2} \gamma^4 \Bigr) \, e^{+\frac{i\phi}{2} \gamma^5} \eta_0^1 
    \equiv \check{M}_{\Sphere^3} \eta_0^1 ,
  \end{aligned}
\end{equation}
where $\eta_0^I$ are constant spinors with two complex components.

Let us consider the first of these solutions. We note that
\begin{equation}
\label{eq:orth-rot-s3}
  \hat{M}_{\Sphere^3}^{-1} \, \gamma_A \, \hat{M}_{\Sphere^3} \, E_m^A = \gamma_A \, \hat{\mathcal{M}}^A{}_B E_m^B ,\qquad
  \check{M}_{\Sphere^3}^{-1} \, \gamma_A \, \check{M}_{\Sphere^3} \, E_m^A = \gamma_A \, \check{\mathcal{M}}^A{}_B E_m^B ,
\end{equation}
where $\hat{\mathcal{M}}^{AB}$ and  $\check{\mathcal{M}}^{AB}$ are orthogonal matrices. Using these matrices we can introduce  new dreibeins, obtained by a rotation in tangent space
\begin{equation}
  \hat{K}_m^A = \hat{\mathcal{M}}^A{}_B \, E_m{}^B \,, \qquad 
  \check{K}_m{}^A = \check{\mathcal{M}}^A{}_B E_m{}^B \,.
\end{equation}
The components of the inverse dreibein are given by a fairly compact expression and can be written in the factorized form
\begin{equation}
  \begin{aligned}
\label{eq:E-components-S3}
  \hat{K}_A{}^m &=
  \begin{pmatrix}
    + \cos\phi & + \sin\phi & 0 \\
    - \sin\phi & + \cos\phi & 0 \\
    0 & 0 & 1
  \end{pmatrix}
  \begin{pmatrix}
    1 + \frac{y_3^2 - y_4^2}{4} & + \frac{y_3 y_4}{2} & -\frac{y_4}{1 - \frac{y_3^2 + y_4^2}{4}} \\
    + \frac{y_3 y_4}{2} & 1 - \frac{y_3^2 - y_4^2}{4} & +\frac{y_3}{1 - \frac{y_3^2 + y_4^2}{4}} \\
    + y_4 & - y_3 & 1
  \end{pmatrix} \,,
\\
  \check{K}_A{}^m &=
  \begin{pmatrix}
    + \cos\phi & - \sin\phi & 0 \\
    + \sin\phi & + \cos\phi & 0 \\
    0 & 0 & 1
  \end{pmatrix}
  \begin{pmatrix}
    1 + \frac{y_3^2 - y_4^2}{4} & + \frac{y_3 y_4}{2} & +\frac{y_4}{1 - \frac{y_3^2 + y_4^2}{4}} \\
    + \frac{y_3 y_4}{2} & 1 - \frac{y_3^2 - y_4^2}{4} & -\frac{y_3}{1 - \frac{y_3^2 + y_4^2}{4}} \\
    - y_4 & + y_3 & 1
  \end{pmatrix} .
  \end{aligned}
\end{equation}
Like all vielbeins, the $\hat{K}_n{}^A$ are covariantly constant
\begin{equation}\label{eq:covariant-constant-E}
  \hat{D}_m \hat{K}_n{}^A =
  \partial_m \hat{K}_n{}^A - \Gamma^k_{mn} \hat{K}_k{}^A + \hat{\omega}_m{}^A{}_B \hat{K}_n^B = 0 \,.
\end{equation}
Here, $\hat{\omega}_{mAB}$ is the spin connection in the rotated tangent space. $\hat{K}_m{}^A$ and $\hat{\omega}_{mAB}$ further satisfy the relation
\begin{equation}\label{eq:symmetric-relation-E-omega}
  \hat{\omega}_m{}^A{}_B \hat{K}_n{}^B + \hat{\omega}_n{}^A{}_B \hat{K}_m{}^B = 0 \,,
\end{equation}
which is \emph{not} true for generic vielbeins and spin-connections.
Equations~\eqref{eq:covariant-constant-E} and~\eqref{eq:symmetric-relation-E-omega} together give the relation
\begin{equation}\label{eq:killing-vector-s3}
  \hat{D}_m \hat{K}_n{}^A + \hat{D}_n \hat{K}_m{}^A
  = \partial_m \hat{K}_n{}^A + \partial_n \hat{K}_m{}^A - 2 \Gamma^k_{mn} \hat{K}_k{}^A = 0 .
\end{equation}
This is the Killing vector equation; similar equations hold for $\check{K}_n{}^A$.

The vectors $\hat{K}_m{}^A$ and $\check{K}_m{}^A$ together generate the $\algSO(4) = \algSU(2) \oplus \algSU(2)$ isometry algebra of $\Sphere^3$,
\begin{equation}\label{eq:s3-isometries}
  \comm{\hat{K}_A}{\hat{K}_B} = +2\epsilon_{AB}{}^C \hat{K}_C , \qquad
  \comm{\check{K}_A}{\check{K}_B} = -2\epsilon_{AB}{}^C \check{K}_C , \quad
  \comm{\hat{K}_A}{\check{K}_B} = 0 .
\end{equation}

\paragraph{Killing spinors on $\AdS_3$.}

We consider the $\AdS_3$ metric
\begin{equation}
  ds^2_{\AdS^3} = -\Bigl(\frac{1 + \frac{z_1^2 + z_2^2}{4}}{1 - \frac{z_1^2 + z_2^2}{4}}\Bigr)^2 dt^2 + \Bigl(\frac{1}{1 - \frac{z_1^2 + z_2^2}{4}}\Bigr)^2 ( dz_1^2 + dz_2^2 ) ,
\end{equation}
with the diagonal dreibein\footnote{%
  For $\AdS_3$ we denote the tangent space directions $A=0,1,2$.
}%
\begin{equation}
\label{eq:diag-ads3-vielbein}
  E_m{}^A =
  \begin{pmatrix}
    \frac{1 + \frac{z_1^2 + z_2^2}{4}}{1 - \frac{z_1^2 + z_2^2}{4}} & 0 & 0 \\
    0 & \frac{1}{1 - \frac{z_1^2 + z_2^2}{4}} & 0 \\
    0 & 0 & \frac{1}{1 - \frac{z_1^2 + z_2^2}{4}}
  \end{pmatrix} .
\end{equation}
The $\AdS_3$ Killing spinors satisfy\cite{Lu:1996rhb,Lu:1998nu}
\begin{equation}
  \label{eq:AdS3-Killing-spinor-eq}
  \partial_\alpha \epsilon_{\AdS_3}^I + \frac{1}{4} \omega_\alpha^{AB} \gamma_{AB} \epsilon_{\AdS_3}^I + \frac{1}{2} E_\alpha{}^A \gamma_A \sigma_3^{IJ} \epsilon_{\AdS_3}^J = 0,
\end{equation}
These equations have the solutions
\begin{equation}
  \begin{aligned}
    \tilde{\epsilon}_{\AdS_3}^1 &=
    \frac{1}{\sqrt{1 - \frac{z_1^2 + z_2^2}{4}}} \Bigl( 1 - \frac{z_1}{2} \gamma^1 - \frac{z_2}{2} \gamma^2 \Bigr) \, e^{+\frac{t}{2} \gamma^0} \epsilon_0^1 
    \equiv \hat{M}_{\AdS_3} \epsilon_0^1 ,
    \\
    \tilde{\epsilon}_{\AdS_3}^2 &=
    \frac{1}{\sqrt{1 - \frac{z_1^2 + z_2^2}{4}}} \Bigl( 1 + \frac{z_1}{2} \gamma^1 + \frac{z_2}{2} \gamma^2 \Bigr) \, e^{-\frac{t}{2} \gamma^0} \epsilon_0^2
    \equiv \check{M}_{\AdS_3} \epsilon_0^2 ,
  \end{aligned}
\end{equation}
where $\epsilon_0^I$ are constant spinors.

Like we did for the $\Sphere^3$ we introduce matrices $\hat{\mathcal{M}}^A{}_B$ and $\check{\mathcal{M}}^A{}_B$
\begin{equation}
\label{eq:orth-rot-ads3}
  \hat{M}_{\AdS_3}^{-1} \, \gamma_A \, \hat{M}_{\AdS_3} \, E_m^A = \gamma_A \, \hat{\mathcal{M}}^A{}_B E_m^B ,
  \qquad
  \check{M}_{\AdS_3}^{-1} \, \gamma_A \, \check{M}_{\AdS_3} \, E_m^A = \gamma_A \, \check{\mathcal{M}}^A{}_B E_m^B ,
\end{equation}
These matrices are orthogonal with respect to a metric of signature $(-1,+1,+1)$. 
The rotated dreibeins are defined by
\begin{equation}
  \hat{K}_m{}^A = \hat{\mathcal{M}}^A{}_B E_m{}^B ,
  \qquad
  \check{K}_m{}^A = \check{\mathcal{M}}^A{}_B E_m{}^B ,
\end{equation}
are their inverses can be written in components as
\begin{equation}\label{eq:E-components-AdS3}
  \begin{aligned}
    \hat{K}_A{}^m &=
    \begin{pmatrix}
      1 & 0 & 0 \\
      0 & + \cos t & + \sin t \\
      0 & - \sin t & + \cos t
    \end{pmatrix}
    \begin{pmatrix}
      + 1 & + z_2 & - z_1 \\
      + \frac{z_2}{1 + \frac{z_1^2 + z_2^2}{4}} & 1 - \frac{z_1^2 - z_2^2}{4} & - \frac{z_1 z_2}{2} \\
      - \frac{z_1}{1 + \frac{z_1^2 + z_2^2}{4}} & - \frac{z_1 z_2}{2} & 1 + \frac{z_1^2 - z_2^2}{4}
    \end{pmatrix} ,
    \\
    \check{K}_A{}^m &=
    \begin{pmatrix}
      1 & 0 & 0 \\
      0 & + \cos t & - \sin t \\
      0 & + \sin t & + \cos t
    \end{pmatrix}
    \begin{pmatrix}
      + 1 & - z_2 & + z_1 \\
      - \frac{z_2}{1 + \frac{z_1^2 + z_2^2}{4}} & 1 - \frac{z_1^2 - z_2^2}{4} & - \frac{z_1 z_2}{2} \\
      + \frac{z_1}{1 + \frac{z_1^2 + z_2^2}{4}} & - \frac{z_1 z_2}{2} & 1 + \frac{z_1^2 - z_2^2}{4}   
    \end{pmatrix} .
  \end{aligned}
\end{equation}
Like in the $\Sphere^3$ case these dreibeins are Killing vectors satisfying the $\algSO(2,2) = \algSL(2) \oplus \algSL(2)$ algebra
\begin{equation}\label{eq:ads3-isometries}
  \comm{\hat{K}_A}{\hat{K}_B} = +2\epsilon_{AB}{}^C \hat{K}_C ,
  \qquad
  \comm{\check{K}_A}{\check{K}_B} = -2\epsilon_{AB}{}^C \check{K}_C ,
  \qquad
  \comm{\hat{K}_A}{\check{K}_B} = 0 .
\end{equation}

\paragraph{Killing spinors on $\AdS_3 \times \Sphere^3 \times \Torus^4$.}

The above constructions of Killing spinors on $\AdS_3$ and $\Sphere^3$ can readily be used to show that the spinors~\eqref{eq:full-ks-soln} satisfy the Killing spinor equations~\eqref{eq:AdS3-S3-T4-Killing-spinor-eq}. Writing the Killing spinors in the penta-spinor notation used to define the gamma matrices in appendix~\ref{sec:gamma-matrix-conventions}, we have
\begin{equation}
  \begin{aligned}
    \varepsilon^1
    &=
    \begin{pmatrix} 1 \\ 0 \end{pmatrix}
    \otimes
    \begin{pmatrix} 0 \\ 1 \end{pmatrix}
    \otimes \tilde{\epsilon}_{\AdS_3}^1
    \otimes \tilde{\eta}_{\Sphere^3}^1
    \otimes \psi_0^1
    =
    \hat{M} \Bigl[
    \begin{pmatrix} 1 \\ 0 \end{pmatrix}
    \otimes
    \begin{pmatrix} 0 \\ 1 \end{pmatrix}
    \otimes \epsilon_0^1
    \otimes \eta_0^1
    \otimes \psi_0^1
    \Bigr] ,
    \\
    \varepsilon^2
    &=
    \begin{pmatrix} 1 \\ 0 \end{pmatrix}
    \otimes
    \begin{pmatrix} 0 \\ 1 \end{pmatrix}
    \otimes \tilde{\epsilon}_{\AdS_3}^2
    \otimes \tilde{\eta}_{\Sphere^3}^2
    \otimes \psi_0^2
    =
    \check{M} \Bigl[
    \begin{pmatrix} 1 \\ 0 \end{pmatrix}
    \otimes
    \begin{pmatrix} 0 \\ 1 \end{pmatrix}
    \otimes \epsilon_0^2
    \otimes \eta_0^2
    \otimes \psi_0^2
    \Bigr] ,
  \end{aligned}
\end{equation}
where $\epsilon_0^I$, $\eta_0^I$ and $\psi_0^I$ are constant two-component spinors. In this basis, the matrices $\hat{M}$ and $\check{M}$ are given by
\begin{equation}
  \hat{M} = \mathds{1} \otimes \mathds{1} \otimes \hat{M}_{\AdS_3} \otimes \hat{M}_{\Sphere^3} \otimes \mathds{1} ,
  \qquad
  \check{M} = \mathds{1} \otimes \mathds{1} \otimes \check{M}_{\AdS_3} \otimes \check{M}_{\Sphere^3} \otimes \mathds{1} ,
\end{equation}
and satisfy
\begin{equation}
  \hat{M}^t T = \hat{M}^{-1} \, T , \quad
  \check{M}^t T = \check{M}^{-1} \, T , \quad
  \hat{M}^\dag \Gamma^0 = \hat{M}^{-1} \, \Gamma^0 , \quad
  \check{M}^\dag \Gamma^0 = \check{M}^{-1} \, \Gamma^0 .
\end{equation}
One can easily check that the Killing spinors $\varepsilon^I$ are chiral in the 5+1 and 9+1 dimensional sense
\begin{equation}
  \frac{1}{2} ( 1 + \Gamma^{012345} ) \tilde{\epsilon}^I = 0 ,
  \qquad
  \frac{1}{2} ( 1 - \Gamma^{0123456789} ) \tilde{\epsilon}^I = 0 \,.
\end{equation}

Again we introduce orthogonal matrices $\hat{\mathcal{M}}^{AB}$ and $\check{\mathcal{M}}^{AB}$ satisfying
\begin{equation}\label{eq:MN-identities}
  \hat{M}^{-1} \Gamma^A \hat{M} = \Gamma^B \hat{\mathcal{M}}_B{}^A ,
  \qquad
  \check{M}^{-1} \Gamma^A \check{M} = \Gamma^B \check{\mathcal{M}}_B{}^A .
\end{equation}
These matrices are block diagonal,
\begin{equation}
\label{eq:orth-rot-ads3-s3}
  \hat{\mathcal{M}} = \hat{\mathcal{M}}_{\AdS_3} \oplus \hat{\mathcal{M}}_{\Sphere^3} \oplus \mathds{1}_4 ,
  \qquad
  \check{\mathcal{M}} = \check{\mathcal{M}}_{\AdS_3} \oplus \check{\mathcal{M}}_{\Sphere^3} \oplus \mathds{1}_4 \,,
\end{equation}
and are used in equation~\eqref{eq:def-of-hat-check-vielbeins} to define a preferred set of vielbeins $\hat{K}_m{}^A $ and $\check{K}_m{}^A$ used in much of the paper.
The tangent space rotations above only affect the $\AdS_3 \times \Sphere^3$ directions $A=0,\dotsc,5$ but leave the $\Torus^4$ directions $i=6,\dotsc,9$ untouched.
It is therefore convenient 
to introduce the contractions
\begin{equation}
  \label{eq:various-vielbeins}\!\!
  \slashed{\bar{E}}_m = \sum_{A=0}^5 E_m{}^A \Gamma_A , \quad
  \slashed{\hat{K}}_m = \sum_{A=0}^5 \hat{K}_m{}^A \Gamma_A , \quad
  \slashed{\check{K}}_m = \sum_{A=0}^5 \check{K}_m{}^A \Gamma_A , \quad
  \slashed{\dot{E}}_m = \sum_{i=6}^9 E_m{}^i \Gamma_i , \!
\end{equation}
of the various vielbeins in the two subspaces.

After performing the tangent space rotations we find new spin connections $\hat{\omega}_{mAB}$ and $\check{\omega}_{mAB}$, and corresponding covariant derivatives $\hat{D}_m$ and $\check{D}_m$ satisfying
\begin{equation}
  \hat{D}_m 
  \equiv \hat{M}^{-1} \, D_m \, \hat{M}
  = \hat{M}^{-1} \bigl( \partial_m + \frac{1}{4} \omega_m^{AB} \Gamma_{AB} \bigr) \hat{M}
  = \partial_m + \frac{1}{4} \hat{\omega}_m^{AB} \Gamma_{AB} ,
\end{equation}
and similar for $\check{D}_m$. The spin connections can be written as
\begin{equation}
  \begin{aligned}
    \frac{1}{4} \slashed{\wh}_m &= - \frac{1}{4} \slashed{\hat{K}}_m ( \Gamma^{012} + \Gamma^{345} ) - \frac{1}{4} ( \Gamma^{012} + \Gamma^{345} ) \slashed{\hat{K}}_m  ,
    \\
    \frac{1}{4} \slashed{\wc}_m &= + \frac{1}{4} \slashed{\check{K}}_m ( \Gamma^{012} + \Gamma^{345} ) + \frac{1}{4} ( \Gamma^{012} + \Gamma^{345} ) \slashed{\check{K}}_m  ,
  \end{aligned}
\end{equation}
which leads to
\begin{equation}
  \begin{aligned}
    \partial_m + \frac{1}{4} \slashed{\omega}_m + \frac{1}{24} \slashed{F} \slashed{E}_m
    &=
    \hat{M} \Bigl(
    \partial_m - \frac{1}{4}  \bigl( \slashed{\hat{K}}_m + \slashed{\dot{E}}_m \bigl) \Gamma^{012} \bigl( 1 + \Gamma^{012345} \bigr)
    \Bigr) \hat{M}^{-1} ,
    \\
    \partial_m - \frac{1}{4} \slashed{\omega}_m + \frac{1}{24} \slashed{F} \slashed{E}_m
    &=
    \check{M} \Bigl(
    \partial_m + \frac{1}{4}  \bigl ( \slashed{\check{K}}_m + \slashed{\dot{E}}_m \bigr) \Gamma^{012} \bigl( 1 + \Gamma^{012345} \bigr)
    \Bigr) \check{M}^{-1} .
  \end{aligned}
\end{equation}

\section{Proof of the identity~(\ref{eq:epsilon-NME})}
\label{app:proof-wz-identity}

To prove the identity~\eqref{eq:epsilon-NME} we write
\begin{equation}
  \epsilon^{\alpha\beta} \partial_\alpha \bigl(
  \check{M}^{-1} \hat{M} \slashed{\hat{K}}_\beta
  \bigr) (1 - \Gamma^{012345})
  =
  \epsilon^{\alpha\beta}
  \partial_\alpha X^m \partial_\beta X^n \Bigl[
  \partial_m \bigl(\check{M}^{-1} \slashed{E}_n \hat{M} \bigr)
  \Bigr] (1 - \Gamma^{012345}) .
\end{equation}
The expression in the square brackets is antisymmetrized in $m$ and $n$. By multiplying it from the left and the right by $\check{M}$ and $\hat{M}^{-1}$ we get
\begin{equation}
  \check{M} \partial_{[m} \bigl( \check{M}^{-1} \slashed{\bar{E}}_{n]} \hat{M} \bigr) \hat{M}^{-1}
  =
  \partial_{[m} \slashed{\bar{E}}_{n]}
  + \bigl( \check{M} \partial_{[m} \check{M}^{-1} \bigr) \slashed{\bar{E}}_{n]}
  + \slashed{\bar{E}}_{[m} \bigl( \hat{M} \partial_{m]} \hat{M}^{-1} \bigr) .
\end{equation}
Using the relations
\begin{equation}
  \frac{1}{4} \slashed{\omega}_m 
  = \frac{1}{4} \hat{M} \slashed{\wh}_m \hat{M}^{-1} + \hat{M} \partial_m \hat{M}^{-1}
  = \frac{1}{4} \check{M} \slashed{\wc}_m \check{M}^{-1} + \check{M} \partial_m \check{M}^{-1}
\end{equation}
we find
\begin{equation}
  \begin{aligned}
    \check{M} \partial_{[m} \bigl( \check{M}^{-1} \slashed{\bar{E}}_{n]} \hat{M} \bigr) \hat{M}^{-1}
    &=
    \partial_{[m} \slashed{\bar{E}}_{n]}
    + \frac{1}{4} \bigl( \slashed{\omega}_{[m} \slashed{\bar{E}}_{n]} + \slashed{\bar{E}}_{[m} \slashed{\omega}_{n]} \bigr)
    \\ &\phantom{{}= \partial_{[m} \slashed{\bar{E}}_{n]}}
    - \frac{1}{4} \check{M} \slashed{\wc}_{[m} \slashed{\check{K}}_{n]} \check{M}^{-1}
    - \frac{1}{4} \hat{M} \slashed{\hat{K}}_{[m} \slashed{\wh}_{n]} \hat{M}^{-1} .
  \end{aligned}
\end{equation}
The first term on the right-hand side can be shown to be zero using the covariant constancy of the vielbein.
Using the expressions for the spin connection in section~\ref{app:Killing-spinors} we can check that the second term is proportional to the projector $(1 + \Gamma^{0123456})$. Hence the expression appearing in~\eqref{eq:epsilon-NME} vanishes.


\section{Useful identities for \texorpdfstring{$\algSO(4)$}{SO(4)} gamma matrices}
\label{sec:SO4-gamma-identities}

The following identities involving blocks of the $\algSO(4)$ gamma matrices $\hat{\gamma}^{\underline{i}}$ and $\hat{\tau}^i$ are useful
\begin{equation*}
  \begin{aligned}
    \gamma^{\underline{i}} \tilde{\gamma}^{\underline{j}} &= + \delta^{\underline{ij}} + \gamma^{\underline{ij}} , \qquad &
    \tau^i \tilde{\tau}^j &= - \delta^{ij} + \tau^{ij} ,
    \\
    \tilde{\gamma}^{\underline{i}} \gamma^{\underline{j}} &= + \delta^{\underline{ij}} + \gamma^{\underline{ij}} , \qquad &
    \tilde{\tau}^i \tau^j &= - \delta^{ij} + \tau^{ij} ,
  \end{aligned}
\end{equation*}
\begin{equation}
  \label{eq:useful-cg-relations}
  \begin{aligned}
    \gamma^{\underline{i}} \tilde{\gamma}^{\underline{j}} \gamma^{\underline{k}} &= + \epsilon^{\underline{ijkl}} \gamma_{\underline{l}} + \delta^{\underline{ij}} \gamma^{\underline{k}} - \delta^{\underline{ik}} \gamma^{\underline{j}} + \delta^{\underline{jk}} \gamma^{\underline{i}} ,
    \\
    \tilde{\gamma}^{\underline{i}} \gamma^{\underline{j}} \tilde{\gamma}^{\underline{k}} &= - \epsilon^{\underline{ijkl}} \tilde{\gamma}_{\underline{l}} + \delta^{\underline{ij}} \tilde{\gamma}^{\underline{k}} - \delta^{\underline{ik}} \tilde{\gamma}^{\underline{j}} + \delta^{\underline{jk}} \tilde{\gamma}^{\underline{i}} ,
  \end{aligned}
\end{equation}
\begin{equation*}
  \begin{aligned}
    \gamma^{\underline{i}} \tilde{\gamma}^{\underline{j}} \gamma^{\underline{k}} \tilde{\gamma}^{\underline{l}} &= \delta^{\underline{jk}} \delta^{\underline{il}} - \delta^{\underline{jl}} \delta^{\underline{ik}} + \delta^{\underline{kl}} \delta^{\underline{ij}} + \epsilon^{\underline{ijkl}}
    - \epsilon^{\underline{jklm}} \gamma^{\underline{i}}{}_{\underline{m}} + \delta^{\underline{jk}} \gamma^{\underline{il}} - \delta^{\underline{jl}} \gamma^{\underline{ik}} + \delta^{\underline{kl}} \gamma^{\underline{ij}} ,
    \\
    \tilde{\gamma}^{\underline{i}} \gamma^{\underline{j}} \tilde{\gamma}^{\underline{k}} \gamma^{\underline{l}} &= \delta^{\underline{jk}} \delta^{\underline{il}} - \delta^{\underline{jl}} \delta^{\underline{ik}} + \delta^{\underline{kl}} \delta^{\underline{ij}} - \epsilon^{\underline{ijkl}}
    + \epsilon^{\underline{jklm}} \tilde{\gamma}^{\underline{i}}{}_{\underline{m}} + \delta^{\underline{jk}} \tilde{\gamma}^{\underline{il}} - \delta^{\underline{jl}} \tilde{\gamma}^{\underline{ik}} + \delta^{\underline{kl}} \tilde{\gamma}^{\underline{ij}} ,
  \end{aligned}
\end{equation*}
\begin{equation*}
  \tilde{\tau}^k \tau^{ij} = \epsilon^{kijl} \tilde{\tau}_l - \delta^{ki} \tilde{\tau}^j + \delta^{kj} \tilde{\tau}^i \,.
\end{equation*}
We also use the relations
\begin{equation}
\label{eq:useful-taus}
  (\tilde{\tau}^i)^{\dot{a}}{}_a \epsilon^{ab} (\tilde{\tau}^j)^{\dot{b}}{}_b
  =
  - \delta^{ij} \epsilon^{\dot{a}\dot{b}} + (\tilde{\tau}^{ij})^{\dot{a}}{}_{\dot{d}} \, \epsilon^{\dot{d}\dot{b}} ,
  \qquad
  (\tilde{\tau}^i)^{\dot{a}}{}_a \, (\tilde{\tau}^i)^{\dot{b}}{}_b
  =
  + 2 \epsilon^{\dot{a}\dot{b}} \epsilon_{ab} .
\end{equation}

\section{Relations between \texorpdfstring{$\Gamma^A$}{Gamma} and \texorpdfstring{$\algSO(4)_1\oplus \algSO(4)_2$}{SO(4) + SO(4)} gamma matrices}
\label{sec:relations-large-small-gamma-matrices}

Let us relate the action on the fermions of the ten dimensional gamma matrices of appendix~\ref{sec:gamma-matrix-conventions} with the action of the $\algSO(4)_1 \oplus \algSO(4)_2$ gamma matrices introduced in section~\ref{sec:gf-action-with-bisponors}.
The chiral spinors $\eta_I$ and $\chi_I$ have eigenvalue $+1$ under $\Gamma^{0123456789}$. Here we write the action of the gamma matrices that preserve this chirality. Performing the change of basis~\eqref{eq:gamma-matrix-basis-change} we find the decomposition
\begin{equation}
  \newcommand{\I}{\mathrlap{\,\mathds{1}}\hphantom{\sigma_a}}
  \begin{aligned}
    T \Gamma^0 
    &= - \phantom{i} \sigma_3 \otimes \I \otimes \sigma_1 \otimes \sigma_2 \otimes \sigma_2
    \to - \begin{pmatrix} + \epsilon & 0 \\ 0 & + \epsilon \end{pmatrix} \otimes \begin{pmatrix} + \epsilon & 0 \\ 0 & + \epsilon \end{pmatrix}
    \\
    T \Gamma^1
    &= + i \sigma_3 \otimes \I \otimes \sigma_3 \otimes \sigma_2 \otimes \sigma_2
    \to - i \begin{pmatrix} 0 & +\epsilon \gamma^1 \\ -\epsilon \tilde{\gamma}^1 & 0 \end{pmatrix} \otimes \begin{pmatrix} + \epsilon & 0 \\ 0 & + \epsilon \end{pmatrix} ,
    \\
    T \Gamma^2
    &= - \phantom{i} \sigma_3 \otimes \I \otimes \I \otimes \sigma_2 \otimes \sigma_2
    \to - i \begin{pmatrix} 0 & +\epsilon \gamma^2 \\ -\epsilon \tilde{\gamma}^2 & 0 \end{pmatrix} \otimes \begin{pmatrix} + \epsilon & 0 \\ 0 & + \epsilon \end{pmatrix} ,
    \\
    T \Gamma^3
    &= + \phantom{i} \sigma_3 \otimes \sigma_3 \otimes \sigma_2 \otimes \sigma_3 \otimes \sigma_2
    \to - i \begin{pmatrix} 0 & +\epsilon \gamma^3 \\ -\epsilon \tilde{\gamma}^3 & 0 \end{pmatrix} \otimes \begin{pmatrix} + \epsilon & 0 \\ 0 & - \epsilon \end{pmatrix} ,
    \\
    T \Gamma^4
    &= + i \sigma_3 \otimes \sigma_3 \otimes \sigma_2 \otimes \I \otimes \sigma_2
    \to - i \begin{pmatrix} 0 & +\epsilon \gamma^4 \\ -\epsilon \tilde{\gamma}^4 & 0 \end{pmatrix} \otimes \begin{pmatrix} + \epsilon & 0 \\ 0 & - \epsilon \end{pmatrix} ,
    \\
    T \Gamma^5
    &= - \phantom{i} \sigma_3 \otimes \sigma_3 \otimes \sigma_2 \otimes \sigma_1 \otimes \sigma_2
    \to + \begin{pmatrix} +\epsilon & 0 \\ 0 & -\epsilon \end{pmatrix} \otimes \begin{pmatrix} + \epsilon & 0 \\ 0 & - \epsilon \end{pmatrix} ,
    \\
    T \Gamma^6
    &= - \phantom{i} \sigma_3 \otimes \sigma_1 \otimes \sigma_2 \otimes \sigma_2 \otimes \sigma_3
    \to - i \begin{pmatrix} +\epsilon \gamma^{34} & 0 \\ 0 & -\epsilon \tilde{\gamma}^{34} \end{pmatrix} \otimes \begin{pmatrix} 0 & - \epsilon \tau^6  \\ + \epsilon \tilde{\tau}^6 & 0 \end{pmatrix} ,
    \\
    T \Gamma^7
    &= - i \sigma_3 \otimes \sigma_1 \otimes \sigma_2 \otimes \sigma_2 \otimes \I
    \to - i \begin{pmatrix} +\epsilon \gamma^{34} & 0 \\ 0 & -\epsilon \tilde{\gamma}^{34} \end{pmatrix} \otimes \begin{pmatrix} 0 & - \epsilon \tau^7  \\ + \epsilon \tilde{\tau}^7 & 0 \end{pmatrix} ,
    \\
    T \Gamma^8
    &= + \phantom{i} \sigma_3 \otimes \sigma_1 \otimes \sigma_2 \otimes \sigma_2 \otimes \sigma_1
    \to - i \begin{pmatrix} +\epsilon \gamma^{34} & 0 \\ 0 & -\epsilon \tilde{\gamma}^{34} \end{pmatrix} \otimes \begin{pmatrix} 0 & - \epsilon \tau^8  \\ + \epsilon \tilde{\tau}^8 & 0 \end{pmatrix} ,
    \\
    T \Gamma^9
    &= + i \I \otimes \sigma_2 \otimes \sigma_2 \otimes \sigma_2 \otimes \sigma_2
    \to - i \begin{pmatrix} +\epsilon \gamma^{34} & 0 \\ 0 & -\epsilon \tilde{\gamma}^{34} \end{pmatrix} \otimes \begin{pmatrix} 0 & - \epsilon \tau^9  \\ + \epsilon \tilde{\tau}^9 & 0 \end{pmatrix} ,
  \end{aligned}
\end{equation}
where the arrows indicate a restriction to the upper left $16 \times 16$ block.
We further find
\begin{equation}
  \newcommand{\I}{\mathrlap{\,\mathds{1}}\hphantom{\sigma_a}}
  \begin{aligned}
    T \Gamma^{012}
    &= - \sigma_3 \otimes \I \otimes \sigma_2 \otimes \sigma_2 \otimes \sigma_2
    \to + \begin{pmatrix} +\epsilon \gamma^{34} & 0 \\ 0 & -\epsilon \tilde{\gamma}^{34} \end{pmatrix} \otimes \begin{pmatrix} +\epsilon & 0 \\ 0 & + \epsilon \end{pmatrix} ,
    \\
    T \Gamma^{345}
    &= - \sigma_3 \otimes \sigma_3 \otimes \sigma_2 \otimes \sigma_2 \otimes \sigma_2
    \to + \begin{pmatrix} +\epsilon \gamma^{34} & 0 \\ 0 & -\epsilon \tilde{\gamma}^{34} \end{pmatrix} \otimes \begin{pmatrix} +\epsilon & 0 \\ 0 & - \epsilon \end{pmatrix} ,
      \end{aligned}
\end{equation}
and
\begin{equation}
  \newcommand{\I}{\mathrlap{\,\mathds{1}}\hphantom{\sigma_a}}
  \begin{aligned}
    \Gamma^{1234}
    &= - \I \otimes \I \otimes \sigma_3 \otimes \sigma_3 \otimes \I
    \to + \sigma_3 \otimes \I \otimes \I \otimes \I ,
    \\
    \Gamma^{6789}
    &= + \sigma_3 \otimes \sigma_3 \otimes \I \otimes \I \otimes \I
    \to +  \I \otimes \I \otimes \sigma_3 \otimes \I ,
    \\
    \Gamma^{05}
    &= - \I \otimes \sigma_3 \otimes \sigma_3 \otimes \sigma_3 \otimes \I
    \to +  \sigma_3 \otimes \I \otimes \sigma_3 \otimes \I .
  \end{aligned}
\end{equation}
It is also useful to note that
\begin{equation}
  \begin{aligned}
    \Gamma^{12} &\to +\hat{\gamma}^{12} \otimes \mathds{1} , \\
    \Gamma^{\underline{ij}} &\to +\hat{\gamma}^{\underline{ij}} \otimes \hat{\tau}^{6789} , && \text{for $\underline{i}=1,2$, $\underline{j}=3,4$} , \\
    \Gamma^{34} &\to +\hat{\gamma}^{34} \otimes \mathds{1} , \\
    \Gamma^{ij} &\to - \mathds{1} \otimes \hat{\tau}^{ij} \,.
  \end{aligned}
\end{equation}

\section{\texorpdfstring{$\lagr_{\text{WZ}}$}{LWZ} in \texorpdfstring{$\algSO(4)_1\oplus \algSO(4)_2$}{SO(4) + SO(4)} components}
\label{sec:Lwz-components}

In this appendix we write down the expression for $\mathcal{L}_{\text{WZ}}$ in terms of  $\algSO(4)_1\oplus \algSO(4)_2$ bispinors introduced in section~\ref{sec:gf-action-with-bisponors}. In order to do this notice first that the matrix $M_0$, when written after the basis change~\eqref{eq:gamma-matrix-basis-change}, is of the form
\begin{equation}
M_0= \mathds{1}_2\otimes m_0\otimes \mathds{1}_4\,,
\end{equation}
where $m_0$ is a $4\times 4$ matrix. In other words it acts non-trivially only on  $\algSO(4)_1$. Explicitly we find
\begin{equation}
  m_0 =
  \frac{1}{\sqrt{\bigl(1 - \frac{z^2}{4} \bigr)\bigl(1 + \frac{y^2}{4}\bigr)}}
  \biggl( 1 -
  \frac{i \epsilon^{\underline{ij}} z_{\underline{i}}}{2}
  \begin{pmatrix}
    0 & + \gamma_{\underline{j}} \\ - \tilde{\gamma}_{\underline{j}} & 0
  \end{pmatrix}
  \biggr)
  \biggl( 1 -
  \frac{i \epsilon^{\underline{kl}} y_{\underline{k}}}{2} 
  \begin{pmatrix}
    0 & +\gamma_{\underline{l}} \\ +\tilde{\gamma}_{\underline{l}} & 0
  \end{pmatrix}
  \biggr)
\end{equation}
and it is also useful to note
\begin{equation}
  m_0^2 = 
  \begin{pmatrix}
    \frac{\bigl(1 + \frac{z^2}{4} \bigr)\bigl(1 - \frac{y^2}{4}\bigr) - \epsilon^{\underline{ij}} \epsilon^{\underline{kl}} z_{\underline{i}} y_{\underline{k}} \gamma_{\underline{jl}}}{\bigl(1 - \frac{z^2}{4} \bigr)\bigl(1 + \frac{y^2}{4}\bigr)} &
    -i \frac{\bigl( 1 + \frac{z^2}{4} \bigr) \epsilon^{\underline{ij}} y_{\underline{i}} \gamma_{\underline{j}} + \bigl( 1 - \frac{y^2}{4} \bigr) \epsilon^{\underline{ij}} z_{\underline{i}} \gamma_{\underline{j}}}{\bigl(1 - \frac{z^2}{4} \bigr)\bigl(1 + \frac{y^2}{4}\bigr)} \\
    -i \frac{\bigl( 1 + \frac{z^2}{4} \bigr) \epsilon^{\underline{ij}} y_{\underline{i}} \tilde{\gamma}_{\underline{j}} - \bigl( 1 - \frac{y^2}{4} \bigr) \epsilon^{\underline{ij}} z_{\underline{i}} \tilde{\gamma}_{\underline{j}}}{\bigl(1 - \frac{z^2}{4} \bigr)\bigl(1 + \frac{y^2}{4}\bigr)} &
    \frac{\bigl(1 + \frac{z^2}{4} \bigr)\bigl(1 - \frac{y^2}{4}\bigr) + \epsilon^{\underline{ij}} \epsilon^{\underline{kl}} z_{\underline{i}} y_{\underline{k}} \tilde{\gamma}_{\underline{jl}}}{\bigl(1 - \frac{z^2}{4} \bigr)\bigl(1 + \frac{y^2}{4}\bigr)}
  \end{pmatrix} .
\end{equation}
The above matrices are written in block form, with the blocks having the same index structure as that given in equation~\eqref{eq:so(4)-gamma-matrices}.  Using this block structure we define
\begin{equation}
  m_0^2\equiv\begin{pmatrix}
    (m_0^2)^{\underline{a}}{}_{\underline{b}} & (m_0^2)^{\underline{a}}{}_{\underline{\dot{b}}} \\ 
    (m_0^2)^{\underline{\dot{a}}}{}_{\underline{b}} & (m_0^2)^{\underline{\dot{a}}}{}_{\underline{\dot{b}}} 
  \end{pmatrix}\,.
\end{equation}
It is straightforward to check that
\begin{equation}
  m_0^{-2}\equiv\begin{pmatrix}
    \phantom{+} (m_0^2)^{\underline{a}}{}_{\underline{b}} & -(m_0^2)^{\underline{a}}{}_{\underline{\dot{b}}} \\ 
    -(m_0^2)^{\underline{\dot{a}}}{}_{\underline{b}} & \phantom{+} (m_0^2)^{\underline{\dot{a}}}{}_{\underline{\dot{b}}} \end{pmatrix}\,.
\end{equation}

Equipped with these observations, it is now straightforward though laborious to express 
$\mathcal{L}_{\text{WZ}}$ given in equation~\eqref{eq:lwzgfix}, in terms of $\algSO(4)_1\oplus \algSO(4)_2$ bispinor fermions,
\begin{align}\label{eq:L-wz-bispinor}
  \mathcal{L}_{\text{WZ}} =  \epsilon^{\alpha\beta} 
  \Bigl(
  & 2i\check{E}^+_\alpha\bar{\eta}_2 m_0^2 \partial_\beta \eta_1+ 2i\hat{E}^+_\alpha\bar{\eta}_1 m_0^2 \partial_\beta \eta_2
  +2i\check{E}^+_\alpha\bar{\chi}_2 m_0^2 \partial_\beta \chi_1+ 2i\hat{E}^+_\alpha\bar{\chi}_1 m_0^2 \partial_\beta \chi_2
  \nonumber \\
  &+\!\sum_{\underline{i}=1,2}\bigl(
  \hat{E}_\alpha^{\underline{i}} \bar{\eta}_1 \tilde{\gamma}^{\underline{i}} m_0^2 \partial_\beta \eta_2
  -\check{E}_\alpha^{\underline{i}} \bar{\eta}_2 \tilde{\gamma}^{\underline{i}} m_0^2 \partial_\beta \eta_1
  +\check{E}_\alpha^{\underline{i}} \bar{\chi}_2 \gamma^{\underline{i}} m_0^2 \partial_\beta \chi_1
  -\check{E}_\alpha^{\underline{i}} \bar{\chi}_1 \gamma^{\underline{i}} m_0^2 \partial_\beta \chi_2
  \bigr)
  \nonumber \\
  &
  +\!\sum_{\underline{i}=3,4}\bigl(
  \check{E}_\alpha^{\underline{i}} \bar{\eta}_2 \tilde{\gamma}^{\underline{i}} m_0^2 \partial_\beta \eta_1
  -\hat{E}_\alpha^{\underline{i}} \bar{\eta}_1 \tilde{\gamma}^{\underline{i}} m_0^2 \partial_\beta \eta_2
  +\check{E}_\alpha^{\underline{i}} \bar{\chi}_2 \gamma^{\underline{i}} m_0^2 \partial_\beta \chi_1
  -\check{E}_\alpha^{\underline{i}} \bar{\chi}_1 \gamma^{\underline{i}} m_0^2 \partial_\beta \chi_2
  \bigr)
  \nonumber \\
  &-2
  E_\alpha^i \bigl(\bar{\chi}_2\gamma^{34}m_0^2\tau^i\partial_\beta\eta_1
  -\bar{\chi}_1 \gamma^{34}m_0^2\tau^i\partial_\beta\eta_2
  \bigr)
  \nonumber \\
  &-2i\bigl(
  \hat{E}_\alpha^+\bar{\chi}_1 m_0^2\gamma^{34}\chi_2
  -\check{E}_\alpha^+\bar{\chi}_2 m_0^2\gamma^{34}\chi_1
  \bigr)\partial_\beta x^-
  \nonumber \\
  &-2i\bigl(
  \hat{E}_\alpha^+\bar{\eta}_1 m_0^2\tilde{\gamma}^{34}\eta_2
  -\check{E}_\alpha^+\bar{\eta}_2 m_0^2\tilde{\gamma}^{34}\eta_1
  \bigr)\partial_\beta x^+
  \nonumber \\
  &+\bigl(
  \check{E}_\alpha^{\underline{i}} \bar{\chi}_2 \gamma^{\underline{i}} m_0^2\gamma^{34}\chi_1
  +
  \hat{E}_\alpha^{\underline{i}} \bar{\chi}_1\gamma^{\underline{i}} m_0^2\gamma^{34}\chi_2
  \bigr)\partial_\beta x^-
  \nonumber \\
  &-\sum_{\underline{i}=1,2}\bigl(
  \check{E}_\alpha^{\underline{i}} \bar{\eta}_2\tilde{\gamma}^{\underline{i}} m_0^2\tilde{\gamma}^{34}\eta_1
  +\hat{E}_\alpha^{\underline{i}} \bar{\eta}_1\tilde{\gamma}^{\underline{i}} m_0^2\tilde{\gamma}^{34}\eta_2
  \bigr)\partial_\beta x^+
  \nonumber \\
  &+\sum_{i=3,4}\bigl(
  \check{E}_\alpha^{\underline{i}} \bar{\eta}_2 \tilde{\gamma}^{\underline{i}} m_0^2\tilde{\gamma}^{34}\eta_1
  +\hat{E}_\alpha^{\underline{i}} \bar{\eta}_1 \tilde{\gamma}^{\underline{i}} m_0^2\tilde{\gamma}^{34}\eta_2
  \bigr)\partial_\beta x^+
  \nonumber \\
  &
  -2E_\alpha^i\bigl(
  \bar{\chi}_2\gamma^{34}m_0^2\tilde{\gamma}^{34}\tau^i\eta_1
  +\bar{\chi}_1\gamma^{34}m_0^2\tilde{\gamma}^{34}\tau^i\eta_2
  \bigr)\partial_\beta x^+
  \nonumber \\
  &
  +\frac{i}{2}E_\alpha^iE_\beta^j\bigl(
  \bar{\chi}_2\gamma^{34}m_0^2\tau^{ij}\chi_1
  -\bar{\chi}_1\gamma^{34}m_0^2\tau^{ij}\chi_2
  \bigr)\partial_\beta x^+
  \nonumber \\
  &-2i\check{E}_\alpha^+\check{E}_\beta^-\bar{\chi}_2\gamma^{34}m_0^2\chi_1
  +2i\hat{E}_\alpha^+\hat{E}_\beta^-\bar{\chi}_1\gamma^{34}m_0^2\chi_2
  \nonumber \\
  &+2\bigl(\check{E}_\alpha^+\check{E}_\beta^i
  \bar{\chi}_2\gamma^i\tilde{\gamma}^{34}m_0^2\chi_1
  -\hat{E}_\alpha^+\hat{E}_\beta^i
  \bar{\chi}_1\gamma^i\tilde{\gamma}^{34}m_0^2\chi_2\bigr)
  \nonumber \\
  &-\frac{i}{2}\bigl(\check{E}_\alpha^{\underline{i}}\check{E}_\beta^{\underline{j}} \bar{\chi}_2\gamma^{\underline{ij}}\gamma^{34}m_0^2\chi_1
  -\hat{E}_\alpha^{\underline{i}} \hat{E}_\beta^{\underline{j}} \bar{\chi}_1 \gamma^{\underline{ij}}\gamma^{34}m_0^2\chi_2\bigr)
  \Bigr) \,.
\end{align}
As expected, many of the terms above break the $\algSO(4)_1$ explicitly.

\section{Equations of motion}
\label{sec:eoms}

In this appendix we write down the bosonic and fermionic equations of motion for the physical fields of the fully gauge-fixed theory to leading order in fermionic fields and sub-leading order in bosonic fields. The bosonic equations of motion are
\begin{align}
  \ddot{z}_{\underline{i}} &= \ppri{z}_{\underline{i}} - z_{\underline{i}} + (z^2-y^2) \ppri{z}_{\underline{i}} + (2 z \cdot \pri{z} - y \cdot \pri{y}) \pri{z}_{\underline{i}} - y \cdot \dot{y} \, \dot{z}_{\underline{i}}
  \nonumber \\ &\qquad
  +\frac{1}{2}\bigl(y^2-\dot{y}^2-\pri{y}^2-2\pri{z}^2-\dot{x}^2-\pri{x}^2
  \bigr)z_{\underline{i}} +\dotsb\,,\\
  \ddot{y}_{\underline{i}}&=\ppri{y}_{\underline{i}}-y_i+(z^2-y^2)\ppri{y}_{\underline{i}}+(z \cdot \pri{z} - 2 y \cdot \pri{y}) \pri{y}_{\underline{i}} + z \cdot \dot{z} \, \dot{y}_{\underline{i}}
  \nonumber \\ &\qquad
  +\frac{1}{2}\bigl(\dot{z}^2+\pri{z}^2-z^2+2\pri{y}^2+\dot{x}^2+\pri{x}^2
  \bigr)y_{\underline{i}} +\dotsb\,,\\
  \ddot{x}_i&=\ppri{x}_i+(z^2-y^2)\ppri{x}_i+(z\cdot \pri{z}-y\cdot\pri{y})\pri{x}_i+(z\cdot\dot{z}-y\cdot\dot{y})\dot{x}_i +\dotsb \,.
\end{align}
The fermionic equations of motion are
\begin{align}
  2\dot{\eta}_1=
  &-2\pri{\eta}_2 - 2\tilde{\gamma}^{34}\eta_1
  +\epsilon^{\underline{ij}}(z_{\underline{i}} \pri{z}_{\underline{j}} - y_{\underline{i}} \pri{y}_{\underline{j}} ) \pri{\eta}_1 + \bigl(y\cdot\pri{y}-z\cdot\pri{z}+(y_{\underline{i}} \pri{z}_{\underline{j}} - z_{\underline{i}} \pri{y}_{\underline{j}})\tilde{\gamma}^{\underline{ij}}\bigr)\eta_2
  \nonumber \\
  &+\bigl(y^2-z^2+(\dot{z}_{\underline{i}} + \dot{y}_{\underline{i}})(z_{\underline{j}} - y_{\underline{j}})\tilde{\gamma}^{34}\tilde{\gamma}^{\underline{i}}\gamma^{\underline{j}}\bigr)\pri{\eta}_2
  +\bigl(y\cdot\pri{y}-z\cdot\pri{z}+(z_{\underline{i}} \pri{y}_{\underline{j}} - y_{\underline{i}} \pri{z}_{\underline{j}})\tilde{\gamma}^{34}\tilde{\gamma}^{\underline{ij}}\bigr)\dot{\eta}_2
  \nonumber \\
  &+(z_{\underline{i}} - y_{\underline{i}}) \pri{x}_i \tilde{\gamma}^{\underline{i}} \tilde{\tau}^i\dot{\chi}_2
  -(z_{\underline{i}} - y_{\underline{i}}) \dot{x}_i \tilde{\gamma}^{\underline{i}} \tilde{\tau}^i\pri{\chi}_2
  \nonumber \\
  &
  +\bigl((\dot{z}_{\underline{i}} - \dot{y}_{\underline{i}}) \pri{x}_i - (\pri{z}_{\underline{i}} - \pri{y}_{\underline{i}}) \dot{x}_i + (z_{\underline{i}} - y_{\underline{i}}) \dot{x}_i \tilde{\gamma}^{34}
  \bigr)\tilde{\gamma}^{\underline{i}} \tilde{\tau}^i\chi_2+\dotsb\,,
  \\
  2\dot{\eta}_2=&-2\pri{\eta}_1+2\tilde{\gamma}^{34}\eta_2
  -\epsilon^{\underline{ij}}(z_{\underline{i}} \pri{z}_{\underline{j}} - y_{\underline{i}} \pri{y}_{\underline{j}})\pri{\eta}_2 + \bigl(y\cdot\pri{y}-z\cdot\pri{z} + (y_{\underline{i}} \pri{z}_{\underline{j}} - z_{\underline{i}} \pri{y}_{\underline{j}}) \tilde{\gamma}^{\underline{ij}}\bigr)\eta_1
  \nonumber \\
  &+\bigl(y^2-z^2-(\dot{z}_{\underline{i}}+\dot{y}_{\underline{i}})(z_{\underline{j}} - y_{\underline{j}})\tilde{\gamma}^{34}\tilde{\gamma}^{\underline{i}}\gamma^{\underline{j}}\bigr)\pri{\eta}_1
  -\bigl(y\cdot\pri{y}-z\cdot\pri{z}+(z_{\underline{i}} \pri{y}_{\underline{j}} - y_{\underline{i}} \pri{z}_{\underline{j}})\tilde{\gamma}^{34}\tilde{\gamma}^{\underline{ij}} \bigr)\dot{\eta}_1
  \nonumber \\
  &-(z_{\underline{i}} - y_{\underline{i}}) \pri{x}_i \tilde{\gamma}^{\underline{i}} \tilde{\tau}^i\dot{\chi}_1
  +(z_{\underline{i}} - y_{\underline{i}}) \dot{x}_i \tilde{\gamma}^{\underline{i}} \tilde{\tau}^i\pri{\chi}_1
  \nonumber \\
  &
  -\bigl((\dot{z}_{\underline{i}} - \dot{y}_{\underline{i}}) \pri{x}_i - (\pri{z}_{\underline{i}} - \pri{y}_{\underline{i}}) \dot{x}_i - (z_{\underline{i}} - y_{\underline{i}}) \dot{x}_i \tilde{\gamma}^{34}
  \bigr)\tilde{\gamma}^{\underline{i}}\tilde{\tau}^i\chi_1+ \dotsb \,,
  \\
  4\dot{\chi}_1=&-4\pri{\chi}_2
  +2\epsilon^{\underline{ij}}(y_{\underline{i}} - z_{\underline{i}}) \pri{x}_i \gamma^{\underline{j}} \tau^i\eta_2
  -2(y_{\underline{i}} - z_{\underline{i}}) \pri{x}_i \gamma^{\underline{j}} \tau^i\dot{\eta}_2
  +2(y_{\underline{i}} - z_{\underline{i}}) \dot{x}_i \gamma^{\underline{i}} \tau^i\pri{\eta}_2
  \nonumber \\
  &+\epsilon^{\underline{ij}}(z_{\underline{i}} \pri{z}_{\underline{j}} - y_{\underline{i}} \pri{y}_{\underline{j}}) \pri{\chi}_1
  +(\dot{x}^2+\dot{y}^2+\dot{z}^2-\pri{x}^2-\pri{y}^2-\pri{z}^2-y^2-z^2)\gamma^{34}\chi_1
  \nonumber \\
  &+2\bigl(\dot{x}_i\pri{x}_j\gamma^{34}\tau^{ij}
  +\tfrac{1}{2}\epsilon^{\underline{ij}}(\dot{y}_{\underline{i}} \pri{y}_{\underline{j}} - \dot{z}_{\underline{i}} \pri{z}_{\underline{j}})
  +(y_{\underline{i}} - z_{\underline{i}})(\pri{y}_{\underline{j}} + \pri{z}_{\underline{j}})\gamma^{\underline{j}}\tilde{\gamma}^{\underline{i}}
  -(\dot{y}_{\underline{i}} \pri{z}_{\underline{j}} + \dot{z}_{\underline{i}} \pri{y}_{\underline{j}})\gamma^{34}\gamma^{\underline{ji}} \bigr)\chi_2
  \nonumber \\
  &
  +2\bigl(y^2-z^2-(z_{\underline{i}} - y_{\underline{i}})(\dot{z}_{\underline{j}} - \dot{y}_{\underline{j}})\gamma^{34}\gamma^{\underline{j}} \tilde{\gamma}^{\underline{i}}\bigr)\pri{\chi}_2
  \nonumber \\
  &
  +\bigl(y\cdot\pri{y}+z\cdot\pri{z}-2( z_{\underline{i}} \pri{y}_{\underline{j}} + \pri{z}_{\underline{j}} y_{\underline{i}} )
  \gamma^{\underline{ij}} \bigr)\gamma^{34}\dot{\chi}_2+\dotsb\,,
  \\
  4\dot{\chi}_2=&-4\pri{\chi}_1
  +2\epsilon^{\underline{ij}}(y_{\underline{i}} - z_{\underline{i}}) \pri{x}_i \gamma^{\underline{j}} \tau^i\eta_1
  +2(y_{\underline{i}} - z_{\underline{i}}) \pri{x}_i \gamma^{\underline{j}} \tau^i\dot{\eta}_1
  -2(y_{\underline{i}} - z_{\underline{i}}) \dot{x}_i \gamma^{\underline{i}} \tau^i\eta'_1
  \nonumber \\
  &-\epsilon^{\underline{ij}} (z_{\underline{i}} \pri{z}_{\underline{j}} - y_{\underline{i}} \pri{y}_{\underline{j}})\pri{\chi}_2
  -(\dot{x}^2+\dot{y}^2+\dot{z}^2-\pri{x}^2-\pri{y}^2-\pri{z}^2-y^2-z^2)\gamma^{34}\chi_2
  \nonumber \\
  &-2\bigl(\dot{x}_i\pri{x}_j\gamma^{34}\tau^{ij}
  +\tfrac{1}{2}\epsilon^{\underline{ij}}(\dot{y}_{\underline{i}} \pri{y}_{\underline{j}} - \dot{z}_{\underline{i}} \pri{z}_{\underline{j}})
  -(y_{\underline{i}} - z_{\underline{i}})(\pri{y}_{\underline{j}} + \pri{z}_{\underline{j}})\gamma^{\underline{j}} \tilde{\gamma}^{\underline{i}}
  -(\dot{y}_{\underline{i}} \pri{z}_{\underline{j}} + \dot{z}_{\underline{i}} \pri{y}_{\underline{j}})\gamma^{34}\gamma^{\underline{ji}} \bigr)\chi_1
  \nonumber \\
  &
  +2\bigl(y^2-z^2+(z_{\underline{i}} - y_{\underline{i}})(\dot{z}_{\underline{j}} - \dot{y}_{\underline{j}})\gamma^{34}\gamma^{\underline{j}}\tilde{\gamma}^{\underline{i}}\bigr)\pri{\chi}_1
  \nonumber \\
  &
  -\bigl(y\cdot\pri{y}+z\cdot\pri{z}-2(z_{\underline{i}} \pri{y}_{\underline{j}} + \pri{z}_{\underline{j}} y_{\underline{i}})
  \gamma^{\underline{ij}}\bigr)\gamma^{34}\dot{\chi}_1+\dotsb\,.
\end{align}

\section{Poisson bracket for \texorpdfstring{$\eta_I$}{eta} and \texorpdfstring{$\chi_I$}{chi}}
\label{sec:app-poisson-b-fermions}

In this appendix we compute the Poisson brackets for $\eta_I$ and $\chi_I$. To do this let us write all the terms in the action that have a $\tau$-derivative acting on the fermions up to quadratic order in transverse bosons. Keeping only the terms up to the order in the number of fields that we require,  $\mathcal{L}_{\text{\scriptsize kin}}$ gives rise to two such terms
\begin{equation}
  \begin{aligned}
    -2i \gamma^{00} \bigl( 
    \bar{\eta}_1 \hat{E}_\tau^+ \Gamma^- \dot{\eta}_1
    + \bar{\chi}_1 \hat{E}_\tau^+ \Gamma^- \dot{\chi}_1
    \bigr) 
    &\approx
    2i \bigl( 
    \bar{\eta}_1  \dot{\eta}_1 + \bar{\chi}_1 \dot{\chi}_1
    \bigr) \bigl( 1 + \tfrac{1}{2} ( \epsilon^{\underline{ij}} z_{\underline{i}} \dot{z}_{\underline{j}} - \epsilon^{\underline{ij}} y_{\underline{i}} \dot{y}_{\underline{j}} ) \bigr) ,
    \\
    -2i \gamma^{00} \bigl( 
    \bar{\eta}_2 \check{E}_\tau^+ \Gamma^- \dot{\eta}_2
    + \bar{\chi}_2 \check{E}_\tau^+ \Gamma^- \dot{\chi}_2
    \bigr) 
    &\approx
    2i \bigl( 
    \bar{\eta}_2 \dot{\eta}_2 + \bar{\chi}_2 \dot{\chi}_2
    \bigr) \bigl( 1 - \tfrac{1}{2} ( \epsilon^{\underline{ij}} z_{\underline{i}} \dot{z}_{\underline{j}} - \epsilon^{\underline{ij}} y_{\underline{i}} \dot{y}_{\underline{j}} ) \bigr) .
  \end{aligned}
\end{equation}
From $\mathcal{L}_{\text{\scriptsize WZ}}$ we get additional terms 
\begin{equation}
  \begin{aligned}
    -i \bar{\eta}_1 \slashed{\check{E}}_\sigma M_0^{-2} \dot{\eta}_2
    &\approx
    + i \bar{\eta}_1  \tilde{\gamma}^{34} \dot{\eta}_2 ( z \cdot \pri{z} - y \cdot \pri{y} )
    - i \bar{\eta}_1 \tilde{\gamma}^{34} \tilde{\gamma}^{\underline{ij}} \dot{\eta}_2 ( z_{\underline{i}} \pri{y}_{\underline{j}} + z_{\underline{i}} \pri{y}_{\underline{j}} ) ,
    \\
    -i \bar{\eta}_2 \slashed{\hat{E}}_\sigma M_0^{+2} \dot{\eta}_1
    &\approx
    - i \bar{\eta}_2 \tilde{\gamma}^{34} \dot{\eta}_1 ( z \cdot \pri{z} - y \cdot \pri{y} )
    + i \bar{\eta}_2 \tilde{\gamma}^{34} \tilde{\gamma}^{\underline{ij}} \dot{\eta}_1 ( z_{\underline{i}} \pri{y}_{\underline{j}} + z_{\underline{i}} \pri{y}_{\underline{j}} ),
    \\
    -i \bar{\chi}_1 \slashed{\hat{E}}_\sigma M_0^{-2} \dot{\chi}_2 ,
    &\approx
    - i \bar{\chi}_1 \gamma^{34} \dot{\chi}_2 ( z \cdot \pri{z} + y \cdot \pri{y} )
    - i \bar{\chi}_1 \gamma^{34} \gamma^{\underline{ij}}  \dot{\chi}_2 ( z_{\underline{i}} \pri{y}_{\underline{j}} - \pri{z}_{\underline{i}} y_{\underline{j}} ),
    \\
    -i \bar{\chi}_2 \slashed{\check{E}}_\sigma M_0^{+2} \dot{\chi}_1 ,
    &\approx
    + i \bar{\chi}_2 \gamma^{34} \dot{\chi}_1 ( z \cdot \pri{z} + y \cdot \pri{y} )
    + i \bar{\chi}_2 \gamma^{34} \gamma^{\underline{ij}}  \dot{\chi}_1 ( z_{\underline{i}} \pri{y}_{\underline{j}} - \pri{z}_{\underline{i}} y_{\underline{j}} ) ,
  \end{aligned}
\end{equation}
as well as terms that mix massless and massive fields,
\begin{equation}
  \begin{aligned}
    -i \bar{\eta}_1 \slashed{\dot{E}}_\sigma M_0^{-2} \dot{\chi}_2
    &\approx
    -i \bar{\eta}_1  \tilde{\gamma}^{\underline{i}} \tilde{\tau}^i  \dot{\chi}_2 ( z_{\underline{i}} - y_{\underline{i}} ) \pri{x}^i ,
    \\
    -i \bar{\eta}_2 \slashed{\dot{E}}_\sigma M_0^{+2} \dot{\chi}_1
    &\approx
    +i \bar{\eta}_2  \tilde{\gamma}^{\underline{i}} \tilde{\tau}^i  \dot{\chi}_1 ( z_{\underline{i}} - y_{\underline{i}} ) \pri{x}^i ,
    \\
    -i \bar{\chi}_1 \slashed{\dot{E}}_\sigma M_0^{-2} \dot{\eta}_2
    &\approx
    -i \bar{\chi}_1  \epsilon \gamma^{\underline{i}}  \tau^i  \dot{\eta}_2 ( z_{\underline{i}} - y_{\underline{i}} ) \pri{x}^i ,
    \\
    -i \bar{\chi}_2 \slashed{\dot{E}}_\sigma M_0^{+2} \dot{\eta}_1
    &\approx
    +i \bar{\chi}_2  \gamma^{\underline{i}} \tau^i  \dot{\eta}_1 ( z_{\underline{i}} - y_{\underline{i}} ) \pri{x}^i .
  \end{aligned}
\end{equation}
These terms can be diagonalised by a field redefinition. Let us introduce a new set of spinors $\psi_{\mathrm{a}}$, $\mathrm{a}=1,\dotsc,4$ by
\begin{equation}
  \eta_1 = \psi_1 + \tfrac{1}{2} A_{1\mathrm{a}} \psi_{\mathrm{a}} , \quad
  \eta_2 = \psi_2 + \tfrac{1}{2} A_{2\mathrm{a}} \psi_{\mathrm{a}} , \quad
  \chi_1 = \psi_3 + \tfrac{1}{2} A_{3\mathrm{a}} \psi_{\mathrm{a}} , \quad
  \chi_2 = \psi_4 + \tfrac{1}{2} A_{4\mathrm{a}} \psi_{\mathrm{a}} .
\end{equation}
The kinetic term then take the canonical form
\begin{equation}
  2 i \psi_{\mathrm{a}} \epsilon \epsilon \dot{\psi}_{\mathrm{a}} ,
\end{equation}
provided the coefficients $A_{\mathrm{a}\mathrm{b}}$ are chosen as
\begin{equation}
\label{eq:PB-A-coeffs}
  \begin{aligned}
    +A_{11} = -A_{22} = +A_{33} = -A_{44} &= -\tfrac{1}{2} \epsilon^{\underline{ij}} ( z_{\underline{i}} \dot{z}_{\underline{j}} - y_{\underline{i}} \dot{y}_{\underline{j}} ) , \\
    +A_{12} = -A_{21} &= -\tfrac{1}{2} \tilde{\gamma}^{34} ( z \cdot \pri{z} - y \cdot \pri{y} ) + \tfrac{1}{2} \tilde{\gamma}^{34} \tilde{\gamma}^{\underline{ij}} ( z_{\underline{i}} \pri{y}_{\underline{j}} + \pri{z}_{\underline{i}} y_{\underline{j}} ) , \\
    +A_{34} = -A_{43} &= +\tfrac{1}{2} \gamma^{34} ( z \cdot \pri{z} + y \cdot \pri{y} ) + \tfrac{1}{2} \gamma^{34} \gamma^{\underline{ij}} ( z_{\underline{i}} \pri{y}_{\underline{j}} - \pri{z}_{\underline{i}} y_{\underline{j}} ) , \\
    A_{13} = A_{31} = A_{24} = A_{42} &= 0 , \\
    +A_{14} = -A_{23} &= +\tfrac{1}{2} \tilde{\gamma}^{\underline{i}} \tilde{\tau}^i ( z_{\underline{i}} - y_{\underline{i}} ) \pri{x}_i , \\
    +A_{41} = -A_{32} &= -\tfrac{1}{2} \gamma^{\underline{i}} \tau^i ( z_{\underline{i}} - y_{\underline{i}} ) \pri{x}_i .
  \end{aligned}
\end{equation}
It is useful to note that the coefficients satisfy
\begin{equation}
  A_{\mathrm{a}\mathrm{b}} = (\epsilon\epsilon) A_{\mathrm{b}\mathrm{a}}^t (\epsilon\epsilon) .
\end{equation}
The fermions $\psi_{\mathrm{a}}$ have canonical Poisson brackets
\begin{equation}
  \acommPB{\psi_{\mathrm{a}}}{\psi_{\mathrm{b}}} = - \frac{i}{4} \delta_{\mathrm{a}\mathrm{b}} \epsilon \epsilon .
\end{equation}
The Poisson brackets for the $\eta_i$ and $\chi_I$ then follow immediately and are given in equation~\eqref{eq:pb-between-etas-and-chis}.

\section{Some Poisson brackets used in section~\ref{sec:massless-A-calculation}}
\label{app:mless-pbs}

To calculate the Poisson bracket between the exponential factor in $\mathcal{Q}$ and $\ham$ we need the relations
\begin{equation}
  \begin{aligned}
    \commPB{x^-(\sigma)}{x^i(\sigma')}
    &= + \frac{1}{p_-} \int_{-\infty}^{\sigma} ds \, \pri{x}^i(s) \delta(s-\sigma') \\
    &= + \frac{1}{p_-} \pri{x}^i(\sigma') \epsilon(\sigma - \sigma') ,
    \\
    \commPB{x^-(\sigma)}{\pri{x}^i(\sigma')}
    &= + \frac{1}{p_-} \int_{-\infty}^{\sigma} ds \, \partial_{\sigma'} ( \pri{x}^i(s) \delta(s-\sigma') \\
    &= + \frac{1}{p_-} \ppri{x}^i(\sigma') \epsilon(\sigma-\sigma') 
    - \frac{1}{p_-} \pri{x}^i(\sigma') \delta(\sigma-\sigma') ,
    \\
    \commPB{x^-(\sigma)}{p_i(\sigma')}
    &= -\frac{1}{p_-} \int_{-\infty}^{\sigma} ds \, p_i(s) \partial_s \delta(s-\sigma') \\
    &= + \frac{1}{p_-} \pri{p}_i(\sigma') \epsilon(\sigma-\sigma') - \frac{1}{p_-} p_i(\sigma) \delta(\sigma-\sigma') .
  \end{aligned}
\end{equation}
Using these relations we find
\begin{equation}
  \begin{split}
    \commPB{ x^-(\sigma) }{ \tfrac{1}{2} ( p^2(\sigma') + \pri{x}^2(\sigma') ) }
    =
    &+ \frac{1}{p_-^2} \partial_{\sigma'} \bigl[ \bigl( p^2(\sigma') + \pri{x}^2(\sigma') \bigr) \epsilon(\sigma-\sigma') \bigr] \\
    &- \frac{1}{p_-^2} \bigl( p^2(\sigma') + \pri{x}^2(\sigma') \bigr) \delta (\sigma-\sigma')
  \end{split}
\end{equation}
For the Poisson bracket between the Hamiltonian and the quadratic supercharge we note
\begin{equation}
  \begin{aligned}
    \commPB{ p_i(\sigma) \tilde{\tau}^i \chi_1(\sigma) }{ \tfrac{1}{2} \pri{x}^2(\sigma') }
    &=
    - \pri{x}_i(\sigma') \tilde{\tau}^i \chi_1(\sigma) \partial_{\sigma'} \delta(\sigma-\sigma') ,
    \\
    \commPB{ \pri{x}_i(\sigma) \tilde{\tau}^i \chi_2(\sigma) }{ \tfrac{1}{2} p^2(\sigma') }
    &=
    + p_i(\sigma') \, \tilde{\tau}^i \chi_2(\sigma) \partial_{\sigma} \delta(\sigma-\sigma') ,
    \\
    \commPB{ p_i(\sigma) \tilde{\tau}^i \chi_1(\sigma) }{ \chi_1(\sigma') \epsilon \epsilon \pri{\chi}_2 ( \sigma' ) }
    &=
    - \frac{i}{2p_-} p_i(\sigma) \tilde{\tau}^i \pri{\chi}_2(\sigma') \delta(\sigma-\sigma') ,
    \\
    \commPB{ p_i(\sigma) \tilde{\tau}^i \chi_1(\sigma) }{ \chi_2(\sigma') \epsilon \epsilon \pri{\chi}_1 ( \sigma' ) }
    &=
    + \frac{i}{2p_-} p_i(\sigma) \tilde{\tau}^i \chi_2(\sigma') \partial_{\sigma'} \delta(\sigma-\sigma') , 
  \end{aligned}
\end{equation}
and
\begin{equation}
  \begin{aligned}
    \commPB{ p_i \tilde{\tau}^i \chi_1 }{ \chi_1 (\epsilon\gamma^{34})\epsilon \chi_1 }
    &=
    - \frac{i}{p_-} p_i \gamma^{34} \tilde{\tau}^i \chi_1 ,
    \\
    \commPB{ \pri{x}_i \tilde{\tau}^i\chi_2 }{ \chi_2 (\epsilon\gamma^{34})\epsilon \chi_2 }
    &=
    - \frac{i}{p_-} \pri{x}_i \gamma^{34} \tilde{\tau}^i \chi_2 ,
    \\
    \commPB{ p_k \tilde{\tau}^k \chi_1 }{ \chi_1 (\epsilon\gamma^{34})(\epsilon\tau^{ij}) \chi_2 }
    &=
    - \frac{i}{2p_-} p_k \gamma^{34} \tilde{\tau}^k \tau^{ij} \chi_2 ,
    \\
    \commPB{ \pri{x}_k \tilde{\tau}^k \chi_2 }{ \chi_2 (\epsilon\gamma^{34})(\epsilon\tau^{ij}) \chi_1 }
    &=
    - \frac{i}{2p_-} \pri{x}_k \gamma^{34} \tilde{\tau}^k \tau^{ij} \chi_1 .
  \end{aligned}
\end{equation}
 
We also collect here some formulas that are useful in deriving equation~\eqref{eq:conc-suchg-massless}.
First, note that for a constant $\alpha$
\begin{equation}
  \begin{aligned}
    e^{\alpha x^-(\sigma)} \commPB{ \mathcal{Q}_2(\sigma) }{ H }
    =
    &+ \frac{2}{p_-} \partial_\sigma \bigl( e^{\alpha x^-} ( \pri{x}_i \tilde{\tau}^i \chi_1 - p_i \tilde{\tau}^i \chi_2 ) \bigr) \\
    &- \frac{2}{p_-^2} e^{\alpha x^-} \bigl( p_i \pri{x}^i \gamma^{34} + \alpha p_- \pri{x}^- \bigr) ( \pri{x}_i \tilde{\tau}^i \chi_1 - p_i \tilde{\tau}^i \chi_2 ) \\
    &+ \frac{1}{p_-^2} e^{\alpha x^-} ( p^2 + \pri{x}^2 ) \gamma^{34}  ( p_i \tilde{\tau}^i \chi_1 - \pri{x}_i \tilde{\tau}^i \chi_2 )
  \end{aligned}
\end{equation}
and
\begin{equation}
  \begin{aligned}
    \mathcal{Q}_2(\sigma)
    \commPB{ e^{\alpha x^-(\sigma)} }{ H }
    =
    &- \frac{1}{p_-^2} e^{\alpha x^-(\sigma)} \alpha ( p^2(\sigma) + \pri{x}^2(\sigma) ) ( p_i \gamma^{34} \tilde{\tau}^i \chi_1 - \pri{x}_i \tilde{\tau}^i \chi_2 ) .
  \end{aligned}
\end{equation}
Putting this together we find,
up to total derivatives of expressions that vanish at $\pm \infty$,
\begin{equation}
  \begin{aligned}
    \commPB{e^{\alpha x^-}  \mathcal{Q}_2}{H}
    =
    &- \frac{2}{p_-^2} \int_{-\infty}^{+\infty} d\sigma \,
    e^{\alpha x^-} p_i \pri{x}^i ( \gamma^{34} - \alpha )
    ( \pri{x}_j \tilde{\tau}^j \chi_1 - p_j \tilde{\tau}^j \chi_2 ) \\
    &- \frac{1}{p_-^2} \int_{-\infty}^{+\infty} d\sigma
    e^{\alpha x^-} ( \pri{x}^2 + p^2 ) ( \gamma^{34} - \alpha ) ( p_i \tilde{\tau}^i \chi_1 - \pri{x}_i \tilde{\tau}^i \chi_2 ) .
  \end{aligned}
\end{equation}
Hence, if we set
\begin{equation}
  \alpha =  \gamma^{34}
\end{equation}
the expression $e^{\alpha x^-} \mathcal{Q}_2$ gives a conserved charge.

\section{Derivation of equation~(\ref{eq:pb-of-sucharges-for-centr-ext})}
\label{app:der-of-centr-ext-eq}

In this appendix we give the details of the how equation~\eqref{eq:pb-of-sucharges-for-centr-ext} is derived. 
The various terms in $\acommPB{j_1}{j_2}$ are given by
\begin{equation}
  \begin{aligned}
   2ip_- \acommPB{j_{1,\text{massless}}}{j_{2,\text{massless}}}
    =&
    +2 \dot{x} \cdot \pri{x} \epsilon\epsilon \\ &
    + \epsilon^{\underline{ij}} ( z_{\underline{i}} \dot{z}_{\underline{j}} - y_{\underline{i}} \dot{y}_{\underline{j}} ) \dot{x}_i \pri{x}_j \epsilon\tilde{\tau}^{ij} \epsilon \\ &
    - \tfrac{1}{2} ( z \cdot \pri{z} + y \cdot \pri{y} ) ( \dot{x}^2 - \pri{x}^2 )  \gamma^{34} \epsilon  \epsilon  \\ &
    - \tfrac{1}{2} ( z_{\underline{i}} \pri{y}_{\underline{j}} - \pri{z}_{\underline{i}} y_{\underline{j}} ) ( \dot{x}^2 - \pri{x}^2 )  \gamma^{\underline{ij}} \gamma^{34} \epsilon  \epsilon  ,
    \\
    2ip_- \acommPB{j_{1,\text{massless}}}{j_{2,\text{massive}}}
    =&
    - \tfrac{1}{2} (z_{\underline{i}} - y_{\underline{i}}) (\dot{z}_{\underline{j}} - \dot{y}_{\underline{j}}) \dot{x}_i \pri{x}_j  \gamma^{34} \gamma^{\underline{i}} \tilde{\gamma}^{\underline{j}} \epsilon  \tilde{\tau}^i \tau^j \epsilon  \\ &
    - \tfrac{1}{2} (z_{\underline{i}} - y_{\underline{i}}) (z_{\underline{j}} + y_{\underline{j}}) \dot{x}_i \pri{x}_j  \gamma^{34} \gamma^{\underline{i}} \tilde{\gamma}^{\underline{j}} \gamma^{34} \epsilon  \tilde{\tau}^i \tau^j \epsilon  \\ &
    - \tfrac{1}{2} (z_{\underline{i}} - y_{\underline{i}}) (\pri{z}_{\underline{j}} - \pri{y}_{\underline{j}}) \pri{x}^2  \gamma^{34} \gamma^{\underline{i}} \tilde{\gamma}^{\underline{j}} \epsilon  \epsilon  ,
    \\
    2ip_- \acommPB{j_{1,\text{massive}}}{j_{2,\text{massless}}}
    =&
    - \tfrac{1}{2} (\dot{z}_{\underline{i}} - \dot{y}_{\underline{i}}) (z_{\underline{j}} - y_{\underline{j}}) \pri{x}_i \dot{x}_j  \gamma^{\underline{i}} \tilde{\gamma}^{\underline{j}} \gamma^{34} \epsilon  \tilde{\tau}^i \tau^j \epsilon  \\ &
    - \tfrac{1}{2} (z_{\underline{i}} + y_{\underline{i}}) (z_{\underline{j}} - y_{\underline{j}}) \pri{x}_i \dot{x}_j  \gamma^{34} \gamma^{\underline{i}} \tilde{\gamma}^{\underline{j}} \gamma^{34} \epsilon  \tilde{\tau}^i \tau^j \epsilon  \\ &
    - \tfrac{1}{2} (\pri{z}_{\underline{i}} - \pri{y}_{\underline{i}}) (z_{\underline{j}} - y_{\underline{j}}) \pri{x}^2  \gamma^{\underline{i}} \tilde{\gamma}^{\underline{j}} \gamma^{34} \epsilon  \epsilon  ,
    \\
    2ip_- \acommPB{j_{1,\text{massive}}}{j_{2,\text{mixed}}}
    =&
    - \tfrac{1}{2} (\dot{z}_{\underline{i}} - \dot{y}_{\underline{i}}) (z_{\underline{j}} - y_{\underline{j}}) \dot{x} \cdot \pri{x}  \gamma^{\underline{i}} \tilde{\gamma}^{\underline{j}} \gamma^{34} \epsilon  \epsilon  \\ &
    - \tfrac{1}{2} (z_{\underline{i}} + y_{\underline{i}}) (z_{\underline{j}} - y_{\underline{j}}) \dot{x} \cdot \pri{x}  \gamma^{34} \gamma^{\underline{i}} \tilde{\gamma}^{\underline{j}} \gamma^{34} \epsilon  \epsilon  \\ &
    + \tfrac{1}{4} (\pri{z}_{\underline{i}} - \pri{y}_{\underline{i}}) (z_{\underline{j}} - y_{\underline{j}}) (\dot{x}^2 + \pri{x}^2)  \gamma^{\underline{i}} \tilde{\gamma}^{\underline{j}} \gamma^{34} \epsilon  \epsilon  ,
    \\
    2ip_- \acommPB{j_{1,\text{mixed}}}{j_{\text{2,massive}}}
    =&
    - \tfrac{1}{2} (z_{\underline{i}} - y_{\underline{i}}) (\dot{z}_{\underline{j}} - \dot{y}_{\underline{j}}) \dot{x} \cdot \pri{x}  \gamma^{34} \gamma^{\underline{i}} \tilde{\gamma}^{\underline{j}} \epsilon  \epsilon  \\ &
    - \tfrac{1}{2} (z_{\underline{i}} - y_{\underline{i}}) (z_{\underline{j}} + y_{\underline{j}}) \dot{x} \cdot \pri{x}  \gamma^{34} \gamma^{\underline{i}} \tilde{\gamma}^{\underline{j}} \gamma^{34} \epsilon  \epsilon  \\ &
    + \tfrac{1}{4} (z_{\underline{i}} - y_{\underline{i}}) (\pri{z}_{\underline{j}} - \pri{y}_{\underline{j}}) (\dot{x}^2 + \pri{x}^2)  \gamma^{34} \gamma^{\underline{i}} \tilde{\gamma}^{\underline{j}} \epsilon  \epsilon  ,
    \\
    2ip_- \acommPB{j_{1,\text{massless}}}{j_{2,\text{mixed}}}
    =&
    - (z^2 - y^2) \dot{x}_i \pri{x}_j \epsilon \tilde{\tau}^{ij} \epsilon \\ &
    + z_{\underline{i}} y_{\underline{j}} \dot{x}_i \pri{x}_j \gamma^{\underline{ij}} \epsilon \tilde{\tau}^i \tau^j \epsilon ,
    \\
    2ip_- \acommPB{j_{1,\text{mixed}}}{j_{2,\text{massless}}}
    =&
    + (z_2 - y^2) \dot{x}_i \pri{x}_i \epsilon \tilde{\tau}_{ab} \epsilon \\ &
    - z_{\underline{i}} y_{\underline{j}} \pri{x}_i \dot{x}_j \gamma^{\underline{ij}} \epsilon \tilde{\tau}^i \tau^j \epsilon ,
    \\
    2ip_- \acommPB{j_{1,\text{massive}}}{j_{2,\text{massive}}}
    =&
    + 2 ( \dot{z} \cdot \pri{z} + \dot{y} \cdot \pri{y} ) \epsilon\epsilon \\ &
    + 2 ( z \cdot \pri{z} - y \cdot \pri{y} ) \gamma^{34} \epsilon\epsilon \\ &
    - 2 ( z_{\underline{i}} \pri{y}_{\underline{j}} + \pri{z}_{\underline{i}} y_{\underline{j}} )  \gamma^{34} \gamma^{\underline{ij}} \epsilon\epsilon .
  \end{aligned}
\end{equation}

Putting this together we find
\begin{equation}
  \begin{aligned}
    2ip_- \acommPB{j_1}{j_2}
    =
    + 2 e^{+\gamma^{34} x^-} \Bigl( 
    &( \dot{z} \cdot \pri{z} + \dot{y} \cdot \pri{y} + \dot{z} \cdot \pri{z} ) \epsilon \epsilon 
    + 
    ( z \cdot \pri{z} - y \cdot \pri{y} ) \gamma^{34} \epsilon\epsilon 
    \\
    & - 
    ( z_{\underline{i}} \pri{y}_{\underline{j}} + \pri{z}_{\underline{i}} y_{\underline{j}} ) \gamma^{34} \gamma^{\underline{ij}} \epsilon \epsilon
    \Bigr) e^{-\gamma^{34} x^-}
    \\
    =
    + e^{+\gamma^{34} x^-} \Bigl( 
    & 2 ( \dot{z} \cdot \pri{z} + \dot{y} \cdot \pri{y} + \dot{z} \cdot \pri{z} ) \epsilon \epsilon 
    +  \partial_\sigma ( z^2 - y^2 ) \gamma^{34} \epsilon\epsilon \\
    &-  \partial_\sigma ( z_{\underline{i}} y_{\underline{j}} ) \gamma^{34} \gamma^{\underline{ij}} \epsilon\epsilon
    \Bigr) e^{-\gamma^{34} x^-} .
  \end{aligned}
\end{equation}


\section{Oscillator algebra}
\label{app:algebra}
As a preliminary step to facilitate the study of non-perturbative representations, let us rewrite
the supercurrents in components and expand them in terms of
oscillators in a momentum-basis. In this way, it will be easier to read off the form of the representations, and to deform their dependence on the momentum. In order to elucidate the structure of the representations it will be sufficient to  consider their leading order in a field expansion.

We introduce complex coordinates
\begin{equation}
Z=-z_2+i\,z_1\,,\qquad
\bar{Z}=-z_2-i\,z_1\,,\qquad\qquad
Y=y_3+i\,y_4\,,\qquad
\bar{Y}=y_3-i\,y_4\,,
\end{equation}
and the corresponding conjugate momenta~$P_{Z},P_{\bar{Z}}$
and~$P_{Y},P_{\bar{Y}}$ so that the variables satisfy canonical
commutation relations
\begin{equation}
\begin{aligned}
&[Z(\sigma_1),P_{\bar{Z}}(\sigma_2)]=[\bar{Z}(\sigma_1),P_{Z}(\sigma_2)]=i\,\delta(\sigma_1-\sigma_2)\,,\\
&[Y(\sigma_1),P_{\bar{Y}}(\sigma_2)]=[\bar{Y}(\sigma_1),P_{Y}(\sigma_2)]=i\,\delta(\sigma_1-\sigma_2)\,.
\end{aligned}
\end{equation}
Similarly, for $\Torus^4$ coordinates we introduce
\begin{equation}
X^{12}=- x_8 + i \,x_9 \,,
\quad
X^{21}= x_8 +i \,x_9\,,
\qquad
X^{11}= x_6 - i\, x_7\,,
\quad
X^{22}= x_6 + i\, x_7\,,
\end{equation}
and canonical momenta satisfying
\begin{equation}
[X^{{\dot{a}a}}(\sigma_1), P_{{\dot{b}b}}(\sigma_2)]=i\,\delta^{{\dot{a}}}_{\
{\dot{b}}}\,\delta^{{a}}_{\
{b}}\,\delta(\sigma_1-\sigma_2)\,.
\end{equation}
It will also be useful to expand the fermions in components. For the
massive ones we have
\begin{equation}
\big(\eta_1\big)_{\underline{\dot{a}}\dot{{a}}}=\left(
\begin{array}{cc}
-e^{+i\pi/4}\,\bar{\eta}_{\smallL 1} & -e^{+i\pi/4}\,
\bar{\eta}_{\smallL 2} \\
\phantom{-}e^{-i\pi/4}\,\eta_{\smallL}^{\ 2} &
-e^{-i\pi/4}\,\eta_{\smallL}^{\ 1}
\end{array}
\right),
\qquad
\big(\eta_2\big)_{\underline{\dot{a}}\dot{{a}}}=\left(
\begin{array}{cc}
-e^{-i\pi/4}\,\eta_{\smallR 2} &
\phantom{-}e^{-i\pi/4}\,\eta_{\smallR}^{\ 1}\\
-e^{+i\pi/4}\,\bar{\eta}_{\smallR}^{\ 1} & -e^{+i\pi/4}\,
\bar{\eta}^{\smallR 2}
\end{array}
\right),
\end{equation}
where we introduced a rotation of~$e^{\pm i\pi/4}$ for later
convenience. Similarly, we have
\begin{equation}
\big(\chi_1\big)_{\underline{a}a}=\left(
\begin{array}{cc}
\phantom{-}e^{+i\pi/4}\bar{\chi}_{-2} & -e^{+i\pi/4}
\bar{\chi}_{-1} \\
\phantom{-}e^{-i\pi/4}\chi_{-}^{\ 1} & \phantom{-}e^{-i\pi/4}\chi_{-}^{\ 2}
\end{array}
\right), \qquad
\big(\chi_2\big)_{\underline{a}a}=\left(
\begin{array}{cc}
-e^{-i\pi/4}\chi_{+}^{\ 1} &
-e^{-i\pi/4} \chi_{+}^{\ 2} \\
\phantom{-}e^{+i\pi/4}\bar{\chi}_{+2} &
-e^{+i\pi/4}\bar{\chi}_{+1}
\end{array}
\right).
\end{equation}
On the right-hand side of the above expressions for $\eta_I$ (respectively $\chi_I$) the indices~$1,2$
correspond to the unbroken~$\alg{su}(2)_{\bullet}$ (respectively
$\su(2)_{\circ}$), and complex conjugation is indicated by a bar.
We write superscript~$\alg{su}(2)_{\bullet}$ indices for left massive fermions
and subscript indices for right massive fermions (the opposite for their
conjugates) because they transform in the
fundamental and anti-fundamental representations, respectively.
The canonical anti-commutation relations take the form
\begin{equation}
\begin{aligned}
&\ \{\bar{\eta}_{\smallL{\dot{a}}}(\sigma_1),{\eta}_{\sL}^{\ {\dot{b}}}(\sigma_2)\}=
\{\bar{\eta}_{\smallR}^{\ {\dot{b}}}(\sigma_1),{\eta}_{\smallR{\dot{a}}}(\sigma_2)\}=\delta_{{\dot{a}}}^{\
{\dot{b}}}\,\delta(\sigma_1-\sigma_2),\\
&\{\bar{\chi}_{+{a}}(\sigma_1),{\chi}_{+}^{{b}}(\sigma_2)\}=
\{\bar{\chi}_{-{a}}(\sigma_1),\chi_{-}^{{b}}(\sigma_2)\}=\delta_{{a}}^{\
{b}}\,\delta(\sigma_1-\sigma_2),
\end{aligned}
\end{equation}

Using these expressions, we can write the leading order expression of the charges, recalling that we take $\epsilon^{12}=-\epsilon_{12}=+1$,
\begin{equation}
\begin{aligned}
&\gen{Q}_{\smallL}^{\ {\dot{a}}}=e^{-\frac{\pi}{4}i}\int\de\sigma\Bigg(
 \frac{1}{2}P_{Z}\eta^{\ {\dot{a}}}_{\smallL}-i Z'\bar{\eta}^{\ {\dot{a}}}_{\smallR}+  iZ\eta^{\ {\dot{a}}}_{\smallL}
-\epsilon^{{\dot{a}\dot{b}}}\,
\big(\frac{i}{2} P_{\bar{Y}} \bar{\eta}_{\smallL{\dot{b}}}- {\bar{Y}'}\eta_{\smallR {\dot{b}}}+ \bar{Y}\bar{\eta}_{\smallL {\dot{b}}}\big)\\
&
\qquad\qquad\qquad\qquad\qquad
\qquad\qquad\qquad\qquad\qquad
-\frac{1}{2}\epsilon^{{\dot{a}\dot{b}}} P_{{\dot{b}a}}\chi_{+}^{\ {a}}
-i(X^{{\dot{a}a}})'\, \bar{\chi}_{-{a}} \Bigg),\\
&\gen{Q}_{\smallR {\dot{a}}}=e^{-\frac{\pi}{4}i}\int\de\sigma\Bigg(
\frac{1}{2}P_{\bar{Z}}\eta_{\smallR {\dot{a}}}-i\bar{Z}'\bar{\eta}_{\smallL {\dot{a}}}+i\bar{Z}\eta_{\smallR {\dot{a}}}
+\epsilon_{{\dot{a}\dot{b}}}\,\big(\frac{i}{2} {P}_Y\bar{\eta}^{\ {\dot{b}}}_{\smallR}-{Y'}\eta^{\ {\dot{b}}}_{\smallL}+ Y\bar{\eta}^{\ {\dot{b}}}_{\smallR}\big)\\
&\qquad\qquad\qquad\qquad\qquad\qquad\qquad\qquad\qquad\qquad
+\frac{1}{2}P_{{\dot{a}a}}\chi_{-}^{\ {a}}
-i \epsilon_{{\dot{a}\dot{b}}}(X^{{\dot{b}a}})'\, \bar{\chi}_{+{a}}\Bigg),
\end{aligned}
\end{equation}
while their Hermitian conjugates can be found directly by
\begin{equation}
\overline{\gen{Q}}{}_{\smallL {\dot{a}}}= (\gen{Q}_{\smallL}^{\ {\dot{a}}})^\dagger,
\qquad\qquad
\overline{\gen{Q}}{}_{\smallR}^{\ {\dot{a}}}= (\gen{Q}_{\smallR {\dot{a}}})^\dagger.
\end{equation}
The identification with the supercharges in \eqref{eq:total-current-tau} goes as follows
\begin{equation}
\gen{Q}_{\smallL}^1 = -(\mathcal{Q}_1)^{21}, \quad
\gen{Q}_{\smallL}^2 = (\mathcal{Q}_1)^{22}, \quad
\gen{Q}_{\smallR {1}} = (\mathcal{Q}_2)^{12}, \quad
\gen{Q}_{\smallR {2}} = (\mathcal{Q}_2)^{11},
\end{equation}
where for simplicity we have written $\mathcal{Q}_I = \int \de \sigma \,  j_I^\tau$.

We can now introduce ladder operators satisfying canonical
(anti-)commutation relations. Let us define the wavefunction parameters
\begin{equation}
\omega(p,m)=\sqrt{m^2+p^2},\quad
f(p,m)=\sqrt{\frac{\omega(p,m)+|m|}{2}},\quad
g(p,m)=-\frac{p}{2f(p,m)},
\end{equation}
satisfying
\begin{equation}
\omega(p,m)=f(p,m)^2+g(p,m)^2,\qquad
|m|=f(p,m)^2-g(p,m)^2,
\end{equation}
and the short-hand notation
\begin{equation}\begin{aligned}
\omega_p=\omega(p,\pm1)\qquad
f_p=f(p,\pm1),\qquad
g_p=g(p,\pm1),\\
\tilde{\omega}_p=\omega(p,0),\qquad\ \ 
\tilde{f}_p=f(p,0),\qquad\ \ 
\tilde{g}_p=g(p,0).\ 
\end{aligned}
\end{equation}
Then, for the massive bosons we have
\begin{equation}
\begin{aligned}
a_{\sL z} (p) &= \frac{1}{\sqrt{2\pi}} \int
\frac{\de\sigma}{\sqrt{\omega_p}} \left(\omega_p \bar{Z} +\frac{i}{2} P_{\bar{Z}}
\right) e^{-i p \sigma},\\
a_{\sR z} (p)&= \frac{1}{\sqrt{2\pi}} \int \frac{\de\sigma}{\sqrt{\omega_p}}
\left(\omega_p Z + \frac{i}{2} P_{Z} \right) e^{-i p \sigma}, \\
\\
a_{\sL y}(p) &= \frac{1}{\sqrt{2\pi}} \int
\frac{\de\sigma}{\sqrt{\omega_p}} \left(\omega_p \bar{Y} + \frac{i}{2} P_{\bar{Y}}
\right) e^{-i p \sigma},\\
a_{\sR y}(p) &= \frac{1}{\sqrt{2\pi}} \int \frac{\de\sigma}{\sqrt{\omega_p}}
\left(\omega_p Y + \frac{i}{2} P_{Y} \right) e^{-i p \sigma}, \\
\end{aligned}
\end{equation}
while for the massive  fermions
\begin{equation}
\label{eq:fermion-par-massive}
\begin{aligned}
d_{\sL {\dot{a}}}(p) &=+ \frac{e^{+i \pi / 4}}{\sqrt{2\pi}} \int
\frac{\de\sigma}{\sqrt{\omega_p}} \ \epsilon_{{\dot{a}\dot{b}}}\left( f_p \,
{\eta}^{\ {\dot{b}}}_{\smallL}
+i g_p \, \bar{\eta}^{\ {\dot{b}}}_{\smallR} \right) e^{-i p \sigma}, \\
d_{\sR}^{\ {\dot{a}}}(p) &=- \frac{e^{+i \pi / 4}}{\sqrt{2\pi}} \int
\frac{\de\sigma}{\sqrt{\omega_p}} \ \epsilon^{{\dot{a}\dot{b}}} \left( f_p \, \eta_{\smallR{\dot{b}}}
+i g_p \, \bar{\eta}_{\smallL{\dot{b}}} \right) e^{-i p \sigma}. \\
\end{aligned}
\end{equation}
The corresponding creation operators are found by taking the
complex conjugate of the above expressions, and raising and lowering the $\alg{su}(2)$ indices by the tensor~$\epsilon$, yielding
\begin{equation}
\begin{gathered}
[a^{\dagger}_{\sL\, z}(p_1),a_{\sL\, z}(p_2)]=
[a^{\dagger}_{\sR\, z}(p_1),a_{\sR\, z}(p_2)]=
\delta(p_1-p_2)\,,\\
[a^{\dagger}_{\sL\, y}(p_1),a_{\sL\, y}(p_2)]=
[a^{\dagger}_{\sR\, y}(p_1),a_{\sR\, y}(p_2)]=
\delta(p_1-p_2)\,,\\
\{d_{\sL}^{\ {\dot{a}}\,\dagger}(p_1),d_{\sL {\dot{b}}}(p_2)\}=
\{d_{\sR {\dot{b}}}^{\,\dagger}(p_1),d_{\sR}^{\ {\dot{a}}}(p_2)\}=\delta_{{\dot{b}}}^{\ {\dot{a}}}\,\delta(p_1-p_2)\,.
\end{gathered}
\end{equation}
For the massless bosons we have
\begin{equation}
\begin{aligned}
a_{{\dot{a}a}} (p) &= \frac{1}{\sqrt{2\pi}} \int
\frac{\de\sigma}{\sqrt{\tilde{\omega}_p}} \left(\tilde{\omega}_p X_{{\dot{a}a}} + \frac{i}{2}
P_{{\dot{a}a}} \right) e^{-i p \sigma}.
\end{aligned}
\end{equation}
and for the massless fermions
\begin{equation}
\label{eq:fermion-par-massless}
\begin{aligned}
\tilde{d}_{{a}}(p)&= \frac{e^{-i \pi / 4}}{\sqrt{2\pi}} \int
\frac{\de\sigma}{\sqrt{\tilde{\omega}_p}} \left( \tilde{f}_p  \bar{\chi}_{+{a}} +i
\tilde{g}_p \, \epsilon_{{ab}}
{\chi}_{-}^{\ {b}} \right) e^{-i p \sigma},
\\
{d}_{{a}}(p) &= \frac{e^{+i \pi / 4}}{\sqrt{2\pi}} \int
\frac{\de\sigma}{\sqrt{\tilde{\omega}_p}} \left( \tilde{f}_p\, \epsilon_{{ab}}
{\chi}_{+}^{\ {b}} +i \tilde{g}_p\, \bar{\chi}_{-{a}}\right) e^{-i p \sigma}.
\end{aligned}
\end{equation}
The commutators for the creation and annihilation operators are then
\begin{equation}
\begin{gathered}
[a^{\dagger}_{{\dot{a}a}}(p_1),a^{{\dot{b}b}}(p_2)]=\delta_{{\dot{a}}}^{\ {\dot{b}}}\,\delta_{{a}}^{\ {b}}\,
\delta(p_1-p_2)\,,\\
\{\tilde{d}^{{a}\,\dagger}(p_1),\tilde{d}_{{b}}(p_2)\}=
\{d^{\ {a}\,\dagger}(p_1),d_{{b}}(p_2)\}=\delta_{{b}}^{\ {a}}\,\delta(p_1-p_2)\,.
\end{gathered}
\end{equation}

We define states by acting with the creation operators on the vacuum. We have eight massive excitations
\begin{equation}
\ket{Z^{\sL,\sR}}=a^\dagger_{\sL,\sR\, z}\ket{0},\quad
\ket{Y^{\sL,\sR}}=a^\dagger_{\sL,\sR\, y}\ket{0},
\qquad
\ket{\eta_{\sL}^{\ {\dot{a}}}}=d^{{\dot{a}} \dagger}_{\sL}\ket{0},\qquad
\ket{\eta_{\sR {\dot{a}}}}=d^{\dagger}_{\sR {\dot{a}}}\ket{0},
\end{equation}
and eight massless ones
\begin{equation}
\ket{T^{{\dot{a}a}}}=a^{{{\dot{a}a}}\,\dagger}\ket{0},\qquad
\ket{\chi^{{a}}}=d^{{a}\,\dagger}\ket{0},\qquad
\ket{\widetilde{\chi}^{{a}}}=\tilde{d}^{{a}\,\dagger}\ket{0}.
\end{equation}
The same notation is used in the main text for the states on which the non-perturbative S matrix acts.

Finally, the supercharges in terms of ladder operators take the form
\begin{equation}
\begin{aligned}
&\gen{Q}_{\smallL}^{\ {\dot{a}}}= \int \de p \ \Bigg[
 (d_{\sL}^{\ {\dot{a}}\,\dagger} a_{\sL y} + \epsilon^{{\dot{a}\dot{b}}}\, a_{\sL
z}^\dagger  d_{\sL {\dot{b}}})f_p
+ (a_{\sR y}^\dagger  d_{\sR}^{\ {\dot{a}}} +\epsilon^{{\dot{a}\dot{b}}}\, d_{\sR
{\dot{b}}}^\dagger  a_{\sR z})g_p \\
&\qquad\qquad\qquad\qquad\qquad\qquad\qquad\qquad\qquad\qquad
+ \left( \epsilon^{{\dot{a}\dot{b}}}\, \tilde{d}^{{a}\,\dagger}a_{{\dot{b}a}}+
a^{{\dot{a}a}\,\dagger}d_{{a}}\right)\tilde{f}_p\Bigg],\\
&\gen{Q}_{\smallR {\dot{a}}}=\int \de p \ \Bigg[
 (d_{\sR {\dot{a}}}^\dagger  a_{\sR y} -\epsilon_{{\dot{a}\dot{b}}}\, a_{\sR z}^\dagger
d_{\sR}^{\ {\dot{b}}})f_p
+ (a_{\sL y}^\dagger  d_{\sL {\dot{a}}} -\epsilon_{{\dot{a}\dot{b}}}\,  d_{\sL}^{\
{\dot{b}}\,\dagger} a_{\sL z})g_p\\
&\qquad\qquad\qquad\qquad\qquad\qquad\qquad\qquad\qquad\qquad
 + \left( d^{{a}\,\dagger}a_{{\dot{a}a}}-
\epsilon_{{\dot{a}\dot{b}}}\, a^{{\dot{b}a}\,\dagger}\tilde{d}_{{a}}\right)\tilde{g}_p\Bigg].
\end{aligned}
\end{equation}
Note that these supercharges are manifestly covariant under~$\alg{so}(4)_2=\alg{su}(2)_{\bullet}\oplus\alg{su}(2)_{\circ}$, and furthermore enjoy a discrete ``left/right'' symmetry under
\begin{equation}
\label{eq:LR-oscillator-massive}
a_{\sL z}\longleftrightarrow a_{\sR z},\qquad
a_{\sL y}\longleftrightarrow a_{\sR y},\qquad
d_{\sL{\dot{a}}} \longleftrightarrow d_{\sR}^{\ {\dot{a}}},
\end{equation}
and
\begin{equation}
\label{eq:LR-oscillator-massless}
a_{{a\dot{a}}}\longleftrightarrow a^{{a\dot{a}}},
\qquad
\tilde{d}_{{a}}\longleftrightarrow +\frac{\tilde{g}_p}{\tilde{f}_p} d^{{a}},
\qquad	
{d}_{{a}}\longleftrightarrow -\frac{\tilde{f}_p}{\tilde{g}_p} \tilde{d}^{{a}},
\end{equation}
where we also used the fact that~$\tilde{f}^2=\tilde{g}^2$.


\section{Parametrisation of \texorpdfstring{$\alg{su}(1|1)^2_{\ce}$}{su(1|1)**2 c.e.} S-matrix elements}
\label{app:smat-param}

Here we give an explicit parametrisation of the $\alg{su}(1|1)^2_{\ce}$ invariant S-matrix elements. We use the Zhukovski variables introduced in equation~\eqref{eq:zhukovski}. In the left-left sector we have
\begin{equation}
\begin{aligned}
A^{\sL\sL}_{pq} &= 1, &
\qquad
B^{\sL\sL}_{pq} &= \phantom{-}\left( \frac{x^-_p}{x^+_p}\right)^{1/2} \frac{x^+_p-x^+_q}{x^-_p-x^+_q}, \\
C^{\sL\sL}_{pq} &= \left( \frac{x^-_p}{x^+_p} \frac{x^+_q}{x^-_q}\right)^{1/2} \frac{x^-_q-x^+_q}{x^-_p-x^+_q} \frac{\eta_p}{\eta_q}, 
\qquad &
D^{\sL\sL}_{pq} &= \phantom{-}\left(\frac{x^+_q}{x^-_q}\right)^{1/2}  \frac{x^-_p-x^-_q}{x^-_p-x^+_q}, \\
E^{\sL\sL}_{pq} &= \frac{x^-_p-x^+_p}{x^-_p-x^+_q} \frac{\eta_q}{\eta_p}, 
\qquad &
F^{\sL\sL}_{pq} &= - \left(\frac{x^-_p}{x^+_p} \frac{x^+_q}{x^-_q}\right)^{1/2} \frac{x^+_p-x^-_q}{x^-_p-x^+_q},
\end{aligned}
\end{equation}
while in the left-right sector
\begin{equation}
\begin{aligned}
 A^{\sL\sR}_{pq} &= \left(\frac{x^+_p}{x^-_p} \right)^{1/2} \frac{1-\frac{1}{x^+_p x^-_q}}{1-\frac{1}{x^-_p x^-_q}}, 
 \qquad &
 C^{\sL\sR}_{pq} &= \phantom{-}1, \\
B^{\sL\sR}_{pq} &= -\frac{2i}{h} \, \left(\frac{x^-_p}{x^+_p}\frac{x^+_q}{x^-_q} \right)^{1/2} \frac{\eta_{p}\eta_{q}}{ x^-_p x^+_q} \frac{1}{1-\frac{1}{x^-_p x^-_q}},  
\qquad &
D^{\sL\sR}_{pq} &=\phantom{-}\left(\frac{x^+_p}{x^-_p}\frac{x^+_q}{x^-_q} \right)^{1/2} \frac{1-\frac{1}{x^+_p x^+_q}}{1-\frac{1}{x^-_p x^-_q}}, \\
F^{\sL\sR}_{pq} &= \frac{2i}{h} \left(\frac{x^+_p}{x^-_p}\frac{x^+_q}{x^-_q} \right)^{1/2}  \frac{\eta_{p}\eta_{q}}{ x^+_p x^+_q} \frac{1}{1-\frac{1}{x^-_p x^-_q}},
\qquad &
 E^{\sL\sR}_{pq} &= - \left(\frac{x^+_q}{x^-_q} \right)^{1/2} \frac{1-\frac{1}{x^-_p x^+_q}}{1-\frac{1}{x^-_p x^-_q}}.
\end{aligned}
\end{equation}
Note that for convenience we normalise these S-matrices in a different way from what was done in~\cite{Borsato:2012ud}. In particular, in order to satisfy unitarity one would need to multiply them by an appropriate scalar factor.

\section{Explicit form of the S-matrix elements}
\label{app:explicitSmatelements}
For the reader's convenience we write the explicit form of the S-matrix in the different sectors. To this end, we introduce the graded tensor product~$\check{\otimes}$ for the matrices~$\mathcal{S}$ so that up to prefactors
\begin{equation}
\mathcal{S}_{\alg{psu}(1|1)^4}\approx\mathcal{S}_{\alg{su}(1|1)^2} \;\check{\otimes}\; \mathcal{S}_{\alg{su}(1|1)^2},
\end{equation}
defined by
\begin{equation}
\left( \mathcal{A}\,\check{\otimes}\,\mathcal{B} \right)_{MM',NN'}^{KK',LL'} = (-1)^{\epsilon_{M'}\epsilon_{N}+\epsilon_{L}\epsilon_{K'}} \ \mathcal{A}_{MN}^{KL} \  \mathcal{B}_{M'N'}^{K'L'}\,,
\end{equation}
see~\eqref{eq:gradedtensorpr} for comparison with the case of~$\mathbf{S}$.

\subsection{The mixed-mass sector}
In the following we write the S-matrix in the case of a massive excitation that scatters with a massless excitation. We write only the matrix part of it. These elements need to be multiplied by the appropriate scalar factors introduced in (\ref{eq:norm-sect}).
In the case of left massive excitations that scatter with massless excitations transforming in the $\varrho_{\sL} \otimes \widetilde{\varrho}_{\sL}$ representation of $\alg{psu}(1|1)^4_{\ce}$ we find
\begingroup
\addtolength{\jot}{1ex}
\begin{equation}\label{eq:mixed-S-matrix-L}
\begin{aligned}
\mathcal{S}^{\sL\sL}\check{\otimes}\mathcal{S}^{\sL\tilde{\sL}} \ket{Z^{\sL}_p T^{\dot{a}a}_q} =& - F^{\sL\sL}_{pq}D^{\sL\sL}_{pq} \ket{T^{\dot{a}a}_q Z^{\sL}_p } - F^{\sL\sL}_{pq}E^{\sL\sL}_{pq} \ket{\widetilde{\chi}^{a}_q \eta^{\sL \dot{a}}_p}, \\
\mathcal{S}^{\sL\sL}\check{\otimes}\mathcal{S}^{\sL\tilde{\sL}} \ket{Y^{\sL}_p T^{\dot{a}a}_q} =& + A^{\sL\sL}_{pq}B^{\sL\sL}_{pq} \ket{T^{\dot{a}a}_q Y^{\sL}_p } - A^{\sL\sL}_{pq}C^{\sL\sL}_{pq} \ket{ \chi^{a}_q \eta^{\sL \dot{a}}_p }, \\
\mathcal{S}^{\sL\sL}\check{\otimes}\mathcal{S}^{\sL\tilde{\sL}} \ket{\eta^{\sL \dot{a}}_p \widetilde{\chi}^{a}_q} =& + F^{\sL\sL}_{pq}B^{\sL\sL}_{pq} \ket{\widetilde{\chi}^a_q \eta^{\sL \dot{a}}_p } + F^{\sL\sL}_{pq}C^{\sL\sL}_{pq} \ket{ T^{\dot{a}a}_q Z^{\sL}_p }, \\
\mathcal{S}^{\sL\sL}\check{\otimes}\mathcal{S}^{\sL\tilde{\sL}} \ket{\eta^{\sL \dot{a}}_p \chi^a_q} =& -A^{\sL\sL}_{pq}D^{\sL\sL}_{pq} \ket{\chi^a_q \eta^{\sL \dot{a}}_p } + A^{\sL\sL}_{pq}E^{\sL\sL}_{pq} \ket{ T^{\dot{a}a}_q Y^{\sL}_p }, \\
\mathcal{S}^{\sL\sL}\check{\otimes}\mathcal{S}^{\sL\tilde{\sL}} \ket{Z^{\sL}_p \widetilde{\chi}^a_q} =& + F^{\sL\sL}_{pq}F^{\sL\sL}_{pq} \ket{\widetilde{\chi}^a_q Z^{\sL}_p }, \\
\mathcal{S}^{\sL\sL}\check{\otimes}\mathcal{S}^{\sL\tilde{\sL}} \ket{Y^{\sL}_p \chi^a_q} =& + A^{\sL\sL}_{pq}A^{\sL\sL}_{pq} \ket{\chi^a_q Y^{\sL}_p }, \\
\mathcal{S}^{\sL\sL}\check{\otimes}\mathcal{S}^{\sL\tilde{\sL}} \ket{Z^{\sL}_p \chi^a_q} =& + D^{\sL\sL}_{pq}D^{\sL\sL}_{pq} \ket{\chi^a_q Z^{\sL}_p } + E^{\sL\sL}_{pq}E^{\sL\sL}_{pq} \ket{\widetilde{\chi}^a_q Y^{\sL}_p } + D^{\sL\sL}_{pq}E^{\sL\sL}_{pq} \, \epsilon_{\dot{a}\dot{b}} \ket{ T^{\dot{a}a}_q \eta^{\sL \dot{b}}_p}, \\
\mathcal{S}^{\sL\sL}\check{\otimes}\mathcal{S}^{\sL\tilde{\sL}} \ket{Y^{\sL}_p \widetilde{\chi}^a_q} =& + B^{\sL\sL}_{pq}B^{\sL\sL}_{pq} \ket{ \widetilde{\chi}^a_q Y^{\sL}_p } + C^{\sL\sL}_{pq}C^{\sL\sL}_{pq} \ket{\chi^a_q Z^{\sL}_p } + B^{\sL\sL}_{pq}C^{\sL\sL}_{pq} \, \epsilon_{\dot{a}\dot{b}} \ket{ T^{\dot{a}a}_q \eta^{\sL \dot{b}}_p}, \\
\mathcal{S}^{\sL\sL}\check{\otimes}\mathcal{S}^{\sL\tilde{\sL}} \ket{ \eta^{\sL \dot{a}}_p T^{\dot{b}a}_q} =& + D^{\sL\sL}_{pq}B^{\sL\sL}_{pq} \ket{ T^{\dot{a}a}_q \eta^{\sL \dot{b}}_p } - E^{\sL\sL}_{pq}C^{\sL\sL}_{pq} \ket{ T^{\dot{b}a}_q \eta^{\sL \dot{a}}_p } \\[-1ex]
& + D^{\sL\sL}_{pq}C^{\sL\sL}_{pq} \, \epsilon^{\dot{a}\dot{b}} \ket{ \chi_q^{a} Z^{\sL}_p} + E^{\sL\sL}_{pq}B^{\sL\sL}_{pq} \, \epsilon^{\dot{a}\dot{b}} \ket{ \widetilde{\chi}_q^{a} Y^{\sL}_p}.
\end{aligned}
\end{equation}
\endgroup
When we scatter a right excitation with a massless one we can write the S-matrix elements as\footnote{To be rigorous we should write the right massive fermion with a lower $\alg{su}(2)$ index, since the identification with right $\alg{psu}(1|1)^2$ representations is correctly implemented only in that case, see equation~\eqref{eq:mv-tensor}. To write the S-matrix we decide to raise this index with $\epsilon^{ab}$ to have a better notation.}
\begingroup
\addtolength{\jot}{1ex}
\begin{equation}\label{eq:mixed-S-matrix-R}
\begin{aligned}
\mathcal{S}^{\sR\sL}\check{\otimes}\mathcal{S}^{\sR\tilde{\sL}} \ket{Z^{\sR}_p T^{\dot{a}a}_q} =& -D^{\sL\sR}_{pq}E^{\sL\sR}_{pq} \ket{T^{\dot{a}a}_q Z^{\sR}_p } + D^{\sL\sR}_{pq}F^{\sL\sR}_{pq} \ket{\chi^a_q \eta^{\sR \dot{a}}_p}, \\
\mathcal{S}^{\sR\sL}\check{\otimes}\mathcal{S}^{\sR\tilde{\sL}} \ket{Y^{\sR}_p T^{\dot{a}a}_q} =& + A^{\sL\sR}_{pq}C^{\sL\sR}_{pq} \ket{T^{\dot{a}a}_q Y^{\sR}_p } -B^{\sL\sR}_{pq}C^{\sL\sR}_{pq}  \ket{ \widetilde{\chi}^a_q \eta^{\sR \dot{a}}_p }, \\
\mathcal{S}^{\sR\sL}\check{\otimes}\mathcal{S}^{\sR\tilde{\sL}} \ket{\eta^{\sR \dot{a}}_p \chi^a_q} =& -D^{\sL\sR}_{pq}A^{\sL\sR}_{pq} \ket{\chi^a_q \eta^{\sR \dot{a}}_p } + D^{\sL\sR}_{pq}B^{\sL\sR}_{pq} \ket{ T^{\dot{a}a}_q Z^{\sR}_p }, \\
\mathcal{S}^{\sR\sL}\check{\otimes}\mathcal{S}^{\sR\tilde{\sL}} \ket{\eta^{\sR \dot{a}}_p \widetilde{\chi}^a_q} =& + E^{\sL\sR}_{pq}C^{\sL\sR}_{pq} \ket{\widetilde{\chi}^a_q \eta^{\sR \dot{a}}_p } - F^{\sL\sR}_{pq}C^{\sL\sR}_{pq} \ket{ T^{\dot{a}a}_q Y^{\sR}_p }, \\
\mathcal{S}^{\sR\sL}\check{\otimes}\mathcal{S}^{\sR\tilde{\sL}} \ket{Z^{\sR}_p \chi^a_q} =& + D^{\sL\sR}_{pq}D^{\sL\sR}_{pq} \ket{\chi^a_q Z^{\sR}_p }, \\
\mathcal{S}^{\sR\sL}\check{\otimes}\mathcal{S}^{\sR\tilde{\sL}} \ket{Y^{\sR}_p \widetilde{\chi}^a_q} =& + C^{\sL\sR}_{pq}C^{\sL\sR}_{pq} \ket{\widetilde{\chi}^a_q Y^{\sR}_p }, \\
\mathcal{S}^{\sR\sL}\check{\otimes}\mathcal{S}^{\sR\tilde{\sL}} \ket{Z^{\sR}_p \widetilde{\chi}^a_q} =& + E^{\sL\sR}_{pq}E^{\sL\sR}_{pq} \ket{\widetilde{\chi}^a_q Z^{\sR}_p }  -F^{\sL\sR}_{pq}F^{\sL\sR}_{pq} \ket{\chi^a_q Y^{\sR}_p } +F^{\sL\sR}_{pq}E^{\sL\sR}_{pq}  \, \epsilon_{\dot{a}\dot{b}} \ket{ T^{\dot{a}a}_q \eta^{\sR \dot{b}}_p}, \\
\mathcal{S}^{\sR\sL}\check{\otimes}\mathcal{S}^{\sR\tilde{\sL}} \ket{Y^{\sR}_p \chi^a_q} =& + A^{\sL\sR}_{pq}A^{\sL\sR}_{pq} \ket{ \chi^a_q Y^{\sR}_p } - B^{\sL\sR}_{pq}B^{\sL\sR}_{pq}\ket{\widetilde{\chi}^a_q Z^{\sR}_p } - B^{\sL\sR}_{pq}A^{\sL\sR}_{pq} \, \epsilon_{\dot{a}\dot{b}} \ket{ T^{\dot{a}a}_q \eta^{\sR \dot{b}}_p}, \\
\mathcal{S}^{\sR\sL}\check{\otimes}\mathcal{S}^{\sR\tilde{\sL}} \ket{ \eta^{\sR \dot{a}}_p T^{\dot{b}a}_q} =& + B^{\sL\sR}_{pq}F^{\sL\sR}_{pq} \ket{ T^{\dot{a}a}_q \eta^{\sR \dot{b}}_p } - A^{\sL\sR}_{pq}E^{\sL\sR}_{pq} \ket{ T^{\dot{b}a}_q \eta^{\sR \dot{a}}_p } \\[-1ex]
& - B^{\sL\sR}_{pq}E^{\sL\sR}_{pq} \, \epsilon^{\dot{a}\dot{b}} \ket{ \widetilde{\chi}^{a}_q Z^{\sR}_p} + A^{\sL\sR}_{pq}F^{\sL\sR}_{pq} \, \epsilon^{\dot{a}\dot{b}} \ket{ \chi^{a}_q Y^{\sR}_p}.
\end{aligned}
\end{equation}
\endgroup
After taking into account a proper normalisation (see (\ref{eq:norm-sect})), the S-matrix elements for left-massless and right-massless scattering can be related by LR symmetry. In order to do so one needs to implement it on massive and massless excitations as in equations~\eqref{eq:LR-massive} and~\eqref{eq:LR-massless}.

\subsection{The massless sector}
We write the non-vanishing entries of the two-particle S~matrix in the massless sector. First we focus on the structure fixed by the $\alg{psu}(1|1)^4$ invariance. For this reason we omit the indices corresponding to $\alg{su}(2)_{\circ}$.
\begingroup
\addtolength{\jot}{1ex}
\begin{equation}\label{eq:massless-S-matrix}
\begin{aligned}
\mathcal{S}^{\sL\sL}\check{\otimes}\mathcal{S}^{\tilde{\sL}\tilde{\sL}} \ket{T^{\dot{a}\ }_p T^{\dot{b}\ }_q} =& -C^{\sL\sL}_{pq} E^{\sL\sL}_{pq} \ket{T^{\dot{a}\ }_q T^{\dot{b}\ }_p} +B^{\sL\sL}_{pq}D^{\sL\sL}_{pq} \ket{T^{\dot{b}\ }_q T^{\dot{a}\ }_p} \\[-1ex]
& +\epsilon^{\dot{a}\dot{b}} \left(C^{\sL\sL}_{pq} D^{\sL\sL}_{pq} \ket{\chi^{\ }_q \widetilde{\chi}^{\ }_p} + B^{\sL\sL}_{pq}E^{\sL\sL}_{pq}  \ket{\widetilde{\chi}^{\ }_q \chi^{\ }_p}\right), \\
\mathcal{S}^{\sL\sL}\check{\otimes}\mathcal{S}^{\tilde{\sL}\tilde{\sL}} \ket{T^{\dot{a}\ }_p \widetilde{\chi}^{\ }_q} =&-B^{\sL\sL}_{pq}F^{\sL\sL}_{pq} \ket{\widetilde{\chi}^{\ }_q T^{\dot{a}\ }_p} - C^{\sL\sL}_{pq}F^{\sL\sL}_{pq} \ket{T^{\dot{a}\ }_q \widetilde{\chi}^{\ }_p}, \\
\mathcal{S}^{\sL\sL}\check{\otimes}\mathcal{S}^{\tilde{\sL}\tilde{\sL}} \ket{\widetilde{\chi}^{\ }_p T^{\dot{a}\ }_q} =&-F^{\sL\sL}_{pq}D^{\sL\sL}_{pq}\ket{T^{\dot{a}\ }_q \widetilde{\chi}^{\ }_p} - F^{\sL\sL}_{pq}E^{\sL\sL}_{pq}\ket{\widetilde{\chi}^{\ }_q T^{\dot{a}\ }_p}, \\
\mathcal{S}^{\sL\sL}\check{\otimes}\mathcal{S}^{\tilde{\sL}\tilde{\sL}} \ket{T^{\dot{a}\ }_p \chi^{\ }_q} =&-B^{\sL\sL}_{pq}F^{\sL\sL}_{pq} \ket{\chi^{\ }_q T^{\dot{a}\ }_p} - C^{\sL\sL}_{pq}F^{\sL\sL}_{pq} \ket{T^{\dot{a}\ }_q \chi^{\ }_p}, \\
\mathcal{S}^{\sL\sL}\check{\otimes}\mathcal{S}^{\tilde{\sL}\tilde{\sL}} \ket{\chi^{\ }_p T^{\dot{a}\ }_q} =&-F^{\sL\sL}_{pq}D^{\sL\sL}_{pq}\ket{T^{\dot{a}\ }_q \chi^{\ }_p} - F^{\sL\sL}_{pq}E^{\sL\sL}_{pq}\ket{\chi^{\ }_q T^{\dot{a}\ }_p}, \\
\mathcal{S}^{\sL\sL}\check{\otimes}\mathcal{S}^{\tilde{\sL}\tilde{\sL}} \ket{\widetilde{\chi}^{\ }_p \widetilde{\chi}^{\ }_q} =& -A^{\sL\sL}_{pq}A^{\sL\sL}_{pq} \ket{\widetilde{\chi}^{\ }_q \widetilde{\chi}^{\ }_p}, \\
\mathcal{S}^{\sL\sL}\check{\otimes}\mathcal{S}^{\tilde{\sL}\tilde{\sL}} \ket{\chi^{\ }_p \chi^{\ }_q} =& -A^{\sL\sL}_{pq}A^{\sL\sL}_{pq} \ket{\chi^{\ }_q \chi^{\ }_p}, \\
\mathcal{S}^{\sL\sL}\check{\otimes}\mathcal{S}^{\tilde{\sL}\tilde{\sL}} \ket{\widetilde{\chi}^{\ }_p \chi^{\ }_q} =& -D^{\sL\sL}_{pq}D^{\sL\sL}_{pq}\ket{\chi^{\ }_q \widetilde{\chi}^{\ }_p} - E^{\sL\sL}_{pq}E^{\sL\sL}_{pq}\ket{\widetilde{\chi}^{\ }_q \chi^{\ }_p} - E^{\sL\sL}_{pq} D^{\sL\sL}_{pq} \epsilon_{\dot{a}\dot{b}} \ket{T^{\dot{a}\ }_q T^{\dot{b}\ }_p}, \\
\mathcal{S}^{\sL\sL}\check{\otimes}\mathcal{S}^{\tilde{\sL}\tilde{\sL}} \ket{\chi^{\ }_p \widetilde{\chi}^{\ }_q} =& -D^{\sL\sL}_{pq}D^{\sL\sL}_{pq}\ket{\widetilde{\chi}^{\ }_q \chi^{\ }_p} - E^{\sL\sL}_{pq}E^{\sL\sL}_{pq}\ket{\chi^{\ }_q \widetilde{\chi}^{\ }_p} + E^{\sL\sL}_{pq} D^{\sL\sL}_{pq} \epsilon_{\dot{a}\dot{b}} \ket{T^{\dot{a}\ }_q T^{\dot{b}\ }_p}.
\end{aligned}
\end{equation}
\endgroup
The structure fixed by the $\alg{su}(2)_{\circ}$ symmetry is as follows
\begin{equation}
\mathcal{S}_{\alg{su}(2)} \ket{\mathcal{X}^a_p \mathcal{Y}^b_q} = \frac{1}{1+\varsigma_{pq}} \left( \varsigma_{pq} \ket{\mathcal{Y'}^b_q \mathcal{X'}^a_p} + \ket{\mathcal{Y'}^a_q \mathcal{X'}^b_p}\right),
\end{equation}
where we use $\mathcal{X},\mathcal{Y},\mathcal{X'},\mathcal{Y'}$ to denote any of the excitations that appear above.
The full S~matrix in the massless sector is then found by combining the structures fixed by $\alg{psu}(1|1)^4_{\ce}$ and $\alg{su}(2)_{\circ}$ and multiplying each element by the scalar factor as in (\ref{eq:norm-sect}).
This S-matrix automatically satisfies the LR-symmetry, where this is implemented on massless excitations as in (\ref{eq:LR-massless}).

\section{Normalization of S-matrix elements}
\label{app:dressingfactors}
In order to obtain the normalisation of section~\ref{sec:normalisation} we can can multiply each block of the S~matrix by the following prefactors
\begin{equation}\label{eq:norm-sect}
\begin{aligned}
\text{LL, RR:} \qquad & \frac{x^+_p}{x^-_p} \, \frac{x^-_q}{x^+_q} \, \frac{x^-_p - x^+_q}{x^+_p - x^-_q} \, \frac{1-\frac{1}{x^-_p x^+_q}}{1-\frac{1}{x^+_p x^-_q}} \, \frac{1}{\left(\sigma^{\bullet\bullet}_{pq} \right)^2 }, \\
\text{LR, RL:} \qquad & \left(\frac{x^+_q}{x^-_q} \right)^{-1}  \, \frac{1-\frac{1}{x^-_p x^+_q}}{1-\frac{1}{x^+_p x^-_q}} \, \zeta_{pq}^2 \, \frac{1}{\left(\tilde{\sigma}^{\bullet\bullet}_{pq}\right)^2}  , 
\end{aligned}
\end{equation}
\begin{equation}
\begin{aligned}
\bullet\circ: \qquad & \left( \frac{x^+_p}{x^-_p} \right)^{-1/2}\, \left( \frac{1-\frac{1}{x^-_p x^+_q}}{1-\frac{1}{x^+_p x^-_q}} \right)^{1/2} \, \zeta_{pq} \, \frac{1}{\left(\sigma^{\bullet\circ}_{pq} \right)^2 }, \\
\circ\bullet: \qquad & \left( \frac{x^+_q}{x^-_q} \right)^{1/2}\, \left( \frac{1-\frac{1}{x^-_p x^+_q}}{1-\frac{1}{x^+_p x^-_q}} \right)^{1/2} \, \zeta_{pq}^{-1} \, \frac{1}{\left(\sigma^{\circ\bullet}_{pq} \right)^2 },
\end{aligned}
\end{equation}
\begin{equation}
\begin{aligned}
\circ\circ: \qquad & \left( \frac{x^+_p}{x^-_p} \, \frac{x^-_q}{x^+_q} \right)^{1/2}\, \frac{x^-_p - x^+_q}{x^+_p - x^-_q} \, \frac{1}{\left(\sigma^{\circ\circ}_{pq} \right)^2 }, 
\end{aligned}
\end{equation}
where we defined
\begin{equation}
 \zeta_{pq} = \left(\frac{1-\frac{1}{x^-_p x^-_q}}{1-\frac{1}{x^+_p x^+_q}} \right)^{1/2}.
\end{equation}

\section{Crossing equations for the S-matrix elements}
\label{app:crossing}
For completeness we present the crossing equations also in terms of the S-matrix elements.
To obtain simpler expressions we choose to focus on processes involving the highest weight states of the left, right and massless
modules, which are enough to constrain all of the dressing factors.
We then rewrite the crossing equations in terms of these scattering processes and the ones with the conjugates of the highest weight states.

For example, in the massive sector we can consider
\begin{equation}
\begin{aligned}
\mathcal{A}^{\bullet\bullet}_{pq} &\equiv \bra{ {Y}^{\sL}_q
{Y}^{\sL}_p } \mathcal{S} \ket{{Y}^{\sL}_p {Y}^{\sL}_q},
\qquad
\mathcal{B}^{\bullet\bullet}_{pq} &\equiv  \bra{{Y}^{\sL}_q
{Y}^{\sR}_p} \mathcal{S} \ket{{Y}^{\sR}_p {Y}^{\sL}_q},
\\
\widetilde{\mathcal{A}}^{\bullet\bullet}_{pq} &\equiv \bra{
{Z}^{\sR}_q {Y}^{\sL}_p } \mathcal{S} \ket{{Y}^{\sL}_p {Z}^{\sR}_q},
\qquad
\widetilde{\mathcal{B}}^{\bullet\bullet}_{pq} &\equiv
\bra{{Z}^{\sR}_q {Y}^{\sR}_p} \mathcal{S} \ket{{Y}^{\sR}_p
{Z}^{\sR}_q}.
\end{aligned}
\end{equation}
The two crossing equations are then equivalent to imposing
\begin{equation}
\mathcal{A}^{\bullet\bullet}_{pq}
\mathcal{B}^{\bullet\bullet}_{\bar{p}q} =1, \qquad
\widetilde{\mathcal{A}}^{\bullet\bullet}_{pq}
\widetilde{\mathcal{B}}^{\bullet\bullet}_{\bar{p}q} =1.
\end{equation}
Similarly, in the mixed-mass sector one can look at the processes
\begin{equation}
\begin{aligned}
\mathcal{A}^{\bullet\circ}_{pq} &\equiv \bra{ \chi^a_q {Y}^{\sL}_p }
\mathcal{S} \ket{{Y}^{\sL}_p \chi^a_q},
\qquad
\mathcal{B}^{\bullet\circ}_{pq} &\equiv  \bra{\chi^a_q {Y}^{\sR}_p}
\mathcal{S} \ket{{Y}^{\sR}_p \chi^a_q},
\\
\mathcal{A}^{\circ\bullet}_{pq} &\equiv \bra{{Y}^{\sL}_q \chi^a_p }
\mathcal{S} \ket{ \chi^a_p {Y}^{\sL}_q},
\qquad
\mathcal{B}^{\circ\bullet}_{pq} &\equiv  \bra{{Y}^{\sL}_q
\tilde{\chi}_{a \, p} } \mathcal{S} \ket{ \tilde{\chi}_{a \, p}
{Y}^{\sL}_q},
\end{aligned}
\end{equation}
and obtain the corresponding crossing equations by imposing
\begin{equation}
\mathcal{A}^{\bullet\circ}_{pq} \mathcal{B}^{\bullet\circ}_{\bar{p}q} =1, \qquad
\mathcal{A}^{\circ\bullet}_{pq} \mathcal{B}^{\circ\bullet}_{\bar{p}q} =1.
\end{equation}
To conclude, in the massless sector we can choose
\begin{equation}
\begin{aligned}
\mathcal{A}^{\circ\circ}_{pq} &\equiv \bra{\chi^a_q \chi^a_p}
\mathcal{S} \ket{\chi^a_p \chi^a_q}, \qquad
\mathcal{B}^{\circ\circ}_{pq} &\equiv  \bra{\chi^a_q \tilde{\chi}_{a
\, p}} \mathcal{S} \ket{\tilde{\chi}_{a \, p} \chi^a_q} ,
\end{aligned}
\end{equation}
and rewrite the crossing equations for $\sigma^{\circ\circ}_{pq}$ and $w_p$ as
\begin{equation}
\mathcal{A}^{\circ\circ}_{pq} \mathcal{B}^{\circ\circ}_{\bar{p}q} =1.
\end{equation}

\bibliographystyle{nb}
\bibliography{massless}

\begin{thebibliography}{100}
\ifx\href\asklfhas\newcommand{\href}[2]{#2}\fi
\ifx\arxivref\asklfhas\newcommand{\arxivref}[2]{\href{http://arxiv.org/abs/#1}{#2}}\fi
\ifx\doiref\asklfhas\newcommand{\doiref}[2]{\href{http://dx.doi.org/#1}{#2}}\fi
\raggedright
\small
\parskip 0pt

\bibitem{Maldacena:1997re}
J.~M.~Maldacena,
\textit{``The large {N} limit of superconformal field theories and
  supergravity''},
\textsf{Adv.~Theor.~Math.~Phys.~2,~231~(1998)},
\texttt{\arxivref{hep-th/9711200}{hep-th/9711200}}.

\bibitem{Witten:1998qj}
E.~Witten,
\textit{``Anti-de {S}itter space and holography''},
\textsf{Adv.~Theor.~Math.~Phys.~2,~253~(1998)},
\texttt{\arxivref{hep-th/9802150}{hep-th/9802150}}.

\bibitem{Gubser:1998bc}
S.~S.~Gubser, I.~R.~Klebanov and A.~M.~Polyakov,
\textit{``Gauge theory correlators from non-critical string theory''},
\textsf{\doiref{10.1016/S0370-2693(98)00377-3}{Phys.~Lett.~B428,~105~(1998)}},
\texttt{\arxivref{hep-th/9802109}{hep-th/9802109}}.

\bibitem{'tHooft:1973jz}
G.~'t~Hooft,
\textit{``A Planar Diagram Theory for Strong Interactions''},
\textsf{\doiref{10.1016/0550-3213(74)90154-0}{Nucl.Phys.~B72,~461~(1974)}}.

\bibitem{Arutyunov:2009ga}
G.~Arutyunov and S.~Frolov,
\textit{``Foundations of the {$\AdS_5 \times \Sphere^5$} Superstring. Part
  {I}''},
\textsf{\doiref{10.1088/1751-8113/42/25/254003}{J.Phys.A~A42,~254003~(2009)}},
\texttt{\arxivref{0901.4937}{arxiv:0901.4937}}.

\bibitem{Beisert:2010jr}
N.~Beisert et~al.,
\textit{``Review of {AdS/CFT} Integrability: An Overview''},
\textsf{\doiref{10.1007/s11005-011-0529-2}{Lett.Math.Phys.~99,~3~(2012)}},
\texttt{\arxivref{1012.3982}{arxiv:1012.3982}}.

\bibitem{Aharony:2008ug}
O.~Aharony, O.~Bergman, D.~L.~Jafferis and J.~Maldacena,
\textit{``{$\superN = 6$} superconformal {C}hern-{S}imons-matter theories,
  {M2}-branes and their gravity duals''},
\textsf{\doiref{10.1088/1126-6708/2008/10/091}{JHEP~0810,~091~(2008)}},
\texttt{\arxivref{0806.1218}{arxiv:0806.1218}}.

\bibitem{Klose:2010ki}
T.~Klose,
\textit{``Review of {AdS/CFT} Integrability, {C}hapter {IV.3}: {$\superN = 6$}
  {C}hern-{S}imons and Strings on {$\AdS_4 \times \CP^3$}''},
\textsf{\doiref{10.1007/s11005-011-0520-y}{Lett.Math.Phys.~99,~401~(2010)}},
\texttt{\arxivref{1012.3999}{arxiv:1012.3999}}.

\bibitem{Brown:1986nw}
J.~D.~Brown and M.~Henneaux,
\textit{``Central Charges in the Canonical Realization of Asymptotic
  Symmetries: An Example from Three-Dimensional Gravity''},
\textsf{\doiref{10.1007/BF01211590}{Commun.~Math.~Phys.~104,~207~(1986)}}.

\bibitem{Banados:1992wn}
M.~Ba{\~n}ados, C.~Teitelboim and J.~Zanelli,
\textit{``The Black hole in three-dimensional space-time''},
\textsf{\doiref{10.1103/PhysRevLett.69.1849}{Phys.Rev.Lett.~69,~1849~(1992)}},
\texttt{\arxivref{hep-th/9204099}{hep-th/9204099}}.

\bibitem{Banados:1992gq}
M.~Ba{\~n}ados, M.~Henneaux, C.~Teitelboim and J.~Zanelli,
\textit{``Geometry of the (2+1) black hole''},
\textsf{\doiref{10.1103/PhysRevD.48.1506}{Phys.Rev.~D48,~1506~(1993)}},
\texttt{\arxivref{gr-qc/9302012}{gr-qc/9302012}}.

\bibitem{Strominger:1997eq}
A.~Strominger,
\textit{``Black hole entropy from near horizon microstates''},
\textsf{\doiref{10.1088/1126-6708/1998/02/009}{JHEP~9802,~009~(1998)}},
\texttt{\arxivref{hep-th/9712251}{hep-th/9712251}}.

\bibitem{Strominger:1996sh}
A.~Strominger and C.~Vafa,
\textit{``Microscopic origin of the {B}ekenstein-{H}awking entropy''},
\textsf{\doiref{10.1016/0370-2693(96)00345-0}{Phys.Lett.~B379,~99~(1996)}},
\texttt{\arxivref{hep-th/9601029}{hep-th/9601029}}.

\bibitem{Seiberg:1999xz}
N.~Seiberg and E.~Witten,
\textit{``The {D1/D5} system and singular {CFT}''},
\textsf{\doiref{10.1088/1126-6708/1999/04/017}{JHEP~9904,~017~(1999)}},
\texttt{\arxivref{hep-th/9903224}{hep-th/9903224}}.

\bibitem{Boonstra:1998yu}
H.~J.~Boonstra, B.~Peeters and K.~Skenderis,
\textit{``Brane intersections, anti-de {S}itter spacetimes and dual
  superconformal theories''},
\textsf{\doiref{10.1016/S0550-3213(98)00512-4}{Nucl.~Phys.~B533,~127~(1998)}},
\texttt{\arxivref{hep-th/9803231}{hep-th/9803231}}.

\bibitem{Giveon:1998ns}
A.~Giveon, D.~Kutasov and N.~Seiberg,
\textit{``Comments on string theory on {$\AdS_3$}''},
\textsf{Adv.~Theor.~Math.~Phys.~2,~733~(1998)},
\texttt{\arxivref{hep-th/9806194}{hep-th/9806194}}.

\bibitem{deBoer:1998pp}
J.~de~Boer, H.~Ooguri, H.~Robins and J.~Tannenhauser,
\textit{``String theory on {$\AdS_3$}''},
\textsf{\doiref{10.1088/1126-6708/1998/12/026}{JHEP~9812,~026~(1998)}},
\texttt{\arxivref{hep-th/9812046}{hep-th/9812046}}.

\bibitem{Elitzur:1998mm}
S.~Elitzur, O.~Feinerman, A.~Giveon and D.~Tsabar,
\textit{``String theory on {$\AdS_3 \times \Sphere^3 \times \Sphere^3 \times
  \Sphere^1$}''},
\textsf{\doiref{10.1016/S0370-2693(99)00101-X}{Phys.~Lett.~B449,~180~(1999)}},
\texttt{\arxivref{hep-th/9811245}{hep-th/9811245}}.

\bibitem{Aharony:1999ti}
O.~Aharony, S.~S.~Gubser, J.~M.~Maldacena, H.~Ooguri and Y.~Oz,
\textit{``Large {N} field theories, string theory and gravity''},
\textsf{\doiref{10.1016/S0370-1573(99)00083-6}{Phys.~Rept.~323,~183~(2000)}},
\texttt{\arxivref{hep-th/9905111}{hep-th/9905111}}.

\bibitem{Maldacena:2000hw}
J.~M.~Maldacena and H.~Ooguri,
\textit{``Strings in {$\AdS_3$} and {$\grpSL(2,R)$} {WZW} model. {I}''},
\textsf{\doiref{10.1063/1.1377273}{J.~Math.~Phys.~42,~2929~(2001)}},
\texttt{\arxivref{hep-th/0001053}{hep-th/0001053}}.

\bibitem{Maldacena:2000kv}
J.~M.~Maldacena, H.~Ooguri and J.~Son,
\textit{``Strings in {$\AdS_3$} and the {$\grpSL(2,R)$} {WZW} model. {II}:
  {E}uclidean black hole''},
\textsf{\doiref{10.1063/1.1377039}{J.~Math.~Phys.~42,~2961~(2001)}},
\texttt{\arxivref{hep-th/0005183}{hep-th/0005183}}.

\bibitem{Maldacena:2001km}
J.~M.~Maldacena and H.~Ooguri,
\textit{``Strings in {$\AdS_3$} and the {$\grpSL(2,R)$} {WZW} model.~{III}:
  Correlation functions''},
\textsf{\doiref{10.1103/PhysRevD.65.106006}{Phys.~Rev.~D65,~106006~(2002)}},
\texttt{\arxivref{hep-th/0111180}{hep-th/0111180}}.

\bibitem{Berkovits:1999im}
N.~Berkovits, C.~Vafa and E.~Witten,
\textit{``Conformal field theory of {AdS} background with {R}amond-{R}amond
  flux''},
\textsf{\doiref{10.1088/1126-6708/1999/03/018}{JHEP~9903,~018~(1999)}},
\texttt{\arxivref{hep-th/9902098}{hep-th/9902098}}.

\bibitem{Babichenko:2009dk}
A.~Babichenko, B.~Stefa{\'n}ski,~jr. and K.~Zarembo,
\textit{``Integrability and the {$\AdS_3/\CFT_2$} correspondence''},
\textsf{\doiref{10.1007/JHEP03(2010)058}{JHEP~1003,~058~(2010)}},
\texttt{\arxivref{0912.1723}{arxiv:0912.1723}}.

\bibitem{Sundin:2013uca}
P.~Sundin and L.~Wulff,
\textit{``The low energy limit of the {$\AdS_3 \times \Sphere^3 \times M_4$}
  spinning string''},
\textsf{\doiref{10.1007/JHEP10(2013)111}{JHEP~1310,~111~(2013)}},
\texttt{\arxivref{1306.6918}{arxiv:1306.6918}}.

\bibitem{Cagnazzo:2012se}
A.~Cagnazzo and K.~Zarembo,
\textit{``{B}-field in {$\AdS_3/\CFT_2$} Correspondence and Integrability''},
\textsf{\doiref{10.1007/JHEP11(2012)133,
  10.1007/JHEP04(2013)003}{JHEP~1211,~133~(2012)}},
\texttt{\arxivref{1209.4049}{arxiv:1209.4049}}.

\bibitem{Zarembo:2010sg}
K.~Zarembo,
\textit{``Strings on Semisymmetric Superspaces''},
\textsf{\doiref{10.1007/JHEP05(2010)002}{JHEP~1005,~002~(2010)}},
\texttt{\arxivref{1003.0465}{arxiv:1003.0465}}.

\bibitem{Sundin:2012gc}
P.~Sundin and L.~Wulff,
\textit{``Classical integrability and quantum aspects of the {$\AdS_3 \times
  \Sphere^3 \times \Sphere^3 \times \Sphere^1$} superstring''},
\textsf{\doiref{10.1007/JHEP10(2012)109}{JHEP~1210,~109~(2012)}},
\texttt{\arxivref{1207.5531}{arxiv:1207.5531}}.

\bibitem{Pakman:2009mi}
A.~Pakman, L.~Rastelli and S.~S.~Razamat,
\textit{``A Spin Chain for the Symmetric Product {$\CFT_2$}''},
\textsf{\doiref{10.1007/JHEP05(2010)099}{JHEP~1005,~099~(2010)}},
\texttt{\arxivref{0912.0959}{arxiv:0912.0959}}.

\bibitem{OhlssonSax:2011ms}
O.~Ohlsson~Sax and B.~Stefa{\'n}ski,~jr.,
\textit{``Integrability, spin-chains and the {$\AdS_3/\CFT_2$}
  correspondence''},
\textsf{\doiref{10.1007/JHEP08(2011)029}{JHEP~1108,~029~(2011)}},
\texttt{\arxivref{1106.2558}{arxiv:1106.2558}}.

\bibitem{Sax:2012jv}
O.~Ohlsson~Sax, B.~Stefa{\'n}ski,~jr. and A.~Torrielli,
\textit{``On the massless modes of the {$\AdS_3/\CFT_2$} integrable systems''},
\textsf{\doiref{10.1007/JHEP03(2013)109}{JHEP~1303,~109~(2013)}},
\texttt{\arxivref{1211.1952}{arxiv:1211.1952}}.

\bibitem{Borsato:2012ud}
R.~Borsato, O.~Ohlsson~Sax and A.~Sfondrini,
\textit{``A dynamic {$\algSU(1|1)^2$} {S}-matrix for {$\AdS_3/\CFT_2$}''},
\textsf{\doiref{10.1007/JHEP04(2013)113}{JHEP~1304,~113~(2013)}},
\texttt{\arxivref{1211.5119}{arxiv:1211.5119}}.

\bibitem{Borsato:2012ss}
R.~Borsato, O.~Ohlsson~Sax and A.~Sfondrini,
\textit{``All-loop {B}ethe ansatz equations for {$\AdS_3/\CFT_2$}''},
\textsf{\doiref{10.1007/JHEP04(2013)116}{JHEP~1304,~116~(2013)}},
\texttt{\arxivref{1212.0505}{arxiv:1212.0505}}.

\bibitem{Borsato:2013qpa}
R.~Borsato, O.~Ohlsson~Sax, A.~Sfondrini, B.~Stefa{\'n}ski,~jr. and
  A.~Torrielli,
\textit{``The all-loop integrable spin-chain for strings on {$\AdS_3 \times
  \Sphere^3 \times \Torus^4$}: the massive sector''},
\textsf{\doiref{10.1007/JHEP08(2013)043}{JHEP~1308,~043~(2013)}},
\texttt{\arxivref{1303.5995}{arxiv:1303.5995}}.

\bibitem{Borsato:2013hoa}
R.~Borsato, O.~Ohlsson~Sax, A.~Sfondrini, B.~Stefa{\'n}ski,~jr. and
  A.~Torrielli,
\textit{``Dressing phases of {$\AdS_3/\CFT_2$}''},
\textsf{\doiref{10.1103/PhysRevD.88.066004}{Phys.Rev.~D88,~066004~(2013)}},
\texttt{\arxivref{1306.2512}{arxiv:1306.2512}}.

\bibitem{Rughoonauth:2012qd}
N.~Rughoonauth, P.~Sundin and L.~Wulff,
\textit{``Near {BMN} dynamics of the {$\AdS_3 \times \Sphere^3 \times \Sphere^3
  \times \Sphere^1$} superstring''},
\textsf{\doiref{10.1007/JHEP07(2012)159}{JHEP~1207,~159~(2012)}},
\texttt{\arxivref{1204.4742}{arxiv:1204.4742}}.

\bibitem{Beccaria:2012kb}
M.~Beccaria, F.~Levkovich-Maslyuk, G.~Macorini and A.~A.~Tseytlin,
\textit{``Quantum corrections to spinning superstrings in {$\AdS_3 \times
  \Sphere^3 \times M^4$}: determining the dressing phase''},
\textsf{\doiref{10.1007/JHEP04(2013)006}{JHEP~1304,~006~(2013)}},
\texttt{\arxivref{1211.6090}{arxiv:1211.6090}}.

\bibitem{Beccaria:2012pm}
M.~Beccaria and G.~Macorini,
\textit{``Quantum corrections to short folded superstring in {$\AdS_3 \times
  \Sphere^3 \times M^4$}''},
\textsf{\doiref{10.1007/JHEP03(2013)040}{JHEP~1303,~040~(2013)}},
\texttt{\arxivref{1212.5672}{arxiv:1212.5672}}.

\bibitem{Abbott:2012dd}
M.~C.~Abbott,
\textit{``Comment on Strings in {$\AdS_3 \times \Sphere^3 \times \Sphere^3
  \times \Sphere^1$} at One Loop''},
\textsf{\doiref{10.1007/JHEP02(2013)102}{JHEP~1302,~102~(2013)}},
\texttt{\arxivref{1211.5587}{arxiv:1211.5587}}.

\bibitem{Sundin:2013ypa}
P.~Sundin and L.~Wulff,
\textit{``Worldsheet scattering in {$\AdS_3/\CFT_2$}''},
\texttt{\arxivref{1302.5349}{arxiv:1302.5349}}.

\bibitem{Abbott:2013ixa}
M.~C.~Abbott,
\textit{``The {$\AdS_3 \times \Sphere^3 \times \Sphere^3 \times \Sphere^1$}
  {H}ern{\'a}ndez-{L}{\'o}pez Phases: a Semiclassical Derivation''},
\textsf{\doiref{10.1088/1751-8113/46/44/445401}{J.~Phys.~A46,~445401~(2013)}},
\texttt{\arxivref{1306.5106}{arxiv:1306.5106}}.

\bibitem{Hoare:2013pma}
B.~Hoare and A.~A.~Tseytlin,
\textit{``On string theory on {$\AdS_3 \times \Sphere^3 \times \Torus^4$} with
  mixed 3-form flux: tree-level {S}-matrix''},
\textsf{\doiref{10.1016/j.nuclphysb.2013.05.005}{Nucl.Phys.~B873,~682~(2013)}},
\texttt{\arxivref{1303.1037}{arxiv:1303.1037}}.

\bibitem{Hoare:2013ida}
B.~Hoare and A.~Tseytlin,
\textit{``Massive {S}-matrix of {$\AdS_3 \times \Sphere^3 \times \Torus^4$}
  superstring theory with mixed 3-form flux''},
\textsf{\doiref{10.1016/j.nuclphysb.2013.04.024}{Nucl.Phys.~B873,~395~(2013)}},
\texttt{\arxivref{1304.4099}{arxiv:1304.4099}}.

\bibitem{Hoare:2013lja}
B.~Hoare, A.~Stepanchuk and A.~Tseytlin,
\textit{``Giant magnon solution and dispersion relation in string theory in
  {$\AdS_3 \times \Sphere^3 \times \Torus^4$} with mixed flux''},
\texttt{\arxivref{1311.1794}{arxiv:1311.1794}}.

\bibitem{Engelund:2013fja}
O.~T.~Engelund, R.~W.~McKeown and R.~Roiban,
\textit{``Generalized unitarity and the worldsheet {S} matrix in {$\AdS_n
  \times \Sphere^n \times M^{10-2n}$}''},
\texttt{\arxivref{1304.4281}{arxiv:1304.4281}}.

\bibitem{Sundin:2014sfa}
P.~Sundin,
\textit{``Worldsheet two- and four-point functions at one loop in
  {$\AdS_3/\CFT_2$}''},
\texttt{\arxivref{1403.1449}{arxiv:1403.1449}}.

\bibitem{Babichenko:2014yaa}
A.~Babichenko, A.~Dekel and O.~Ohlsson~Sax,
\textit{``Finite-gap equations for strings on {$\AdS_3 \times \Sphere^3 \times
  \Torus^4$} with mixed 3-form flux''},
\texttt{\arxivref{1405.6087}{arxiv:1405.6087}}.

\bibitem{Bianchi:2014rfa}
L.~Bianchi and B.~Hoare,
\textit{``{$\AdS_3 \times \Sphere^3 \times M^4$} string {S}-matrices from
  unitarity cuts''},
\texttt{\arxivref{1405.7947}{arxiv:1405.7947}}.

\bibitem{Sfondrini:2014via}
A.~Sfondrini,
\textit{``Towards integrability for {$\AdS_3/\CFT_2$}''},
\texttt{\arxivref{1406.2971}{arxiv:1406.2971}}.

\bibitem{Zamolodchikov:1992zr}
A.~B.~Zamolodchikov and A.~B.~Zamolodchikov,
\textit{``Massless factorized scattering and sigma models with topological
  terms''},
\textsf{\doiref{10.1016/0550-3213(92)90136-Y}{Nucl.Phys.~B379,~602~(1992)}}.

\bibitem{Fendley:1993wq}
P.~Fendley, H.~Saleur and A.~B.~Zamolodchikov,
\textit{``Massless flows. 1. The {S}ine-{G}ordon and {$\grpO(n)$} models''},
\textsf{\doiref{10.1142/S0217751X93002265}{Int.J.Mod.Phys.~A8,~5717~(1993)}},
\texttt{\arxivref{hep-th/9304050}{hep-th/9304050}}.

\bibitem{Fendley:1993xa}
P.~Fendley, H.~Saleur and A.~B.~Zamolodchikov,
\textit{``Massless flows, 2. The Exact {S} matrix approach''},
\textsf{\doiref{10.1142/S0217751X93002277}{Int.J.Mod.Phys.~A8,~5751~(1993)}},
\texttt{\arxivref{hep-th/9304051}{hep-th/9304051}}.

\bibitem{David:2014qta}
J.~R.~David and A.~Sadhukhan,
\textit{``Spinning strings and minimal surfaces in {$\AdS_3$} with mixed 3-form
  fluxes''},
\texttt{\arxivref{1405.2687}{arxiv:1405.2687}}.

\bibitem{Ahn:2014tua}
C.~Ahn and P.~Bozhilov,
\textit{``String solutions in {$\AdS_3 \times \Sphere^3 \times \Torus^4$} with
  {NS}-{NS} {B}-field''},
\texttt{\arxivref{1404.7644}{arxiv:1404.7644}}.

\bibitem{Banerjee:2014gga}
A.~Banerjee, K.~L.~Panigrahi and P.~M.~Pradhan,
\textit{``Spiky strings on {$\AdS_3 \times \Sphere^3$} with {NS}-{NS} flux''},
\texttt{\arxivref{1405.5497}{arxiv:1405.5497}}.

\bibitem{David:2010yg}
J.~R.~David and B.~Sahoo,
\textit{``{S}-matrix for magnons in the {D1}-{D5} system''},
\textsf{\doiref{10.1007/JHEP10(2010)112}{JHEP~1010,~112~(2010)}},
\texttt{\arxivref{1005.0501}{arxiv:1005.0501}}.

\bibitem{David:2008yk}
J.~R.~David and B.~Sahoo,
\textit{``Giant magnons in the {D1}-{D5} system''},
\textsf{\doiref{10.1088/1126-6708/2008/07/033}{JHEP~0807,~033~(2008)}},
\texttt{\arxivref{0804.3267}{arxiv:0804.3267}}.

\bibitem{Lloyd:2013wza}
T.~Lloyd and B.~Stefa{\'n}ski,~jr.,
\textit{``{$\AdS_3/\CFT_2$}, finite-gap equations and massless modes''},
\texttt{\arxivref{1312.3268}{arxiv:1312.3268}}.

\bibitem{Borsato:2014exa}
R.~Borsato, O.~Ohlsson~Sax, A.~Sfondrini and B.~Stefanski,~jr.,
\textit{``All-loop worldsheet S matrix for {$\AdS_3 \times \Sphere^3 \times
  \Torus^4$}''},
\texttt{\arxivref{1403.4543}{arxiv:1403.4543}}.

\bibitem{Frolov:2006cc}
S.~Frolov, J.~Plefka and M.~Zamaklar,
\textit{``The {$\AdS_5 \times \Sphere^5$} superstring in light-cone gauge and
  its {B}ethe equations''},
\textsf{\doiref{10.1088/0305-4470/39/41/S15}{J.Phys.~A39,~13037~(2006)}},
\texttt{\arxivref{hep-th/0603008}{hep-th/0603008}}.

\bibitem{Arutyunov:2006ak}
G.~Arutyunov, S.~Frolov, J.~Plefka and M.~Zamaklar,
\textit{``The off-shell symmetry algebra of the light-cone {$\AdS_5 \times
  \Sphere^5$} superstring''},
\textsf{\doiref{10.1088/1751-8113/40/13/018}{J.~Phys.~A40,~3583~(2007)}},
\texttt{\arxivref{hep-th/0609157}{hep-th/0609157}}.

\bibitem{Arutyunov:2006yd}
G.~Arutyunov, S.~Frolov and M.~Zamaklar,
\textit{``The {Z}amolodchikov-{F}addeev algebra for {$\AdS_5 \times \Sphere^5$}
  superstring''},
\textsf{\doiref{10.1088/1126-6708/2007/04/002}{JHEP~0704,~002~(2007)}},
\texttt{\arxivref{hep-th/0612229}{hep-th/0612229}}.

\bibitem{Rahmfeld:1998zn}
J.~Rahmfeld and A.~Rajaraman,
\textit{``The {GS} string action on {$\AdS_3 \times \Sphere^3$} with
  {R}amond-{R}amond charge''},
\textsf{\doiref{10.1103/PhysRevD.60.064014}{Phys.Rev.~D60,~064014~(1999)}},
\texttt{\arxivref{hep-th/9809164}{hep-th/9809164}}.

\bibitem{Park:1998un}
J.~Park and S.-J.~Rey,
\textit{``{G}reen-{S}chwarz superstring on {$\AdS_3 \times \Sphere^3$}''},
\textsf{\doiref{10.1088/1126-6708/1999/01/001}{JHEP~9901,~001~(1999)}},
\texttt{\arxivref{hep-th/9812062}{hep-th/9812062}}.

\bibitem{Metsaev:2000mv}
R.~Metsaev and A.~A.~Tseytlin,
\textit{``Superparticle and superstring in {$\AdS_3 \times \Sphere^3$}
  {R}amond-{R}amond background in light cone gauge''},
\textsf{\doiref{10.1063/1.1377274}{J.Math.Phys.~42,~2987~(2001)}},
\texttt{\arxivref{hep-th/0011191}{hep-th/0011191}}.

\bibitem{Grisaru:1985fv}
M.~T.~Grisaru, P.~S.~Howe, L.~Mezincescu, B.~Nilsson and P.~K.~Townsend,
\textit{``{$\superN=2$} Superstrings in a Supergravity Background''},
\textsf{\doiref{10.1016/0370-2693(85)91071-8}{Phys.~Lett.~B162,~116~(1985)}}.

\bibitem{Wulff:2014kja}
L.~Wulff,
\textit{``Superisometries and integrability of superstrings''},
\texttt{\arxivref{1402.3122}{arxiv:1402.3122}}.

\bibitem{Wulff:2013kga}
L.~Wulff,
\textit{``The type {II} superstring to order {$\theta^4$}''},
\textsf{\doiref{10.1007/JHEP07(2013)123}{JHEP~1307,~123~(2013)}},
\texttt{\arxivref{1304.6422}{arxiv:1304.6422}}.

\bibitem{Cvetic:1999zs}
M.~Cveti{\v c}, H.~L{\"u}, C.~N.~Pope and K.~S.~Stelle,
\textit{``{T}-Duality in the {G}reen-{S}chwarz Formalism, and the
  Massless/Massive {IIA} Duality Map''},
\textsf{\doiref{10.1016/S0550-3213(99)00740-3}{Nucl.~Phys.~B573,~149~(2000)}},
\texttt{\arxivref{hep-th/9907202}{hep-th/9907202}}.

\bibitem{Lu:1996rhb}
H.~L{\"u}, C.~Pope and P.~Townsend,
\textit{``Domain walls from anti-de {S}itter space-time''},
\textsf{\doiref{10.1016/S0370-2693(96)01443-8}{Phys.Lett.~B391,~39~(1997)}},
\texttt{\arxivref{hep-th/9607164}{hep-th/9607164}}.

\bibitem{Lu:1998nu}
H.~L{\"u}, C.~Pope and J.~Rahmfeld,
\textit{``A Construction of Killing spinors on {$\Sphere^n$}''},
\textsf{\doiref{10.1063/1.532983}{J.Math.Phys.~40,~4518~(1999)}},
\texttt{\arxivref{hep-th/9805151}{hep-th/9805151}}.

\bibitem{Alday:2005ww}
L.~F.~Alday, G.~Arutyunov and S.~Frolov,
\textit{``{G}reen-{S}chwarz strings in {TsT}-transformed backgrounds''},
\textsf{\doiref{10.1088/1126-6708/2006/06/018}{JHEP~0606,~018~(2006)}},
\texttt{\arxivref{hep-th/0512253}{hep-th/0512253}}.

\bibitem{Metsaev:2001bj}
R.~Metsaev,
\textit{``Type {IIB} {G}reen-{S}chwarz superstring in plane wave
  {R}amond-{R}amond background''},
\textsf{\doiref{10.1016/S0550-3213(02)00003-2}{Nucl.Phys.~B625,~70~(2002)}},
\texttt{\arxivref{hep-th/0112044}{hep-th/0112044}}.

\bibitem{Arutyunov:2005hd}
G.~Arutyunov and S.~Frolov,
\textit{``Uniform light-cone gauge for strings in {$\AdS_5 \times \Sphere^5$}:
  {S}olving {$\algSU(1|1)$} sector''},
\textsf{\doiref{10.1088/1126-6708/2006/01/055}{JHEP~0601,~055~(2006)}},
\texttt{\arxivref{hep-th/0510208}{hep-th/0510208}}.

\bibitem{Zoubos:2010kh}
K.~Zoubos,
\textit{``Review of {AdS/CFT} Integrability, {C}hapter {IV.2}: Deformations,
  Orbifolds and Open Boundaries''},
\textsf{\doiref{10.1007/s11005-011-0515-8}{Lett.Math.Phys.~99,~375~(2010)}},
\texttt{\arxivref{1012.3998}{arxiv:1012.3998}}.

\bibitem{vanTongeren:2013gva}
S.~J.~van~Tongeren,
\textit{``Integrability of the {$\AdS_5 \times S^5$} superstring and its
  deformations''},
\texttt{\arxivref{1310.4854}{arxiv:1310.4854}}.

\bibitem{Gava:2002xb}
E.~Gava and K.~S.~Narain,
\textit{``Proving the pp-wave / {$\CFT_2$} duality''},
\textsf{\doiref{10.1088/1126-6708/2002/12/023}{JHEP~0212,~023~(2002)}},
\texttt{\arxivref{hep-th/0208081}{hep-th/0208081}}.

\bibitem{Gomis:2002qi}
J.~Gomis, L.~Motl and A.~Strominger,
\textit{``pp-wave / {$\CFT_2$} duality''},
\textsf{\doiref{10.1088/1126-6708/2002/11/016}{JHEP~0211,~016~(2002)}},
\texttt{\arxivref{hep-th/0206166}{hep-th/0206166}}.

\bibitem{Green:1982tk}
M.~B.~Green and J.~H.~Schwarz,
\textit{``Extended Supergravity in Ten-Dimensions''},
\textsf{\doiref{10.1016/0370-2693(83)90781-5}{Phys.Lett.~B122,~143~(1983)}}.

\bibitem{Green:1982tc}
M.~B.~Green and J.~H.~Schwarz,
\textit{``Superstring Interactions''},
\textsf{\doiref{10.1016/0550-3213(83)90475-3}{Nucl.Phys.~B218,~43~(1983)}}.

\bibitem{Green:1983wt}
M.~B.~Green and J.~H.~Schwarz,
\textit{``Covariant Description of Superstrings''},
\textsf{\doiref{10.1016/0370-2693(84)92021-5}{Phys.Lett.~B136,~367~(1984)}}.

\bibitem{Berenstein:2002jq}
D.~E.~Berenstein, J.~M.~Maldacena and H.~S.~Nastase,
\textit{``Strings in flat space and {pp} waves from {$\superN = 4$} super
  {Y}ang {M}ills''},
\textsf{\doiref{10.1088/1126-6708/2002/04/013}{JHEP~0204,~013~(2002)}},
\texttt{\arxivref{hep-th/0202021}{hep-th/0202021}}.

\bibitem{Hofman:2006xt}
D.~M.~Hofman and J.~M.~Maldacena,
\textit{``Giant magnons''},
\textsf{J.~Phys.~A39,~13095~(2006)},
\texttt{\arxivref{hep-th/0604135}{hep-th/0604135}}.

\bibitem{Plefka:2006ze}
J.~Plefka, F.~Spill and A.~Torrielli,
\textit{``On the {H}opf algebra structure of the {$\AdS/\CFT$} {S}-matrix''},
\textsf{\doiref{10.1103/PhysRevD.74.066008}{Phys.~Rev.~D74,~066008~(2006)}},
\texttt{\arxivref{hep-th/0608038}{hep-th/0608038}}.

\bibitem{Torrielli:2011zz}
A.~Torrielli,
\textit{``The Hopf superalgebra of {$\AdS/\CFT$}''},
\textsf{\doiref{10.1016/j.geomphys.2010.10.003}{J.Geom.Phys.~61,~230~(2011)}}.

\bibitem{Torrielli:2011gg}
A.~Torrielli,
\textit{``{Y}angians, {S}-matrices and {$\AdS/\CFT$}''},
\textsf{\doiref{10.1088/1751-8113/44/26/263001}{J.Phys.~A44,~263001~(2011)}},
\texttt{\arxivref{1104.2474}{arxiv:1104.2474}}.

\bibitem{Gross:1974jv}
D.~J.~Gross and A.~Neveu,
\textit{``Dynamical Symmetry Breaking in Asymptotically Free Field Theories''},
\textsf{\doiref{10.1103/PhysRevD.10.3235}{Phys.~Rev.~D10,~3235~(1974)}}.

\bibitem{Fendley:1993jh}
P.~Fendley and H.~Saleur,
\textit{``Massless integrable quantum field theories and massless scattering in
  {$(1+1)$}-dimensions''},
\texttt{\arxivref{hep-th/9310058}{hep-th/9310058}}.

\bibitem{Janik:2006dc}
R.~A.~Janik,
\textit{``The {$\AdS_5 \times \Sphere^5$} superstring worldsheet {S}-matrix and
  crossing symmetry''},
\textsf{\doiref{10.1103/PhysRevD.73.086006}{Phys.~Rev.~D73,~086006~(2006)}},
\texttt{\arxivref{hep-th/0603038}{hep-th/0603038}}.

\bibitem{Arutyunov:2009kf}
G.~Arutyunov and S.~Frolov,
\textit{``The Dressing Factor and Crossing Equations''},
\textsf{\doiref{10.1088/1751-8113/42/42/425401}{J.~Phys.~A42,~425401~(2009)}},
\texttt{\arxivref{0904.4575}{arxiv:0904.4575}}.

\bibitem{Arutyunov:2004vx}
G.~Arutyunov, S.~Frolov and M.~Staudacher,
\textit{``{B}ethe ansatz for quantum strings''},
\textsf{\doiref{10.1088/1126-6708/2004/10/016}{JHEP~0410,~016~(2004)}},
\texttt{\arxivref{hep-th/0406256}{hep-th/0406256}}.

\bibitem{Kazakov:2004qf}
V.~A.~Kazakov, A.~Marshakov, J.~A.~Minahan and K.~Zarembo,
\textit{``Classical/quantum integrability in {AdS/CFT}''},
\textsf{\doiref{10.1088/1126-6708/2004//24}{JHEP~5,~24~(2004)}},
\texttt{\arxivref{hep-th/0402207}{hep-th/0402207}}.

\bibitem{Beisert:2005bm}
N.~Beisert, V.~A.~Kazakov, K.~Sakai and K.~Zarembo,
\textit{``The algebraic curve of classical superstrings on {$\AdS_5 \times
  \Sphere^5$}''},
\textsf{Commun.~Math.~Phys.~263,~659~(2006)},
\texttt{\arxivref{hep-th/0502226}{hep-th/0502226}}.

\bibitem{Kruczenski:2003gt}
M.~Kruczenski,
\textit{``Spin chains and string theory''},
\textsf{\doiref{10.1103/PhysRevLett.93.161602}{Phys.Rev.Lett.~93,~161602~(2004)}},
\texttt{\arxivref{hep-th/0311203}{hep-th/0311203}}.

\bibitem{Kruczenski:2004kw}
M.~Kruczenski, A.~Ryzhov and A.~A.~Tseytlin,
\textit{``Large spin limit of {$\AdS_5 \times \Sphere^5$} string theory and
  low-energy expansion of ferromagnetic spin chains''},
\textsf{\doiref{10.1016/j.nuclphysb.2004.05.028}{Nucl.Phys.~B692,~3~(2004)}},
\texttt{\arxivref{hep-th/0403120}{hep-th/0403120}}.

\bibitem{Hernandez:2004uw}
R.~Hern{\'a}ndez and E.~L{\'o}pez,
\textit{``The {$\grpSU(3)$} spin chain sigma model and string theory''},
\textsf{\doiref{10.1088/1126-6708/2004/04/052}{JHEP~0404,~052~(2004)}},
\texttt{\arxivref{hep-th/0403139}{hep-th/0403139}}.

\bibitem{Stefanski:2004cw}
B.~Stefa{\'n}ski,~Jr. and A.~A.~Tseytlin,
\textit{``Large spin limits of {$\AdS/\CFT$} and generalized
  {L}andau-{L}ifshitz equations''},
\textsf{\doiref{10.1088/1126-6708/2004/05/042}{JHEP~0405,~042~(2004)}},
\texttt{\arxivref{hep-th/0404133}{hep-th/0404133}}.

\bibitem{Stefanski:2005tr}
B.~Stefa{\'n}ski,~Jr. and A.~A.~Tseytlin,
\textit{``Super spin chain coherent state actions and {$\AdS_5 \times
  \Sphere^5$} superstring''},
\textsf{\doiref{10.1016/j.nuclphysb.2005.04.026}{Nucl.Phys.~B718,~83~(2005)}},
\texttt{\arxivref{hep-th/0503185}{hep-th/0503185}}.

\bibitem{Stefanski:2007dp}
B.~Stefa{\'n}ski,~jr.,
\textit{``{L}andau-{L}ifshitz sigma-models, fermions and the {$\AdS/\CFT$}
  correspondence''},
\textsf{\doiref{10.1088/1126-6708/2007/07/009}{JHEP~0707,~009~(2007)}},
\texttt{\arxivref{0704.1460}{arxiv:0704.1460}}.

\bibitem{Ambjorn:2005wa}
J.~Ambj{\o}rn, R.~A.~Janik and C.~Kristjansen,
\textit{``Wrapping interactions and a new source of corrections to the
  spin-chain/string duality''},
\textsf{\doiref{10.1016/j.nuclphysb.2005.12.007}{Nucl.~Phys.~B736,~288~(2006)}},
\texttt{\arxivref{hep-th/0510171}{hep-th/0510171}}.

\bibitem{Arutyunov:2007tc}
G.~Arutyunov and S.~Frolov,
\textit{``On String {S}-matrix, Bound States and {TBA}''},
\textsf{\doiref{10.1088/1126-6708/2007/12/024}{JHEP~0712,~024~(2007)}},
\texttt{\arxivref{0710.1568}{arxiv:0710.1568}}.

\bibitem{Arutyunov:2009zu}
G.~Arutyunov and S.~Frolov,
\textit{``String hypothesis for the {$\AdS_5 \times \Sphere^5$} mirror''},
\textsf{\doiref{10.1088/1126-6708/2009/03/152}{JHEP~0903,~152~(2009)}},
\texttt{\arxivref{0901.1417}{arxiv:0901.1417}}.

\bibitem{Gromov:2009tv}
N.~Gromov, V.~Kazakov and P.~Vieira,
\textit{``Integrability for the Full Spectrum of Planar {AdS/CFT}''},
\textsf{\doiref{10.1103/PhysRevLett.103.131601}{Phys.~Rev.~Lett.~103,~131601~(2009)}},
\texttt{\arxivref{0901.3753}{arxiv:0901.3753}}.

\bibitem{Bombardelli:2009ns}
D.~Bombardelli, D.~Fioravanti and R.~Tateo,
\textit{``Thermodynamic {B}ethe Ansatz for planar {AdS/CFT}: a proposal''},
\textsf{\doiref{10.1088/1751-8113/42/37/375401}{J.~Phys.~A42,~375401~(2009)}},
\texttt{\arxivref{0902.3930}{arxiv:0902.3930}}.

\bibitem{Arutyunov:2009ur}
G.~Arutyunov and S.~Frolov,
\textit{``Thermodynamic {B}ethe Ansatz for the {$\AdS_5 \times \Sphere^5$}
  Mirror Model''},
\textsf{\doiref{10.1088/1126-6708/2009/05/068}{JHEP~0905,~068~(2009)}},
\texttt{\arxivref{0903.0141}{arxiv:0903.0141}}.

\bibitem{Cavaglia:2010nm}
A.~Cavaglia, D.~Fioravanti and R.~Tateo,
\textit{``Extended {Y}-system for the {$\AdS_5/\CFT_4$} correspondence''},
\textsf{\doiref{10.1016/j.nuclphysb.2010.09.015}{Nucl.Phys.~B843,~302~(2011)}},
\texttt{\arxivref{1005.3016}{arxiv:1005.3016}}.

\bibitem{Gromov:2014caa}
N.~Gromov, V.~Kazakov, S.~Leurent and D.~Volin,
\textit{``Quantum spectral curve for arbitrary state/operator in
  {$\AdS_5/\CFT_4$}''},
\texttt{\arxivref{1405.4857}{arxiv:1405.4857}}.

\bibitem{Gukov:2004ym}
S.~Gukov, E.~Martinec, G.~W.~Moore and A.~Strominger,
\textit{``The search for a holographic dual to {$\AdS_3 \times \Sphere^3 \times
  \Sphere^3 \times \Sphere^1$}''},
\textsf{Adv.~Theor.~Math.~Phys.~9,~435~(2005)},
\texttt{\arxivref{hep-th/0403090}{hep-th/0403090}}.

\bibitem{Tong:2014yna}
D.~Tong,
\textit{``The Holographic Dual of {$\AdS_3 \times \Sphere^3 \times \Sphere^3
  \times \Sphere^1$}''},
\texttt{\arxivref{1402.5135}{arxiv:1402.5135}}.

\bibitem{Gaberdiel:2013vva}
M.~R.~Gaberdiel and R.~Gopakumar,
\textit{``Large {$\superN = 4$} Holography''},
\texttt{\arxivref{1305.4181}{arxiv:1305.4181}}.

\bibitem{David:2011iy}
J.~R.~David and A.~Sadhukhan,
\textit{``Classical integrability in the {BTZ} black hole''},
\textsf{\doiref{10.1007/JHEP08(2011)079}{JHEP~1108,~079~(2011)}},
\texttt{\arxivref{1105.0480}{arxiv:1105.0480}}.

\bibitem{David:2012aq}
J.~R.~David, C.~Kalousios and A.~Sadhukhan,
\textit{``Generating string solutions in {BTZ}''},
\textsf{\doiref{10.1007/JHEP02(2013)013}{JHEP~1302,~013~(2013)}},
\texttt{\arxivref{1211.5382}{arxiv:1211.5382}}.

\end{thebibliography}

\end{document}